\documentclass[11pt,a4paper,twoside,openright]{book} 

\usepackage[swedish, english]{babel} 
\usepackage[ansinew]{inputenc} 

\usepackage{mathptmx} 
\usepackage{helvet} 
\usepackage{courier}
\usepackage[T1]{fontenc}
\usepackage{Thesis} 
\ifpdf
   \usepackage[pdftex]{graphicx}
   \usepackage{microtype} 
  \else
    \usepackage[dvips]{graphicx}
\fi 

\def\color[#1]#2{}

\usepackage{bbm,amsmath,amsfonts,amssymb,bm}
\usepackage{boxedminipage}

\def\pp{{+\!\!\!+}}
\def\mm{=}
\def\ppmm{{\mbox{\tiny${}_{\stackrel\pp =}$}}}
 
\def\del{\partial}
\def\e{\mathrm{e}}
\def\abs#1{\left|#1\right|}
\def\dilaton{{\mathit{\phi}}}
\def\ph#1{\phantom{#1}}
\def\phn{{\phantom{0}}}
\def\sign{\mathrm{sign}}

\def\i{\mathrm{i}}

\def\ie{i.\,e.\ }

\def\eg{e.\,g.\ }
\def\cf{cf.\ }

\renewcommand*{\d}{\mathrm{d}}
\renewcommand* {\vec}[1]{\mathbf{#1}}
\newcommand*{\dx}{\d x}
\newcommand*{\dz}{\d z}

\newcommand*{\dt}{\d t}
\newcommand*{\ds}{\d s}

\def\ts#1{{\textstyle #1}}
\def\half{\frac{1}{2}}
\def\smallhalf{\ts{\half}}
\def\tsfrac#1#2{\ts{\frac{#1}{#2}}}

\def\genJ{{\bf J}}
\def\genG{{\bf G}}
\def\genI{{\bf I}}
\def\matrix#1#2{\left(\begin{array}{#1}#2\end{array}\right)}
\def\Litenmatrix#1{\left(\begin{smallmatrix}#1\end{smallmatrix}\right)}
\def\gcgMatrix#1{\matrix{cc}{#1}}

\def\GgenJ#1{\mathbbm{J}^{#1}}
\def\GgenG{\mathbbm{G}}
\def\GGamma{\Gamma\hspace{-5pt}\mathrm{I}} 
\def\GR{\mathbbm{R}}

\let\rund\r

\renewcommand{\AA}{\rund{A}}

\def\a{\alpha}
\def\b{\beta}
\def\g{\gamma}
\def\s{\sigma}
\def\r{\rho}
\def\t{\tau}

\newcommand{\F}{\phi}
\renewcommand{\S}{S}
\newcommand{\eps}{\epsilon}

\newcommand{\p}{{(+)}}

\def\f{\phi}

\def\tVec#1#2#3#4{\Litenmatrix{#1\\#2\\#3\\#4}}

\def\tLambda#1{\tVec{D\f^0_{\ph{#1}}}{D\f^#1_{\ph{#1}}}{S_0^{\ph{#1}}}{S_#1^{\ph{#1}}}}

\def\tMatrix#1{\Litenmatrix{#1}}

\def\Tpairing#1#2{\tMatrix{ 0^{\ph{#1}}_{\ph{#1}}&0^{\ph{#1}}_{\ph{#1}}&1^{\ph{#1}}_{\ph{#1}}&0^{\ph{#1}}_{\ph{#1}} \\ 0^{\ph{#1}}_{\ph{#1}}&0^{\ph{#1}}_{\ph{#1}}&0^{\ph{#1}}_{\ph{#1}}&\delta^#2_#1 \\ 1^{\ph{#1}}_{\ph{#1}}&0^{\ph{#1}}_{\ph{#1}}&0^{\ph{#1}}_{\ph{#1}}&0^{\ph{#1}}_{\ph{#1}} \\ 0^{\ph{#1}}_{\ph{#1}}&\delta^#1_#2&0^{\ph{#1}}_{\ph{#1}}&0^{\ph{#1}}_{\ph{#1}}}}

\def\Tdual#1#2{\tMatrix{0       & b_{0#2}                       & -1                      & 0           \\
                        0       & \delta^#1_#2                  & 0                       & 0           \\
                        -1      & -\frac{g_{0#2}}{g_{00}}       & 0                       & 0           \\
                        B_{0#1} & \frac{B_{0#1}g_{0#2}}{g_{00}} & -\frac{g_{0#1}}{g_{00}} & \delta^#2_#1} }

\def\invTdual#1#2{\tMatrix{0                       & -\frac{g_{0#2}}{g_{00}}       & -1      & 0            \\
                           0                       & \delta^#1_#2                  & 0       & 0            \\
                           -1                      & B_{0#2}                       & 0       & 0            \\
                           -\frac{g_{0#1}}{g_{00}} & \frac{g_{0#1}B_{0#2}}{g_{00}} & B_{0#1} & \delta^#2_#1 } }

\def\TgenJ#1#2{\tMatrix{-J^0_0 & -J^0_#2 & 0 & P^{0#2} \\ -J^#1_0 & -J^#1_#2 & P^{#1 0} & P^{#1#2} \\ 
                        0 & L_{0#2} & J^0_0 & J^#2_0 \\ L_{#1 0} & L_{#1#2} & J^0_#1 & J^#2_#1}}

\newcommand{\authorSurname}{Persson} 
\newcommand{\authorFirstName}{Jonas} 
\newcommand{\authorFirstInitial}{J} 
\newcommand{\authorEmail}{Jonas.Persson\@teorfys.uu.se} 
\newcommand{\dissertationTitle}{Strings as sigma models and in the tensionless limit} 
\let\dissertationSubtitle\empty 
\newcommand{\yearOfPublication}{2007} 
\newcommand{\placeOfDisputation}{H\"{a}ggsalen, \AA ngstr\"{o}m Laboratory} 
\newcommand{\dateOfDisputation}{Friday, April 27, 2007} 
\newcommand{\timeOfDisputation}{13:15} 
\newcommand{\disputationLanguage}{English} 
\newcommand{\numberOfPages}{viii+127} 
\newcommand{\placeOfPublication}{Uppsala} 
\newcommand{\ISBN}{978-91-554-6848-4} 
\newcommand{\series}{Digital Comprehensive Summaries of Uppsala Dissertations from the Faculty of Science and Technology} 
\newcommand{\serialNumber}{286} 
\newcommand{\keywords}{Theoretical Physics, String Theory, Tensionless Strings, Supergravity Backgrounds, Plane Wave, Extended Supersymmetry, Non-Linear Sigma Models, Generalized Complex Geometry, T-duality} 
\newcommand{\department}{Department of Theoretical Physics} 
\newcommand{\departmentaddress}{\AA ngstr\"{o}mlaboratoriet, L\"{a}gerhyddsvägen 1, SE-752 37 Uppsala,
Sweden} 
\newcommand{\ISSN}{1651-6214} 
\newcommand{\urn}{urn:nbn:se:uu:diva-7783} 
\newcommand\abstracttext{%
This thesis considers two different aspects of string theory, the tensionless limit of the string and supersymmetric sigma models with extended supersymmetry. First, the tensionless limit is used to find a IIB supergravity background generated by a tensionless string. The background has the characteristics of a gravitational shock-wave. Then, the quantization of the tensionless string in a pp-wave background is performed and the result is found to agree with what is obtained by taking a tensionless limit directly in the quantized theory of the tensile string. Hence, in the pp-wave background the tensionless limit commutes with quantization. Next, supersymmetric sigma models and the relation between extended world-sheet supersymmetry and target space geometry is studied. The sigma model with $\mathcal{N}=(2,2)$ extended supersymmetry is considered and the requirement on the target space to have a bi-Hermitean geometry is reviewed. The Hamiltonian formulation of the model is constructed and the target space is shown to have generalized K\"{a}hler geometry. The equivalence between bi-Hermitean geometry and generalized K\"{a}hler follows, in this context, from the equivalence between the Lagrangian- and Hamiltonian formulation of the sigma model. Then, T-duality in the Hamiltonian formulation of the sigma model is studied and the explicit T-duality transformation is constructed. It is shown that the transformation is a symplectomorphism, \ie a generalization of a canonical transformation. Under certain assumptions, the amount of extended supersymmetry present in the sigma model is shown to be preserved under the T-duality transformation. Next, extended supersymmetry in a first order formulation of the sigma model is studied. By requiring $\mathcal{N}=(2,2)$ extended world-sheet supersymmetry an intriguing geometrical structure arises and in a special case generalized complex geometry is found to be contained in the new framework.
} 

\renewcommand{\listofpapers}
{\chapter*{List of Papers}
This thesis is based on the following papers, which are referred
to in the text by their Roman numerals. \vspace{13pt}
\begin{romanlist}
 \setlength{\itemsep}{1.5ex}
 \item A.~Bredthauer, U.~Lindström and J.~Persson.
  $SL(2,\mathbb{Z})$ tensionless string backgrounds in IIB string theory.
  {\sl Classical and Quantum Gravity} {\bf 20} (2003) 3081-3086.
  {hep-th/0303225}.
 \item   A.~Bredthauer, U.~Lindström, J.~Persson and L.~Wulff.
  Type IIB tensionless superstrings in a pp-wave background.
  {\sl Journal of High Energy Physics} {\bf 02} (2004) 051. {hep-th/0401159}.
 \item  A.~Bredthauer, U.~Lindström and J.~Persson.
  First-order supersymmetric sigma models and target space geometry.
  {\sl Journal of High Energy Physics} {\bf 01} (2006) 144. {hep-th/0508228}.
 \item  A.~Bredthauer, U.~Lindström, J.~Persson and M.~Zabzine.
  Generalized Kähler geometry from supersymmetric sigma models.
  {\sl Letters in Mathematical Physics} (2006) {\bf 77}:291-308. {hep-th/0603130}.
 \item  J.~Persson.
  T-duality and generalized complex geometry.
  {\sl Journal of High Energy Physics} {\bf 03} (2007) 025. {hep-th/0612034}.
\end{romanlist}
}

\newcommand{\dedication}%
{\cleardoublepage
\thispagestyle{empty}
\vspace*{\stretch{3}}
\begin{flushright}
		
		{\fontfamily{pzc}\Large\selectfont
        \emph{To \rund{A}sa}}

\end{flushright}
\vspace*{\stretch{1}}} 

\begin{document}
	\pagenumbering{roman}
	  \frontmatterDigitalPublishing 
    
    \tableofcontents
    

	\cleardoublepage
	\pagenumbering{arabic}
	\setcounter{page}{1}
\let\r\rund 
\selectlanguage{swedish}

\chapter[Introduction in Swedish]{Svensk introduktion}

Strängteori är en möjlig kandidat till att vara en teori för alla fysikaliska fenomen vi observerar runt oss i världen. Ambitionen med strängteorin är att den ska vara en förenande teori som kan beskriva och förklara alla fysikaliska fenomen, krafter och partiklar vi observerar. Dock är den inte färdigutvecklad och i dagsläget kan vi inte veta om detta verkligen kommer att bli den slutgiltiga teorin för hur den fysikaliska världen fungerar, men det finns ett antal indikationer på att strängteorin verkligen kan beskriva alla de fysikaliska krafterna. Innan vi går in på dessa indikationer behöver vi en liten bakgrund till varför vi letar efter en sådan förenande teori.

De fysikaliska krafterna som vi observerar i vår värld är den elektromagnetiska kraften, den svaga kraften, den starka kraften och gravitationen. 

Den elektromagnetiska kraften beskrivs av en teori som kallas elektromagnetism. Teorin beskriver bland annat hur elektriskt laddade partiklar interagerar med varandra och hur elektromagnetisk strålning, t.ex. ljus eller radiovågor, beter sig. Denna kraft är bl.a. ansvarig för att elektroner hålls kvar kring en atoms kärna. 

Den svaga kraften ger bl.a. upphov till att atomer kan sönderfalla via $\beta$-sönderfall, där en neutron sönderfaller till en proton genom att sända ut en högenergetisk elektron och en anti-elektron-neutrino. Den högenergetiska elektronen är det man brukar kalla för $\beta$-strålning som är ett exempel på joniserande strålning. Den svaga kraften är, jämfört med den starka kraften, väldigt svag och har därav fått sitt namn. 

Den starka kraften verkar mellan kvarkar, vilka är beståndsdelarna av t.ex. protoner och neutroner. Denna kraft är bl.a. ansvarig för att hålla ihop neutroner och protoner så att de bildar atomkärnor. Den starka kraften har fått sitt namn eftersom den är mycket starkare än den elektromagnetiska kraften. Detta är anledningen till att en atomkärna hålls ihop trots att den elektromagnetiska kraften verkar repulsivt mellan protonerna i kärnan.

Gravitationen beskrivs av Einsteins allmänna relativitetsteori. Den är t.ex. ansvarig för att månen hålls kvar i sin bana kring jorden och att ett äpple faller ner mot jorden när det lossnar från sitt träd. Denna teori ger oss en geometrisk formulering av hur gravitationen fungerar. Den beskriver gravitation som en krökning i rum-tiden, där tid och rum behandlas på samma sätt. 

I början av nittonhundratalet formulerades de två teorierna speciell relativitetsteori, som beskriver hur fysiken fungerar vid höga hastigheter och energier, och kvantmekanik, som är ett ramverk för att beskriva hur fysiken fungerar på mycket korta avstånd. Dessa teorier visade sig vara mycket framgångsrika och genom att sammanfoga dessa två i en och samma teori får man en teori som väl beskriver höga energier och korta avstånd. En sådan teori kallas för en relativistisk kvantfältteori. Att göra om en klassisk teori, dvs.\ en teori som inte är formulerad m h a kvantmekanik, till en teori som är formulerad m h a  kvantmekanik kallas att kvantisera teorin. Det går att kvantisera den elektromagnetiska kraften så att den och partiklarna den påverkar beskrivs i termer av en relativistisk kvantfältteori. Den teori vi får kallas kvantelektrodynamik och det visar sig att den är en otroligt exakt teori för att beskriva hur elektromagnetismen beter sig. Två av de andra krafterna, den svaga och den starka kraften, kan också kvantiseras och formuleras som kvantfältteorier. Den förenade teorin för den elektromagnetiska, svaga och starka kraften kallas för partikelfysikens Standardmodell. Denna är en väldigt exakt teori som beskriver alla interaktioner, förutom gravitationen, mellan de allra minsta beståndsdelarna i vår värld.

Motiverad av denna framgång att formulera tre av de fyra krafterna i en förenad teori kan vi fortsätta och försöka förena gravitationen med Standardmodellen. Gör man detta på samma sätt som för de andra krafterna, genom att kvantisera gravitationen, så fungerar det inte. Det man får ut ur teorin då man beräknar sannolikheter för att olika fysikaliska processer ska äga rum är bara nonsens. Det dyker upp oändligheter i beräkningarna som inte går att hantera på ett vettigt sätt. Det fungerar helt enkelt inte och något radikalt nytt måste tillföras teorin för att vi ska kunna formulera en förenad teori för alla de fyra krafterna. 

En alternativ ståndpunkt är att vi helt enkelt kan säga att vi har en bra beskrivning av hur fysiken fungerar på stora avstånd, den allmänna relativitetsteorin, och en bra beskrivning av hur den fungerar på mycket korta avstånd, Standardmodellen, och vara nöjd med det. Dock så blir det problem med de två teorierna under extrema förhållanden. Då vi vill förstå hur ett svart hål fungerar behövs allmän relativitetsteori för att beskriva den extrema rum-tids krökningen i närheten av det svarta hålets centrum och samtidigt behövs kvantmekanik för att beskriva fysiken på de korta avstånden i samma region. Problemet är att dessa teorier inte fungerar tillsammans. 

Ett annat exempel på behovet av en förenande teori för alla krafter är då vi vill förstå hur vårt universum betedde sig kort efter den Stora Smällen (the Big Bang). Återigen har vi att göra med stor rum-tids krökning men nu måste denna kombineras med höga energier. Kvantfältteorier beskriver också höga energier, så vad vi behöver är en teori för kvantiserad gravitation.

\begin{figure}
\begin{centering}
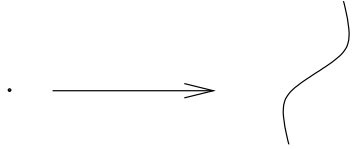
\caption{I strängteorin antar man att matriens minsta beståndsdelar är endimensionella strängar istället för punktpartiklar.}
\label{svensk:punkt-strang}
\end{centering}
\end{figure}

Det är viktigt att poängtera att de kvantfältteorier som beskriver den elektromagnetiska, svaga och starka kraften alla bygger på antagandet att de minsta beståndsdelarna i vår värld beskrivs som nolldimensionella punkter, dvs.\ utan någon som helst utsträckning. Vad strängteorin gör är att helt enkelt anta att istället för att vara nolldimensionella punkter så är de minsta beståndsdelarna endimensionella strängar. Istället för att studera punkter så studerar vi strängar, se figur \ref{svensk:punkt-strang}. Denna till synes enkla generalisering leder till oanade konsekvenser.

Vi kan tänka på denna generalisering på följande sätt: Om vi förstorar punktpartiklarna väldigt mycket så kommer vi att se en sträng. Detta leder till att strängar måste vara mycket små, så små att även i de mest avancerade experiment som vi hittills genomfört med stora partikelacceleratorer kommer en sträng att se ut som en punktpartikel. Låt oss nu fundera på hur stor en sträng kan antas vara. Vi vill att strängen ska beskriva alla de egenskaper vi hittat med hjälp av punktpartikelteorierna men vi vill också gärna att den ska beskriva kvantiserad gravitation. Om vi kombinerar Planks konstant $\hbar$, Newtons gravitationskonstant $G$ och ljushastigheten i vakuum $c$ så att vi får en längd hittar vi den så kallade Plancklängden
\begin{align*}
l_P = \sqrt{\frac{\hbar G}{c^3}} \thickapprox 10^{-35} m.
\end{align*}
Detta är den fundamentala längdskalan där kvantiserad gravitation antas bli nödvändig för att beskriva fysiken. En sträng antas därför vara av samma storleksordning som Planklängden. För att förstå hur ofantligt liten en sträng är kan vi använda oss av följande liknelser. Antag att vi förstorar upp ett äpple till samma storlek som jorden, i den skalan kommer en atom att vara lika stor som ett äpple. Låt oss därefter förstora upp atomen så att dess radie är ungefär tio gånger så stor som hela vårt solsystem, i den skalan är atomens kärna ungefär lika stor som solen. Låt oss vidare förstora upp atomkärnan så att den blir lika stor som hela vårt solsystem, då blir en sträng i denna skala lika lång som några hundra atomer på rad, i vår vanliga skala. I dagens experiment har vi helt enkelt inte tillräcklig upplösning för att ''se'' strängens utsträckning utan ''ser'' den bara som en punktpartikel och därför fungerar punktpartikelteorierna, som inte inkluderar gravitationen, bra.

Strängteorin antar alltså att materiens minsta beståndsdelar är strängar. Dessa strängar kan vibrera och svänga på olika sätt. Beroende på hur en sträng svänger beter den sig som en viss typ av partikel. En sträng som svänger på ett sätt beter sig som en elektron och en sträng som svänger på ett annat beter sig som en foton. På detta sätt får man från en och samma sträng en mängd olika typer av partiklar. En av de upptäckter som gör strängteorin extra intressant är att ett av de sätt en sträng kan svänga på motsvarar precis hur en graviton beter sig. En graviton är partikeln som förmedlar gravitation mellan partiklar. Eftersom strängteorin går att kvantisera blir den därför en naturlig kandidat till kvantgravitation och en förenad teori för alla de fyra krafterna.

Inom strängteorin finns två olika typer av strängar, slutna strängar, som formar en sluten ögla, och öppna strängar vars ändar sitter fast på högredimensionella objekt. Dessa objekt kallas D-bran, eller ibland D$p$-bran där $p$ anger antalet rumsdimensioner som branet har. Det visar sig att de öppna strängarna som sitter fast på D-branet beskriver hur det rör sig. 

Den enklaste strängteorin som går att formulera kallas bosonisk strängteori och kan bara beskriva kraftförmedlingspartiklar, sk. bosoner. I denna teori finns inga materiepartiklar och grundtillståndet för en bosonisk sträng visar sig vara en tachyon, ett tillstånd med imaginär massa vilket innebär att den rör sig fortare än ljuset. Existensen av detta tillstånd är en indikation på att teorin inte är stabil. Dessutom kräver teorin att rum-tiden har 26 dimensioner, en tidsriktning och 25 rumsriktningar, samt att rum-tiden uppfyller en generalisering av Einsteins ekvationer för allmän relativitetsteori. Detta är återigen en indikation på att strängteorin naturligt inkluderar gravitation.

För att få en realistisk fysikalisk modell behöver vi inkludera både bosoner och fermioner, dvs.\ både interaktionspartiklar och materiepartiklar. För att göra detta använder man sig av något som är känt som supersymmetri. Detta är en symmetri mellan bosonerna och fermionerna så att de kommer i par. Till varje boson hör en fermion och vise versa. Supersymmetri är en unik utvidgning av symmetrierna som vi observerar hos fysiken, rum-tids translationerna och Lorentz rotationerna. 

Införandet av supersymmetri i strängteorin leder till inte mindre än fem självmotsägelsefria supersträngteorier. Dessa kallas typ I, typ IIA, typ IIB, heterotisk $E_8\times E_8$ och heterotisk $SO(32)$. Ett specifikt villkor som dyker upp från kvantiseringen av teorierna är att alla dessa kräver en rum-tid med tio dimensioner. Dessutom visar det sig att ingen av dessa teorier har någon tachyon i sitt spektrum men inkluderar gravitonen.

Vad vi observerar runt oss till vardags är en tidsriktning och tre rumsriktningar, totalt har den rum-tid vi observerar fyra dimensioner. Om det verkligen är så att de minsta beståndsdelarna av materien är strängar leder detta därför till frågan var de övriga sex dimensionerna som supersträngteorierna kräver har tagit vägen.

Det finns olika sätt man kan tänka sig att bli kvitt de sex extra dimensionerna. Ett sätt är den s.k. branvärldsmodellen där man antar att vi lever på ett D3-bran, med fyra rum-tidsdimensioner, som i sin tur lever i en tiodimensionell rum-tid. På D3-branet sitter det öppna strängar som beskriver den fysik vi observerar. Dessutom kan man i denna modell tänka sig universa på andra D$p$-bran, med olika antal rum-tidsdimensioner, som existerar parallellt med vårt i den tiodimensionella rum-tiden. Genom att låta två parallella D$p$-bran, som beskriver parallella universum, kollidera kan man modellera Stora Smällen. Huruvida det på detta sätt går att skapa modeller som realistiskt beskriver den Stora Smällen är inte helt utrett.

Ett annat sätt att göra sig av med de extra dimensionerna är att kompaktifiera dem. Detta innebär att vi rullar ihop de extra dimensionerna så att de blir så små att vi inte kan observera dem. För att förstå hur detta går till så tänk dig en myra på ett pappersark, myran kan röra sig framåt - bakåt och höger - vänster längs med pappersarket. Den kan röra sig i två dimensioner. Om vi nu kompaktifierar en dimension, säg höger - vänster, betyder det att vi rullar ihop pappret så att det bildar en tub. Om nu myran går tillräckligt långt åt höger kommer den efter ett tag att komma tillbaka till samma plats som den startade på. Vi har gjort en av de två dimensionerna på pappret kompakt. För att effektivt sett bli av med den kompakta dimensionen måste vi rulla ihop pappret på ett sånt sätt att den resulterande tuben får en väldigt liten radie. Myran på pappret kommer fortfarande att kunna röra sig i de två dimensionerna på tuben men om vi ställer oss långt bort från papperstuben, och eventuellt kisar lite, så kommer den smala tuben att inte längre se ut som en tub utan som en linje, den ser inte längre ut att ha två dimensioner utan bara en. På detta sätt har vi, när vi betraktar det hårt ihoprullade pappret långt ifrån, effektivt sett blivit av med en dimension. Men, den extra dimensionen kommer att påverka fysiken som beskrivs i den lägredimensionella modellen. 

I exemplet har vi bara blivit av med en dimension, men kompaktifiering av sex dimensioner fungerar på samma sätt. Vi rullar ihop dem och säger att de finns där fast vi inte kan se dem, vi kan bara observera fysikaliska effekter av att de extra dimensionerna finns där. Beroende på vilket sätt vi väljer att kompaktifiera de extra dimensionerna kommer vi att få olika effektiva fyrdimensionella modeller. Om vi till exempel vill att den effektiva modellen ska ha supersymmetri måste vi välja att kompaktifiera de sex extra dimensionerna så att de bildar ett sexdimensionellt matematiskt rum som kallas Calabi-Yau rum.

Populärt att nämna i detta sammanhang är att det finns minst $10^{500}$ olika sätt att kompaktifiera de extra dimensionerna. Var och ett av dessa sätt ger upphov till en effektiv fyrdimensionell modell. Det finns till synes ett helt landskap av olika effektiva teorier som dyker upp genom kompaktifiering. Detta landskap av teorier brukar kallas för ''strängteorins landskap''.

Vi sökte från början efter en förenande teori för de fyra fundamentala krafterna och vi har funnit att det finns ett helt landskap av teorier som innehåller kvantiserad gravitation. Man kan nu söka efter olika sätt att dynamiskt välja en teori som beskriver vår värld som vi observerar den. Detta är ett fält inom strängteorin som ännu håller på att utvecklas och den slutgiltiga bilden av hur detta ska gå till är inte färdig.

Låt oss återvända till de fem supersträngteorierna. Vi har alltså fem teorier som alla verkar lovande för att beskriva vår värld som vi observerar den. Det har visat sig att dessa fem teorier tillsammans med en sjätte, kallad elvadimensionell supergravitation, är sammanlänkade via ett nätverk av olika s.k. dualiteter. En dualitet mellan två teorier säger att de beskriver samma fysik fast på olika sätt, ungefär som två olika sidor av samma mynt. En id\'{e} som nätverket av dualiteter har givit upphov till är att alla dessa sex olika teorier troligen är olika gränser av en och samma, mera fundamental elvadimensionell teori kallad M-teori. Denna situation illustreras i figur \ref{I:dualities} i nästa kapitel. Vad M'et står för är oklart men olika förslag är Mother, Mystery, Matrix, ett upp-och-nervänt W för Witten. Också oklart är hur denna M-teori ska formuleras och i termer av vilka fundamentala objekt. 

Framtiden kommer med största sannolikhet att lära oss mer om denna M-teori och om strängteorin. Dessutom kommer det att bli spännande att se vilka typer av experiment som konstrueras för att testa strängteorin. Forskare runt om i hela världen arbetar hela tiden med frågor relaterade till om strängteorin, eller M-teorin, verkligen är den förenande teorin för allt eller om det finns något ännu mer fundamentalt som behövs för att beskriva den fysik vi observerar på ett sätt som förenar de fyra krafterna.

\vspace{2ex}

Den här introduktionen avslutas genom att beskriva de fem artiklar som jag har deltagit i och som ligger till grund för denna avhandling.

I artiklarna [I] och [II] studeras olika aspekter av en specifik gräns av strängteorin, den spänningslösa gränsen. Strängspänningen är strängens viloenergi per enhetslängd och i den spänningslösa gränsen låter man denna gå mot noll. Strängspänningen beskriver hur de olika delarna av strängen påverkar varandra. Man kan tänka på den som spänningen i t.ex. en gitarrsträng. I den spänningslösa gränsen beter sig de olika delarna av strängen som en kontinuerlig fördelning punktpartiklar som klassiskt inte påverkar varandra och som rör sig längs geodeter, raka banor, i rum-tiden. Om vi betraktar en punktpartikel motsvarar den spänningslösa gränsen av strängen gränsen då man låter massan hos partikeln gå mot noll samtidigt som man låter partikelns hastighet gå mot ljushastigheten. Detta är en högenergigräns för partikeln och därför kan den spänningslösa gränsen för strängen ses som en högenergigräns av stränteorin.

I artikel [I] studerade jag tillsammans med Andreas Bredthauer och Ulf Lindström en rum-tid som genereras av en typ IIB sträng. Genom att låta strängen röra sig med ljusets hastighet och samtidigt ha en ändlig energi fann vi en rum-tid som genereras av en spänningslös sträng. Vi verifierade också att energin som finns i denna rum-tid motsvarar precis energin som kommer från en spänningslös sträng.

I artikel [II] studerade jag tillsammans med Andreas Bredthauer, Ulf Lindström och Linus Wulff en annan aspekt av den spänningslösa gränsen av typ IIB strängteori. Vi kvantiserade en spänningslös sträng i en särskild rum-tid, kallad en pp-våg. Vi fann att den kvantiserade spänningslösa strängen även kan erhållas genom att ta en spänningslös gräns av den kvantiserade spänningsfulla IIB strängen på pp-vågen. Anledningen till att kvantiseringen av den spänningslösa strängen är oproblematisk i detta fall är att vi betraktar strängen på den speciella pp-våg rum-tiden. I plant rum är detta inte fallet och kvantisering är problematisk.

I artiklarna [III], [IV] och [V] studeras olika aspekter av utökad supersymmetri på världsytan av en sträng. Att kräva sådan utökad supersymmetri på världsytan innebär att det dyker upp villkor på vilka rum-tider, även kallade målrum, som strängen kan existera i. Anledningen till att det är intressant att studera detta är att vi kan få en klassifiering av de målrum som krävs för olika strängmodeller.

I artikel [III] studerade jag tillsammans med Andreas Bredthauer och Ulf Lindström en första ordningens sigmamodell, som beskriver strängen. Vi undersökte vilka villkor som dyker upp på målrummet som en effekt av att kräva två utökade supersymmetrier. Vi fann en geometrisk struktur som påminner om generaliserad komplex geometri men som verkar innehålla mer än den. Någon exakt geometrisk tolkning av villkoren för utökad supersymmetri kunde vi inte ge eftersom det inte existerar något färdigutvecklat ramverk för att beskriva dessa typer av geometriska strukturer.

I artikel [IV] studerade jag tillsammans med Andreas Bredthauer, Ulf Lindström och Maxim Zabzine Hamiltonformuleringen av en sigmamodell. Genom att kräva utökad supersymmetri i fasrummet fann vi en direkt relation mellan bi-Hermitsk geometri och generaliserad Kähler geometri. Att de två formuleringarna av geometrin är ekvivalenta följer, ur fysikalisk synvinkel, från att Lagrange- och Hamiltonformuleringen av sigmamodellen är ekvivalenta. 

I artikel [V] studerade jag T-dualitet i Hamiltonformuleringen av sigmamodellen. Jag hittade den explicita T-dualitetstransformationen i denna formulering och visade att den är en symplektomorfism, en generalisering av en kanonisk transformation. Under vissa antaganden visade jag att antalet utökade supersymmetrier i modellen bevaras under T-dualitetstransformationen.

\let\rund\r
\def\r{\rho}
\selectlanguage{english}

        \chapter{Introduction}

The key assumption in string theory is that the fundamental objects in the physical world are not point particles, as they are considered to be in \eg the Standard Model of particle physics, but extended one-dimensional objects, strings. These strings can be closed, as a rubber band, or open, as a rubber band with two ends. This small generalization of the point particle model turns out to be very rich in structure. In particular, one single string unifies in an elegant way different types of particles. The different ways the string can oscillate correspond to different particles in the same way as a guitar string gives rise to different notes. In the simplest model of the string, known as bosonic string theory, there are only bosons, \ie interaction particles, present in the string spectrum. One way to include fermions, \ie matter particles, is to introduce a symmetry between the bosons and the fermions known as supersymmetry. When including supersymmetry in the theory we obtain a superstring, which constitutes a more realistic string theory model.

String theory was constructed about forty years ago as a model of the strong nuclear force to describe a phenomenon observed when scattering hadrons. The phenomenon is known as Regge behavior and gives the masses of mesons as $m^2=J/\alpha'$, where $J$ is the spin of the particle and the parameter $\alpha'$ is known as the Regge slope. In 1968 Veneziano constructed an amplitude \cite{Veneziano:1968yb}, known as the Veneziano amplitude, that reproduces this behavior and later it was realized that this amplitude arises from the scattering of four open strings. Thus, strings could be used to describe the scattering of hadrons. However, in 1973-1974 a successful theory known as quantum chromodynamics was invented which in a different way explains the strong force and how the hadrons behave. The string model seemed superfluous.

When Scherk, Schwarz \cite{Scherk:1974ca} and Yoneya \cite{Yoneya:1974jg} in 1974 noted that one of the states in the string spectrum behaves as a graviton, the mediator of gravity, string theory gained a new status as a possible ``theory of everything,'' unifying all the known forces. It was realized that instead of being a model of the strong force and of hadrons, string theory is a theory that naturally includes quantized gravity. Before this, a renormalizable quantized theory of gravity had not been known and the discovery of the graviton in the spectrum together with the discovery \cite{Callan:1985ia} that string theory includes Einsteins theory of gravity in a low energy limit are perhaps the most important motivations for studying string theory.

Since these discoveries string theory has become a whole new field in theoretical physics and developed in its own right. However, interest in string theory was not immense until the ``first superstring revolution'' which was initiated in 1984 when Green and Schwarz showed that certain anomalies cancel in the theory \cite{Green:1984sg}. These anomalies are mathematical inconsistencies that had plagued superstring theory up to that point. This discovery gave the interest in string theory an upswing and the outcome of the ``first superstring revolution'' was the construction of no less then five different consistent superstring theories, called type I, type IIA, type IIB, heterotic $E_8\times E_8$ and heterotic $SO(32)$, each requiring the space-time to be ten-dimensional. The dream of a single unique unifying theory of the physical world seemed shattered.

\begin{figure}
\begin{centering}
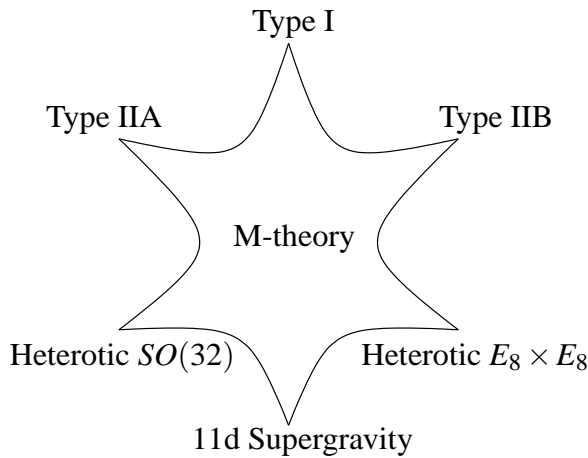
\caption{The five string theories and eleven dimensional supergravity are merely different limits of a more fundamental eleven-dimensional theory, M-theory.}
\label{I:dualities}
\end{centering}
\end{figure}

This situation remained until 1995 when the ``second superstring revolution'' took place. The key observation now was that the five consistent theories of superstrings were all linked to one another via so called dualities \cite{Witten:1995zh, Witten:1995ex}. Further the five theories were also found to be linked by duality to eleven-dimensional supergravity, which is the unique theory of gravity in eleven dimensions that includes supersymmetry. The web of dualities suggested that indeed there may be a more fundamental theory of which the five string theories and eleven-dimensional supergravity are merely different limits. This theory is called M-theory and supposed to be formulated in eleven dimensions. What the M stands for is unclear but different suggestions are Mother, Mystery, Matrix or an up-side-down W for Witten. It is also unclear in terms of which fundamental objects the theory should be formulated. Even though M-theory is not known explicitly, it suggests that string theory, or more precisely M-theory, might actually be a good candidate for being the unique ``theory of everything.'' The relations between the theories are depicted in figure \ref{I:dualities}.

Another outcome of the ``second superstring revolution'' and of the dualities was the discovery that at the non-perturbative level, string theory includes not only strings but also higher dimensional objects, known as D-branes or D$p$-branes. Here $p$ denotes the spatial dimension of the brane and D stands for Dirichlet. These D-branes are subspaces of space-time on which open strings can end. It turns out that D-branes themselves are dynamical objects whose quantum fluctuations are described by the open strings attached to them.


We mentioned earlier that the superstring theories need a ten-dimensional space-time to be consistent. When comparing to our everyday life this presents a problem in that we only observe three spatial directions and one time direction. Our observed space-time has only four dimensions. One way to deal with the six extra dimensions is to consider our world being a D3-brane living in an ambient ten-dimensional space-time, unobservable to us. Models of this type are known as brane-world models. In these models one can imagine other universes existing in parallel to ours and letting two such universes collide gives us models of the Big Bang. Whether or not the brane-world idea can provide realistic models of our universe and the Big Bang is currently under investigation, and no definite conclusion has been reached.

Another way of handling the extra dimensions is to wrap them up into a closed six-dimensional compact space $C_6$, so that the total space-time is given by $M_4\times C_6$, where $M_4$ describes the visible four-dimensional space-time. The idea is then to let the compact dimensions be so small that we in current experiments do not have high enough resolution to discover them. This process of making the extra dimensions compact and small is known as compactification. The $C_6$ becomes an internal space of the effective theory.

However, there is huge number of ways of compactifying the extra dimensions, and to stabilize the geometry of the internal space we need to introduce fluxes. Further, to have a model that resembles the Standard Model of particle physics we also need to include D-branes. The choices of internal manifold, fluxes and D-branes decides the different properties of the effective four dimensional theory, such as which particles are present in the model. It thus seems that we have a huge number of different possible ways of obtaining an effective four-dimensional theory. The set of all the possible effective four-dimensional theories make up what is called the ``string theory landscape.'' This landscape is vast and presently a lot of research is devoted to finding some dynamical selection mechanism that could single out our universe as a possibly unique solution to the theory. Such a mechanism might not exist and maybe we just happen to live in one of many equally probable universes. 

Even though string theory is a fascinating theory that includes quantized gravity there is one main problem. It is that, at present date, there are no direct connections to observations. This is surely a problem, since if string theory is to be a theory of the real world, we have to find a way to relate it to what we observe around us. I hope that this will in the future prove to be the case. Otherwise, I believe that string theory is doomed as a physical theory.

\vspace{2ex}
We conclude this introductory chapter by a summary of the articles on which this thesis is based.

In the articles [I] and [II], different aspects of the tensionless limit of string theory is considered. The string tension, that goes to zero in this limit, is the rest energy of the string per unit length. It describes how the different parts of the string are held together and in the tensionless limit the string effectively behaves as a continuous distribution of point particles, classically not interacting with each other and moving on null geodesics through space-time. The corresponding limit for the point particle is the massless limit where the velocity of the particle is taken to approach the speed of light. At high enough velocity the energy of the particle is dominated by its kinetic energy and its mass can effectively be considered as being zero. In the same way the tensionless limit corresponds to a high energy limit of string theory.

In article [I] together with Andreas Bredthauer and Ulf Lindström, I studied a solution to type IIB supergravity sourced by a type IIB string. By letting the source string move at the speed of light while keeping the total energy finite we found a background that has the structure of a gravitational shock-wave. We verified that the energy content of the background corresponds to the energy of a tensionless string and thus we interpreted the background as being generated by such a string. 


In article [II] together with Andreas Bredthauer, Ulf Lindström and Linus Wulff, I studied another aspect of the tensionless limit of type IIB string theory. We quantized the tensionless string in a pp-wave background. We also found that the quantized tensionless string can be obtained by taking a tensionless limit directly from the quantized tensile IIB string on the pp-wave. The reason why the quantization of the tensionless string is straightforward in this case is related to the existence of a dimensionful parameter in the pp-wave background. This is not the case in a flat background and the quantization is more involved. 


The articles [III], [IV] and [V] consider different aspects of extended supersymmetry on the world-sheet of the string by studying non-linear sigma models. By demanding extended supersymmetry, restrictions on the type of space in which the string may exist arise. The space in which the string is embedded is, in the context of sigma models, known as the target space and the conditions that arise are in general geometrical and it is often possible to formulate them in terms of generalized complex geometry. One reason to why this is interesting to study is to obtain a classification of the possible target space geometries in which strings with different amount of supersymmetry may exist.

In article [III] I studied, together with Andreas Bredthauer and Ulf Lindström, a manifestly $\mathcal{N}=(1,1)$ supersymmetric first order sigma model, describing the string. We examined the conditions that arise on the target space by demanding $\mathcal{N}=(2,2)$ extended supersymmetry. We considered a symplectic sigma model and found a geometric structure that resembles generalized complex geometry. The structure found seems to be one further generalization of generalized complex geometry. By considering a special case we found that generalized complex geometry is contained in the new geometrical structure. However, due to the lack of a proper geometrical framework to describe the type of structures that arise, we were not able to give an exact geometrical interpretation of the conditions for extended supersymmetry.


In article [VI] together with Andreas Bredthauer, Ulf Lindström and Maxim Zabzine, I studied the Hamiltonian formulation of the manifestly $\mathcal{N}=(1,1)$ supersymmetric sigma model. By demanding extended supersymmetry in the phase space we found a direct relation between bi-Hermitean geometry and generalized Kähler geometry. The equivalence between the two formulations is, from a sigma model point of view, due to the equivalence between the Lagrangian and the Hamiltonian formulation. We also used our results to discuss the topological twist in the Hamiltonian setting.


In article [V] I studied T-duality in the Hamiltonian formulation of the $\mathcal{N}=(1,1)$ sigma model. I found the explicit T-duality transformation in this formulation and showed that it is a symplectomorphism, a generalization of a canonical transformation. Under certain assumptions I demonstrated that the amount of extended supersymmetry in the sigma model is preserved under the T-duality transformation. 

        \chapter{Basic string theory}
As seen from the previous chapter, string theory is a vast subject. To describe the topics of this thesis we need a short general introduction to string theory, which is what this chapter provides. Here, we begin by studying the bosonic string in some detail and then go on to see how Einsteins theory of gravity arises out of string theory. We then introduce world-sheet supersymmetry and discuss superstrings.

This entire chapter is based on the standard reference books \cite{Green:1987sp, Polchinski:1998rq, Johnson:2003gi, Zwiebach:2004tj}. The list of references given here is not exhaustive and for more complete set of references the reader is referred to the above books.

\section{Bosonic strings}\label{intro:Bosonic_strings}
\begin{figure}
\begin{centering}
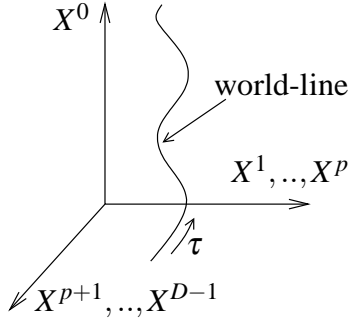
\caption{The world-line of a point particle moving in D-dimensional space-time. The world-line is parametrized by $\tau$.}
\label{img:world-line}
\end{centering}
\end{figure}

The most obvious thing to do when trying to generalize the point particle theory is to replace the point particles with extended objects in space-time. To see how this is done, consider the action for the point particle given by
\begin{align}\label{eqn:particle_action}
S= -m\int\d\tau \sqrt{-g_{\mu\nu}(x)\dot{x}^\mu \dot{x}^\nu}, 
\end{align}
where $x^\mu=x^\mu(\tau)$ and $\dot{x}$ denotes derivation of $x$ with respect to $\tau$.  This action describes the classical propagation of a point particle, of mass $m$, in a curved space-time with metric $g_{\mu\nu}$. The particle traces out a world-line as it moves through the space-time, see fig. \ref{img:world-line}. Extremizing the action provides the classical path of the particle. We say that the action extremizes the length of the path in space-time.

\begin{figure}
\begin{centering}
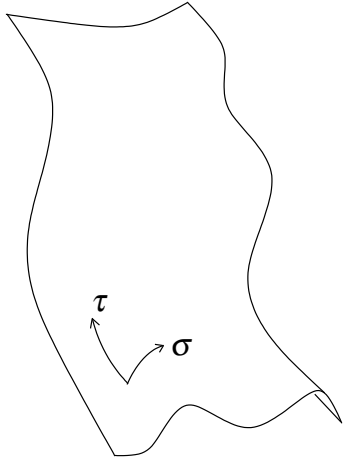
\caption{The string world-sheet. The coordinates used to parametrize the world-sheet are $\sigma$ and $\tau$.}
\label{img:world-sheet}
\end{centering}
\end{figure}

Thus, the most natural generalization from a point particle to a string is to consider the two dimensional world-sheet that the string will trace out when it propagates through space-time, fig. \ref{img:world-sheet}, and to extremize the area of it. To construct an action for the string is not difficult. Note that the volume of a geometrical body must not depend on how we choose to describe the body, \ie the equation we need to solve must have the same form no matter how we choose our coordinates. The action of the string must be invariant under diffeomorphisms.  The general volume element for a p-dimensional manifold that is invariant under diffeomorphisms is given by 
\begin{align}\label{intro:volume_element}
\d V = \sqrt{-\det h_{\alpha\beta}}\,\d^p \sigma,
\end{align}
where $h_{\alpha\beta}$ is the metric of the space. To see that $\d V$ is invariant under a coordinate change $\sigma^{\alpha} \rightarrow \sigma^{\prime\alpha}(\sigma)$, we note that 
\begin{align}
\d^p\sigma' = \left|\det\left(\frac{\partial \sigma^{\prime\alpha}}{\partial \sigma^{\beta}}\right)\right|\, 
               \d^p\sigma
\end{align}
and
\begin{align}
h'_{\alpha\beta} &=\frac{\partial \sigma^{\delta}}{\partial \sigma^{\prime\alpha}} 
         \frac{\partial \sigma^{\gamma}}{\partial \sigma^{\prime\beta}} h_{\delta\gamma}\\
\Longrightarrow \, \det(h'_{\alpha\beta})&= \left(\det\left(
 \frac{\partial \sigma^{\alpha}}{\partial \sigma^{\prime\beta}}\right)\right)^2
 \det(h_{\alpha\beta}).
\end{align}
It follows that the volume element \eqref{intro:volume_element} is invariant under the change of coordinates.

In our case $h_{\alpha\beta}$ is the induced metric on the two dimensional string world-sheet given by $h_{\alpha\beta} = g_{\mu\nu}\partial_\alpha X^\mu\partial_\beta X^\nu$. Now, $\d s^2= h_{\alpha\beta}\d\sigma^\alpha\d\sigma^\beta$ is the square distance between two infinitesimally separated points on the world-sheet, measured along the world-sheet.

Using this induced metric in \eqref{intro:volume_element} and integrating over
the two dimensional world-sheet we obtain the two dimensional volume, the area,
of the world-sheet. Further, we introduce the string tension $T$ that plays a similar role as the mass $m$ in the point particle case. The action obtained in this way is called the Nambu-Goto action and is given by \cite{Nambu:1970,Goto:1971ce}
\begin{align}\label{intro:Nambu-Goto}
S_0=-T\int\d^2\sigma\sqrt{-\det\left( g_{\mu\nu}\partial_\alpha X^\mu\partial_\beta X^\nu \right)},
\end{align}
where $g_{\mu\nu}$ is the metric in the ambient space-time in which the string lives, \ie where the world-sheet is embedded. In the following we will assume that this space-time is flat Minkowski space, \ie $g_{\mu\nu}=\eta_{\mu\nu}$. The coordinates on the world-sheet are $\sigma^\alpha$, where $\alpha\in\{0,1\}$, $\sigma^0=\tau$ and $\sigma^1=\sigma$.  The field $X^\mu = X^\mu(\sigma^\alpha)$ are the coordinates of the world-sheet in space-time. From the world-sheet point of view, the coordinates are just a set of D bosonic scalar fields living on the two dimensional world-sheet.

We have thus found an action that describes the string. There is however a technical difficulty, it contains a square root and is difficult to quantize. To circumvent this difficulty, we introduce a classically equivalent action known as the the Polyakov action \cite{Polyakov:1981rd,Polyakov:1981re}, or perhaps more correctly the Brink - Di Vecchia - Howe - Deser - Zumino action \cite{Brink:1976sc,Deser:1976rb}. It is given by
\begin{align}\label{intro:Polyakov}
S=-\frac{T}{2}\int\d^2\sigma \sqrt{-\det \gamma_{\alpha\beta}}\,\gamma^{\alpha\beta}
\partial_\alpha X^\mu \partial_\beta X^\nu \eta_{\mu\nu},
\end{align}
where the field $\gamma_{\alpha\beta}$ is an independent world-sheet metric. The equation of motion arising by variation of $\gamma_{\alpha\beta}$ is 
\begin{align}\label{intro:eqn_T=0}
T_{\alpha\beta} \equiv \partial_\alpha X^\mu\partial_\beta X^\nu \eta_{\mu\nu} 
  -\frac{1}{2}\gamma_{\alpha\beta}\gamma^{\sigma\delta}
                    \partial_\sigma X^\mu \partial_\delta X^\nu \eta_{\mu\nu}=0,
\end{align}
Taking the square root of minus the determinant of equation \eqref{intro:eqn_T=0} yields \begin{align}
\sqrt{-\det(\eta_{\mu\nu}\partial_\alpha X^\mu\partial_\beta X^\nu)} = \frac{1}{2}\sqrt{-\det{\gamma_{\a\b}}} \gamma^{\sigma\delta} \partial_\sigma X^\mu \partial_\delta X^\nu \eta_{\mu\nu}. \label{intro:relation_NG-P}
\end{align}
This shows that the Polyakov action \eqref{intro:Polyakov} and the Nambu-Goto action \eqref{intro:Nambu-Goto} in flat space-time are classically equivalent. Further, dividing \eqref{intro:eqn_T=0} by \eqref{intro:relation_NG-P} gives a relation between the induced metric $h_{\a\b}=\eta_{\mu\nu}\partial_\alpha X^\mu\partial_\beta X^\nu$ and the world-sheet metric $\gamma_{\a\b}$ as
\begin{align}
h_{\a\b} (-\det(h_{\a\b}))^{-1/2} = \gamma_{\a\b}(-\det(\gamma_{\a\b}))^{-1/2}. \label{intro:relation_h-gamma}
\end{align}

The $T_{\alpha\beta}$ defined in \eqref{intro:eqn_T=0} is the world-sheet energy-momentum tensor. Note that it is symmetric and that its trace vanishes, \ie $\gamma^{\alpha\beta}T_{\alpha\beta}=0$. This implies that there are only two independent components of the world-sheet energy-momentum tensor and \eqref{intro:eqn_T=0} tells us that these components have to be zero.

\subsection{Invariances}\label{intro:invariances}
Invariances under transformations are always important in physics. In the
classical theory the existence of an invariance give rise to a quantity that is
conserved when the system evolves in time. In ordinary quantum mechanics invariance under a transformation implies the existence of an observable that can be
diagonalized simultaneously with the Hamiltonian. Hence the observable specifies a quantum number that is used to characterize the state of the system under
consideration. To specify the state completely we need to find all the
observables that commute with the Hamiltonian (and each other) and hence span
the space of states. This means that finding all invariances of a system that
are independent is important to completely describe the physics. 
Invariance under a transformation that depends on the space-time point is the hallmark of so called gauge theories. When quantizing gauge theories we find that path integrals diverge and give nonsense answers unless we have control over the gauge degrees of freedom.
 
The Polyakov action \eqref{intro:Polyakov} is invariant under the following transformations:
\begin{itemize}
\item Global space-time Poincar\'{e} transformations,
\begin{align}
X^\mu &\rightarrow X^{\prime \mu} = \Lambda^\mu_\nu X^\nu + A^\mu, \label{intro:poincare_inv}\\
\gamma_{\a\b} &\rightarrow \gamma'_{\a\b}=\gamma_{\a\b},
\end{align}
where $\Lambda^\mu_\nu$ is a Lorentz transformation and $A^\mu$ is a space-time translation.
Invariance is explicit since the action is written in covariant form. This tells us that the theory is a relativistic theory, as it certainly should be.

\item Local world-sheet reparametrizations, $\sigma^\alpha \rightarrow \sigma^{\prime\alpha}=f^\alpha(\sigma^\alpha)$. These transformations are also known as diffeomorphisms. Under an infinitesimal transformations, when $f^\alpha(\sigma^\alpha)=\sigma^\alpha + \zeta^\alpha$, the fields in the action transform as
\begin{align}
X^\mu &\rightarrow  X^{\prime \mu} = X^\mu + \zeta^\alpha\partial_\alpha X^\mu, \label{intro:diffeos1}\\
\gamma^{\alpha\beta} &\rightarrow 
     \gamma^{\prime\alpha\beta} =\gamma^{\alpha\beta} 
         + \zeta^\delta\partial_\delta\gamma^{\alpha\beta}
         - \partial_\delta\zeta^\alpha\gamma^{\delta\beta}
         - \partial_\delta\zeta^\beta\gamma^{\alpha\delta}.\label{intro:diffeos2}
\end{align}
This invariance tells us that the physics that we are describing is independent
of the choice of coordinate system on the world-sheet of the string.

\item Local Weyl transformations,
\begin{align}
X^\mu &\rightarrow  X^{\prime \mu} = X^\mu,\\
\gamma_{\alpha\beta} &\rightarrow \gamma'_{\alpha\beta}=\e^{2\omega}\gamma_{\alpha\beta}, \label{intro:weyl_invariance}
\end{align}
for a function $\omega=\omega(\sigma,\tau)$. This symmetry means that two world-sheet metrics that differ only by a Weyl transformation describe the same physical situation.
\end{itemize}

We can use these invariances to put the Polyakov action \eqref{intro:Polyakov} into a nice form, \ie we can choose a convenient gauge. One such choice is the conformal gauge, where we use Weyl invariance and diffeomorphism invariance to gauge fix the world-sheet metric to
\begin{align}
\gamma_{\alpha\beta} = \eta_{\alpha\beta}. \label{intro:conformal_gauge}
\end{align}
Here $\eta_{\alpha\beta}$ is the two dimensional flat Minkowski metric with signature $(-,+)$. This gauge choice simplifies the form of the Polyakov action \eqref{intro:Polyakov} to
\begin{align}
S&=-\frac{T}{2}\int\d^2\sigma \partial_\alpha X^\mu \partial^\alpha X^\nu \eta_{\mu\nu} \label{intro:confAction}\\
 &=-\frac{T}{2}\int\d^2\sigma \,\left(\partial_\sigma X^\mu \partial_\sigma X^\nu
                               - \partial_\tau X^\mu \partial_\tau X^\nu \right)\eta_{\mu\nu}.
\end{align}

Since the vanishing of the world-sheet energy-momentum tensor does not follow as equations of motion from this gauge fixed action we need to impose the condition \eqref{intro:eqn_T=0} by hand. 

The invariance of the Polyakov action \eqref{intro:Polyakov} under diffeomorphisms \eqref{intro:diffeos1}-\eqref{intro:diffeos2} implies that the world-sheet energy-momentum tensor is conserved. In the conformal gauge the conservation equation reads 
\begin{align}
\partial^\alpha T_{\alpha\beta} = 0. \label{intro:T_conserv}
\end{align}
After having chosen the conformal gauge \eqref{intro:conformal_gauge} there is still a residual symmetry that preserves the gauge choice. This symmetry arises as follows; perform a diffeomorphism that changes the world-sheet metric only by a scale factor and then a Weyl transformation that scales the world-sheet metric back to the original form. This residual symmetry turns out to be exactly the two dimensional conformal transformations as can be seen from the following argument. We go to the Euclidean version of the world-sheet by letting $\tau\rightarrow -\i\tau$. We then introduce the new complex coordinate $z=\tau +\i\sigma$, its complex conjugate $\bar{z}=\tau - \i\sigma$ and the derivatives $\partial_z = \frac{1}{2}(\partial_\tau -\i\partial_\sigma)$ and $\partial_{\bar{z}}=\frac{1}{2}(\partial_\tau +\i\partial_\sigma)$. In these coordinates the fact that the world-sheet energy-momentum is traceless translates to $T_{z\bar{z}}=0$. The conservation equation \eqref{intro:T_conserv} now becomes
\begin{align}
\partial_{z} T_{\bar{z}\bar{z}} = \partial_{\bar{z}} T_{zz} = 0.
\end{align}
This implies that the components satisfy $T_{zz}=T_{zz}(z)$ and $T_{\bar{z}\bar{z}}=T_{\bar{z}\bar{z}}(\bar{z})$. Next we introduce the holomorphic function $v(z)$ to form the current $j_\alpha$ with components
\begin{align}
j_z (z) = \i v(z)T_{zz}(z), \;\;\;\;\;\;
j_{\bar{z}}(\bar{z}) = \i v(z)^* T_{\bar{z}\bar{z}}(\bar{z}).
\end{align}
This current is conserved, $\partial^\alpha j_\alpha=0$, and generates the residual symmetry of the gauge fixed action. The symmetry transformation is given by
\begin{align}
\delta X^\mu = -\epsilon v(z)\partial_z X^\mu - \epsilon v(z)^* \partial_{\bar{z}}X^\mu.
\end{align}
This is the infinitesimal version of the two dimensional conformal transformations. Hence, the residual symmetry that here arises from the conservation and tracelessness of the world-sheet energy-momentum tensor is the two dimensional conformal symmetry. We will use this conformal symmetry when we discuss string interactions in section \ref{intro:section_interactions}.

\subsection{Classical equations of motion}
\begin{figure}
\begin{centering}
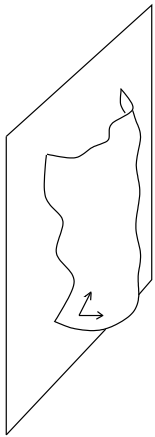
\caption{An open string world-sheet attached to a D1-brane.}
\label{img:Dbrane}
\end{centering}
\end{figure}

The classical equations of motion are found by variation of the fields $X^\mu$ in the gauge fixed action \eqref{intro:confAction},
\begin{align}\label{intro:EOM}
\Box X^\mu=\left(\partial^2_\sigma-\partial^2_\tau\right)X^\mu=0.
\end{align}
This is the wave equation in two dimensions for each $\mu\in\{0,...,D-1\}$. The general solution to this equation is $X^\mu=X^\mu_L(\tau-\sigma) + X^\mu_R(\tau+\sigma)$, meaning that the solution separates into one left moving and one right moving part.  The equations of motion are accompanied by conditions from the boundary terms that arise in the variation of the action. The variation of the field $X^\mu$ in the action
yields the boundary term 
\begin{align}
-T\int\d\tau \left\{\partial_\sigma X_\mu \delta X^\mu|_{\sigma=\pi}
                - \partial_\sigma X_\mu \delta X^\mu|_{\sigma=0} \right\} =0.
\end{align}
To satisfy this equation we must put conditions on the endpoints of the string,
\ie specify boundary conditions for $X^\mu$.  There are different
choices for the boundary conditions:
\begin{itemize}
 \item Periodic boundary conditions, $X^\mu(\sigma+\pi,\tau)=X^\mu(\sigma,\tau)$,
 \item Dirichlet boundary conditions, $\delta X(\sigma=0,\tau) = \delta X(\sigma=\pi,\tau) = 0$,
 \item Neumann boundary conditions, $\partial_\sigma X_\mu|_{\sigma=0}=\partial_\sigma X_\mu|_{\sigma=\pi}=0$. 
\end{itemize}
The first case specifies that the string world-sheet is periodic in the
$\sigma$ direction. This means that imposing this boundary condition give us
the closed strings. The other two possible boundary conditions give us open 
strings and there is a possibility to combine the two in the sense that for some
coordinates $X^\mu$ with $\mu=0,...,p$ we use Neumann boundary conditions and for the other coordinates $\mu=p+1,...,D-1$ we use Dirichlet boundary conditions. In this case
the first set of coordinates are free to vary but the second set is not, these are
fixed to specific values. This means that the string endpoints can move freely on a
$p+1$ dimensional hypersurface but they can not leave this hypersurface, figure \ref{img:Dbrane}. This surface is called a D$p$-brane.

One thing to note about the Neumann boundary condition is that it states that there is no 
flow of momentum off the string. This is not the case for the directions in which the Dirichlet boundary conditions are imposed, hence there can be momentum transfer between the D-brane and the string.

\subsection{The closed string}
The solution to the equations of motion \eqref{intro:EOM} that also respects the periodic boundary conditions are given by $X^\mu=X^\mu_R + X^\mu_L$, where the mode expansions of the right respective the left moving solution are given by
\begin{align}
X^\mu_R(\sigma,\tau) &= \frac{1}{2}x^\mu +\alpha' p^\mu(\tau-\sigma) +\i\sqrt{\frac{\alpha'}{2}}
                            \sum_{n\neq0}\frac{1}{n}\alpha_n^\mu\e^{-2\i n(\tau-\sigma)}, 
                \label{intro:mode_closed1}\\
X^\mu_L(\sigma,\tau) &= \frac{1}{2}x^\mu +\alpha' p^\mu(\tau+\sigma) +\i\sqrt{\frac{\alpha'}{2}}
                            \sum_{n\neq0}\frac{1}{n}\tilde{\alpha}_n^\mu\e^{-2\i n(\tau+\sigma)}.
    \label{intro:mode_closed2}
\end{align}
Here $\sigma\in [0,\pi]$ and $\alpha' =(2\pi T)^{-1}$. The total momentum for the string is given by integrating the conjugate momentum $P^\mu=T \partial_\tau X^\mu$ over the strings spatial extension $\sigma\in[0,\pi]$. The result is that the total momentum of the string is given by $p^\mu$, which motivates the interpretation of the coefficient $p^\mu$ as the center of mass momentum of the string. Further, since the fields $X^\mu$ are coordinates in the target space they must be real. This implies that, after quantization, mode operators have to satisfy ${a^\mu_{-n}=\left(a_n^\mu\right)^\dagger}$ and ${\tilde{a}^\mu_{-n}=\left(\tilde{a}_n^\mu\right)^\dagger}$. For the zero modes $x^\mu$ and $p^\mu$ the reality condition implies that they have to be real.

To quantize the closed string we promote the fields to operators and impose
equal $\tau$ commutation relations between the field $X^\mu$ and its conjugate momentum
$P^\mu$,
\begin{align}
[X^\mu(\sigma,\tau),P^\nu(\sigma',\tau)]&=\i\eta^{\mu\nu}\delta(\sigma-\sigma'),\\
[X^\mu(\sigma,\tau),X^\nu(\sigma',\tau)]&=[P^\mu(\sigma,\tau),P^\nu(\sigma',\tau)]=0 .
\end{align}
These translate into the following commutators for the modes,
\begin{align}
\hspace*{-0.5cm}
\left[x^\mu ,p^\nu\right]=\i \eta^{\mu\nu},\;\;\;\;
\left[\alpha_m^\mu ,\alpha_n^\nu\right]=m \delta_{m+n}\eta^{\mu\nu},\;\;\;\;
\left[\tilde{\alpha}_m^\mu ,\tilde{\alpha}_n^\nu\right]=m \delta_{m+n}\eta^{\mu\nu}, \label{intro:comm_bos_modes}
\end{align}
and all other commutators being zero.  This means that $\alpha_n^\mu$ and $\tilde{\alpha}_n^\mu$ now become creation ($n<0$) and annihilation ($n>0$) operators in the respective sector of the theory. We define the vacuum state of the theory to be annihilated by all the annihilation operators. The commutator $[p^\mu, \alpha^\nu_{-n}]=0$ tells us that the vacuum state has an additional quantum number $k^\mu$ such that $p^\mu |0,k\rangle= k^\mu|0,k\rangle$. Thus, $k^\mu$ is the center of mass momentum of the vacuum state $|0,k\rangle$. We then construct states in a Fock space by acting on a vacuum state with creation operators. To find the physical spectrum of the string, we must not forget the condition of vanishing energy-momentum tensor \eqref{intro:eqn_T=0}.

If we expand the energy-momentum tensor in modes that we conventionally call $L_m$ and express these in terms of the oscillator modes we find 
\begin{align}
L_m = \frac{1}{2}\sum_{n=-\infty}^{\infty} :\alpha^\mu_{m-n}\alpha_{n\mu}:,
\end{align}
where we have defined $\alpha^\mu_0 = \sqrt{\frac{\alpha'}{2}}p^\mu$ for the closed string and $\alpha^\mu_0 = \sqrt{2\alpha'}p^\mu$ for the open string. The $::$ denotes normal ordering which is only an issue for the $L_0$ operator. For the closed string we can write
\begin{align}
L_0 = - a + \frac{\alpha'}{4} p^\mu p_\mu +\sum_{n=1}^{\infty} \alpha^\mu_{-n}\alpha_{n\mu},
\end{align}
where $a$ is a normal ordering constant that remains to be determined.

For the closed string there exist a second set of Fourier modes of the energy-momentum tensor $\tilde{L}_m$ that is defined in the same way but with tildes on the oscillators and $\tilde{\alpha}^\mu_0 = \sqrt{\frac{\alpha'}{2}}p^\mu$.

Each set of Fourier modes of the energy-momentum tensor form a closed algebra known as the Virasoro algebra,
\begin{align}
 [L_m,L_n]=(m-n)L_{m+n} + \left(\frac{1}{12}D(m^3-m) +2m a \right)\delta_{m+n}, \label{intro:Virasoro_algebra}
\end{align}
where $D$ is the dimension of the space-time. In section \ref{intro:differentQuantizations} we will motivate that the values $D=26$ and $a=1$ are required by consistency of the bosonic string theory. For now we just assume that the values for the space-time dimension and the normal ordering constant are the correct ones.

Since we must require the energy-momentum tensor to vanish in the theory we must impose a constraint that assures that $\langle L_m \rangle=0$ for the physical states. If we note that $L_m^\dagger=L_{-m}$ it turns out that it is enough to require
\begin{align}
L_m|phys\rangle = 0, \;\;\;\tilde{L}_m|phys\rangle = 0, \;\;\;\; m= 0,1,2,... \label{intro:closed_phys_cond}
\end{align}
for the physical states. 

The mass spectrum of the closed string arises from the equation $L_0|phys\rangle=0$ and is given by
\begin{align}
- \alpha'p^\mu p_\mu = \alpha' M^2 = -4 + 4\sum_{n=1}^{\infty}\alpha_{-n}^\mu\alpha_{n \mu} 
       = -4 + 4\sum_{n=1}^{\infty}\tilde{\alpha}_{-n}^\mu \tilde{\alpha}_{n\mu}. \label{intro:closed_string_spectrum}
\end{align}
Hence, the vacuum $|0,k\rangle$ has imaginary mass, \ie is a tachyonic state. This signals an instability of the theory. We will not worry about this defect here since there is a more serious flaw with the bosonic string theory; there are no fermions in the spectrum. Thus, the theory can not describe matter particles which are obviously an essential part of the world around us. When we introduce fermions in the theory we will see that the spectrum will not contain any tachyons.

By considering the equation $(L_0-\tilde{L}_0)|phys\rangle=0$ we obtain a condition on the mode operators,
\begin{align}
\sum_{n=1}^{\infty}\alpha^\mu_{-n}\alpha_{n\mu}
       =\sum_{n=1}^{\infty}\tilde{\alpha}^\mu_{-n}\tilde{\alpha}_{n\mu}. \label{intro:level_matching_cond}
\end{align}
This is the only relation between the left and right moving sectors in the closed string. It states that for the physical states, the sum of the mode numbers of the excited modes in the state must be the same in the left and right sector. For example, the states $\alpha^\mu_{-2}\tilde{\alpha}^\nu_{-2} |0,k\rangle$ and
$\alpha^\mu_{-1}\alpha^\mu_{-1}\tilde{\alpha}^\nu_{-2} |0,k\rangle$ are physical states of mass $(4/\alpha')^{1/2}$ that satisfy \eqref{intro:level_matching_cond}.

If we turn to the massless part of the spectrum we find that there is only one possible combination of the creation operators that is physical, namely 
$\alpha^\mu_{-1}\tilde{\alpha}^\nu_{-1}|0,k\rangle$. This state is a space-time tensor, and under Lorentz transformations it reduces into three irreducible representations defining particles. The traceless symmetric part $G_{\mu\nu}$ is the graviton, the antisymmetric part $B_{\mu\nu}$ is the so-called $B$-field and the trace part $\Phi$ is the dilaton.

\subsection{The open string}
We now turn to the open string with Neumann boundary conditions $\partial_\sigma
X^\mu = 0$ at both ends $\sigma=0, \pi$. The mode expansion of
the general solution to the equations of motion \eqref{intro:EOM} satisfying
these boundary conditions is
\begin{align}\label{intro:mode_openNN}
X^\mu(\sigma,\tau) = x^\mu + 2\alpha' p^\mu \tau 
 + \i\sqrt{2\alpha'}\sum_{n\neq0}\frac{1}{n}\alpha_n^\mu\e^{-\i n\tau}\cos(n\sigma).
\end{align}
We note that the open string boundary conditions relate the left and right
moving sectors so that there is just one set of oscillators.

Next we turn to the mode expansion of the coordinates for a string
with Dirichlet conditions at both ends, \ie $\delta X^a|_{\sigma=0,\pi}=0$.
We let one end of the string be attached to a D-brane at position $X^a=x^a_1$
and the other end to an other D-brane at position $X^a=x^a_2$. Solving the
equations of motion \eqref{intro:EOM} with these boundary conditions gives the
mode expansion
\begin{align}\label{intro:mode_openDD}
X^a(\sigma,\tau) = x_1^a + (x_2^a-x_1^a)\frac{\sigma}{\pi}
        + \sqrt{2\alpha'}\sum_{n\neq0}\frac{1}{n}\alpha_n^a \e^{-\i n\tau}\sin(n\sigma).
\end{align}
Note that since the string is attached to D-branes (that are fixed in space-time at given positions) the string will not have any momentum in the
directions normal to the brane, this is the reason why $p^a=0$ in this case. 

Thus for a string attached to D-branes at both ends, some 
of its coordinates will have the mode expansion \eqref{intro:mode_openNN} and
the other coordinates will have mode expansion \eqref{intro:mode_openDD}.

To quantize the theory we proceed as in the case of the closed string,
introducing commutators for the fields and their conjugate momenta. The
Virasoro algebra arises from the mode expansion of the energy-momentum tensor, but in this case there is only one copy of it. The physical state condition is now
\begin{align}
L_m|phys\rangle=0, \;\;\;\; m=0,1,2....  \label{intro:open_phys_cond}
\end{align}
The $L_0$ condition give us the mass spectrum
\begin{align}
M^2 = \left(\frac{x_2^a-x_1^a}{2\pi\alpha'}\right)^2
   -\frac{1}{\alpha'} + \frac{1}{\alpha'}\sum_{n=1}^{\infty} \alpha_{-n}^\mu\alpha_{n\mu}.
\end{align}
Note that since the string tension is given by $T=(2\pi\alpha')^{-1}$, the first term is just the square of classical energy of a string stretched between the two D-branes. This means that if we have a string stretched between two D-branes that are far enough apart there are no tachyons in the spectrum. However if the D-branes are close together the ground state will still be a tachyon, hence indicating an instability of the theory. This is thus not a stable configuration.

If one consider a string with only Neumann boundary conditions or a string that
has both ends attached to the same D-brane, \ie $x^a_1=x^a_2$, the ground state
is a tachyon. For the string attached to a D-brane it has been proposed by
Sen \cite{Sen:2002nu,Sen:2002in} that this tachyon has a physical
interpretation in terms of the decay of the D-brane. The first excited state is
massless and given by $\alpha^\mu_n|0,k\rangle$. This is a massless
gauge boson living in space-time.

\subsection{Different methods of quantization} \label{intro:differentQuantizations}
In the previous sections we quantized the string in a Lorentz covariant fashion. This method gives us a Fock space that needs to be restricted, which is done by imposing the physical state conditions, \eqref{intro:closed_phys_cond} or \eqref{intro:open_phys_cond}. However, it is difficult to prove that there are no negative norm states, so called ghost states, in the quantum theory. So we may ask whether there are other possible quantization schemes that produce Fock spaces that only contain physical states and explicitly no ghost states. It turns out that this is possible.

The different types of quantization schemes we will discuss here are covariant quantization, light-cone quantization and BRST quantization. More thorough introductions to these can be found for example in the books \cite{Peskin:1995ev,Green:1987sp,Polchinski:1998rq}.

\subsubsection{Covariant quantization}
This is the quantization scheme that we used in the previous sections and here
we only briefly summarize the method. In the covariant quantization we impose commutator relations for the fields that imply commutator relations for the modes. The Fock space built by the modes contain ghost states. This is the case since the Minkowski metric has $\eta^{00}=-1$, and hence the open string state $\alpha^0_{-1}|0,k\rangle$ has negative norm. The first step to get rid of these is to impose the condition that arises from the vanishing of the energy-momentum tensor. This means that we impose the conditions, \eqref{intro:closed_phys_cond} or \eqref{intro:open_phys_cond}, on the
states in the Fock space to obtain the physical spectrum.  The critical values
for the dimension $D=26$ and the normal ordering constant $a=1$ are required
since only for these values do the negative norm states decouple from the physical spectrum.

\subsubsection{Light-cone quantization}
Light cone quantization is another method for quantizing the string. In this
method the manifest space-time Lorentz invariance is broken by separating out
two space-time directions, the time direction, $X^0$, and one spacelike direction, \eg $X^{D-1}$. We then define the space-time light-cone coordinates as $X^\pm = \frac{1}{\sqrt{2}}\left(X^0 \pm X^{D-1}\right)$.  It turns out that we can choose $X^+=x^+ + p^+\tau$. This is called the light-cone gauge. In this gauge the condition of vanishing energy-momentum tensor \eqref{intro:eqn_T=0} can be solved and we find that it is possible to express $X^-$ in terms of the coordinates $X^i$ with $i=1,...,D-2$. Writing down the mode expansions for the $X^i$ coordinate fields gives the independent modes. Imposing the commutation relations between the coordinate fields yields the commutators between the modes. These become creation and annihilation operators in the standard way. The crucial difference from the covariant quantization is that the commutators between the modes now have the same sign for all the modes, this follows since the space spanned by the $X^i$'s has an Euclidean metric. Hence the physical state space, built by the creation operators, is manifestly ghost free. 

It turns out that in this light-cone setting it is possible to construct linear independent sets of states that at each mass level are equal in number to the states generated by the covariant mode operators $\alpha^\mu_n$ at the same mass level. Thus, this new set of states is just another basis of the Hilbert space of the string excitations. In this new basis a physical state decomposes into a zero norm physical state that is orthogonal to all physical states, called a spurious state, and a physical state of non-negative norm. Hence, any physical state has non-negative norm and there are no ghost states in the physical spectrum. Since this spectrum is related to the spectrum arising from the covariant quantization by a change of basis in the Hilbert space it follows that the spectrum from the covariant quantization is also free of ghosts. For the details of the no-ghost theorem, see e.g.\ \cite{Green:1987sp}.

This method of quantization started off by making a non-covariant choice when
we selected two space-time directions, this means that the space-time Lorentz
invariance is no longer manifest. However, we want this classical symmetry to be a symmetry also for the quantized theory. This means that we have to require that the Lorentz algebra holds in the quantized theory as well. It turns out that the Lorentz invariance is preserved after quantization precisely if the space-time dimension is twenty-six, $D=26$, and the normal ordering constant arising in the Virasoro algebra is one, $a=1$. 

\subsubsection{BRST quantization}
The last quantization scheme that we will consider is BRST quantization. This differs from the two previous methods in that it uses the path integral as the fundamental object. Furthermore, it takes care of the gauge invariances in the theory in an elegant way.

We start from the path integral formed out of the Polyakov action \eqref{intro:Polyakov} and consider the invariances of the action. In the present case the symmetries we will consider are the two-dimensional diffeomorphism invariance and the Weyl invariance. These symmetries are used to choose a gauge slice in the integration space that fixes the gauge of the world-sheet metric $\gamma_{\alpha\beta}$. In the path integral this gauge slice is represented by delta functions and gauge-fixing determinants.

However, these determinants are difficult to calculate explicitly, but a trick due to Faddeev and Popov \cite{Faddeev:1967fc} makes it possible to proceed. We introduce ghost and anti-ghost fields, called Faddeev-Popov (anti)ghosts, to represent the determinants as integrals. These (anti)ghosts are in fermionic if the gauge fixed fields are bosonic and vice verse. In the present case the Faddeev-Popov (anti)ghosts will be fermionic and in the path integral they are represented by Grassmann variables, \ie anticommuting variables.

This procedure will modify the action of the theory so that the total action of the theory is given by the sum of the Polyakov action \eqref{intro:Polyakov}, a gauge fixing action, which is an integral over a Lagrange multiplier times the gauge fixing conditions, and a ghost action.
When we fix the gauge to be the conformal gauge the ghost action is given by
\begin{align}
S_g = - i \frac{T}{2}\int\d^2\sigma \sqrt{-\det \gamma_{\alpha\beta}}\,\gamma^{\alpha\beta}
      c^\delta \nabla_\alpha b_{\beta\delta}.\label{intro:FP_ghost_action}
\end{align}
Here $c^\delta$ is a vector field representing the Faddeev-Popov ghost field and $b_{\beta\delta}$ is a traceless symmetric tensor field representing the anti-ghost field. Further $\nabla_\alpha$ is the covariant derivative constructed from the world-sheet metric $\gamma_{\alpha\beta}$.

So, we have introduced more fields that represent the gauge degrees of freedom. More fields in the theory and apparently more degrees of freedom. But the ghost fields have the opposite statistics to the coordinate fields and will in fact serve to cancel the unphysical degrees of freedom. We next need to investigate if there are any remaining symmetries of the theory, now with the Faddeev-Popov ghost fields included. It turns out that there is one such remaining symmetry, the BRST symmetry. The fact that such a symmetry exists in a general gauge theory when ghosts fields are used to represent the gauge-fixing determinants was first discovered by Becchi, Rouet, and Stora, \cite{Becchi:1974md} and independently by Tyutin \cite{Tyutin:1975qk} in 1975. The generator $Q$ of this symmetry, called the BRST operator, is in the classical theory nilpotent, $Q^2=0$. However, when going to the quantum theory of the bosonic string the BRST operator is no longer nilpotent unless the dimension of the target space is twenty-six, $D=26$, and the normal ordering constant of the Virasoro generators is one, $a=1$. This is how these critical values appear in the BRST quantization of the bosonic string. We will only consider this critical case.

Since the BRST operator is nilpotent it can be used to define a cohomology. This cohomology defines the physical states of the string. More explicitly, the physical states of bosonic string theory are the ones that that are $Q$-closed, $Q|phys\rangle=0$, but not $Q$-exact, $|phys\rangle\neq Q|\Psi\rangle$ for some state $|\Psi\rangle$. It turns out that we need to make one further restriction to deal with the ghosts, we need to restrict the physical spectrum to states that are annihilated by the zero mode of the anti-ghost field $b$. We think of such a state as not containing any ghosts. It follows that this restriction implies that the physical states satisfy the condition $L_0|phys\rangle=0$, which in turn determines the mass spectrum of the string.

In short, we include ghosts to fix the gauge, discover BRST symmetry and use it to find the set of physical states. These physical states agree with the ones found in light-cone quantization, and the statement that  $Q|phys\rangle=0$ is the same as the usual condition that the Virasoro generators $L_m$ with $m>0$ will annihilate physical states.

Adding the ghost action \eqref{intro:FP_ghost_action} to the theory gives a contribution to the world-sheet energy-momentum tensor \eqref{intro:eqn_T=0}. The mode expansion of the energy-momentum tensor defines the Virasoro generators, which now will contain a ghost part. Hence the central charge, \ie the part in the algebra \eqref{intro:Virasoro_algebra} with $\delta_{m+n}$, in the Virasoro algebra will be modified. The non-vanishing of this central charge represent an anomaly in the Weyl invariance \eqref{intro:weyl_invariance} of the quantum theory. Put differently, the quantum theory is no longer Weyl invariant if the Virasoro algebra has a non-zero central charge. Requiring this anomaly is absent again yields the critical dimension $D=26$ and the normal ordering constant $a=1$. 

\subsection{Oriented vs. unoriented strings}
In the previous sections we have without stating it assumed that there is an
internal direction on the world-sheet, \ie the direction of increasing
$\sigma$. When this is the case the strings that arise are called oriented
strings. However one may consider the case where the strings do not have this
property. This means that the world-sheet, and the theory, is invariant under
the reflection $\sigma\rightarrow -\sigma$ for closed strings and $\sigma
\rightarrow \pi-\sigma$ for the open string. Imposing this symmetry reduces the
spectrum of available physical states.  Performing the reflection explicitly in
the mode expansions (\ref{intro:mode_closed1}, \ref{intro:mode_closed2},
\ref{intro:mode_openNN}, \ref{intro:mode_openDD}) we find that the closed
string modes transform as $\alpha^\mu_n\leftrightarrow\tilde{\alpha}^\mu_n$ and
the open string modes transform as $\alpha^\mu_n\rightarrow
(-1)^n\alpha_n^\mu$. This means that to have an unoriented string the spectrum
should be invariant under these transformations. In particular, for
the open string that the massless photon $A^\mu$ must be removed, but the
tachyon is still present. For the closed string we still have the graviton
$G_{\mu\nu}$ and the dilaton $\Phi$, but the antisymmetric tensor $B_{\mu\nu}$
is no longer present in the spectrum.

\subsection{String interactions and vertex operators} \label{intro:section_interactions}
\begin{figure}
\begin{centering}
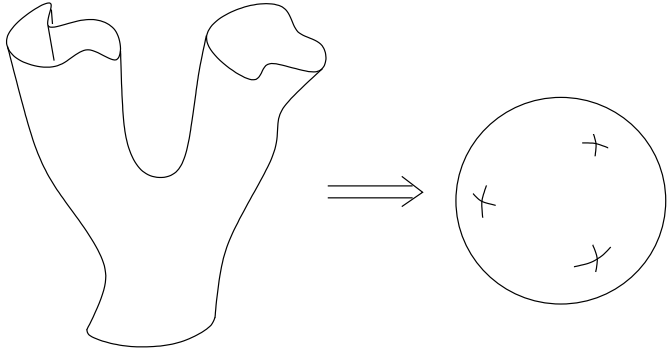
\caption{The pants diagram showing one closed string that splits into two. The
right picture show the same diagram when the world-sheet is mapped to a sphere and vertex operator insertions are used to represent the external states.}
\label{img:pants}
\end{centering}
\end{figure}

String interactions are most naturally described by diagrams of the type in
figure \ref{img:pants}. We interpret this so called ``pants-diagram'' as one
incoming closed string that splits into two outgoing closed strings. In this section we will only consider interactions of the closed string. For a more complete treatment of string interactions see e.g.\ \cite{Green:1987sp, Polchinski:1998rq, Zwiebach:2004tj, Johnson:2003gi}.

To understand how string interactions work we use the residual conformal symmetry, found in section \ref{intro:invariances}, to deform the world-sheet of the string. The conformal symmetry gives us enough freedom to map the closed string world-sheet to, for example, a sphere. Under this rescaling all external states are mapped to points on this sphere where they are represented as vertex operators. To study these operators we introduce new world-sheet coordinates for the string. We begin by going to the Euclidean form of the metric by letting $\tau\rightarrow -i\tau$. We now use the conformal invariance to map the world-sheet to a sphere, as in figure \ref{img:pants}, or to the complex plane. By the change of coordinates $z=\e^{\tau -i\sigma}$ we map the closed string world-sheet to the complex plane, where $z=0$ corresponds to the infinite past of the string, $\tau=-\infty$, and the `point' at infinity to the infinite future of the string, $\tau=\infty$. In these coordinates the Euclidean version of the Polyakov action \eqref{intro:Polyakov} in flat Minkowski space-time becomes
\begin{align}
S_0= \frac{1}{2\pi\alpha'} \int \d z \d\bar{z} \,\partial X^\mu \bar{\partial}X^\nu \eta_{\mu\nu}, \label{intro:C_action}
\end{align} 
where $\bar{z}$ is the complex conjugate of $z$ and $\partial=\frac{\partial}{\partial z}$, $\bar{\partial}=\frac{\partial}{\partial\bar{z}}$.

To insert a vertex operator in a path integral we must integrate over the position of the insertion. This follows since we do not know exactly where an emission or absorption occurs and hence we must take all possibilities into account. The operator obtained in this way by integrating a vertex operator is called an integrated vertex operator. Further, since an integrated vertex operator are to represent a part of a string world-sheet we must require that it respects the conformal symmetry of the theory. For this, note that under a conformal transformation $z=z'(z)$ the measure transforms as
\begin{align}
\d z' \d\bar{z}' = \left(\frac{\partial z'}{\partial z}\right) \left(\frac{\partial\bar{z}'}{\partial\bar{z}}\right)\, \d z \d\bar{z}.
\end{align}
We next define a primary field $\Phi$ of conformal weight $(h,\bar{h})$ as a field that transforms as
\begin{align}
\Phi(z',\bar{z}')= \left(\frac{\partial z'}{\partial z}\right)^{-h} \left(\frac{\partial\bar{z}'}{\partial\bar{z}}\right)^{-\bar{h}} \Phi(z,\bar{z})
\end{align}
under a conformal transformation. This implies that the vertex operators must be primary fields of conformal weight $(1,1)$ so that the integrated vertex operator is a primary field of conformal weight $(0,0)$. Hence, under this condition an integrated vertex operator does preserve the conformal symmetry when it is used to deform the theory. Such an operator is called an exactly marginal operator.

The integrated vertex operator that describes the emission or absorption of a closed string tachyon, with $k^2=4/\alpha'$, is given by
\begin{align}
V_t = \int \d z \d\bar{z}\, :\e^{\i k^\mu X_\mu}:
\end{align}
and the integrated vertex operators describing the emission or absorption of a closed string graviton, with $k^2=0$, is given by 
\begin{align}
V_g = \int \d z \d\bar{z} \, :\xi_{(\mu\nu)} \partial X^\mu \bar{\partial} X^\nu \e^{\i k^\lambda X_\lambda}:.
\end{align} 
The normal ordering, $::$, in these expressions is to say that when the integrands are expanded in terms of the modes we must put all the creation operators $\alpha^\mu_n$, $n<0$ to the left of the annihilation operators $\alpha^\mu_n$, $n>0$.

Consider a closed string propagating in flat space-time that emits a graviton. The Euclidean path integral for this process is given by
\begin{align}
Z = \int \mathcal{D}X\; V_g \e^{- S_0},
\end{align}
where $S_0$ is the action \eqref{intro:C_action} and the integral is over all possible histories of the the string. Next, consider the situation where the string propagates in flat space-time and interacts with a lot of gravitons. The total Euclidean path integral for this process is obtained by the sum over all possible number of closed string graviton insertions on the world-sheet. The amplitude for this process becomes
\begin{align}
Z &= \int\mathcal{D}X \;\left(1 + V_g + \frac{1}{2}(V_g)^2 + \frac{1}{3!}(V_g)^3 + \ldots\right)\e^{- S_0} \cr
  &= \int\mathcal{D}X \; e^{- (S_0 - V_g)},
\end{align}
where the modified action is given by
\begin{align}
S_0-V_g = \frac{1}{2\pi\alpha'}\int\d z \d\bar{z}\, \partial X^\mu \bar{\partial} X^\nu
      \left(\eta_{\mu\nu} - 2\pi\alpha' \xi_{(\mu\nu)}\e^{\i k^\lambda X_\lambda}\right).
\end{align}
This means that by letting the string propagate in this flat background and interact with closed string gravitons we have effectively modified the metric of the background space-time making it non-flat. If we consider a string in a general curved space-time with a metric, the space-time itself can be seen as a coherent state of closed string gravitons. In this sense the strings themselves produce the background in which they live.

\section{Backgrounds}
In 1985 Callan et. al. \cite{Callan:1985ia} showed that Einstein's equations, with small string theoretic corrections, arise as the condition for Weyl invariance to hold in the quantized string theory. This means that string theory, which is a quantum theory, actually contains gravity. This result is a strong indication that string theory might be part of the final theory of everything. In this section we will briefly sketch how Einstein's equations arise out of string theory.

In the previous section we saw that the metric of the background space-time can be thought of as arising out of the massless graviton state of the closed string. However there we considered only the symmetric part of the general massless state. Remember that the general massless state of the closed string also contains an antisymmetric and a trace part. We might expect that these states would play a similar r\^{o}le as the graviton state does for the string. If we incorporate these fields in the action, we obtain
\begin{align}\label{intro:action_BGD}
S = \frac{1}{4\pi\alpha'} \int\d^2\sigma \Big\{& \sqrt{-\det \gamma_{\alpha\beta}} \left(
     \gamma^{\alpha\beta}G_{\mu\nu}\partial_\alpha X^\mu \partial_\beta X^\nu + \alpha' \Phi  R^{(2)}\right)\cr
    & + \epsilon^{\alpha\beta}B_{\mu\nu}\partial_\alpha X^\mu \partial_\beta X^\nu \Big\}.
\end{align}
Here, $G_{\mu\nu}=G_{\mu\nu}(X)$ is a general metric that arise as a condensation of the graviton states, the $B_{\mu\nu}=B_{\mu\nu}(X)$ is antisymmetric and comes from the antisymmetric part of the massless closed string state and $\Phi=\Phi(X)$ is the dilaton field arising from the massless closed string trace part. The $\epsilon^{\alpha\beta}$ is the two dimensional totally antisymmetric tensor and $R^{(2)}$ is the two dimensional scalar curvature of the world-sheet. The $\alpha'$ is present in the last term to make it dimensionless. When we consider $\alpha'$ as a small expansion parameter, we see the presence of the $\alpha'$ in the last term as an indication that the term is a one loop effect.

To consider string interactions we need the action \eqref{intro:action_BGD} to be invariant under conformal transformations. Further, for consistency of the theory we need that it is invariant under Weyl transformations \eqref{intro:weyl_invariance}. We proceed by examining under what circumstances this is the case in the quantized theory.

To investigate the Weyl invariance we note that the non-linear sigma model defined by \eqref{intro:action_BGD} is actually an interacting two dimensional quantum field theory where the fields $G_{\mu\nu}$, $B_{\mu\nu}$ and $\Phi$ are field dependent coupling ``constants''. A Weyl transformation changes the scale of the theory and, as in ordinary quantum field theory, if we want to know how the theory behaves under scaling we construct the $\beta$-functions. These describe how the coupling ``constants'' change with the scale, for details see e.g.\ \cite{Peskin:1995ev}. For the quantum theory to be invariant under Weyl transformations we require that the $\beta$-functions of the theory defined by \eqref{intro:action_BGD} vanish. Note that for string theory the $\beta$-functions are actually functionals since the couplings depend on the fields $X^\mu$ that define the string position in space-time.

The $\beta$-functionals may be calculated using dimensional regularization techniques and the requirement that these should vanish are, to lowest order in $\alpha'$, given by \cite{Callan:1985ia}
\begin{align}
0=\beta^G_{\mu\nu} &= \alpha'\left( R_{\mu\nu} + 2\nabla_\mu\nabla_\nu\Phi
        - \frac{1}{4}H_{\mu\kappa\sigma}H_{\nu}^{\ph{\nu}\kappa\sigma} \right) +O(\alpha^{\prime 2}),\label{intro:betaG}\\
0=\beta^B_{\mu\nu} &= \alpha'\left(-\frac{1}{2}\nabla^\kappa H_{\kappa\mu\nu} 
        + \nabla^\kappa\Phi H_{\kappa\mu\nu} \right)+O(\alpha^{\prime 2}),\\
0=\beta^\Phi_{\ph{\mu\nu}} &= \frac{D-26}{6} + \alpha'\bigg(-\frac{1}{2}\nabla^2\Phi   \cr
 \ph{ 0=\beta^\Phi_{\mu\nu} }& \ph{=}  \ph{D-26}
         + \nabla_\kappa\Phi\nabla^\kappa\Phi -\frac{1}{24}H_{\kappa\mu\nu}H^{\kappa\mu\nu}\bigg)
         +O(\alpha^{\prime 2}),\label{intro:betaPhi}
\end{align} 
where $\nabla_\mu$ is the covariant derivative containing the Levi-Civit\`{a} connection of the space-time metric $G_{\mu\nu}$. Further, $R_{\mu\nu}$ is the space-time Ricci tensor of the metric $G_{\mu\nu}$ and $H_{\mu\nu\kappa}=\partial_{[\mu}B_{\nu\kappa]}$ is the field strength of $B_{\mu\nu}$.

Recall that the vanishing of the trace of the conserved world-sheet energy-momentum tensor gave rise to the conformal symmetry. So, to examine under what conditions the conformal invariance is preserved in the quantum theory we need to consider the trace of the energy-momentum tensor. It may be evaluated to, \cf \eg \cite{Polchinski:1998rq},
\begin{align}
T^\alpha_{\ph{\alpha}\alpha} =& -\frac{1}{2\alpha'}\beta^G_{\mu\nu} \gamma^{\alpha\beta}\partial_\alpha X^\mu \partial_\beta X^\nu 
 -\frac{1}{2\alpha'}\beta^B_{\mu\nu}
     \epsilon^{\alpha\beta}\partial_\alpha X^\mu \partial_\beta X^\nu 
  -\frac{1}{2}\beta^\Phi R^{(2)}.
\end{align}
To have conformal invariance we require this trace to vanish. Hence,we find the same conditions of vanishing $\beta$-functionals as before.

Notice that when the space-time dimension is $26$ and the $B$-field and the dilaton are zero, the conditions \eqref{intro:betaG} - \eqref{intro:betaPhi}, to lowest order in $\alpha'$ reduce to Einstein's equations of gravity in empty space-time, \ie $R_{\mu\nu}=0$. This implies that ordinary general relativity naturally arises out of string theory.

The conditions \eqref{intro:betaG} - \eqref{intro:betaPhi} above can be viewed as equations of motion for the background fields. It turns out that it is possible to give an effective action that governs the dynamics of these fields. It is given by
\begin{align}\label{intro:effective_BG_action}
S&=\frac{1}{2\kappa^2_0}\int\d^D X \sqrt{-G} e^{-2\Phi}\bigg\{
                   R + 4\nabla_\mu\Phi\nabla^\mu\Phi \cr
&\ph{\frac{1}{2\kappa^2_0}\int\d^D X}
 -\frac{1}{12}H_{\mu\nu\lambda}H^{\mu\nu\lambda}
   - \frac{2(D-26)}{3\alpha'} + O(\alpha') \bigg\}.
\end{align}
Here $\kappa_0$ is a normalization constant that is not fixed and can be changed by a redefinition of the dilaton field. This action is the bosonic version of the supergravity action for the superstring. It is the ordinary Einstein-Hilbert action with certain matter fields present. It governs the dynamics of classical gravity interacting with the matter fields.

\section{Superstrings}\label{intro:superstrings}
In this section we will introduce fermions in the theory. We will find that this procedure removes the tachyon from the theory and restricts the number of space-time dimensions to ten. To achieve this we will make use of supersymmetry, a symmetry between bosonic and fermionic degrees of freedom. Studying this symmetry will enable us to write down a manifestly world-sheet supersymmetric action that contains the standard bosonic action \eqref{intro:confAction}.  This procedure introduces world-sheet fermions, which may seem a bit strange since what we are looking for is not world-sheet fermions but rather space-time fermions. We will present the Gliozzi-Scherk-Olive (GSO) projection that is needed in the Ramond-Neveu-Schwarz (RNS) superstring to truncate the spectrum in such a way that it realizes space-time supersymmetry. In this way we obtain space-time fermions.

\subsection{Supersymmetry}

The symmetries we observe, and are required to be present in any relativistic theory are the symmetries under space-time translations, generated by $P_\mu$, and Lorentz rotations, generated by $M_{\mu\nu}$. The generators satisfy the Poincar\'{e} algebra given by
\begin{align}
[P_\mu,P_\nu] =& 0,\label{intro:Poincare1}\\
[M_{\mu\nu},P_\rho] =& \frac{\i}{2}\eta_{\rho[\mu}P_{\nu]},\\
[M_{\mu\nu},M_{\rho\sigma}] =& \frac{\i}{2}\eta_{\rho[\mu}M_{\nu]\sigma} 
- \frac{\i}{2}\eta_{\sigma[\mu}M_{\nu]\rho}\label{intro:Poincare3},
\end{align}
where $A_{[\mu\nu]}$ denotes anti-symmetrization. The Coleman-Mandula no-go theorem \cite{Coleman:1967ad} tells us that the only way to extend the Poincar\'{e} symmetry of a local relativistic quantum field theory is to include an internal symmetry such that the total symmetry group is given by the direct product of the Poincar\'{e} group and the internal symmetry group. This means that the two types of symmetries are combined in a trivial way. 

However, the restrictions from the Coleman-Mandula no-go theorem can be circumvented introducing supersymmetry.  In \cite{Haag:1974qh} it is shown that the Poincar\'{e} algebra has non-trivial extensions if we introduce odd supersymmetry generators $Q_\alpha$.
For simplicity, here we consider only one supersymmetry generator, no central charges and no internal symmetry group. For this case, the above Poincar\'{e} algebra \eqref{intro:Poincare1}-\eqref{intro:Poincare3} is complemented by
\begin{align}
[P_\mu, Q_\alpha]  =& 0, \label{intro:susy_algebraPQ}\\
[M_{\mu\nu}, Q_\a] =& \frac{1}{8}([\Gamma_\mu, \Gamma_\nu])_\alpha^{\ph{\alpha}\beta} Q_\beta,\label{intro:susy_algebraMQ}\\
\{Q_\alpha, Q_\beta\} =& \Gamma^\mu_{\alpha\beta}P_\mu,\label{intro:QQP_def}
\end{align}
where $\Gamma^\mu$ satisfies the Clifford algebra $\{\Gamma^\mu, \Gamma^\nu\}= 2\eta^{\mu\nu}\mathbbm{1}$. The commutators \eqref{intro:susy_algebraPQ} and \eqref{intro:susy_algebraMQ} mean that the supersymmetry generator is invariant under space-time translations and transforms as a spinor under Lorentz-rotations. Further, a property that we will use frequently in later chapters is \eqref{intro:QQP_def}, that the supersymmetry generators anti-commute to a translation.

When considering the internal symmetry it is found that supersymmetry generators transform in a non-trivial representation of this group. For more details on the supersymmetry algebra with non-zero central charges and its relation to the internal symmetry, see \eg \cite{Gates:1983nr,Lindstrom:2002ph}.

\subsection{World-sheet supersymmetry}\label{sec:world-sheet-susy}
To introduce world-sheet fermions we extend the set of coordinates on the world-sheet $(\sigma,\tau)\rightarrow(\sigma,\tau,\theta^1,\theta^2)$. These extra coordinates are taken to be Grassmann odd variables, \ie anticommuting, so that they square to zero and obey $\{\theta^1,\theta^2\}=0$. We group them into a two-dimensional Majorana spinor
\begin{align}
\theta =\left(\begin{array}{c}\theta^1\\ \theta^2\end{array}\right). \label{intro:odd_coords_spinor}
\end{align}
This gives each bosonic coordinate on the world-sheet an anticommuting, or fermionic, partner, enlarging the world-sheet to a superspace. The fields living on this superspace are called superfields and will in general depend both on the original commuting coordinates and the extra anti-commuting coordinates, $\Phi^\mu=\Phi^\mu(\sigma^\alpha,\theta)$. Since the $\theta$'s are anti-commuting any Taylor expansion in these coordinates will terminate after a couple of terms. For example, the superfield that correspond to the position field has the expansion
\begin{align}
\Phi^\mu(\sigma^\alpha, \theta) = X^\mu(\sigma^\alpha)
                                 + \bar{\theta}\psi^\mu(\sigma^\alpha)
                                 +\smallhalf \bar{\theta}\theta F^\mu(\sigma^\alpha), \label{intro:superfield}
\end{align}
where $X^\mu$ is the position field as in the bosonic theory, $\psi^\mu$ is a two-component Majorana fermion and $F^\mu$ will turn out to be an auxiliary field. Further, a bar over any spinor $\psi$ denotes $\bar{\psi} = \psi^\dagger\rho^0$, where $\rho^0$ is one of the two-dimensional Dirac matrices $\rho^\alpha$ that satisfy the algebra $\{\rho^\alpha,\rho^\beta\} = -2\eta^{\alpha\beta}\mathbbm{1}$. In the following we will use the basis
\begin{align}
\rho^0 = \left(\begin{array}{cc} 0&-\i\\ \i&0\\ \end{array}\right),\;\;\;\;\;\;
\rho^1 = \left(\begin{array}{cc} 0& \i\\ \i&0\\ \end{array}\right), \label{intro:2dgamma}
\end{align}
for the two-dimensional Dirac-matrices.

The theory we want to formulate is to have a symmetry between the bosonic and the fermionic degrees of freedom, this is realized in terms of supersymmetry. We thus consider the supersymmetry transformations of the world-sheet coordinates
\begin{align}
\delta(\epsilon) \sigma^\alpha &= \i \bar{\epsilon}\rho^\alpha \theta, \label{intro:susy_trnsf_1}\\
\delta(\epsilon) \theta &= \epsilon, \label{intro:susy_trnsf_2}
\end{align}
where $\epsilon$ is an anticommuting two-component Majorana spinor. The generator of this transformation is given by
\begin{align}
Q=\frac{\partial}{\partial \bar{\theta}} + \i \rho^\alpha\theta\partial_\alpha,
\end{align}
so that $\delta(\epsilon) \sigma^\alpha = \bar{\epsilon}Q\sigma^\alpha$ and
$\delta(\epsilon) \theta=\bar{\epsilon}Q\theta$. 

Next, we investigate how the superfield \eqref{intro:superfield} transforms under the
supersymmetry transformation,
\begin{align}
\delta(\epsilon) \Phi^\mu =& \bar{\epsilon}Q\Phi^\mu \label{intro:superfield_transfn}\\
                =& \bar{\epsilon}\left(\frac{\partial}{\partial \bar{\theta}} 
                                        + \i \rho^\alpha\theta\partial_\alpha \right)
  \left( X^\mu + \bar{\theta}\psi^\mu +\smallhalf \bar{\theta}\theta F^\mu\right)\\
 \equiv & \delta(\epsilon) X^\mu + \bar{\theta}\delta(\epsilon) \psi^\mu +\smallhalf \bar{\theta}\theta \delta(\epsilon) F^\mu.
\end{align}
We use the two-dimensional Fierz relation $\theta_A\bar{\theta}_B = -\frac{1}{2}\delta_{AB}\bar{\theta}_C \theta_C$, where the capital letters denote the components of the spinors, to find the supersymmetry transformation for the component fields
\begin{align}
&\delta(\epsilon) X^\mu = \bar{\epsilon}\psi^\mu, \label{intro:fieldsusy_trnsf_1}\\
&\delta(\epsilon) \psi^\mu = -\i\rho^\alpha \epsilon\partial_\alpha X^\mu + \epsilon F^\mu, \label{intro:fieldsusy_tranf_Psi}\\
&\delta(\epsilon) F^\mu = -\i\bar{\epsilon}\rho^\alpha\partial_\alpha\psi^\mu.\label{intro:fieldsusy_trnsf_2}
\end{align}
The definition of a superfield is a field that transform as $\Phi^\mu$ in \eqref{intro:superfield_transfn}. Note that the $\bar{\theta}\theta$ component of $\bar{\epsilon}Q\Phi^\mu$ is proportional to a total $\sigma^\a$-derivative of the $\bar{\theta}$ component of $\Phi^\mu$, this is important for formulating an invariant action.

Further, using the identity for two-dimensional Majorana spinors $\bar{\epsilon}_1\rho^\alpha\epsilon_2 = - \bar{\epsilon}_2\rho^\alpha\epsilon_1$ it is straightforward to show that 
\begin{align}
[\bar{\epsilon}_1 Q,\bar{\epsilon}_2Q] = - 2\i \bar{\epsilon}_1\rho^\alpha\epsilon_2 \partial_\alpha. \label{intro:susy_comm_2_transl}
\end{align}
Since an infinitesimal translation on the world-sheet $\sigma^\alpha\rightarrow \sigma^\alpha + a^\alpha$ act on the coordinate fields as $\delta X^\mu = a^\alpha\partial_\alpha X^\mu$, the above relation \eqref{intro:susy_comm_2_transl} tells us that the commutator of two supersymmetry transformations is a translation on the world-sheet. Remember that $\epsilon_1$ and $\epsilon_2$ are Grassmann-odd and hence \eqref{intro:susy_comm_2_transl} is nothing but one of the defining properties of supersymmetry \eqref{intro:QQP_def}.

Next we look for a superspace covariant derivative that makes the derivative of a superfield transform in the same way as the superfield. The derivative is given by
\begin{align}
D=\frac{\partial}{\partial \bar{\theta}} - \i \rho^\alpha\theta\partial_\alpha.
\end{align}
If we note that $\{D,Q\}=0$, it is easy to verify the transformation $\delta(\epsilon)(D\Phi^\mu) = \bar{\epsilon}Q(D\Phi^\mu)$ which means that $D\Phi^\mu$ is a superfield.

Further, $\bar{\epsilon}Q$ is a derivation and obeys the Leibniz rule, it follows that a product of superfields is again a superfield. For example, consider the product of two superfields,
\begin{align}
\delta(\epsilon)(\Phi_1 \Phi_2) = (\bar{\epsilon}Q \Phi_1)\Phi_2 + \Phi_1(\bar{\epsilon}Q\Phi_2) 
= \bar{\epsilon}Q(\Phi_1\Phi_2). \label{intro:Leibnitz}
\end{align}

To write an action in terms of superfields we need to define integration over the Grassmann coordinates. This is called the Berezin integral and is defined as
$\int\d^2\theta (a + b \theta_1 + c\theta_2 + d\theta_1\theta_2) = d$. This implies that $\int\d^2\theta \;\bar{\theta}\theta = -2\i$. The integration picks out the coefficient of the $\bar{\theta}\theta$ component of the integrand. As noted above this coefficient of $\bar{\epsilon}Q\mathcal{L}$, where $\mathcal{L}$ is any superfield, is proportional to a total $\sigma^\a$-derivative. Hence, the supersymmetry transformation of the integral 
\begin{align}
\delta(\epsilon)\int\d^2\sigma\d^2\theta \mathcal{L} 
 =\int\d^2\sigma\d^2\theta \bar{\epsilon}Q \mathcal{L} = 0.
\end{align}
This means that an action whose Lagrangian is written as a product of superfields is manifestly invariant under the world-sheet supersymmetry transformation \eqref{intro:susy_trnsf_1} - \eqref{intro:susy_trnsf_2}. Here, we let the background be flat Minkowski space-time and write the string action as 
\begin{align}
S = \frac{\i}{8\pi\alpha'}\int\d^2\sigma\d^2\theta \bar{D}\Phi^\mu D\Phi_\mu. \label{intro:susy_action}
\end{align}
It is manifestly supersymmetric since it is written as an integral over a product of superfields.

We now have a manifestly supersymmetric action \eqref{intro:susy_action} in terms of the superfields \eqref{intro:superfield}. To make contact with the bosonic action that we studied in previous sections, we expand out the superfields and the covariant derivatives in components and integrate over the fermionic directions. The resulting action after this procedure is 
\begin{align}\label{intro:ws_susy_off}
S=-\frac{1}{4\pi\alpha'}\int\d^2\sigma \left(\partial_\alpha X^\mu \partial^\alpha X_\mu
-\i \bar{\psi}^\mu \rho^\alpha\partial_\alpha\psi_\mu - F^\mu F_\mu
\right).
\end{align}
Remember that $T=(2\pi\alpha')^{-1}$. Then we recognize the first term in the action as the gauge fixed version of the Polyakov action \eqref{intro:confAction}. An explicit check verifies that this action is invariant under \eqref{intro:fieldsusy_trnsf_1} - \eqref{intro:fieldsusy_trnsf_2}. The equation of motion for the $\psi^\mu$ field is the two-dimensional massless Dirac equation $\i\rho^\alpha\partial_\alpha\psi^\mu=0$ meaning that we have found the world-sheet fermions. The equations of motion for the field $F^\mu$ sets it to zero, it is an auxiliary field. However, if $F^\mu$ is to be completely removed from the theory the transformation $\delta(\epsilon) F^\mu$ in \eqref{intro:fieldsusy_trnsf_2} must also be zero. This condition is nothing but the equation of motion for $\psi^\mu$ field, implying that we need to use the equations of motion, \ie ``go on shell'', to completely remove the auxiliary $F^\mu$ field. As expected, the supersymmetry transformation, \eqref{intro:fieldsusy_trnsf_1} and \eqref{intro:fieldsusy_tranf_Psi} with $F^\mu=0$, close to world-sheet translations up to equations of motion.

\subsection{Quantization}
As for the bosonic string, quantization of the superstring can be performed in 
several ways. Here, we will only discuss the covariant quantization scheme.
To do this we start by writing down the equations of motion for the fields in the 
theory. From the action \eqref{intro:ws_susy_off} we find that the equation of motion and the
boundary conditions for the $X^\mu$-field are the same as for the bosonic string. The $F$-field equation of motion makes the field vanish. The only new thing, compared to the bosonic string, is the world-sheet fermions, whose equations of motion are given by
\begin{align}
\i\rho^\alpha\partial_\alpha\psi^\mu = 0.
\end{align}
To study this equation we define the components of the Majorana fermion as
\begin{align}
\psi^\mu   = \left(\begin{array}{c} \psi^\mu_{-}\\ \psi^\mu_{+}\\ \end{array} \right). \label{intro:Psi_components}
\end{align}
We use the basis \eqref{intro:2dgamma} for the Dirac-matrices and introduce the world-sheet light-cone coordinates, $\sigma_\ppmm=\tau \pm\sigma$, that imply $\partial_\ppmm=\frac{1}{2}(\partial_\tau \pm \partial_\sigma)$. The equations of motion for the fermion components now become
\begin{align}
\partial_\pp \psi^\mu_-=0, \;\;\;\;\;\;
\partial_\mm \psi^\mu_+=0. \label{intro:eoms_fermion}
\end{align}
These are, for the open string, accompanied by boundary conditions arising from the requirement that the boundary contributions in the variation of the action vanish. For each $\mu = 0,...,D-1$, this requirement reads
\begin{align}
\psi_{\mu+} \delta \psi^\mu_+ - \psi_{\mu-} \delta \psi^\mu_- = 0 \;\;\;\;
\mbox{at}\;\;\;\;
\sigma = 0,\pi.
\end{align}
There are two possibilities to satisfy these open string boundary conditions
\begin{align}
(\mbox{R})\;\; &\psi^\mu_+(0,\tau) = \psi^\mu_-(0,\tau),\;\;\;\;\;\;\;\,
      \psi^\mu_+(\pi,\tau) = \psi^\mu_-(\pi,\tau),  \\
(\mbox{NS})\;\;&\psi^\mu_+(0,\tau) = \psi^\mu_-(0,\tau),\;\;\;\;\;\;\;\,
      \psi^\mu_+(\pi,\tau) = -\psi^\mu_-(\pi,\tau).
\end{align}
The string sector that arise from the first set of boundary condition is called
the Ramond (R) sector and the one arising from the second set is called Neveu-Schwarz (NS) sector.
The mode expansion of the field with Ramond boundary conditions is
\begin{align}
(\mbox{R})\;\;&\psi^\mu_\pm=\frac{1}{\sqrt{2}}\sum_{n\in\mathbb{Z}} d^\mu_n\e^{-\i n\sigma_\ppmm}.
\end{align}
For Neveu-Schwarz boundary conditions it is given by
\begin{align}
(\mbox{NS})\;\;&\psi^\mu_\pm=\frac{1}{\sqrt{2}}\sum_{r\in\mathbb{Z}+\frac{1}{2}} b^\mu_r\e^{-\i r\sigma_\ppmm}.
\end{align}

For the closed string different cases arise depending on the boundary condition. The closed string is essentially two copies of the oscillators of an open string that are unrelated, except for the level matching condition discussed later. So, the possible boundary conditions that apply for each sector is periodicity (R), $\psi^\mu_\pm(\sigma+\pi,\tau)=\psi^\mu_\pm(\sigma,\tau)$ or antiperiodicity (NS), $\psi^\mu_\pm(\sigma+\pi,\tau)=-\psi^\mu_\pm(\sigma,\tau)$. The mode
expansions for the right moving fields are
\begin{align}
(\mbox{R}) \;\; \psi^\mu_- = \sum_{n\in \mathbb{Z}} d^\mu_n\e^{-2\i n\sigma_\mm} 
\;\;\;\;\mbox{or}\;\;\;\; 
(\mbox{NS}) \;\; \psi^\mu_- = \sum_{r\in \mathbb{Z}+\frac{1}{2}} b^\mu_r\e^{-2\i r \sigma_\mm},
\end{align}
and for the left moving fields
\begin{align}
(\mbox{R}) \;\;\psi^\mu_+ = \sum_{n\in \mathbb{Z}} \tilde{d}^\mu_n\e^{-2\i n\sigma_\pp} 
\;\;\;\;\mbox{or}\;\;\;\; 
(\mbox{NS})\;\;\psi^\mu_+ = \sum_{r\in \mathbb{Z}+\frac{1}{2}} 
                              \tilde{b}^\mu_r\e^{-2\i r \sigma_\pp}.
\end{align}
We find that there are four combinations possible; NS-NS, NS-R, R-NS and R-R. 

Next, we quantize the theory by imposing commutation relations between the bosonic fields and anti-commutation relations for the fermionic fields. For the bosonic fields we have the same situation as when quantizing the bosonic string, \ie \eqref{intro:comm_bos_modes}.  For the world-sheet fermions, the result of this procedure is the anti-commutators
\begin{align}
\{b^\mu_r ,b^\nu_s\} = \eta^{\mu\nu}\delta_{r+s},\;\;\;\;\;
\{d^\mu_m ,d^\nu_n\} = \eta^{\mu\nu}\delta_{m+n}. \label{intro:ferm_comm_rel}
\end{align}
For the closed string there is a second set of anticommutators for the modes $\tilde{b}^\mu_r$ and $\tilde{d}^\mu_n$. This implies that we have extended our previous set of creation operators and we can use the modes $d_n$ with $n<0$ in the R sector and $b_r$ with $r<0$ in the NS sector to build the Fock space. As for the bosonic string the Fock space must be restricted to remove unphysical states.

In the bosonic case the Virasoro algebra arises from the energy-momentum tensor. In the present case we have an additional symmetry, the supersymmetry. It gives rise to a current on the world-sheet. If we make this supersymmetry local, \ie dependent on the position on the world-sheet, construct an action $S_{inv}$ that is invariant under this local transformation and find the equations of motion for the additional fields that were needed to make the action invariant, we find the so called super-Virasoro constraints, here written in world-sheet light-cone coordinates,
\begin{align}
J_\pm &\equiv \psi^\mu_\pm \partial_\ppmm X_\mu = 0, \label{intro:susy_J=0}\\
T_{\pm\pm} &\equiv \partial_\ppmm X^\mu\partial_\ppmm X_\mu
               + \frac{\i}{2}\psi^\mu_\pm \partial_\ppmm \psi_{\pm\mu} = 0.
        \label{intro:susy_T=0}
\end{align}
The components $T_{\pm\mp}=0$ vanishes identically. These constraints are needed for the action \eqref{intro:ws_susy_off} to be equivalent to the gauge invariant action $S_{inv}$, in the same way as in the bosonic case where \eqref{intro:eqn_T=0} was required for \eqref{intro:confAction} to be equivalent to the Polyakov action \eqref{intro:Polyakov}.

Notice that the energy-momentum tensor now contains a contribution from the world-sheet fermions. The $J$ can be thought of as the fermionic partner to the energy-momentum tensor. The above conditions are needed to remove the unphysical degrees of freedom in the theory. The modes of $J$ and $T$ form an algebra, the super-Virasoro or super-conformal algebra. Exactly as previously the zero modes suffer from a normal ordering ambiguity, and when we require the algebra, including Faddeev-Popov ghosts, to be anomaly free in the quantized theory, this fixes the normal ordering parameter and determines the dimension of space-time to be ten.

The modes of the two energy-momentum tensors are commonly denoted $L_n$ for the modes of $T$ and $G_r$ for the modes of $J$. Here $n$ is an integer and $r\in\mathbb{Z}$ in the R sector or $r\in\mathbb{Z}+\frac{1}{2}$ in the NS sector. To satisfy the conditions \eqref{intro:susy_J=0} and \eqref{intro:susy_T=0} in the quantized theory we impose the physical state conditions
\begin{align}\label{intro:susy_phys_cond}
G_r|phys\rangle=0, \; r>0;   \;\;\;\;\;\;\; 
L_n|phys\rangle=0, \; n\geq0. 
\end{align}
Level number operators are defined as
\begin{alignat}{2}
&(\mbox{NS}) &\quad\quad\quad 
N_{NS} &= \sum_{n=1}^\infty \alpha^\mu_{-n} \alpha_{n\mu} 
              + \sum_{r=\frac{1}{2}}^\infty r b^\mu_{-r}b_{r\mu},\\
&(\mbox{R}) &
N_{R} &=\sum_{n=1}^\infty \left(\alpha^\mu_{-n} \alpha_{n\mu} 
              + n\, d^\mu_{-n}d_{n\mu}\right),
\end{alignat}
where the $\alpha_n^\mu$'s are the standard modes of the bosonic field $X^\mu$.

The constraint from $L_0$ give us a mass formula and for the open string
NS sector as
\begin{align}\label{intro:NS_mass}
M^2 = \frac{1}{\alpha'}\left(N_{NS} -\frac{1}{2}\right),
\end{align}
and for the R sector
\begin{align}\label{intro:R_mass}
M^2 = \frac{1}{\alpha'} N_{R}.
\end{align}
For the closed string there is, as in the bosonic theory, a level matching condition that arises from $(L_0-\tilde{L}_0)|phys\rangle=0$. Here we get one number operator for the right moving sector and one for the left moving sector. Using these operators the level matching conditions for the different sectors read
\begin{alignat}{2}
&(\mbox{NS-NS}) & \quad\quad\quad N_{NS} &= \tilde{N}_{NS},\\
&(\mbox{NS-R})  & N_{NS} - \smallhalf &= \tilde{N}_{R},\\
&(\mbox{R-NS})  & N_{R} &= \tilde{N}_{NS} - \smallhalf,\\
&(\mbox{R-R})   & N_{R} &= \tilde{N}_{R}.
\end{alignat}
As for the bosonic string these are the only relations between the right and left moving 
sectors.

\subsection{Space-time supersymmetry}
As mentioned previously, we are not actually looking for two dimensional 
world-sheet fermions but rather ten-dimensional space-time ones. The field $\psi^\mu$ transforms as a space-time vector, \ie as a boson. So what we found in the previous section is really a ten-dimensional boson whose components behave like world-sheet fermions. This may seem a bit awkward. However, it turns out that the spectrum that arises from this theory can consistently be truncated so that each mass level fills out an irreducible representation of the ten-dimensional supersymmetry algebra \cite{Gliozzi:1976qd}.

\subsubsection{GSO projection}
We find from \eqref{intro:NS_mass} that the ground state in the NS sector still is a tachyon. However, one of the motivations for introducing fermions was that we wanted to find a way to remove the bosonic tachyon, and now we find that there is still a tachyon in the spectrum. What to do? We can focus on the R sector and note that it does not contain a tachyon. So at least we have one sector that seems to make sense. However, since we have not found the required space-time fermions, it means that we have to do better than to just throw away the NS sector.

For this we need to study the ground states of the two sectors of the open string. The NS sector ground state, denoted $|NS\rangle$, is non-degenerate and, as mentioned, a tachyonic state.

For the R sector the situation is a little different. Here we have the mode operators $d^\mu_0$. These do not change the mass of the state on which they act, hence, the R sector ground state will be degenerate. The anti-commutation relations \eqref{intro:ferm_comm_rel} for the $d^\mu_0$'s become $\{d^\mu_0,d^\nu_0\}=\eta^{\mu\nu}$. The ten mode operators can be split into five creation and five annihilation operators by defining
\begin{align}
d^\pm_i &= \frac{1}{\sqrt{2}}\left(d^{2i}_0 \pm \i d^{2i+1}_0\right), \;\;\; i=1,...,4\\
d^\pm_0 &= \frac{1}{\sqrt{2}}\left(d^1_0 \mp d^0_0\right),
\end{align}
which satisfy $\{d^{+}_i,d^-_j\}=\delta_{ij}$. Assuming a unique vacuum $|0\rangle$, we find that the ground state is $2^5=32$ fold degenerate. This means that the R sector ground state has the same number of independent components as a SO(9,1) Majorana fermion in ten dimensions, namely 32 real components. The physical state conditions \eqref{intro:susy_phys_cond} reduce the independent components to 16. Further the ground state can be split into a state $|R_1\rangle$ with positive chirality and which has an even number of creation operators $d^+_i$ acting on $|0\rangle$ and a state $|R_2\rangle$ with negative chirality and which has an odd number of creation operators $d^+_i$ acting on $|0\rangle$. These two states have 8 independent components each.

The full ground state of the superstring is then given by the tensor product of the bosonic ground state $|0,k\rangle$ with the ground state of the fermionic sector under consideration. The spectrum of the superstring is built on this state by acting with the creation operators $\alpha_{-n}^\mu$, $d^\mu_{-n}$ or $b^\mu_{-r}$, where $n>0$ and $r>0$. Further, we must impose the the physical state conditions \eqref{intro:susy_phys_cond}.

Moreover, to get rid of the unwanted tachyon we impose the so called GSO projection \cite{Gliozzi:1976qd}. For the NS sector it instructs us to keep only states with an odd number of creation operators $b^\mu_{-r}$ acting on the ground state $|NS\rangle$. Hence the tachyon is projected out of the spectrum. For the R sector we can choose which of the two states $|R_1\rangle$ or $|R_2\rangle$ we should consider to be the ground state. If we choose $|R_1\rangle$ as the ground state the GSO projection instructs us to keep states built on $|R_1\rangle$ with an even number of creation operators $d^\mu_{-n}$ and states built on $|R_2\rangle$ with an odd number of creation operators $d^\mu_{-n}$. A physically equivalent situation is obtained if we instead choose $|R_2\rangle$ as the ground state, then we are instructed to keep states built on $|R_2\rangle$ with an even number of creation operators and states built on $|R_1\rangle$ with an odd number of creation operators.

The state of lowest mass that survives the GSO projection in the NS sector of the open string is the state $b^\mu_{-1/2}|NS\rangle$. It corresponds to a massless vector field, this is the open string photon state. This ten-dimensional massless particle state transforms under Lorentz transformations in the eight-dimensional vector representation $\mathbf{8}_v$ of $SO(8)$.

In the R sector the GSO projection selects the state $|R_1\rangle$ as the lowest mass state. This zero mass state transforms in the eight-dimensional spinor representation $\mathbf{8}_s$ of $SO(8)$. If on the other hand we would have chosen $|R_2\rangle$ as the ground state we would have obtained the eight-dimensional conjugate spinor representation $\mathbf{8}_c$ of $SO(8)$.

The complete ground state spectrum of the open superstring is given by the direct sum of the massless states of the two sectors. With the choice $|R_1\rangle$ as the ground state it thus transforms as  $\mathbf{8}_v \oplus \mathbf{8}_s$ which is nothing but a massless vector multiplet of ten-dimensional $\mathcal{N}=1$ space-time supersymmetry. Hence, we have found that at the massless level the spectrum exactly fills out a supersymmetry multiplet. This is a first indication that the GSO projection apart from removing the tachyon from the spectrum realizes space-time supersymmetry. This open string theory is not by itself complete in that when we include interactions the two ends of the string can join together to form a closed string. However, we here use it as a building block for the closed string theory.

The closed string is built by a tensor product of the left- and right-moving sectors with the level matching condition taken into account. Meaning that we find the spectrum by taking a tensor product of two open strings. However we can now choose the same or different ground states for the Ramond sector for the left- and right-moving sector. If we choose different ground states for the left and right moving sector we obtain type IIA string theory, and if we choose the same ground state in both the left- and right-moving sector we obtain type IIB string theory. At the massless level the states in the respective theory transform under $SO(8)$ as:
\begin{align}
\mbox{Type IIA:} \;\;\;\;\ &(\mathbf{8}_v\oplus\mathbf{8}_s)\otimes(\mathbf{8}_v\oplus\mathbf{8}_c)\\
\mbox{Type IIB:} \;\;\;\;\ &(\mathbf{8}_v\oplus\mathbf{8}_s)\otimes(\mathbf{8}_v\oplus\mathbf{8}_s)
\end{align}
Here we have considered that the Ramond sector ground state in the type IIA theory is given by $|R_1\rangle$ in the left moving sector and by $|R_2\rangle$ in the right moving sector. For the type IIB theory we have used $|R_1\rangle$ as the Ramond sector ground state for both the left and right moving sectors.

For both type IIA and type IIB string theory the massless state in the NS-NS sector is given by $b^\mu_{-1/2}|NS\rangle\otimes\tilde{b}^\nu_{-1/2}|NS\rangle$. This state decomposes into irreducible representations of $SO(8)$ as
\begin{align}
\mathbf{8}_v\otimes\mathbf{8}_v = \mathbf{1} \oplus \mathbf{28} \oplus \mathbf{35}.
\end{align}
This implies that the state separates into the dilaton $\Phi$, the antisymmetric field $B_{\mu\nu}$ and the graviton $G_{\mu\nu}$. 

In type IIA string theory, the massless state in the NS-R sector is given by $b^\mu_{-1/2}|NS\rangle\otimes|R_2\rangle$ and in the R-NS sector by $|R_1\rangle\otimes\tilde{b}^\mu_{-1/2}|NS\rangle$. These two states decompose under $SO(8)$ as 
\begin{align}
\mathbf{8}_v\otimes\mathbf{8}_c =& \mathbf{8}_s \oplus \mathbf{56}_c\\
\mathbf{8}_s\otimes\mathbf{8}_v =& \mathbf{8}_c \oplus \mathbf{56}_v \label{intro:8sx8v}
\end{align}
which are a spinors and gravitinos. The gravitino is the superpartner of the graviton. For type IIB the massless states in the NS-R and R-NS sectors are, with our choice of Ramond ground state, given by two copies of \eqref{intro:8sx8v}.

In the R-R sector the massless state decomposes under $SO(8)$ as 
\begin{align}
\mbox{Type IIA:} \;\;\;\;\ &\mathbf{8}_s\otimes\mathbf{8}_c = \mathbf{8}_s\oplus \mathbf{56}_t\\
\mbox{Type IIB:} \;\;\;\;\ &\mathbf{8}_s\otimes\mathbf{8}_s = \mathbf{1}\oplus \mathbf{28}\oplus \mathbf{35}
\end{align}
which are two different consistent sets of antisymmetric tensor fields. Type IIA contains the R-R fields $C^{(1)}_\mu$ and $C^{(3)}_{\mu\nu\gamma}$ and type IIB contains the R-R fields $C^{(0)}$, $C^{(2)}_{\mu\nu}$ and $C^{(4)}_{\mu\nu\gamma\kappa}$. Further, the Hodge duals of these antisymmetric fields can be constructed which gives that 
the type IIA theory contains antisymmetric $C^{(p)}$ fields with $p$ odd, and type IIB contains fields with $p$ even. It turns out that all these antisymmetric fields couple to D$p$-branes via a coupling of the form
\begin{align}
\int_{M_{p+1}}C^{(p+1)},
\end{align}
where $M_{p+1}$ is the $(p+1)$-dimensional world-volume of the D$p$-brane. Hence, the D$p$-branes that couple to the IIA string are those with $p$ even and the ones that couple to the IIB string are those with $p$ odd.

The massless states found above in type IIA and type IIB string theory are the on-shell multiplets of ten-dimensional IIA and IIB supergravity respectively. In this sense type IIA and type IIB supergravity is the low energy limit of the respective string theory.
Further, it means that at the massless level ten-dimensional supersymmetry is realized.

It has been shown that the GSO projection realizes space-time supersymmetry at every mass level of the RNS superstring spectrum \cite{Gliozzi:1976qd}.

\subsection{Five different but equal theories}
Two different string theories were briefly mentioned in the previous section, type IIA and type IIB. It turns out that the above construction and the requirements give three different string theories. Different in the sense that they have different physical spectrum. The three theories that arise are; type I that contains unoriented open and closed strings and have ten-dimensional $\mathcal{N}=1$ supersymmetry, since the left and right moving sectors are related, type IIA and type IIB that only contain closed strings and have ten-dimensional $\mathcal{N}=2$ supersymmetry. The type IIA theory contains only non-chiral space-time fields and type IIB contains only chiral space-time fields.

There are two further possibilities to construct a consistent string theory, the heterotic strings \cite{Gross:1985fr}.  These are closed string hybrids, the left moving sector is taken to be the same as for a bosonic closed string while the right moving sector is chosen from a closed superstring, \ie from a type II theory. The right moving sector gives the space-time supersymmetry while the left moving sector is interpreted as 10 ordinary bosonic coordinate fields $X^\mu$ and 32 world-sheet Majorana fermions. This combination of the bosonic string and the superstring is only consistent if the theory has $E_8\times E_8$ or $SO(32)$ space-time gauge symmetry. These two theories are called Heterotic $E_8\times E_8$ and Heterotic $SO(32)$.

This means that it has emerged no less than five different consistent string theories out of the urge to create a unique theory of everything. This was the situation until Edward Witten in 1995 pointed out that all the five theories and 11 dimensional supergravity are linked together by a web of dualities \cite{Witten:1995zh, Witten:1995ex}. Duality is a map between two different string theories that tells us that the two theories describe merely two sides of the same physics. The central idea that sprung from these facts is that these theories might just be different limits of a underlying, more general, theory called M-theory, \cf figure \ref{I:dualities}. In chapter \ref{T-dual:chapter} we will study one of these dualities, T-duality, in some detail.

\subsection{Green-Schwarz superstrings}\label{intro:Green-Schwarz}
Another way to introduce space-time fermions is to construct a manifestly space-time supersymmetric action. This formulation of the superstring was introduced by Green and Schwarz in 1984 \cite{Green:1983wt}, and these strings are therefore called Green-Schwarz superstrings. 

The advantage of this method is that we, already from the start, have space-time fermions. There is no need for a GSO like projection. However there is one major drawback, no-one knows how to quantize the action in a covariant manner.  What can be done is to go to light-cone gauge where manifest Lorentz invariance is lost, but the theory is possible to quantize. 

The manifestly supersymmetric theory in $D$ space-time dimensions is most conveniently formulated on $D$-dimensional superspace. The coordinates of this space are the standard bosonic coordinate fields $X^\mu = X^\mu(\sigma^\alpha)$ and $\mathcal{N}$ independent $SO(9,1)$ spinor coordinate fields $\theta^A=\theta^A(\sigma^\alpha)$ with $A=1..\mathcal{N}$. This superspace is used to formulate a theory with $\mathcal{N}$ manifest supersymmetries. However, if the theory should describe the correct number of propagating degrees of freedom in ten dimensions, it turns out that $\mathcal{N}\leq 2$, and the $\theta^A$ must be Majorana-Weyl spinors. Otherwise it is not possible to write down an action that has an extra local fermionic symmetry, known as the $\kappa$-symmetry, which ensures the correct number of propagating degrees of freedom. The $\kappa$-symmetric action that describes the superstring in ten dimensions is called the Green-Schwarz action \cite{Green:1983wt}. However, we will not go into detail of this model until chapter \ref{Tnollquant} where we study the Green-Schwarz string in a specific non-flat background, the pp-wave. For the flat background case we refer to the book \cite{Green:1987sp}. 

Let us here just count the number of degrees of freedom in the model. In even dimensions a Dirac spinor has $2^{D/2}$ independent complex components. In ten dimensions the spinor $\theta^A$ has 32 complex components. The Majorana condition reduce the number of independent components to 32 real. The chirality condition, \ie the Weyl condition, is imposed to further reduce the number of degrees of freedom to 16 real. For the case $\mathcal{N}=2$ there are two possible choices when imposing the chirality condition. Either we choose the two $\theta^1$ and $\theta^2$ to have the opposite chirality or the same chirality. Choosing opposite chirality for the spinors lead us to the type IIA string and choosing the same chirality to the type IIB string.

Further, the $\kappa$-symmetry can be gauge fixed to remove the unphysical degrees of freedom. The gauge fixing removes one half of the remaining degrees of freedom of the spinors. Thus, each $\theta^A$ has eight real physical degrees of freedom. Comparing to the bosonic $X^\mu$ coordinates, which in the light-cone gauge has eight left moving and eight right moving physical degrees of freedom, we find that for $\mathcal{N}=2$ in ten dimensions the number of fermionic degrees of freedom matches the number of bosonic degrees of freedom. 

        \chapter{The tensionless limit of string theory}

In this chapter we will study some aspects of what happens to string theory when the string tension goes to zero. This tensionless limit was first studied in \cite{Schild:1976vq} and later a renewed interest in the subject aroused from \cite{Karlhede:1986wb}.

One motivation to study this limit is that the tensionless string is the string theory analogue of the massless point particle. Since the massless point particle is the high energy limit of the point particle the analogue suggests that we can consider the tensionless limit as the high energy limit, or equivalently the short distance limit, of string theory. One might further expect that the limit should expose high energy symmetries that at lower energies are broken and give the properties of the tensile string. 
Unbroken symmetries of string theory at high energies was first discussed in \cite{Gross:1988ue}. The appearance of new symmetries in the tensionless limit, taken in a flat background, has been shown in \cite{Karlhede:1986wb,Isberg:1992ia,Isberg:1993av, Gustafsson:1994kr}. It has also been shown that massless higher spin fields appear in the tensionless limit of the string both in a flat and in an $AdS$-background \cite{Sundborg:2000wp,Lindstrom:2003mg}. This reflects the fact that the symmetry is enhanced in the limit.

Another motivation to study the tensionless limit is the behavior of tensile strings close to space-time singularities. The string length scales as $\alpha^{\prime 1/2}\propto T^{-1/2}$ and by denoting the effective radius of curvature of the space-time by $R_c$, we find that the dimensionless combination $R_c T^{1/2}$, that compares the radius of curvature to the string length, goes to zero as we approach a space-time singularity, since $R_c$ is zero at the singularity. Note that we find the same situation by letting the string tension go to zero while keeping the radius of curvature fixed. Thus, the tensile string close to a space-time singularity effectively behaves as a tensionless string \cite{deVega:1994hu}, and hence, to understand this situation we need to study the tensionless limit of the string.

In this chapter we begin with a brief discussion of the massless limit of the point particle. After this, we turn to the string case and review the results of article [I] to present a supergravity background generated by a tensionless string source. For an extensive introduction to the tensionless limit of string theory see \eg \cite{Isberg:1993av}.

\section{Massive and massless relativistic particles}
It is well known that the relativistic point particle dynamics is described by extremizing the particle's path length in space-time. In flat Minkowski space-time the action to be extremized follows from \eqref{eqn:particle_action} and is given by 
\begin{eqnarray}\label{tl:startS} 
S = -m \int \d\tau \sqrt{-\dot{x}^2}
\end{eqnarray}
where $m$ is the mass and $x^\mu=x^\mu(\tau)$ is the position of the particle, parametrized by the particle's eigentime. Hence $\dot{x}^2=\frac{\d x^\mu}{\d\tau}\frac{\d x_\mu}{\d\tau}$. On one hand this action is useful for describing massive particles, for which $m\neq 0$. On the other hand for massless particles this action vanishes and does not tell us anything. The solution is to rewrite this action in a form that is equivalent when $m\neq 0$ but which also has a sensible limit $m\rightarrow 0$.

To find this equivalent action we start by calculating the conjugate momenta to the position. It is found to be, 
\begin{eqnarray}\label{tl:conjmom}
p_\mu = \frac{m \dot{x}_\mu}{\sqrt{-\dot{x}^2}}.
\end{eqnarray} 
Note that the Hamiltonian for the system vanishes identically, \ie $H = p_\mu\dot{x}^\mu - L = 0$. However, from \eqref{tl:conjmom} we find that there is a constraint in the system, $p^\mu p_\mu + m^2=0$. We introduce this constraint in the Hamiltonian via a Lagrange multiplier $e/2$. Using this Hamiltonian to define the phase space Lagrangian we find 
\begin{eqnarray}
L = p_\mu \dot{x}^\mu - \frac{e}{2}\left(p^\mu p_\mu + m^2\right).
\end{eqnarray}
The equation of motion for the momenta is $p_\mu= e^{-1} \dot{x}_\mu$. Using this equation to eliminate $p_\mu$ from the phase space Lagrangian leaves us with a more standard Lagrangian independent of the conjugate momentum. Writing the action for the system as $S=\int L$ gives us 
\begin{eqnarray}\label{tl:reparinvS}
S= \frac{1}{2}\int\d\tau \left(\frac{\dot{x}^2}{e} - e m^2\right).
\end{eqnarray}
Note that in this action it is possible to set $m=0$ and still have something interesting left. This means that the action \eqref{tl:reparinvS} is useful for describing massless particles. Note also that if we integrate out the Lagrange multiplier $e$ we recover the original action \eqref{tl:startS} that we started with.

The action \eqref{tl:reparinvS} is invariant under local reparametrization transformations
\begin{eqnarray}
\delta x^\mu &=& \xi \dot{x}^\mu,\\
\delta e &=& \frac{\d}{\d\tau} (\xi e).
\end{eqnarray}

\section{The tensionless limit of string theory}
Comparing the point particle action \eqref{tl:startS} and the Nambu-Goto action \eqref{intro:Nambu-Goto} we find that the tension $T$ of the string plays the same r\^{o}le as the mass of the point particle. In particular, the tensionless string $T=0$ can not be described using this action. This means that we could try to follow the same line of reasoning as in the previous section to obtain an action for the tensionless string. This approach turns out to yield a sensible action.

To start, we consider the Nambu-Goto action \eqref{intro:Nambu-Goto} with the metric being the flat Minkowski metric,
\begin{eqnarray}\label{tl:Nambu-Goto}
S_0=-T\int\d^2\sigma\sqrt{-\det\left( \eta_{\mu\nu}\partial_\alpha X^\mu\partial_\beta X^\nu \right)}.
\end{eqnarray}
Note that $\det\left( \eta_{\mu\nu}\partial_\alpha X^\mu\partial_\beta X^\nu \right) = \dot{X}^2 X^{\prime 2} - (\dot{X}^\mu X'_\mu)^2$. Further we find the conjugate momenta to the $X^\mu$ field to be,
\begin{eqnarray}
P_\mu = T \frac{\dot{X}_\mu X^{\prime 2}- X'_{\mu}(\dot{X}^\nu X'_\nu)}
                    {\left(-\dot{X}^2 X^{\prime 2} + (\dot{X}^\nu X'_\nu)^2\right)^{1/2}}.
\end{eqnarray}
Just as for the point particle the Hamiltonian constructed by $H=P_\mu \dot{X}^\mu-L$ vanishes.  We thus want to find constraints in the theory so that we can define a more general Hamiltonian with Lagrangian multipliers. It turns out that for the string there are two constraints, $P_\mu X^{\prime\mu}=0$ and
$P^2 + T^2X^{\prime 2}=0$, and hence we introduce the two Lagrange multipliers $\lambda$ and $\rho$ to write the phase space Lagrangian as $L=-\lambda(P_\mu X^{\prime\mu}) - \frac{\rho}{2}(P^2 + T^2X^{\prime 2})$. The equation of motion for $P_\mu$ now yields $P_\mu = \rho^{-1}\dot{X}_\mu -\lambda\rho^{-1}X'_\mu$, using this to eliminate $P_\mu$ give us the action
\begin{eqnarray}
S=\int\d^2\sigma\left(\frac{1}{2\rho}\dot{X}^2 - \frac{\lambda}{\rho} \dot{X}^\mu X'_\mu +
        \frac{\lambda^2-T^2\rho^2}{2\rho}X^{\prime 2}\right).\label{tl:T=0_possible_action}
\end{eqnarray}
Integrating out the Lagrange multipliers while assuming that $T\neq0$ give us back the Nambu-Goto action. In the form \eqref{tl:T=0_possible_action} the action does not vanish if we take the limit $T\rightarrow 0$, so it is appropriate to use it to describe the tensionless string. Taking the limit and defining a world sheet vector
density $V^\alpha$, with components $V^0=\rho^{-1/2}$ and $V^1=-\lambda\rho^{-1/2}$, we find that the action can be written as
\begin{eqnarray}\label{tl:tensionless_action}
S = \frac{1}{2}\int\d^2\sigma V^\alpha V^\beta \partial_\alpha X^\mu\partial_\beta X^\nu \eta_{\mu\nu}.
\end{eqnarray}
This action is the starting point for discussing the bosonic tensionless string. The quantization of the action and the implications is discussed in \cite{Isberg:1992ia, Isberg:1993av}. In the next chapter we will see how \eqref{tl:tensionless_action} may be generalized to incorporate fermions and to describe a tensionless string on a pp-wave background.

The effect the presence of a string has on space-time is described by the space-time energy-momentum tensor. In \cite{Gurses:1974cm} this tensor for a string moving in $D$ dimensional space-time was found to be 
\begin{eqnarray}\label{tl:tensile_space-time_energy_mom}
T_{\mu\nu}(x^\mu) = \int \d^2\sigma \, T \sqrt{h}h^{\alpha\beta} 
         \partial_\alpha X_\mu\partial_\beta X_\nu \,\delta^{D}(x^\mu -X^\mu(\sigma,\tau)).
\end{eqnarray}
Here, $X^\mu(\sigma,\tau)$ is the position of the string world sheet in space-time and $x^\mu$ is the position in space time at which the energy momentum tensor is evaluated.  The delta-function in this relation reflects the fact that the energy density in space-time that arise from the presence of the
string only is nonzero at the position of the string world-sheet. For the Polyakov action \eqref{intro:Polyakov}, with $\eta_{\mu\nu}$ replaced by a general metric $g_{\mu\nu}$, we note that \eqref{tl:tensile_space-time_energy_mom} follows from
\begin{eqnarray}
T_{\mu\nu}(x^\mu) = \int \d^2\sigma \,\frac{\delta S}{\delta g^{\mu\nu}}
\,\delta^{D}(x^\mu -X^\mu(\sigma,\tau)).\label{tl:def_Energy_momentum_tensor}
\end{eqnarray}
To find the corresponding space-time energy-momentum tensor for the tensionless string described by the action \eqref{tl:tensionless_action}, we replace the flat Minkowski metric in \eqref{tl:tensionless_action} by a general space-time metric and perform the variation with respect to it. Via \eqref{tl:def_Energy_momentum_tensor} we find
\begin{eqnarray}\label{tl:tensionless_EM_tensor_0}
T_{\mu\nu}(x^\mu) = \int \d^2\sigma \, V^\alpha V^\beta 
         \partial_\alpha X_\mu\partial_\beta X_\nu \, \delta^{D}(x^\mu -X^\mu(\sigma,\tau)),
\end{eqnarray}
where we have absorbed one minus sign in the $V^\alpha$'s. To simplify
\eqref{tl:tensionless_EM_tensor_0} we use reparametrizations and
diffeomorphisms to chose the transverse gauge \cite{Isberg:1993av}, $V^\alpha=(1,0)$. In this gauge the space-time energy-momentum tensor for the tensionless string becomes
\begin{eqnarray}\label{tl:tensionless_EM_tensor_1}
T_{\mu\nu}(x^\mu) = \int \d^2\sigma \, 
        \partial_\tau X_\mu\partial_\tau X_\nu \, \delta^{D}(x^\mu -X^\mu(\sigma,\tau)).
\end{eqnarray} 
In the classical approximation we may use \eqref{tl:tensionless_EM_tensor_1} to study how the presence of a tensionless string curves space time. In section \ref{tl:sec_TLbackground} we will find a background that describes the space-time around a tensionless string and reproduces this energy-momentum tensor.

\section{Gravitational field of a massless relativistic point particle}\label{tl:aichelburg-sexl}
In the following sections we will present the ideas that lead to the construction of a gravitational background for the tensionless string in [I]. To do this we begin by studying how the background of a massless particle is obtained by taking a limit in a background of a massive point particle.

In \cite{Aichelburg:1970dh} a derivation is presented that produces the gravitational background of a massless relativistic point particle. A massive point particle with mass $m$ in four dimensions generates a curved space time described by the Schwarzschild metric. In isotropic coordinates the
line element reads
\begin{eqnarray}
\d s^2= -\frac{(1-A)^2}{(1+A)^2}\d t^2 + (1+A)^4(\d x^2+\d y^2+\d z^2),
\end{eqnarray}
where $A=\frac{m}{2r}$ and $r=(x^2+y^2+z^2)^{1/2}$. A Lorentz boost in the $x$-direction is given by $t'=\gamma(t+vx)$, $x'=\gamma(x+vt)$, where $\gamma=(1-v^2)^{-1/2}$. Thus, performing the boost the metric, for an observer at rest, looks like 
\begin{eqnarray}\label{tl:AS_metric1}
\d s^2 &=& (1+A)^2(-\d t^2+ \d x^2 +\d y^2 + \d z^2) \cr 
 &&\ph{(1+A)} + \gamma\left((1+A)^4-\frac{(1-A)^2}{(1+A)^2}\right)(\d t -v\d x)^2,
\end{eqnarray}
where we have dropped the primes of the new coordinates and where $A$ is boosted to
\begin{eqnarray}
A=\frac{\gamma^{-1} m}{2\sqrt{(x-vt)^2+\gamma^{-2}(y^2+z^2)}}.
\end{eqnarray}
When we take the limit $v\rightarrow 1$, the last term in \eqref{tl:AS_metric1} becomes infinite, which is a reflection of the fact that it takes infinite energy to boost a particle to the speed of light.  This is clearly a problem if we want to use the metric to study massless particles, which travels at the speed of light. To circumvent the problem we keep the energy finite under the boost by rescaling the particle mass as $m=\gamma^{-1} p$ and keep $p$ constant while $v\rightarrow 1$. This means that $m\rightarrow 0$ under the infinite boost.  Thus, we expect that the resulting metric should describe the background of a massless particle. Calculating the limit carefully produces the result 
\begin{eqnarray}\label{tl:AS_metric2}
&&\hspace*{-1cm}\d s^2 = -\d t^2 + \d x^2 +\d y^2 + \d z^2 \cr
&& + 4p \left(|t-x|^{-1} -2\delta(t^2-x^2) \ln(y^2+z^2)^{1/2}\right)(\d t -\d x)^2.
\end{eqnarray}
This metric is divergent when $x=t$ and $y^2+z^2=0$. However this is not a problem since the divergence is located on the world line of the particle. This behavior is something we should have expected. Moreover, this metric has the form of a plane fronted gravitational shock-wave.

To investigate the classical matter content of the space-time described by
\eqref{tl:AS_metric2}, we use Einstein's equation,
$R_{\mu\nu}-g_{\mu\nu}R=8\pi\, T_{\mu\nu}$. By calculating the Ricci tensor and
scalar from the metric we can read off the energy momentum tensor from the
right side of the equation. It is found to be
\begin{eqnarray}
T^{\mu\nu}= p \delta(t-x)\delta(y)\delta(z)(\delta^\mu_0+\delta^\mu_1)(\delta^\nu_0+\delta^\nu_1).
\end{eqnarray}
This is the energy-momentum tensor for a massless particle in four dimensional space-time meaning that it is consistent to interpret the metric \eqref{tl:AS_metric2} as describing the space-time around a massless particle.

To summarize, we start with a known background for a massive particle, then we perform an infinite Lorentz-boost while keeping the energy of the particle fixed. This forces the mass of the particle to go to zero. Finally to check consistency the matter content of the new space time is investigated.

\section{A background for type IIB string theory}
\label{tl:section_schwarz_Background}

To find a consistent background for the tensionless string we will follow the same path as we used in the last section for the point particle. To do this, we need to decide which string background to boost. In this section we present a suitable background.

Type IIB string theory contains two different three-form field strengths, $H^{(i)}=dB^{(i)}$ $i=1,2$, one belonging to the R-R sector and the other belonging to NS-NS sector. We combine these into a vector ${\bf H}=\d {\bf B}=(H^{(1)},H^{(2)})^t$. Further IIB contains the graviton $g_{\mu\nu}$, a five-form field strength $F_5$ and two scalar fields, the NS-NS dilaton $\Phi$ and a R-R scalar $\chi$. The scalars are combined into one complex field $\lambda=\chi+\i\e^{-\Phi}$. In the present discussion all fermionic fields are taken to be zero.  The charges that couple to the five-form field strength are naturally carried by D3-branes and since, here, we are only interested in the behavior of the string background we put $F_5$ to zero. That is to say that there are no branes present. All of the above fields are massless, and together they form a consistent background for the IIB string to propagate in.

As for the bosonic case the equations of motion for these background fields arise from demanding that the quantized string theory living in this background is scale invariant. This produces the vanishing of the $\beta$-functionals, which is interpreted as equations of motion for the background fields. An action that is written in a covariant manner and that gives rise to the low-energy field equations for the
above bosonic supergravity fields is given by \cite{Hull:1995nu}
\begin{align}
  S_{\mathrm{IIB}}=\frac{1}{2\kappa^2}\int\d^{10}x\sqrt{-g}\left[
  R+\frac{1}{4}\mbox{tr}(\partial \mathbbm{M}\partial \mathbbm{M}^{-1})-
  \frac{1}{12}{\bf H}^t \mathbbm{M}{\bf H}\right], \label{tl:SUGRA_action}
\end{align}
where 
\begin{align}
  \mathbbm{M}=\e^\dilaton\left(\begin{array}{cc}
    \abs{\lambda}^2&\chi\\
    \chi&1
   \end{array}\right) \in SL(2,\mathbb{R}).
\end{align}
The global $SL(2,\mathbb{R})$ transformation $\mathbbm{M}\rightarrow\Lambda \mathbbm{M}\Lambda^t$, ${\bf B}\rightarrow(\Lambda^t)^{-1}{\bf B}$, leaves the action invariant. This is a sign of the fact that type IIB string theory is self-dual under S-duality, a duality between strong and weak couplings. In \cite{Schwarz:1995dk} a solution to the supergravity equations that follow from the action is presented. The solution gives the metric as 
\begin{eqnarray}\label{tl:schwarz_metric}
\ds^2 = A_q^{-3/4}\left(-\dt^{\,2} + (\dx^1)^2 \right)
        + A_q^{1/4}\,\dx\cdot\dx ,
\end{eqnarray}
where
\begin{eqnarray}
A_q = 1 + \frac{\Delta_q^{1/2} Q}{3r^6},\hspace{0.5cm}
\Delta_q^{1/2} = \vec{q}^t \mathbbm{M}^{-1}\vec{q},\hspace{0.5cm}
 \vec{q}=\left(
  \begin{array}{c}q_1\\q_2\end{array}\right),
\end{eqnarray}
$x = \left(x^2, \ldots, x^9\right)$ and $r=|x|$. The $Q$ is the fundamental charge for the coupling of the string to the $B^{(i)}$ fields. The charge of the string under this coupling is given by $(q_1,q_2)$. The solutions for the ${\bf B}$-field and the scalar field $\lambda$ are given by \cite{Schwarz:1995dk}
\begin{eqnarray}\label{tl:Bequation}
&&{\bf B}_{01} = \mathbbm{M}^{-1}\vec{q}\, \Delta_q^{-1/2} A_q^{-1},\\
&&\lambda = \frac{q_1\chi_0- q_2 |\lambda_0|^2 + \i q_1 \e^{-\phi_0}A_q^{1/2}}
                 {q_1 - q_2\chi_0 +\i q_2\e^{-\phi_0}A_q^{1/2}}.
\end{eqnarray}
All other components of ${\bf B}$ are zero.  Note that the background is divergent at $r=0$. This is interpreted as having a string located at this position, \ie the world-sheet lies in the $t-x^1$ plane. By studying the content of this background, \cite{deAlwis:1996ze} found that the action for the string that sources the background is given by 
\begin{eqnarray}\label{tl:schwarz_sourceaction}
S = -\frac{T_q}{2}\int \d^2\xi \left[\Delta^{1/2}_q
     \partial^a X^\mu \partial^\phn_a X^\nu G^\phn_{\mu\nu}
     + \epsilon^{ab}\partial^\phn_a X^\mu \partial^\phn_b X^{\nu}
     {\bf B}^t_{\mu\nu}\vec{q}\right].
\end{eqnarray}
where the string tension is
\begin{eqnarray}\label{tl:string_tension}
T_q=\Delta_q^{1/2} Q.
\end{eqnarray}

The above supergravity background provides the starting point for finding a background for the tensionless string, which is what we will do next.

\section{A background of a tensionless string}\label{tl:sec_TLbackground}
We now follow the lines of \cite{Aichelburg:1970dh} to Lorentz-boost the solution \eqref{tl:schwarz_metric} while keeping the energy finite. We review the derivation in article [I].  We expect the procedure to yield a background for the tensionless string since, as we have mentioned, the tension for the string plays the same r\^{o}le as the mass for the point particle. And, as we saw in section \ref{tl:aichelburg-sexl}, the condition to keep the energy finite under the infinite boost makes the particle mass go to zero. We thus expect the tension to go to zero in the same manner.

We want to boost the metric \eqref{tl:schwarz_metric} in a direction orthogonal to the string extension. Since the string world sheet is extended in the $t-x^1$ plane we choose to perform a Lorentz transformation in the $x^9$-direction. In the following we will call the direction of the boost $z$, \ie $z\equiv x^9$. The form of the transformation is as before the usual $t'= \gamma (t+vz)$, $z'=\gamma(x+vt)$ where $\gamma = (1-v^2)^{-1/2}$.  We introduce the notation $\tilde{x}= \left(x^2, \ldots x^8\right)$, and boost the metric \eqref{tl:schwarz_metric} to obtain
\begin{eqnarray}\label{tl:boosted_metric}
\ds^{\prime 2} &=& \frac{A_q^{\prime -3/2}\Delta_q^{1/2} Q \gamma^2}{3r^{\prime 6}}
         \left(\dt'-v\dz'\right)^2  +A_q^{\prime -3/4}(\dx^1)^2 \cr
      && +A_q^{\prime 1/4}\left(-(\dt')^2 + (\d z)^2 +\d\tilde{x}\cdot\d\tilde{x}\right).
\end{eqnarray}
To find this form of the transformed metric we have added and subtracted $A_q^{\prime 1/4} \gamma^2 (\dt'-v\dz')^2$, used the identity 
\begin{align}
\left(\dz' -v\dt'\right)^2 - \left(\dt'-v\dz'\right)^2 =\gamma^{-2} \left(\dz^{\prime 2} -\dt^{\prime 2} \right)
\end{align}
and finally set $A'_q - 1 = \frac{\Delta_q^{1/2} Q}{3r^{\prime 6}}$. In the above expressions $r'$ is defined as $r^{\prime 6} = \left(\gamma^2(z'-vt')^2 + \tilde{x}\cdot\tilde{x}\right)^3$.

We can now take the limit where the string moves at the speed of light, \ie $v \rightarrow 1$. In this limit the energy of the string diverges, $E_0\rightarrow\infty$. To keep the energy finite we rescale the energy as $E = \gamma^{-1}E_0$. Note that in this limit the background fields $\lambda\rightarrow\lambda_0$, $\mathbbm{M}\rightarrow \mathbbm{M}_0$, \ie go to their background expectation values. This implies that $\Delta_{q}\rightarrow \Delta_{q,0}$.

Since the classical energy of the string is given by $E_0=L T$, where $L$ is the length of the string and $T$ is the string tension, we find the rescaled energy to be $E=\gamma^{-1}L T = L\Delta_{q,0}^{1/2}\gamma^{-1}Q_0 =L\Delta_{q,0}^{1/2}Q$ where we used \eqref{tl:string_tension}. This means that we equivalently can rescale the fundamental charge as $Q=\gamma^{-1}Q_0$. This implies that the string tension is $T = \Delta_{q,0}^{1/2}Q = \gamma^{-1}\Delta_{q,0}^{1/2}Q_0$ and goes to zero as $v\rightarrow 1$. Hence, we interpret the background in this limit as being generated by a tensionless string.

To perform the limit we write 
\begin{eqnarray}
r^{\prime -6} = \frac{\left(1-v^2\right)^3}{\left(\left(z'-vt'\right)^2 + 
\tilde{\rho}^2\right)^3},
\end{eqnarray}
where $\tilde{\rho}^2 = \gamma^{-2} \tilde{x}\cdot\tilde{x}$. Next we define 
the integral $I$ as
\begin{eqnarray}
I = \int_{-\infty}^{z'}\frac{\d \zeta}{\left(\left(\zeta-vt'\right)^2 
+ \tilde{\rho}^2\right)^3}
\end{eqnarray}
and note that that we can express $r^{\prime -6}$ as proportional to a derivative of I;
\begin{eqnarray}
\left(1-v^2\right)^3 \frac{\d}{\dz'} I = r^{\prime -6}.
\end{eqnarray}
Hence, to evaluate the limit $v\rightarrow 1$ we can evaluate the expression
with $r^{\prime -6}$ replaced by $\left(1-v^2\right)^3 I = \gamma^{-6}I$ and
simplify. In the end we take the derivative with respect to $z'$ to obtain the
limit of interest.

Let us first consider the limit of $A'_q$,
\begin{eqnarray}\label{tl:limAq}
A'_q = 1 + \frac{\Delta_q^{1/2}Q_0 \gamma^{-1}}{3r^{\prime 6}} 
     = 1 + \frac{\Delta_q^{1/2}Q_0}{3} \frac{\d}{\dz'}\left(\gamma^{-7}I\right)
\end{eqnarray}
and the limit of
\begin{eqnarray}\label{tl:limFraction}
\frac{\Delta_q^{1/2}Q_0\gamma^{-1}\gamma^2}{3r^{\prime 6}} = 
                        \frac{\Delta_q^{1/2}Q_0}{3}
                        \frac{\d}{\dz'}\left(\gamma^{-5}I\right).
\end{eqnarray}
To find these limits we need to evaluate the integral $I$, we find it
to be
\begin{eqnarray}
I &=& \frac{z'-vt'}{4\tilde{\rho}^2\left(\left(z'-vt'\right)^2 
      + \tilde{\rho}^2\right)^2} 
      + \frac{3}{8\tilde{\rho}^4}\frac{z'-vt'}{\left(z'-vt'\right)^2 
      +\tilde{\rho}^2} \\
   && + \frac{3}{8\tilde{\rho}^5}\left(\arctan 
      \left(\frac{z'-vt'}{\tilde{\rho}}\right) 
      + \frac{\pi}{2}\right).
\end{eqnarray}
Next we calculate the limit 
\begin{align}
\lim_{v\rightarrow 1} \gamma^{-7}I =& \lim_{v\rightarrow 1}\Bigg(
                     \frac{\gamma^{-7}}{4 \tilde{\rho}^2}
                      \frac{z'-vt'}{\left(\left(z'-vt'\right)^2 
                      + \tilde{\rho}^2\right)^2} 
                      + \frac{3\gamma^{-7}}{8\tilde{\rho}^4}
                      \frac{z'-vt'}{\left(z'-vt'\right)^2 +\tilde{\rho}^2}\cr
&\ph{\lim_{v\rightarrow 1}\Bigg(} + \frac{3\gamma^{-7}}{8\tilde{\rho}^5}
        \left(\arctan \left(\frac{z'-vt'}{\tilde{\rho}}\right) 
      + \frac{\pi}{2}\right) \Bigg). \label{tl:lim_gamma-7_I}
\end{align}
Remember that $\tilde{\rho}^2 = \gamma^{-2}\tilde{x}\cdot\tilde{x}$, so 
the first term of \eqref{tl:lim_gamma-7_I} becomes
\begin{eqnarray}
&&\lim_{v\rightarrow 1} \frac{\gamma^{-7}}{4\gamma^{-2}\tilde{x}\cdot\tilde{x}} 
                     \frac{z'-vt'}{\left(\left(z'-vt'\right)^2 + 
                     \gamma^{-2}\tilde{x}\cdot\tilde{x}\right)^2}\cr
&&\hspace*{1cm}= \lim_{v\rightarrow 1}
        \frac{z'-vt'}{\left(\tilde{x}\cdot\tilde{x}\right)\left(\gamma^{5/2}
                     \left(z'-vt'\right)^2 + 
                     \gamma^{1/2}\tilde{x}\cdot\tilde{x}\right)^2} 
                     = 0,
\end{eqnarray}
since $\gamma \rightarrow \infty$ as $v\rightarrow 1$. The second term in \eqref{tl:lim_gamma-7_I} becomes
\begin{eqnarray}
&&\lim_{v\rightarrow 1}
\frac{3\gamma^{-7}}{8\gamma^{-4}\left(\tilde{x}\cdot\tilde{x}\right)^2}
 \frac{z'-vt'}{\left(z'-vt'\right)^2 +\gamma^{-2}\tilde{x}\cdot\tilde{x}}\cr
&&\hspace{1cm}=\lim_{v\rightarrow 1}
\frac{3}{8\left(\tilde{x}\cdot\tilde{x}\right)^2}
                \frac{z'-vt'}{\gamma^3\left(z'-vt'\right)^2
                              +\gamma^1\tilde{x}\cdot\tilde{x}} = 0,
\end{eqnarray}
and the third term
\begin{align}
&\lim_{v\rightarrow 1} \frac{3\gamma^{-7}}{8\gamma^{-5}
\left(\tilde{x}\cdot\tilde{x}\right)^{5/2}}\left(\arctan 
\left(\frac{z'-vt'}{\gamma^{-1}
\left(\tilde{x}\cdot\tilde{x}\right)^{1/2}}\right) + \frac{\pi}{2}\right)\cr
&\hspace*{.5cm}= \lim_{v\rightarrow 1} \frac{1}{\gamma^2}\frac{3}{8
\left(\tilde{x}\cdot\tilde{x}\right)^{5/2}}\left(\arctan
\left(\frac{z'-vt'}{\gamma^{-1}
\left(\tilde{x}\cdot\tilde{x}\right)^{1/2}}\right) + \frac{\pi}{2}\right)
=0.
\end{align}
This since $\arctan(\cdot)$ is bounded and the factor in front goes to zero.
We have thus found
\begin{eqnarray}\label{tl:gamma7I}
\lim_{v\rightarrow 1} \gamma^{-7}I = 0
\end{eqnarray}
and hence the derivative of \eqref{tl:gamma7I} with respect to $z'$ is 
zero. This means that the limit of \eqref{tl:limAq} is
\begin{eqnarray}\label{tl:limAqFinal}
\lim_{v\rightarrow 1} A'_q = 1.
\end{eqnarray}
The next limit we need to consider is the limit of the fraction in the
metric \eqref{tl:limFraction}. To evaluate it we begin by calculating 
\begin{align}
\lim_{v\rightarrow 1} \gamma^{-5}I =& \lim_{v\rightarrow 1}\Bigg(
                     \frac{\gamma^{-5}}{4 \tilde{\rho}^2}
                      \frac{z'-vt'}{\left(\left(z'-vt'\right)^2 
                      + \tilde{\rho}^2\right)^2} 
                      + \frac{3\gamma^{-5}}{8\tilde{\rho}^4}
                      \frac{z'-vt'}{\left(z'-vt'\right)^2 +\tilde{\rho}^2}\cr
&\ph{\lim_{v\rightarrow 1}\Bigg(}
      + \frac{3\gamma^{-5}}{8\tilde{\rho}^5}\left(\arctan 
        \left(\frac{z'-vt'}{\tilde{\rho}}\right) 
       + \frac{\pi}{2}\right) \Bigg).
\end{align}
The two first terms in this limit become zero by the same arguments as used above. 
The third term, however, becomes
\begin{align}
&\lim_{v\rightarrow 1} \frac{\gamma^{-5}}{\gamma^{-5}}\frac{3}{8 
              \left(\tilde{x}\cdot\tilde{x}\right)^{5/2}}
              \left(\arctan \left(\frac{z'-vt'}{\gamma^{-1}
              \left(\tilde{x}\cdot\tilde{x}\right)^{1/2}}\right) 
                +\frac{\pi}{2}\right)\cr
&  = \left\{\begin{array}{lll}
        0 & & \mbox{if } z' < t'\\
        \frac{3\pi}{8} \left(\tilde{x}\cdot\tilde{x}\right)^{-5/2} 
        & &\mbox{if } z'> t'\\
        \end{array} \right.
     =  \frac{3\pi}{8} \left(\tilde{x}\cdot\tilde{x}\right)^{-5/2} 
        \Theta\left(z'-t'\right),
\end{align}
where $\Theta$ is the Heaviside function. Thus we find
the limit of \eqref{tl:limFraction} to be
\begin{align}\label{tl:limFractionFinal}
\lim_{v\rightarrow 1}\frac{\Delta_q^{1/2}Q_0\gamma^{-1}\gamma^2}{3r^{\prime 6}} 
  &= \frac{\Delta_{q,0}^{1/2}Q_0}{3} \frac{\d}{\dz'}\lim_{v\rightarrow 1}
      \gamma^{-5}I \cr
  &= \frac{\pi\Delta_{q,0}^{1/2}Q_0}{8} 
      \left(\tilde{x}\cdot\tilde{x}\right)^{-5/2} \delta\left(z'-t'\right).
\end{align}

We are now in a position to write down the limit of the metric given by equation \eqref{tl:boosted_metric}. Defining $\rho = \left(\tilde{x}\cdot\tilde{x}\right)^{1/2}$ and using the results \eqref{tl:limAqFinal} and \eqref{tl:limFractionFinal}, the limit of the metric
becomes 
\begin{align}
\ds^{\prime 2} =& \frac{\pi\Delta_{q,0}^{1/2} Q_0}{8\rho^5} \delta\left(z'-t'\right)
\left(\dt'-\dz'\right)^2 \cr
&-(\dt')^2 +(\dx^1)^2 +(\dz')^2 + \d \tilde{x}\cdot\d \tilde{x}.
\end{align}
Next we perform a change of variables in the metric to put it into a nicer form. For this we define $u=z'-t'$, $v=z'+t'$ and write the directions transverse to the string in spherical coordinates. The metric now becomes
\begin{eqnarray}\label{tl:boostedmetric}
\ds^{\prime 2} = \d u \d v + (\dx^1)^2 + \d\rho^2 + \rho^2\d\Omega 
          + \frac{\pi\Delta_{q,0}^{1/2} Q_0}{8\rho^5} \delta\left(u\right) \d u^2.
\end{eqnarray}
This metric has the structure of a gravitational shock-wave and also of the typical plane-fronted wave with parallel propagation, or pp-wave, see \eg \cite{Plefka:2003nb},
\begin{eqnarray}
  \ds^2 = \d u \d v + K(x^2, \ldots, x^8, u)\d u^2 + \sum_{i=1}^8 (\d x^i)^2.
  \label{tl:pp-wave_metric}
\end{eqnarray}
In the next chapter another example of this kind of metric will appear as a Penrose-G\"{u}ven limit of the $AdS_5\times S^5$ space-time.

To have the complete supergravity background for the string we also need to consider the limit of ${\bf B}$ given in \eqref{tl:Bequation}. Since the action only depends on ${\bf B}$ through $\mathbbm{H} = \d{\bf B}$, there is a gauge freedom in the choice of ${\bf B}$. Since $\d^2=0$ any two fields ${\bf B}$ and $\tilde{\bf B}$ are physically equivalent if they at most differ by an exact term $\d \lambda$. We use this freedom to make the choice $\d\lambda = -\mathbbm{M}^{-1}\vec{q}\Delta_q^{-1/2}$. The limit $v\rightarrow 1$ of the gauge transformed non zero component ${\bf B}_{01}$ is easily found, 
\begin{eqnarray}
\lim_{v\rightarrow 1}{\bf B}_{01}
   &=&\lim_{v\rightarrow 1}\left(\mathbbm{M}^{-1}
                 \vec{q}\Delta_q^{-1/2}A_q^{-1} 
            - \mathbbm{M}^{-1}\vec{q}\Delta_q^{-1/2}\right)\cr
 &=& \lim_{v\rightarrow 1} \mathbbm{M}^{-1}
                  \vec{q}\Delta_q^{-1/2}(A_q^{-1}-1) = 0
\end{eqnarray}
since $A_q\rightarrow 1$. This means that the ${\bf B}$ field vanishes in the limit.

\begin{figure}
\begin{centering}
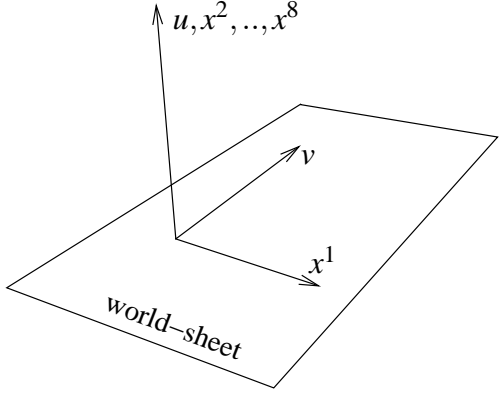
\caption{The world-sheet of the tensionless string in space time.}
\label{img:world-sheet_config}
\end{centering}
\end{figure}

Looking at the metric given by \eqref{tl:boostedmetric} we notice that if $u\neq 0$ the space time is just flat Minkowski space. At $u=0$ the space has a shock wave singularity. We interpret this as the position of the string, see figure \ref{img:world-sheet_config}.

To study the energy content of the space-time defined by the metric \eqref{tl:boostedmetric} we calculate the Ricci tensor. The only non zero component is given by
\begin{eqnarray}
  R_{uu}=-\frac{1}{2}\Delta_7\left(\frac{\pi\Delta_{q,0}^{1/2} Q_0}{8\rho^5}\right),
\end{eqnarray}
where $\Delta_7$ is the seven dimensional Laplacian. Since $1/\rho^5$ is the 
Green's function for this Laplacian, \ie$\Delta_7\rho^{-5}=-16/3\pi^3\delta(\rho)$,
we find
\begin{eqnarray}
  R_{uu}= \frac{1}{3}\pi^4\Delta^{1/2}_{q,0} Q_0^\phn\delta(\rho)\delta(u).
\end{eqnarray}
Further, since this is the only nonzero component of the Ricci tensor, the Ricci scalar vanishes, and from the Einstein's equations we read off the energy-momentum tensor of this space
\begin{eqnarray}\label{tl:Tmunu}
T_{uu}=\frac{1}{24}\pi^3\Delta^{1/2}_{q,0} Q_0^\phn \delta(\rho)\delta(u),
\end{eqnarray}
with all other components being zero. This shows that the energy-momentum of the space-time is located at $\rho=0$ and $u=0$, the position of the string world-sheet, see figure \ref{img:world-sheet_config}. It also implies that the metric \eqref{tl:boostedmetric} solves Einstein's equations in empty space except at the position of the string.

To see if this energy momentum tensor is reproducible by a tensionless string we return to the expression \eqref{tl:tensionless_EM_tensor_1}, which we repeat here,
\begin{eqnarray}\label{tl:tensionless_EM_tensor_2}
T_{\mu\nu}(x^\mu) = \int \d^2\sigma \, 
        \partial_\tau X_\mu\partial_\tau X_\nu \, \delta^{D}(x^\mu -X^\mu(\sigma,\tau)).
\end{eqnarray}
In the classical configuration that we found, figure \ref{img:world-sheet_config}, the world-sheet coordinate $\sigma$ is proportional to $x^1$ and the world-sheet coordinate $\tau$ is proportional to $v$. This enables us to perform the integration in \eqref{tl:tensionless_EM_tensor_2}, it makes $X^1=\alpha \sigma$ and $X^v= \beta \tau$. The other coordinates $X^2,..X^8, X^u$ are fixed to zero by the
delta function. The only coordinate field that is dependent on $\tau$ is $X_u=G_{uv}X^v=\beta\tau$. This means that the only non-vanishing component is 
\begin{eqnarray}
  T_{uu}=\del_\tau X_u\del_\tau X_u\delta(\rho)\delta(u)=
  \beta^2\delta(\rho)\delta(u).
\end{eqnarray}
Comparing this energy momentum tensor with the one found from the metric \eqref{tl:Tmunu}, we find that they have the same structure.  From these expressions we can identify $\beta^2 =\pi^3\Delta^{1/2}_{q,0} Q_0/24$.

We conclude that the energy content of the space-time defined by \eqref{tl:boostedmetric} is the same as that of a tensionless string in a certain configuration. This confirms the claim that this space-time is
generated by a tensionless string source. And finally, we have indeed found a supergravity background generated by tensionless string.

To continue, we could study the behavior of ordinary tensile strings in the new background or we could study the tensionless string in the quantized theory. In the next chapter we will turn to the second of these questions and study the quantized tensionless string in a pp-wave background. The first question is partly answered in \cite{Amati:1988ww} where they study strings in a shock-wave background. This background has similarities with \eqref{tl:boostedmetric} in that it contains a $\delta(u)$ factor, but there are some subtleties that remain to be explored.

        \chapter{Strings in a pp-wave background}\label{Tnollquant}
In the last chapter we considered the tensionless limit of string theory and constructed a background that had the characteristic pp-wave form, \eqref{tl:pp-wave_metric}. The background was obtained by taking a limit of a known supergravity solution. In this chapter we will consider tensile and tensionless strings on a background that also can be obtained as a limit. This background is pp-wave background supported by a R-R five-form flux \cite{Blau:2001ne}. It is a maximally supersymmetric solution to IIB supergravity and in \cite{Blau:2002dy} it is shown that this background arises through a Penrose-G\"{u}ven limit \cite{Penrose:1976,Gueven:2000ru} of $AdS_5\times S^5$ space. 

We will first review the quantization of the tensile string in this pp-wave background and then consider the tensionless case. We will find that the procedure of quantizing the tensionless string simplifies in the pp-wave background, as compared to quantization in flat Minkowski background, discussed in \cite{Isberg:1993av}.

\section{The pp-wave background as a limit}
Here we will present the string background of this chapter, the pp-wave background. 

When we in section \ref{tl:section_schwarz_Background} gave the IIB supergravity action \eqref{tl:SUGRA_action} we used the assumption that there were no D-branes present to set the five-form field strength to zero. In \cite{Horowitz:1991cd} a different solution to the IIB supergravity equations is found. The solution has a non-zero five-form field strength but the three-form field strengths are put to zero, this since they couple to strings that can be disregarded. The resulting space-time is interpreted as being generated by a $N$ coincident D3-branes \cite{Polchinski:1995mt} and is described by \cite{Johnson:2003gi} 
\begin{align}\label{q:3branesol}
\d s^2 =& H^{-1/2}_3 \eta_{\mu\nu}\d x^\mu\d x^\nu + H^{1/2}_3 \delta_{ij}\d x^i \d x^j, \\
\e^{2\Phi} =& g_s^2, \\
C^{(4)} =& (H^{-1}_3 - 1)g_s^{-1} \d x^0 \wedge ...\wedge \d x^3,
\end{align}
where 
\begin{align}
H_3 = 1 + \frac{4\pi g_s N \alpha'^2}{r^4} \;\;\;\mbox{with}\;\;r^2=\delta_{ij} x^i x^j,
\end{align}
and $\mu,\nu = 0,...,3$ and $i,j=4,...,9$. Further, $g_s$ is the string coupling, $\Phi$ is the dilaton and $C^{(4)}$ is the R-R four-form field, \ie $F^{(5)}=\d C^{(4)}+ *\d C^{(4)}$. Note that this space-time has a singularity at $r=0$ with a horizon of a finite size given by the radius
\begin{align}
R_3 = (4\pi g_s\alpha^{\prime 2}N)^{1/4}. \label{q:horizon_radius}
\end{align}
Note also the resemblance between this space-time and the one we studied in the last chapter \eqref{tl:schwarz_metric}. The essential structure is the same only that the present case we have three coordinates grouped together with the time coordinate, corresponding to the 3+1 coordinates of the coincident D3-brane world-volumes. Remember that in \eqref{tl:schwarz_metric} we had only one coordinate grouped with the time coordinate.

We follow the lines of \cite{Maldacena:1997re} and study the geometry in the near-horizon region of the space-time \eqref{q:3branesol}. By letting $\alpha'\rightarrow 0$ while keeping $u=r/\alpha'$ fixed we obtain the near-horizon limit and find that it is described by the space $AdS_5\times S^5$ supported by a five-form field strength and a constant dilaton,
\begin{align}
\d s^2 =& (4\pi g_s \alpha^{\prime 2} N)^{-1/2} u^2 \eta_{\mu\nu}\d x^\mu \d x^\nu \cr 
& + (4\pi g_s \alpha^{\prime 2} N)^{1/2} (u^{-2}\d u^2 + \d\Omega^2_5),\\
F^{(5)}=& 16\pi\alpha^{\prime 2}N(\epsilon_{(5)} + *\epsilon_{(5)}),\\
\Phi =& \mbox{constant}.
\end{align}
Here, $\epsilon_{(5)}$ is the volume form on $S^5$. The radius of the $S^5$ and 
the radius of curvature of the $AdS_5$ are equal and both given by the horizon
radius $R_3$ in \eqref{q:horizon_radius}.
 
Next we want to perform a Penrose-G\"{u}ven limit on this supergravity solution. Such a limit can be thought of as ``blowing up'' the vicinity of a null geodesic in such a way that the null geodesic remains invariant.  A nice review of the limit is given in \eg \cite{Blau:2002mw}. In \cite{Penrose:1976} it is shown that this limit, for any initial space-time, produces plane-fronted waves with parallel propagation, or pp-waves for short. In \cite{Gueven:2000ru} the limit is extended to incorporate not only the metric but also the other supergravity fields. It can be showed that the limit preserves at least half of the supersymmetries of the initial background. In the present case we choose a null-geodesic along a great circle on the $S^5$, this limit can be thought of as having a string rotating very quickly around the $S^5$. When we ``zoom in and blow up'' the space we obtain a maximally supersymmetric pp-wave \cite{Blau:2001ne,Blau:2002dy}, \ie a pp-wave that preserves all 32 supersymmetries present in the $AdS_5\times S^5$ space. In Brinkmann coordinates this pp-wave metric is given by
\begin{align}\label{q:ppwave_metric}
\d s^2 = 2\d x^+\d x^- - \mu^2 x^i x_i (\d x^+)^2 +\d x^i\d x_i,
\end{align}
where $x^{\pm}=2^{-1/2}(x^9\pm x^0)$ and $i=1,...,8$. Here we have also
defined $\mu$ as the inverse of the common radius of the $AdS_5$ and the $S^5$, \ie $\mu=(R_3)^{-1}$.

The Penrose-G\"{u}ven limit of $F^{(5)}$ and $\Phi$ tells us that this space is
supported by a constant R-R five-form field strength and a constant dilaton,
\begin{align}
F^{(5)}_{+1234} = F^{(5)}_{+5678} =& 2\mu, \\
\Phi =& \mbox{constant}.
\end{align}
There are two tings to note about this supergravity background. Firstly the presence of the five-form field strength breaks the manifest $SO(8)$ invariance of \eqref{q:ppwave_metric} to $SO(4)\times SO(4)$, and secondly, when we take the radius of curvature, or the horizon radius, $R_3$ to infinity we obtain flat Minkowski space-time. This means that the string theory living on this pp-wave background must reduce to the standard flat space theory when the common radius of $AdS_5$ and $S^5$ is taken to infinity.

\section{Ordinary string in a pp-wave background}
The standard tensile string in the pp-wave background given above was first studied by in \cite{Metsaev:2001bj}. There, the string is studied in the Green-Schwarz formulation, see section \ref{intro:Green-Schwarz}, and the action describing the system in light-cone coordinates is found. This theory was, for closed strings, successfully quantized in \cite{Metsaev:2002re}.

Here we will follow the lines of \cite{Metsaev:2002re} to review the quantization of the closed superstring in a pp-wave background. The setup is an action with the standard ten dimensional superspace coordinates $X^\mu(\sigma,\tau)$ and $\theta^{A a}(\sigma,\tau)$, where $\mu=0,...,9$, $A=1,2$ and $a=1,..16$.  The two $\theta^A$'s are ten dimensional Majorana-Weyl spinors of the same chirality, \ie they are real and obey $\Gamma^{11} \theta^A= \theta^A$. This choice of chirality implies that we study type IIB strings. The $a$-index labels the 16 independent components of each spinor.

We now introduce the 10-dimensional $32\times 32$ gamma matrices obeying the 10-dimensional Dirac algebra $\{\Gamma^\mu,\Gamma^\nu\}=2 \eta^{\mu\nu} \mathbbm{1}_{32\times 32}$. The above mentioned operator $\Gamma^{11}$ is defined as $\Gamma^{11}=\Gamma^0...\Gamma^9$. In the chiral representation of the $\Gamma$-matrices they are given by 
\begin{eqnarray}
\Gamma^\mu = \left(\begin{array}{cc}0&\gamma^\mu\\\bar{\gamma}^\mu&0\end{array}\right),
\end{eqnarray}
where the $\gamma$ and $\bar{\gamma}$ are $16\times 16$ matrices obeying 
$\gamma^\mu\bar{\gamma}^\nu+\gamma^\nu\bar{\gamma}^\mu=2\eta^{\mu\nu}\mathbbm{1}_{16\times 16}$. Another object that we will need is the fermionic mass operator $\Pi$ that is given by $\Pi=\gamma^1\bar{\gamma}^2\gamma^3\bar{\gamma}^4$ and obeys $\Pi^2=1$.

The $\kappa$-symmetry fixed action that describes the string in the pp-wave background \eqref{q:ppwave_metric} is given by \cite{Metsaev:2001bj}
\begin{align}
S= -T \int\d^2 \sigma \Big\{&\sqrt{-\det(\gamma_{\alpha\beta})}\gamma^{\alpha\beta} h_{\alpha\beta}\cr
& -\i\epsilon^{\alpha\beta}\partial_\alpha X^+\left(
  \theta^{1}\bar{\gamma}^-\partial_\beta\theta^{1} - 
  \theta^{2}\bar{\gamma}^-\partial_\beta\theta^{2}
   \right) \Big\}, \label{q:original_action}
\end{align}
where the induced metric is given by
\begin{align}
h_{\alpha\beta} =&  2\partial_\alpha X^+ \left(\partial_\beta X^- + \i\theta^1 \bar{\gamma}^- \partial_\beta \theta^1 +\i \theta^2 \bar{\gamma}^- \partial_\beta\theta^2 \right) \cr
& -\left(\mu^2 X^I X_I + 4\i \mu \theta^1 \bar{\gamma}^- \Pi \theta^2\right) \partial_\alpha X^+ \partial_\beta X^+ + \partial_\alpha X^I \partial_\beta X_I.
\end{align}
The gauge we have chosen to fix the $\kappa$-symmetry is the, so called, fermionic light-cone gauge. This gauge imposes the condition $\Gamma^+\theta^A=0$ on the fermionic coordinate fields, where $\Gamma^+$ is defined as $\Gamma^+=\frac{1}{\sqrt{2}}(\Gamma^9+\Gamma^0)$.

The action \eqref{q:original_action} is invariant under world-sheet diffeomorphisms and Weyl rescalings which let us choose the conformal gauge for the world sheet metric. However, when choosing the conformal gauge we must remember that we have to impose the Virasoro conditions, arising from variation of the world-sheet metric, by hand. After choosing the conformal gauge there is still enough symmetry left to allow the light-cone choice, $X^+ =  p^+ \tau/T$. In this gauge the Virasoro conditions read
\begin{align}
2 p^+\dot{X}^- + T\dot{X}^I\dot{X}_I + T X^{I\prime}X'_I - T m^2 X^I X_I \hspace*{2.5cm}&\cr
   + 2\i p^+ \left(
    \theta^1 \bar{\gamma}^-\dot{\theta}^1 
    + \theta^2 \bar{\gamma}^-\dot{\theta}^2 
    - 2 m \theta^1\bar{\gamma}^-\Pi\theta^2 \right) =& 0,\label{q:tensile_virasoro1}\\
p^+ X^{-\prime} + T X'_I \dot{X}^I 
    +\i p^+\left(\theta^1 \bar{\gamma}^-\theta^{1\prime} 
    + \theta^2 \bar{\gamma}^-\theta^{2\prime} \right) =& 0,\label{q:tensile_virasoro2}
\end{align}
where a dot over a field denotes a derivative with respect to $\tau$ and a prime denotes a derivative with respect to $\sigma$.

Fixing the above gauges in \eqref{q:original_action} gives us the $\kappa$-symmetry fixed action in the light-cone gauge as \cite{Metsaev:2002re}
\begin{align}\label{q:tensile_action}
  S = \frac{T}{2} & \int \d^2\sigma \bigg\{
       \partial_+ X^I\partial_- X_I - m^2 X^I X_I\cr
    &+ \frac{2\i p^+}{T} \left( \theta^1\bar{\gamma}^-\partial_+\theta^1
     + \theta^2\bar{\gamma}^-\partial_-\theta^2 \right)
     - \frac{4\i m p^+}{T} \theta^1\bar{\gamma}^- \Pi\theta^2 \bigg\}.
\end{align}
where $m= p^+ \mu/T$ and $\partial_\pm = \partial_\tau
\pm\partial_\sigma$.  The $\bar{\gamma}^-$ is defined as an eigenmatrix of the
$\gamma^0\bar{\gamma}^9$ operator with eigenvalue $-1$.

The classical equations of motion that follow from this action read
\begin{align}
\partial_+\partial_- X^I + m^2 X^I &= 0, \\
\partial_+\theta^1 - m\Pi\theta^2 &= 0,  \\
\partial_-\theta^2 + m\Pi\theta^1 &= 0.    
\end{align}
In the following we let the world sheet $\sigma$-coordinate take values between $0$ and $1$. The general solution for the bosonic coordinate field that satisfy the closed string boundary conditions is
\begin{align}
X^I(\sigma,\tau)=& x_0^I \cos(m\tau) + \frac{p_0^I}{m T}\sin(m\tau) \cr
&        +\i\sum_{n\neq0}
         \frac{1}{\omega_n}\e^{-\i\omega_n\tau}\left(\alpha^{1I}_n \e^{2\pi\i n\sigma} 
          +\alpha^{2I}_n \e^{- 2\pi\i n\sigma} \right),
\end{align}
with $\omega_n =\sign(n)\sqrt{4\pi^2 n^2+ m^2}$. And for the two fermionic coordinate fields the solutions are
\begin{align}
\theta^1(\sigma,\tau) =& \cos(m\tau)\theta^1_0 +\sin(m\tau)\Pi\theta^2_0 \cr
&    +\sum_{n\neq 0}c_n \e^{-\i\omega_n\tau}
        \left( \theta^1_n \e^{2\pi\i n\sigma } +\i\frac{\omega_n -2\pi n}{m}\Pi\theta^2 \e^{-2\pi\i n \sigma}
         \right),  \\
\theta^2(\sigma,\tau) =& \cos(m\tau)\theta^2_0 -\sin(m\tau)\Pi\theta^1_0 \cr
&      +\sum_{n\neq 0} c_n \e^{-\i\omega_n\tau}
        \left( \theta^2_n \e^{2\pi\i n\sigma} -\i\frac{\omega_n -2\pi n}{m}\Pi\theta^1 \e^{-2\pi\i n\sigma}
         \right),
\end{align}
where 
\begin{align}
c_n = \frac{1}{\sqrt{1+(\frac{\omega_n- 2\pi n}{m})^2}}.
\end{align}
When quantizing we need the conjugate momenta of the coordinate fields. The bosonic conjugate momenta is found to be $P^I = T \dot{X}^I$. We impose the standard equal $\tau$ commutation relations $[P_I(\sigma, \tau) ,X^J(\sigma',\tau)]= -\i \delta_I^J\delta(\sigma-\sigma')$. This translates into the following relations for the bosonic modes
\begin{eqnarray}
[p^I_0,x^J_0]=-\i\delta^{IJ}, \hspace{1.5cm} 
[\alpha^{A I}_m,\alpha^{B J}_n]=\frac{\omega_m}{2 T}\delta_{m+n,0}\delta^{IJ}\delta^{AB},
\end{eqnarray}
where $A,B = 1,2$.

Turning to the conjugated momentum for the fermionic coordinates we find that it is given by $\pi^A = -\i p^+ \bar{\gamma}^-\theta^A$ for $A=1,2$. The absence of a $\tau$-derivative means that the system has the two second class constraints $\chi^{A}=\pi^{A} +\i p^+\bar{\gamma}^-\theta^{A} = 0$. Dirac taught us how to quantize systems with these kinds of constraints, instead of basing the quantum commutators on the Poisson brackets we must use a generalization, called the Poisson-Dirac bracket. This bracket is defined as \cite{Dirac:1950pj}
\begin{eqnarray}
[\,\cdot,\cdot\,]_{P.D}\equiv [\,\cdot,\cdot\,]_P -[\,\cdot ,\chi_a]_P C^{ab}[\chi_b,\cdot\,]_P.
\end{eqnarray}
where $[\,\cdot,\cdot\,]_P$ is the standard Poisson bracket and $C^{ab}$ is the inverse of $[\chi_a,\chi_b]_P$. We begin by assuming that the equal $\tau$ anti-Poisson bracket for a fermionic coordinate field given by
\begin{eqnarray}
\{\pi_b^{A}(\sigma,\tau),\theta^{B a} (\sigma',\tau)\}_P 
        = \frac{1}{2} (\gamma^+\bar{\gamma}^-)^a_{\ph{a}b} 
                \delta(\sigma-\sigma')\delta^{IJ}\delta^{AB},
\end{eqnarray}
where $\frac{1}{2}\gamma^+\bar{\gamma}^-$ is a projector that is present to respect the fermionic light-cone gauge. Calculating the corresponding anti-Poisson-Dirac bracket with the aid of several relations for the $\gamma$-matrices, translating the brackets for the fields to brackets for the modes and quantizing by $\{\,\cdot,\cdot\,\}\rightarrow \i\{\,\cdot,\cdot\,\}$, we find the commutators for the fermionic modes as
\begin{align}
\{\theta^{A a}_m,\theta^{B b}_n\} 
          = \frac{1}{4 p^+} (\gamma^+)^{a b}\delta^{AB}
    \delta_{m+n,0}.
\end{align}

Now when we have the commutators we continue to the construction of the physical spectrum of the theory. For this we need to construct the light-cone Hamiltonian, also called the light-cone energy operator. To get a more readable form of the end result we define new mode operators as 
\begin{align}
\begin{array}{lll}
a_0^I=\sqrt{\frac{T}{2m}}(\frac{p^I_0}{T} - \i m x_0^I),& \hspace{1cm}& 
\bar{a}_0^I=\sqrt{\frac{T}{2m}}(\frac{p^I_0}{T} + \i m x_0^I),\\
\eta_0 = \sqrt{\frac{p^+}{2}}(\theta_0^1 - \i\theta_0^2),& &
\bar{\eta}_0 = \sqrt{\frac{p^+}{2}}(\theta_0^1 + \i\theta_0^2),\\
a_n^{AI}=\sqrt{\frac{2T}{\omega_n}}\alpha^{AI}_n, &&  
\eta^A_n = \sqrt{2p^+} \theta^A_n,
\end{array}
\end{align}
for $n\neq0$ and $A=1,2$. The (anti)commutators for these mode operators become
\begin{align}
\begin{array}{lll}
[a^I_0,\bar{a}^J_0]= \delta^{IJ}, &\hspace{0.5cm}&
[a^{AI}_m, a^{BJ}_n] = \delta_{m+n,0}\delta^{IJ}\delta^{AB},\\
\{\eta^a_0,\bar{\eta}^b_0\} = \frac{1}{4}(\gamma^+)^{ab},&&
\{\eta^{Aa}_m, \eta^{Bb}_n\} = \frac{1}{2}(\gamma^+)^{ab}\delta_{m+n,0}\delta^{AB}.
\end{array}
\end{align}

When writing the light-cone Hamiltonian, normal ordering must be considered for the zero modes. For the other modes, normal ordering constants simply cancel since there are equal numbers of bosonic and fermionic oscillators for each $n$. The light-cone Hamiltonian is found to be
\begin{align}\label{q:Tneq0_lightconeHamiltonian}
H_{LC} = &
  m\left(\bar{a}_0^I a_{0I} + 2 \bar{\eta}_0\bar{\gamma}^-\Pi\eta_0 + 4\right)\cr
  &+\sum_{A = 1,2} \sum_{n=1}^{\infty}
     \sqrt{(2\pi n)^2 + m^2}
     \left(a^{A I}_{-n} a^{A}_{In} 
         + \eta^{A}_{-n}\bar{\gamma}^- \eta^{A}_n\right),
\end{align} 
where we recall that $m=\mu p^+/T$. Further, the modes with label $-n$ are creation operators and the ones with label $n$ are annihilation operators. These are used when constructing the spectrum and for defining the vacuum state. The vacuum is defined to be annihilated by $a_0^I$, $\theta^a_0$, $a^{AI}_n$ and $\eta^{Aa}_n$ for $n>0$. The Fock space is now built by acting on the vacuum state with the operators $\bar{a}_0^I$, $\bar{\theta}^a_0$, $a^{AI}_{-n}$ and $\eta^{Aa}_{-n}$ for $n>0$.

As for the bosonic string in flat space the Virasoro condition \eqref{q:tensile_virasoro2} is used to express $X^-$ in terms of the transverse coordinates. To this end we integrate the expression over the range of $\sigma$ and find a constraint. This implies that to get the space of physical states, we must restrict the Fock space to states that fulfill the constraint. In terms of the mode operators the constraint is given by
\begin{align}\label{q:tensile_lvlmatch}
N^1|phys\rangle = N^2|phys\rangle\;\;\mbox{ with }\;\; 
N^A = \sum_{n=1}^{\infty} n \left(a^{A I}_{-n} a^{A}_{In} 
         + \eta^{A}_{-n}\bar{\gamma}^-\eta^{A}_n\right),
\end{align}
where there is no summation over $A$. This is the level matching condition. 

The energy spectrum of the theory is found from \eqref{q:Tneq0_lightconeHamiltonian} as $\frac{T}{p^+} H_{LC}$, and given in terms on two parameters, the string tension $T$ and the inverse radius of curvature of the $AdS_5\times S^5$ space $\mu$. Taking the radius of curvature to infinity we obtain the flat space limit, which is the same as taking $\mu\rightarrow 0$. In this limit the quantized spectrum goes into its flat space form.

It can be shown \cite{Metsaev:2002re} that the supersymmetry algebra is
realized in terms of the super-Poincar\'{e} generators and that the spectrum
obtained above reproduce the correct supersymmetry multiplets and in particular that the massless IIB supergravity multiplet is present.

\subsubsection{The AdS/CFT correspondence}
The main importance of the string in this pp-wave background is that
it provides an exactly solvable sector of string theory in $AdS_5\times S^5$.

String theory near the stack of D3-branes, or more precisely strings living in the near horizon limit of the geometry generated by the stack of D3-branes, is claimed to be a dual description of the $\mathcal{N}=4$ Super Yang-Mills theory living on the stack of D3-branes. This is the famous AdS/CFT duality that was conjectured by Maldacena \cite{Maldacena:1997re} and later made precise in \cite{Witten:1998qj, Gubser:1998bc}. Since the near horizon limit of the D3-brane solution is $AdS_5\times S^5$ the conjecture states that string theory living on this space is dual to the four-dimensional conformal field theory $\mathcal{N}=4$ Super Yang-Mills. 

However it has not yet been possible to quantize the string on $AdS_5\times S^5$. So being an exactly solvable limit of this space, strings on a pp-wave provide an important setting where the conjecture can be tested. The dual description of the pp-wave string is the so called BMN sector \cite{Berenstein:2002jq}, which will not be described here.

\section{Tensionless string in a pp-wave background}
Now let us turn to the main issue of this chapter, the tensionless superstring
in a pp-wave background. We review the results of article [II].

To find the action for a closed tensionless string in a pp-wave background we follow
the same lines as we did in the previous chapter when we found an action for
the massless particle and for the tensionless string in a flat Minkowski
background.  Thus, we start from the action for the ordinary tensile string in a pp-wave background, given in \eqref{q:original_action}. We then integrate out $\gamma_{\alpha\beta}$ to put the action in Nambu-Goto form, from which we calculate the conjugate momenta to the fields and note that there are two bosonic constraints. The fermionic momenta are found to be completely fixed by the theory, and hence considered as constraints. When constructing the naive Hamiltonian we find it to be zero, just as before. We incorporate the constraints into the Hamiltonian by the use of four Lagrange multipliers. Then, we construct the phase space Lagrangian and integrate out all momenta. This procedure puts the Lagrangian in a form in which it is possible to take the tensionless limit $T\rightarrow 0$. Taking this limit gives the action for the tensionless string in a pp-wave background as,
\begin{align}
S = \frac{1}{2}\int\d^2\sigma\; V^\alpha V^\beta \Big\{&
  2\partial_\alpha X^+ \partial_\beta X^- + \partial_\alpha X^I \partial_\beta X_I\cr
&+ 2\i \partial_\alpha X^+\left(\theta^1 \bar{\gamma}^-\partial_\beta\theta^1
                             +   \theta^2 \bar{\gamma}^-\partial_\beta\theta^2\right)\cr
&+\left(\mu^2 X^I X_I + 4\i\mu \theta^1\bar{\gamma}^-\Pi\theta^2 \right)
           \partial_\alpha X^+\partial_\beta X^+ \Big\}. \label{q:tensionless_initial_action}
\end{align}
This action is the generalization of the action \eqref{tl:tensionless_action} that we found in the previous chapter for a bosonic tensionless string in flat space. Further, \eqref{q:tensionless_initial_action} is invariant under world-sheet reparametrizations where $V^\alpha$ transforms as a vector density
\begin{align}
\delta(\epsilon) X^\mu &= \epsilon^\alpha \partial_\alpha X^\mu,\\
\delta(\epsilon) \theta^{A a} &= \epsilon^\alpha \partial_\alpha \theta^{A a},\\
\delta(\epsilon) V^\alpha &= -V^\beta \partial_\beta \epsilon^\alpha 
                             + \epsilon^\beta\partial_\beta V^\alpha 
                             + \frac{1}{2}\partial_\beta \epsilon^\beta V^\alpha.
\end{align}
We now use this reparametrization invariance to go to the transverse gauge, $V^\alpha=(v,0)$ \cite{Isberg:1993av}. After fixing this gauge there is still a residual symmetry left which allows us to go to the light-cone gauge, $X^+=\frac{p^+}{v^2}\tau$. The action now becomes
\begin{align}\label{q:tensionless_action}
S=\int\d^2\sigma \Big\{&
  \frac{v^2}{2}\left(\dot{X}^I\dot{X}_I-m_0^2 X^I X_I\right)\cr
&+ \i p^+ \left(\theta^1 \bar{\gamma}^-\dot{\theta}^1 
    + \theta^2 \bar{\gamma}^-\dot{\theta}^2 
    - 2 m_0 \theta^1\bar{\gamma}^-\Pi\theta^2 \right)\Big\},
\end{align}
where $m_0=\frac{\mu p^+}{v^2}$. The most remarkable feature of this action is that it does not contain any derivatives with respect to $\sigma$. This is a sign of the fact that the tensionless string almost behaves as a collection of point particles moving on light like world-lines that together make up the world-sheet. Note also that this action is just the tensile action \eqref{q:tensile_action} with $T\rightarrow v^2$ and all $\sigma$-derivatives of the fields taken to be zero. More precisely, the limit to take in the action \eqref{q:tensile_action} to obtain \eqref{q:tensionless_action} is $T\rightarrow0$ while keeping $v^2=\frac{T}{\tau}$ fixed. In particular, a result of this limit is that the $\sigma$-derivatives drop out. This can be seen as an indication that the tensionless string effectively behaves as a collection of point particles moving along null-geodesics \cite{Karlhede:1986wb}.

While fixing the gauge simplifies the action we must not forget that the variation of the $V^\alpha$ gives equations of motions that need to be taken as constraints if the gauge fixed action is to be equivalent with the initial one. The the constraints arising from the equations of motion for $V^\alpha$ written in light-cone gauge are
\begin{align}\label{q:virasoro_1}
2 p^+\dot{X}^- +v^2\dot{X}^I\dot{X}_I - v^2 m_0^2 X^I X_I \hspace*{3.5cm}&\cr
   + 2\i p^+ \left(
    \theta^1 \bar{\gamma}^-\dot{\theta}^1 
    + \theta^2 \bar{\gamma}^-\dot{\theta}^2 
    - 2 m_0 \theta^1\bar{\gamma}^-\Pi\theta^2 \right) =& 0,\\
p^+ X^{-\prime} + v^2 X'_I \dot{X}^I 
    +\i p^+\left(\theta^1 \bar{\gamma}^-\theta^{1\prime} 
    + \theta^2 \bar{\gamma}^-\theta^{2\prime} \right) =& 0,\label{q:virasoro_2}
\end{align}
where primes denote derivatives with respect to $\sigma$. Peeking forward to
the equations of motion for the fermions \eqref{q:eom_ferm} we note that in
\eqref{q:virasoro_1} the parenthesis containing the fermions vanishes on shell.
The condition can then be solved to express $X^-$ in terms of the transverse
coordinates. Later we will see that the second constraint \eqref{q:virasoro_2} 
produces the level matching conditions. Note also that this is the only place in
the theory where $\sigma$ derivatives of fields enter.

The equations of motion for the bosonic fields that follow from \eqref{q:tensionless_action} is
\begin{align}
\ddot{X}^I+m_0^2 X^I = 0.
\end{align}
Letting $\sigma\in[0,1]$, taking into account the closed string boundary condition
$X^I(\sigma +1,\tau)= X^I(\sigma,\tau)$ and writing the general solution in a
form that resembles the tensile case as much as possible we find
\begin{align}
 X^I(\sigma,\tau) &= x_0^I\cos(m_0\tau) +\frac{p_0^I}{m_0 v^2} \sin(m_0\tau) \cr
& +\frac{\i}{m_0}\sum_{n\neq 0}\sign(n)\e^{-\i\,\sign(n)m_0\tau} 
     \left(\alpha^{1I}_n\e^{2\pi\i n\sigma} +\alpha^{2I}_n \e^{-2\pi\i n\sigma}\right).
\end{align}

The equations of motion for the fermionic fields are found to be
\begin{align}\label{q:eom_ferm}
\dot{\theta}^1-m_0\Pi\theta^2=0,\;\;\;\;\;\dot{\theta}^2+m_0\Pi\theta^1=0.
\end{align}
Solving these while taking the closed string boundary conditions into account
produces the solutions
\begin{align}
\theta^1(\sigma,\tau)=& \theta^1_0\cos(m_0\tau) +\Pi\theta^2_0\sin(m_0\tau)\cr
&+\frac{1}{\sqrt{2}}\sum_{n\neq0}\e^{-\i\,\sign(n)m_0\tau}
  \left(\theta^1_n \e^{2\pi\i n\sigma} +\i\Pi\theta^2_n \e^{-2\pi\i n\sigma} \right)\\
\theta^2(\sigma,\tau)=& \theta^2_0\cos(m_0\tau) -\Pi\theta^1_0\sin(m_0\tau)\cr
& +\frac{1}{\sqrt{2}}\sum_{n\neq0}\e^{-\i\,\sign(n)m_0\tau}
  \left(\theta^2_n \e^{2\pi\i n\sigma} -\i\Pi\theta^1_n \e^{-2\pi\i n\sigma} \right).
\end{align}
Having the general solutions, we proceed in the same manner as for the tensile string by looking at the momenta. For the bosonic fields we find $P_I=v^2\dot{X}_I$ and for the fermionic fields $\pi^A=-\i p^+\theta^A \bar{\gamma}^-$, for $A = 1,2$. This means that for the fermionic fields we need, as in the tensile case, to use the Poisson-Dirac bracket when quantizing.

In line with the tensile case we introduce new mode operators defined by
\begin{align}
\begin{array}{lll}
a^I_0 =\sqrt{\frac{v^2}{2m_0}}\left(\frac{p^I_0}{v^2}-\i m_0 x_0^I\right),& \hspace{0.5cm} &
\bar{a}^I_0 = \sqrt{\frac{v^2}{2m_0}}\left(\frac{p^I_0}{v^2}+\i m_0 x_0^I\right),\\
\theta_0 = \sqrt{\frac{p^+}{2}}\left(\theta^1_0 - i\theta^2_0\right),&&
\bar{\theta}_0 = \sqrt{\frac{p^+}{2}}\left(\theta^1_0 + i\theta^2_0\right), \\
a^{AI}_n = \sqrt{\frac{2 v^2}{m_0}} \alpha^{AI}_n, &&
\eta^{Aa}_n = \sqrt{2 p^+} \theta^{Aa}_n.
\end{array}
\end{align}
With these definitions the (anti)commutators for the new mode operators are found to be
\begin{align}
\begin{array}{lll}
[a_0^I,\bar{a}_0^J] = \delta^{IJ}, &\hspace{0.5cm}&
[a^{A I}_m, a^{B J}_{n}] 
     = \sign(m)\delta_{m+n,0}\delta^{IJ}\delta^{AB},\\
\{\eta^a_0, \bar{\eta}^b_0\} = \frac{1}{2}(\gamma^+)^{ab},&&
\{\eta_m^{A a},\eta_{n}^{B b}\}
 = \frac{1}{2}(\gamma^+)^{ab} \delta_{m+n,0}\delta^{IJ}\delta^{AB},
\end{array}
\end{align}
for $m,n \neq0$ and $A,B = 1,2$. This means that the oscillators $\bar{\alpha}^I_0$, $\bar{\eta}^a_0$, $a^{A I}_{-n}$ and $\eta_{-n}^{A a}$ with $n>0$ are creation operators and $a^I_0$, $\eta^a_0$, $a^{A I}_{n}$ and $\eta_{n}^{A a}$ with $n>0$ are annihilation operators. The vacuum of the theory is defined as the state annihilated by all annihilation operators and the Fock space of the theory is built by acting on the vacuum with the creation operators. To get the physical spectrum we must impose a level matching condition. Expressing \eqref{q:virasoro_2} in terms of the modes and integrating over $\sigma$ produce the level matching condition
\begin{align}
N^1|phys\rangle = N^2|phys\rangle, \;\;\;\;
N^{A} = \sum_{n=1}^{\infty}n\left(a^{A I}_{-n}a^{A }_{nI}
                        + \eta^{A}_{-n}\bar{\gamma}^-\eta^{A}_n \right).
\end{align}
Note that this is the same level matching condition as found in the tensile case \eqref{q:tensile_lvlmatch}.

The operator that gives the energies of the states is the light-cone Hamiltonian and in terms of the modes it is given by
\begin{align}
H_{LC} =& m_0 \left(4 + \bar{a}^I_0 a_{I0}  
                    + 2 \bar{\eta}_0 \bar{\gamma}^- \Pi \eta_0 \right)  \cr
        & + m_0 \sum_{n=1}^{\infty}\left\{
             a^{1I}_{-n}a^{1}_{nI} + a^{2I}_{-n}a^{2}_{nI} 
           +  \eta^1_{-n}\bar{\gamma}^-\eta^1_n 
           +  \eta^2_{-n}\bar{\gamma}^-\eta^2_n
           \right\}.
\end{align}
We have normal ordered the expression and just as in the tensile case the only normal ordering constant that does not cancel comes from the bosonic zero mode part. Note the absence of the factor $n$ in the sum. This means that, \eg for each $n$ the state $a^{1I}_{-n}a^{2J}_{-n}|0\rangle$ is a state in the spectrum with energy  independent of $n$, implying that the spectrum is infinitely degenerated at this energy level. More generally, every energy level is infinitely degenerated as soon as the energy level contains states built by $a^{A I}_{-n}$ or $\eta^{Aa}_{-n}$. The only energy levels that does not have an infinite degeneracy are the vacuum, $a^I_0|0\rangle$ and $\eta^a_0|0\rangle$.

In article [II] we show that the above spectrum also arises if we take a specific limit directly in the quantized theory of the ordinary string. Hence this limit is called the tensionless limit. This means that when we want to study the tensionless string in the pp-wave background we can choose to take the tensionless limit in the classical theory or we can take the limit after the theory has been quantized. The following diagram commutes.
\begin{align*}
\begin{array}{ccc}
\mbox{Classical tensile string} & \longrightarrow & \mbox{Quantized tensile string}\\
\downarrow && \downarrow\\
\mbox{Classical tensionless string} & \longrightarrow & \mbox{Quantized tensionless string}\\
\end{array}
\end{align*}
Since $T=(2\pi\alpha')^{-1}$, the tensionless limit is the same as taking $\alpha'\rightarrow\infty$. Thus, we might expect the tensionless limit to be well behaved, in the sense that the limit commutes with quantization, if the background is a solution of the string theory to all orders in $\alpha'$, \ie not only valid for small $\alpha'$. This is the case for the pp-wave background since it is an exact solution to type IIB string theory, see \eg \cite{Sadri:2003pr}. The flat Minkowski space is also an exact solution, however, there is one important difference between the pp-wave and flat Minkowski space. For the string in a pp-wave there are two scales, one given by the string tension $T$ and one given by the inverse curvature of the space $\mu$, while for the string in flat Minkowski space there is only the one, given by $T$. This means that when we take the tensionless limit of the string in the pp-wave background we still have one energy scale left to relate to, which is obviously not the case for the string in flat space. The tensionless limit in a flat Minkowski background is in some sense less controlled.

Another observation is that the tensile theory on a pp-wave may be written in a form so that the dependence on the string tension appears in the combination $m=\frac{\mu p^+}{T}$, meaning that the tensionless limit is in some sense equivalent to taking $\mu\rightarrow\infty$, \ie to taking the curvature of the space to infinity. As mentioned in the last chapter, the tensile string close to a space-time singularity behaves as a tensionless string \cite{deVega:1994hu}. In \cite{Metsaev:2002re}, Metsaev and Tseytlin comment on this infinite curvature limit and note that the theory is well behaved in the limit and that the spectrum gets infinitely degenerated. They also comment that $\mu$ can be viewed as a regulator to get a non-trivial tensionless limit of strings in flat space. However, in [II] it is shown that the same spectrum indeed arises from an action describing the tensionless string in a pp-wave background.

        \chapter{Extended supersymmetry and geometry}\label{extended_susy}
In this chapter we consider extended world-sheet supersymmetry for the closed string. We will find that requiring more than one supersymmetry in each of the left and right moving sectors of the closed string imposes conditions on the geometry in which the string lives. To study and interpret the requirements of extended supersymmetry we introduce elements of complex geometry and generalized complex geometry. We will also construct the Hamiltonian formulation of the string sigma model and investigate the requirements of extended supersymmetry in this setting. 

The motivation to why we study the relation between extended supersymmetry and geometry is, apart from gaining a deeper understanding of string theory, to gain insight into the classification of the different target spaces allowed for a string with different amount of extended supersymmetry.

\section{The $\mathcal{N}=(p,q)$ formulation}
In flat space-time, the bosonic and fermionic fields on the world-sheet of a closed string separate into two uncoupled parts, one left and one right moving part. This enables us to have different amount of supersymmetry in the left moving sector and in the right moving sector. The notation is $\mathcal{N}=(p,q)$, where $p$ and $q$ denotes the number of different supersymmetry transformations in the left and right moving sector respectively.

The action for a string with world-sheet supersymmetry is given in section \ref{sec:world-sheet-susy} as \eqref{intro:ws_susy_off}. In conformal gauge, having removed the auxiliary field $F^\mu$ and for ease of notation set $4\pi\alpha'=2$, the action \eqref{intro:ws_susy_off} reads
\begin{align}\label{action:N=11SUSY}
S=- \frac{1}{2}\int \d^2 \sigma\left\{\partial_\a X^\mu \partial^\a X_\mu 
          -\i \bar{\psi}^\mu \rho^\a \partial_\a \psi_\mu \right\}.
\end{align}
In this action the world-sheet fermion fields $\psi^\mu$ are two-component Majorana world-sheet spinors and $\rho^\a$ are the two-dimensional Dirac matrices. We now use the specific basis for the Dirac matrices \eqref{intro:2dgamma} and the definition of the components of the fermion field introduced in \eqref{intro:Psi_components}. We also use world-sheet light-cone coordinates, $\sigma^\ppmm=\tau \pm\sigma$, which imply $\partial_\ppmm=\frac{1}{2}(\partial_\tau \pm \partial_\sigma)$. The equation of motion for the coordinate fields is the wave equation, so a general solution may be written as $X^\mu(\sigma,\tau)=X^\mu_\pp(\sigma^\pp)+X^\mu_\mm(\sigma^\mm)$. Using the above we find that the action written in terms of the components is 
\begin{align}
S= \int \d^2\sigma \left\{
        \partial_\pp X^\mu \partial_\mm X_\mu 
        +\i \psi^\mu_{+}\partial_\mm\psi_{+\mu}
        +\i \psi^\mu_{-}\partial_\pp\psi_{-\mu}
         \right\}. \label{susy:comp_action}
\end{align}
Varying $\psi_\pm$ we find that the equations of motion for the fermion fields, as given in \eqref{intro:eoms_fermion}, dictate that $\psi^\mu_\pm = \psi^\mu_\pm(\sigma^\ppmm)$. Hence the two-component spinors $\psi^\mu$ split into one left and one right moving part.

As discussed in section \ref{sec:world-sheet-susy} the action \eqref{action:N=11SUSY} is invariant under the supersymmetry transformation given by
\begin{align}
\delta X^\mu &= \bar{\varepsilon}\psi^\mu, \\
\delta \psi^\mu &= -\i\rho^\alpha\partial_\alpha X^\mu \varepsilon,
\end{align}
where the parameter $\varepsilon$ is a Grassmann valued Majorana spinor with components $\varepsilon=(-\varepsilon_+,\varepsilon_-)^t$.

Writing the supersymmetry transformation in components we find for the $X^\mu$ part
$\delta X^\mu = \i\varepsilon_- \psi_-^\mu +\i\varepsilon_+\psi_+^\mu$. Since
$\psi^\mu_\pm$ only depend on $\sigma^\ppmm$, and 
$\delta X^\mu$ can be written as $\delta X^\mu_\pp(\sigma^\pp) +\delta X^\mu_\mm(\sigma^\mm)$, we find that the transformation separates into two parts
\begin{align}
&\left\{\begin{array}{l}
\delta X_\pp^\mu =  \i\varepsilon_+\psi_+^\mu\\
\delta \psi_+^\mu = -\varepsilon_+\partial_\pp X^\mu_\pp
\end{array}\right.\\
&\left\{\begin{array}{l}
\delta X_\mm^\mu = \i\varepsilon_- \psi_-^\mu \\
\delta \psi_-^\mu = -\varepsilon_-\partial_\mm X^\mu_\mm
\end{array}\right.
\end{align}
one left moving $(+,\pp)$ and one right moving $(-,\mm)$ supersymmetry. We have two identical and independent copies of the supersymmetry transformations. In this sense it is possible for the closed string in flat space to have different amount of supersymmetry in the left and right moving sectors. The case considered above is denoted by $\mathcal{N}=(1,1)$ and the same kind of reasoning applies to the general case of $\mathcal{N}=(p,q)$ supersymmetry. Whenever $p>1$ or $q>1$ we say that we have extended supersymmetry.

\section{A non-linear sigma model}

A sigma model is defined as a set of $D$ maps $\Phi^\mu$, $\mu=0,...,D-1$, from a parameter space $\Sigma$ into a $D$-dimensional target space $M$,
\begin{align}
\Phi^\mu:\Sigma \rightarrow M,
\end{align}
and an action giving the dynamics of the model. For a non-linear sigma model the action is non-linear. We let the $d$-dimensional parameter space have coordinates $\sigma^\alpha$, $\alpha=0,...,d-1$. The maps  $\Phi^\mu=\Phi^\mu(\sigma^\alpha)$ embed the parameter space into the target space. For the string, the parameter space $\Sigma$ is the two-dimensional world-sheet and the target space $M$ is the $D$-dimensional space-time in which the string lives.

We will use the manifest formulation of $\mathcal{N}=(1,1)$ supersymmetry and consider, as in section \ref{sec:world-sheet-susy}, the world-sheet as a two-dimensional superspace. Here, however, we chose a basis for the Grassmann odd coordinates as $\theta^\pm$. In terms of these coordinates the  supersymmetry generators $Q_\pm$ are given by
\begin{align}
Q_\pm = \i \frac{\partial}{\partial\theta^\pm} + \theta^\pm \partial_\ppmm .
\end{align}
We further introduce spinor derivatives $D_\pm$ that obey the relations
\begin{align}
D^2_\pm = \i \partial_\ppmm, \;\;\;\;\;
\{D_+,D_-\} = 0, \;\;\;\;\;
Q_\pm = \i D_\pm + 2\theta^\pm \partial_\ppmm .\label{susy:N11_odd_derivative}
\end{align}

The non-linear sigma model that we will study in this chapter is defined by the coordinate $\mathcal{N}=(1,1)$ superfields $\Phi^\mu$ and the second order action
\begin{align}\label{susy:S_second_order}
S_{2nd} = \int \d^2 \sigma \d^2 \theta \; D_+\Phi^\mu D_-\Phi^\nu (E^{-1}(\Phi))_{\mu\nu}.
\end{align}
Here, $E^{-1}$ is the sum of the background fields $(E^{-1})_{\mu\nu}=g_{\mu\nu}+B_{\mu\nu}$. Observe that both the metric and the $B$-field depend on the position in space-time, hence $E$ depends on the coordinates $\Phi^\mu$. By defining the components of the superfield as 
\begin{align}
X^\mu  \equiv \Phi^\mu|_{\theta^\pm=0},\;\;\;\;
\psi^\mu_\pm \equiv D_+ \Phi|_{\theta^\pm=0}, \;\;\;\;
F^\mu \equiv D_+ D_- \Phi|_{\theta^\pm=0}
\end{align}
and using the integration measure for the spinor coordinates 
\begin{align}
\int \d^2\theta (\cdot) = D_+ D_- (\cdot)|_{\theta^\pm=0}
\end{align}
we find the component form of the action \eqref{susy:S_second_order}. When we specify the target space to be Minkowski without a $B$-field and integrate out the $F$-field we recover the action \eqref{susy:comp_action}.

\section{Extended supersymmetry}\label{susy:sec:extended}
In 1984 the second order action \eqref{susy:S_second_order} was found to admit extended $\mathcal{N}=(2,2)$ supersymmetry under the condition that the target manifold is a complex bi-Hermitean manifold \cite{Gates:1984nk}. In the following we will sketch how the restrictions on the target space arise.

We start with the manifestly $\mathcal{N}=(1,1)$ action \eqref{susy:S_second_order}, and write $(E^{-1})_{(\mu\nu)}=g_{\mu\nu}$ and $(E^{-1})_{[\mu\nu]} = B_{\mu\nu}$,
\begin{align}
S_{2nd} = \int \d^2 \sigma \d^2 \theta \; D_+\Phi^\mu D_-\Phi^\nu 
             \left(g_{\mu\nu}+B_{\mu\nu}\right). \label{susy:S_2nd_order_explicit}
\end{align}
Note that, even though $B$ is used explicitly to write the action, the theory depends only on the closed 3-form field strength $H=\d B$. This can be seen by reducing the action to its non-manifest form and noting that the part of the reduced action that contains the $B$-field explicitly, is a Wess-Zumino term. Using Stoke's theorem this term may be written as
\begin{align}
\int_{\Sigma_2= \partial\Sigma_3} X^*(B) = \int_{\Sigma_3} X^*(\d B) = \int_{\Sigma_3} X^*(H),
\end{align}
where $X^*(B)$ denotes the pull-back of $B$ to the world-sheet by the bosonic component of the superfield $\Phi$ and $\Sigma_3$ is a three-dimensional manifold whose boundary is the two-dimensional world-sheet. Thus, written in terms of $H$ the invariance $B\rightarrow B+\d \Lambda$, for some one-form $\Lambda$, is manifest and hence the theory only depends on $H=\d B$. It is also possible to consider cases when $H$ is closed but not exact. In local coordinates $H$ is given explicitly by $H_{\rho\s\nu}=\smallhalf(B_{\rho\s,\nu}+B_{\s\nu,\rho}+B_{\nu\rho,\s})$.

Since the action \eqref{susy:S_2nd_order_explicit} is written in terms of $\mathcal{N}=(1,1)$ superfields it is manifestly invariant under the $\mathcal{N}=(1,1)$ transformations
\begin{align}
\delta^\pm_1(\epsilon^\pm) \Phi^\mu &= -\i\epsilon^\pm Q_\pm \Phi^\mu, \label{susy:N11_manifest_transfn}
\end{align}
where $\epsilon^\pm$ are two independent constant Grassmann valued transformation parameters and $Q_{\pm}$ are the generators for supersymmetry transformation in the left respective right sector of the theory. These transformations satisfy the algebra
\begin{align}
\left[\delta^\pm_1 (\epsilon^\pm_1) , \delta^\mp_1(\epsilon^\pm_2)\right]\Phi^\mu =& 0,\\
\left[\delta^\pm_1 (\epsilon^\pm_1) , \delta^\pm_1(\epsilon^\pm_2)\right]\Phi^\mu =& 
-2\i\epsilon^\pm_1 \epsilon^\pm_2 \partial_\ppmm \Phi^\mu. \label{susy:N11_manifest_transfn2}
\end{align}

We next make an ansatz for the extended supersymmetry. By dimensional arguments the ansatz is unique and reads
\begin{align}
\delta_2^\pm(\epsilon^\pm) \Phi^\mu &= \epsilon^\pm D_\pm\Phi^\nu J^{(\pm)\mu}_\nu. \label{susy:N11_extended_susy_transfn}
\end{align}
Demanding that this transformation is a supersymmetry transformation will give conditions on the undetermined tensors $J^{(\pm)\mu}_\nu$. As we shall see in the following, it is possible to interpret these conditions geometrically.

Fist we must require that the transformation \eqref{susy:N11_extended_susy_transfn} is a symmetry of the action \eqref{susy:S_2nd_order_explicit}. The invariance of the action requires that that the two $J^{(\pm)\mu}_\nu$ satisfy
\begin{align}
J^{(\pm)\rho}_{\mu} g^{\ph{(\pm)\rho}}_{\rho\nu} = - J^{(\pm)\rho}_{\nu} g^{\ph{(\pm)\rho}}_{\rho\mu}, \label{susy:bi-herm-cond}
\end{align}
and that 
\begin{align}
\nabla^{(\pm)}_\s J^{(\pm)\mu}_\nu = J^{(\pm)\mu}_{\nu,\s} 
        +\Gamma^{(\pm)\mu}_{\s\rho}J^{(\pm)\rho}_{\nu}
        -\Gamma^{(\pm)\rho}_{\s\nu}J^{(\pm)\mu}_{\rho} = 0. \label{susy:cov_deriv_Jpm}
\end{align}
The last condition means that the $J^{(\pm)}$'s are covariantly constant with respect to two different torsionful connections defined by $\Gamma^{(\pm)\mu}_{\nu\s}=\Gamma^{(0)\mu}_{\nu\s}\pm g^{\mu\rho}H_{\rho\nu\s}$. Here, $\Gamma^{(0)}$ is the Levi-Civit\'{a} connection and $H$ is the closed 3-form field strength defined above.

Secondly we must require that the transformation commute with the transformation for the manifest supersymmetry
\begin{align}
\left[\delta_1^\pm(\epsilon_1^\pm), \delta_2^\pm(\epsilon_2^\pm)\right] \Phi^\mu &= 0, \label{susy:12_comm:1}\\
\left[\delta_1^\pm(\epsilon_1^\pm), \delta_2^\mp(\epsilon_2^\mp)\right] \Phi^\mu &= 0, \label{susy:12_comm:2}
\end{align}
that the transformations in the left and right sectors commute
\begin{align}\label{susy:2pm_comm}
\left[\delta_2^\pm(\epsilon^\pm_1),\delta_2^\mp(\epsilon^\mp_2)\right]\Phi^\mu = 0,
\end{align}
and that the supersymmetry algebra closes to a translation on the world
sheet, \ie obey the same algebra as the manifest supersymmetry \eqref{susy:N11_manifest_transfn2},
\begin{align}\label{susy:2pp_comm}
\left[\delta_2^\pm(\epsilon^\pm_1),\delta_2^\pm(\epsilon^\pm_2)\right]\Phi^\mu 
      = -2\i\epsilon^\pm_1 \epsilon^\pm_2 \partial_\ppmm \Phi^\mu.
\end{align}

When these commutators are calculated we find that the conditions
\eqref{susy:12_comm:1} and \eqref{susy:12_comm:2} are satisfied identically since the second supersymmetry transformation is written in terms of $\mathcal{N}=(1,1)$ superfields. The commutators \eqref{susy:2pm_comm} and \eqref{susy:2pp_comm} are found to be
\begin{align}
\big[\delta_2^\pm(\epsilon^\pm_1),\delta_2^\pm(&\epsilon^\pm_2)\big] \Phi^\mu = 
  2\epsilon^\pm_1\epsilon^\pm_2 D^2_\pm\Phi^\alpha J^{(\pm)\mu}_\rho J^{(\pm)\rho}_\alpha \cr
& + \epsilon^\pm_1\epsilon^\pm_2 D_\pm \Phi^\alpha D_\pm \Phi^\beta \Big(
      J^{(\pm)\rho}_{~~[\beta}J^{(\pm)\mu}_{~~\alpha],\rho}
    - J^{(\pm)\mu}_{~~\rho} J^{(\pm)\rho}_{~~[\alpha,\beta]}
    \Big),\label{susy:calculated_2pp_comm}\\
\big[\delta_2^\pm(\epsilon^\pm_1),\delta_2^\mp(&\epsilon^\mp_2)\big] \Phi^\mu =
\epsilon^\pm_1\epsilon^\mp_2 D_\pm D_\mp\Phi^\alpha \Big(
   J^{(\pm)\mu}_{~~\rho} J^{(\mp)\rho}_{~~\alpha} - J^{(\mp)\mu}_{~~\rho} J^{(\pm)\rho}_{~~\alpha}
   \Big)\cr
&+\epsilon^\pm_1\epsilon^\mp_2 D_\pm\Phi^\alpha D_\mp\Phi^\beta \Big(
              -J^{(\pm)\nu}_{~~\alpha,\beta} J^{(\mp)\mu}_{~~\nu}
              -J^{(\pm)\rho}_{~~\alpha} J^{(\mp)\mu}_{~~\beta,\rho}\cr
&\ph{+\epsilon^\pm_1\epsilon^\mp_2 D_+\Phi^\alpha D_-\Phi^\beta \Big(}
             +J^{(\mp)\nu}_{~~\beta,\alpha} J^{(\pm)\mu}_{~~\nu}
             +J^{(\mp)\rho}_{~~\beta}J^{(\pm)\mu}_{~~\alpha,\rho}
             \Big).\label{susy:pm_offshellcomm}
\end{align}
It is possible to simplify \eqref{susy:pm_offshellcomm} by the use of the equations
of motion,
\begin{align}
D_+ D_-\Phi^\mu + \Gamma^{(-)\mu}_{\lambda\kappa} D_+\Phi^\lambda D_-\Phi^\kappa = 0.
\end{align}
By doing this we ``go on shell'' meaning that we use properties of the action 
to make the algebra close. Hence, it will not be possible to rewrite the action
in a manifestly invariant way. However, by using the equations of motion and the conditions \eqref{susy:cov_deriv_Jpm} we find that the commutators \eqref{susy:pm_offshellcomm} vanish.

We find that we must impose conditions on the $J^{(\pm)}$ tensors so that the algebra behaves as we want it to. To understand these conditions we will make a
small detour to discuss some properties of complex geometry.

\section{Complex geometry}
In this section we will introduce some elements of complex geometry. For a more 
thorough introduction to the subject see \eg \cite{Nakahara:1990th}.

An almost complex structure is defined as a map from the tangent space of a complexified manifold $M$ at a point $p$ to itself, $J:T_p M^{\mathbb{C}} \rightarrow T_p M^{\mathbb{C}}$, that obeys $J^2=-1$. A manifold $M$ that can be equipped with an almost complex structure is called an almost complex manifold. 

The property that $J$ squares to minus one implies that $J$ has two eigenvalues $\pm\i$. This gives us the possibility to divide the tangent space $T_p M^\mathbb{C}$ into two disjoint vector subspaces,
\begin{eqnarray}
T_p M^\mathbb{C} = T_p M^+ \oplus T_p M^-,
\end{eqnarray}
where 
\begin{eqnarray}
T_p M^\pm = \{Z\in T_p M^\mathbb{C} \;|\; J Z = \pm \i Z \}.
\end{eqnarray}
Now, every vector $X$ in the tangent space can be written as a linear combination
$X=X_1 + \bar{X}_2$, where $X_1\in T_p M^+$ and $\bar{X}_2\in T_p M^-$. 
We define the projectors $\mathcal{P}^\pm: T_p M^\mathbb{C} \rightarrow T_p M^\pm$ as
\begin{eqnarray}\label{susy:defproj}
\mathcal{P}^\pm = \smallhalf (1 \mp \i J)
\end{eqnarray}
and say that a distribution $T_p M^{\pm}$ defined by $\mathcal{P}^\pm$ is integrable if
\begin{eqnarray}
X,Y\in T_p M^\pm \;\;\mbox{implies}\;\; [X,Y] \in T_p M^\pm,
\end{eqnarray}
where the bracket is the ordinary Lie bracket. In words, integrability means that the Lie bracket of two vectors in one of the above subspaces lies in the same subspace. If this is the case for a subspace we say that it is involutive under the Lie bracket.

We now use the projectors to write conditions for integrability.
Let $X,Y\in T_p M^\mathbb{C}$,
\begin{eqnarray}\label{susy:integrabilitycond}
\mathcal{P}^{\mp}[\mathcal{P}^{\pm}X,\mathcal{P}^{\pm}Y]=0.
\end{eqnarray}
These conditions are quite natural, we project $X$ and $Y$ to the space $T_p M^\pm$ and take the Lie bracket between the projections. If this bracket is to lie entirely in the space $T_p M^\pm$, it can not have any components in the other space, $T_p M^\mp$, \ie the projection to the space $T_p M^\mp$ must be zero, which is exactly what we demand in \eqref{susy:integrabilitycond}.

Simplifying the conditions \eqref{susy:integrabilitycond} using the definition of the projectors \eqref{susy:defproj} we find 
\begin{eqnarray}
&&\hspace{-43pt} 0 = \mathcal{P}^{\mp}[\mathcal{P}^{\pm}X,\mathcal{P}^{\pm}Y] = 
      \smallhalf (1 \pm \i J)\left[ \smallhalf (1 \mp \i J)X,\smallhalf (1 \mp \i J)Y\right] \cr  
&&= \tsfrac{1}{4} \mathcal{P}^\mp \left( [X,Y] + J[JX,Y] + J[X,JY] - [JX,JY]\right)            \cr
&&= \tsfrac{1}{4} \mathcal{P}^\mp N(X,Y), \label{susy:intcond2}
\end{eqnarray}
where have defined the Nijenhuis tensor $N(X,Y)$ as
\begin{eqnarray}\label{susy:Nijenhuis_def}
N(X,Y) = [X,Y] + J[JX,Y] + J[X,JY] - [JX,JY].
\end{eqnarray}
The two conditions in \eqref{susy:intcond2} together with $N(X,Y)= (\mathcal{P}^+ + \mathcal{P}^-)N(X,Y)$ tell us that $N(X,Y)$ must vanish. The integrability condition thus becomes $N(X,Y)=0$.

It should be noted that in these considerations we do not make use of the component form of the Lie bracket. We will see later that a Nijenhuis type tensor based on more general brackets is also possible to construct.

We now turn to the expression of the Nijenhuis tensor in local coordinates.
In local coordinates the vectors and the $J$ tensor are given by, $X=X^\mu\partial_\mu$, $Y=Y^\mu\partial_\mu$ and $J=J^\mu_\nu \dx^\nu\otimes\partial_\mu$. Further, the Lie bracket between two vectors is given by 
$[X,Y]=(X^\nu\partial_\nu Y^\mu -Y^\nu\partial_\nu X^\mu)\partial_\mu$.
Using the above in the definition of the Nijenhuis tensor \eqref{susy:Nijenhuis_def} we find
\begin{eqnarray}
&&\hspace{-45pt}N(X,Y) = X^\kappa Y^\nu 
                  \left(- J^\mu_\lambda \partial_\nu  J^\lambda_\kappa 
                        + J^\mu_\lambda \partial_\kappa J^\lambda_\nu
                        - J^\lambda_\kappa\partial_\lambda J^\mu_\nu
                        + J^\lambda_\nu \partial_\lambda J^\mu_\kappa \right)\partial_\mu.
\end{eqnarray}
Thus the components of the Nijenhuis tensor are
\begin{eqnarray}
N^\mu_{\kappa\nu} = J^\gamma_\kappa J^\mu_{[\nu,\gamma]} -J^\gamma_\nu J^\mu_{[\kappa,\gamma]},
\end{eqnarray}
where we use antisymmetrization without combinatorial factor, \eg $J^\mu_{[\nu,\gamma]} = J^\mu_{\nu,\gamma}-J^\mu_{\gamma,\nu}$.
In the end we find the that the integrability condition is the same as requiring the vanishing of the tensor $N^\mu_{\kappa\nu}$. 

An almost complex structure $J$ is a complex structure if it defines integrable subspaces, or equivalently its Nijenhuis tensor is zero. If the manifold $M$ admits a complex structure it is a complex manifold.

Two examples of complex manifolds are the complex projective space $\mathbbm{CP}^1$ and the two-dimensional sphere $S^2$. A manifold that admits an almost complex structure is $S^6$, but evidence has been put forward, \cf \cite{Marshakov:2005fn}, that it does not admit a complex structure.

\section{Geometrical interpretation}\label{susy:geom_interpret}

In this section we will explore how the conditions that we found in section \ref{susy:sec:extended} can be given a geometrical interpretation in terms of the above complex structures. The conditions for on-shell closure of the algebra, as follows from \eqref{susy:calculated_2pp_comm}, are given by
\begin{eqnarray}
-1&=& J^{(\pm)\mu}_\rho J^{(\pm)\rho}_\alpha, \label{susy:cond_1}\\
 0&=& J^{(\pm)\rho}_{~~[\beta}J^{(\pm)\mu}_{~~\alpha],\rho} 
    - J^{(\pm)\mu}_{~~\rho} J^{(\pm)\rho}_{~~[\alpha,\beta]}. \label{susy:cond_2}
\end{eqnarray}
The first condition, \eqref{susy:cond_1}, tells us that $J^{(\pm)}$ are two
almost complex structures. Using \eqref{susy:cond_1} to rewrite \eqref{susy:cond_2} yields $N^{(\pm)\mu}_{\beta\alpha}=0$, here $N^{(\pm)}$ is the Nijenhuis tensor built with $J^{(\pm)}$. Hence, for these conditions to be satisfied $J^{(\pm)}$ must be two complex structures on the target space of the sigma model. The condition \eqref{susy:bi-herm-cond} tells us that the metric is Hermitean with respect to both complex structures. Further, the vanishing of the Nijenhuis tensors $N^{(\pm)}=0$ and the conditions \eqref{susy:cov_deriv_Jpm} imply that $J^{(\pm)}$ preserves the torsion \cite{Lyakhovich:2002kc}
\begin{align}
H_{\mu\nu\rho} = J^{(\pm)\lambda}_\mu J^{(\pm)\s}_\nu H_{\lambda\s\rho}
+ J^{(\pm)\lambda}_\rho J^{(\pm)\s}_\mu H_{\lambda\s\nu}
+ J^{(\pm)\lambda}_\nu J^{(\pm)\s}_\rho H_{\lambda\s\mu}.
\end{align}
This relation is often useful when performing explicit calculations.

We conclude that on-shell extended $\mathcal{N}=(2,2)$ supersymmetry in this second order non-linear sigma model requires that the target space must have two covariantly constant complex structures and the metric must be Hermitean with respect to both. This space is called a bi-Hermitean manifold.

If we do not use the equations of motion, the complex structures must obey two
additional conditions coming from \eqref{susy:pm_offshellcomm}.  The extra
conditions for off-shell closure read
\begin{align}
0&= J^{(\pm)\mu}_{~~\rho} J^{(\mp)\rho}_{~~\alpha} - J^{(\mp)\mu}_{~~\rho} J^{(\pm)\rho}_{~~\alpha},
             \label{susy:cond_3}\\
0&= -J^{(\pm)\nu}_{~~\alpha,\beta} J^{(\mp)\mu}_{~~\nu}
             -J^{(\pm)\rho}_{~~\alpha} J^{(\mp)\mu}_{~~\beta,\rho}
             +J^{(\mp)\nu}_{~~\beta,\alpha} J^{(\pm)\mu}_{~~\nu}
             +J^{(\mp)\rho}_{~~\beta}J^{(\pm)\mu}_{~~\alpha,\rho}. \label{susy:cond_4}
\end{align}
The first condition, \eqref{susy:cond_3}, tells us that the two complex structures must commute. Further, if the two complex structures commute, are integrable and covariantly constant \eqref{susy:cov_deriv_Jpm}, the condition \eqref{susy:cond_4} is automatically satisfied \cite{Maes:2006bm}. This off-shell closure means that the model can be written in a manifestly $\mathcal{N}=(2,2)$ supersymmetric way and in \cite{Gates:1984nk} Gates, Hull and Ro\v{c}ek wrote down the manifestly supersymmetric action in terms of chiral and twisted chiral $\mathcal{N}=(2,2)$ superfields.

\section{Generalized complex geometry} \label{susy:GCG}
It turns out that there exist a more natural framework to describe the above 
geometry. This framework is called generalized complex geometry and was introduce by Hitchin \cite{Hitchin:2004ut} and developed by Gualtieri \cite{Gualtieri:2003dx}. In \cite{Gualtieri:2003dx} it is shown that the bi-Hermitean geometry found by Gates, Hull and Ro\v{c}ek \cite{Gates:1984nk} fit nicely into this framework of generalized complex geometry.

One main difference between generalized complex geometry and ordinary complex
geometry is that the generalized formulation incorporates the co-tangent bundle
$T^* M$ in a natural way, while the ordinary complex geometry only considers the tangent bundle. This is encoded in the generalized complex structure which is a map from the complexified direct sum of the tangent bundle and the cotangent bundle, $(T M \oplus T^* M)\otimes\mathbbm{C}$, to itself.

Here, we give a brief introduction to the relevant parts of generalized complex 
geometry. This formulation incorporates the standard complex geometry and the
Poisson geometry as special cases. For a more exhaustive introduction to generalized complex geometry, see \eg \cite{Zabzine:2006uz}.

As mentioned, the construction of generalized complex geometry is based on the complex vector bundle $(TM\oplus T^* M)\otimes\mathbbm{C}$, where $M$ is a manifold of real dimension $d$. A section of the bundle is denoted by $X+\xi\in\Gamma((TM\oplus T^* M)\otimes\mathbbm{C})$, where $X$ is a vector field and $\xi$ is a one-form. The natural pairing on $(TM\oplus T^*M)\otimes\mathbbm{C}$ is given by 
\begin{align}
\langle X+\xi, Y+\eta\rangle = \frac{1}{2}(i_X\eta + i_Y\xi), \label{susy:GCG_pairing}
\end{align} 
where $i_X\eta$ is the contraction of the vector $X$ and the one-form $\eta$ which in local coordinates reads $i_X\eta= X^\mu\eta_\mu$. In the local coordinates $(\partial_\mu, \d x^\mu)$ this natural pairing can be written as
\begin{align}
\langle X+\xi, Y+\eta \rangle = \frac{1}{2}\matrix{cc}{X&\xi}\matrix{cc}{0&\mathbbm{1}\\\mathbbm{1}&0} \matrix{c}{Y\\ \eta},
\end{align}
which defines the matrix
\begin{align}
\genI = \frac{1}{2}\matrix{cc}{0&\mathbbm{1}\\\mathbbm{1}&0}
\end{align}
as a metric on $(TM\oplus T^* M)\otimes\mathbbm{C}$ with signature $(2d,2d)$. Hence elements of $O(2d,2d)$ preserves the natural pairing.

Another important object is the $H$-twisted Courant bracket \cite{courant},
which is a bracket operation on the space $TM\oplus T^*M$. Here, $H$ is a closed three-form, but not the same $H$ as in previous sections. The bracket is given by
\begin{align}
[X+\xi, Y+\eta]_{H} = [X,Y] + \mathcal{L}_X \eta - \mathcal{L}_Y \xi 
                -\tsfrac{1}{2} \d (i_X\eta-i_Y\xi) +i_X i_Y H,
\end{align}
with $X+\xi, Y+\eta\in \Gamma((TM\oplus T^* M)\otimes\mathbbm{C})$. Here, $\mathcal{L}$ is the Lie derivative, $[\,\cdot,\cdot\,]$ is the Lie bracket and $\d$ is the exterior derivative. When $H$ is zero the bracket is simply called the Courant bracket. Note that when the sections do not have components in $T^* M$ and when $H$ is zero, the Courant bracket reduces to the Lie bracket.  In the local coordinates $(\partial_\mu, \d x^\mu)$ the $H$-twisted bracket reads
\begin{align}\label{susy:courant_bracket}
[X+\xi, Y+\eta]_H =& \big\{X^\alpha\partial_\alpha Y^\mu 
                       - Y^\alpha\partial_\alpha X^\mu
                       \big\}\partial_\mu \cr
        & +\big\{X^\alpha\partial_\alpha\eta_\mu 
           +\frac{1}{2}\eta_\alpha\partial_\mu X^\alpha
           -\frac{1}{2}X^\alpha\partial_\mu\eta_\alpha \cr
        & -Y^\alpha\partial_\alpha\xi_\mu 
           -\frac{1}{2}\xi_\alpha\partial_\mu Y^\alpha
           +\frac{1}{2}Y^\alpha\partial_\mu\xi_\alpha \cr
        & - X^\nu Y^\sigma H_{\mu\nu\sigma}\big\} \d x^\mu.
\end{align}
This form of the bracket is useful when doing explicit calculations.

The automorphisms of the $H$-twisted Courant bracket are the diffeomorphisms of $M$ and the $b$-transformations, defined as
\begin{align}
\e^b(X+\xi) = X + \xi + i_X b, \label{susy:b-transform}
\end{align}
where $b$ is a closed two-form. Note that $\e^{-b}$ is the inverse of the $b$-transformation. For non-closed $b$ the $H$-twisted Courant bracket transforms as \cite{Gualtieri:2003dx}
\begin{align}
[\e^b(X+\xi),\e^b(Y+\eta)]_H = \e^b[X+\xi,Y+\eta]_{H+db}. \label{susy:b-transformed_HCourant}
\end{align}

The next important object we introduce is the generalized almost complex structure
$\genJ$. It is an endomorphism of the space $(TM\oplus T^* M)\otimes\mathbbm{C}$,
\begin{align}
\genJ: (TM\oplus T^* M)\otimes\mathbbm{C} \rightarrow (TM\oplus T^* M)\otimes\mathbbm{C}
\end{align}
that satisfies 
\begin{align}
\genJ^2=-\mathbbm{1} \;\;\;\mbox{and}\;\;\; \genJ^t \genI \genJ = \genI. \label{susy:cond_aGCS}
\end{align}
Just as for the ordinary complex structure, the space $(TM\oplus T^*M)\otimes\mathbbm{C}$ can be divided into two parts corresponding to the two the eigenvalues of $\genJ$ as $(TM\oplus T^*M)\otimes\mathbbm{C} = \mathbbm{L}_+ \oplus \mathbbm{L}_-$. We define the projectors $\Pi_{\pm}=\frac{1}{2}(\mathbbm{1}\mp \i\genJ)$ to the $\pm \i$ eigenbundles $\mathbbm{L}_\pm$. If 
\begin{align}
\Pi_\mp[\Pi_\pm(X+\xi),\Pi_\pm(Y+\eta)]_H = 0,\label{susy:GCG_integ_cond_proj}
\end{align}
for any $X+\xi, Y+\eta\in \Gamma((TM \oplus T^* M)\otimes{C})$, the subbundles $\mathbbm{L}_\pm$ are involutive with respect to the $H$-twisted Courant bracket. We say that a $H$-twisted generalized almost complex structure is a $H$-twisted generalized complex structure if it defines $\mathbbm{L}_\pm$ that are involutive with respect to the $H$-twisted Courant bracket. The equation \eqref{susy:GCG_integ_cond_proj} is the integrability condition for a $H$-twisted generalized complex structure.

We next investigate how the integrability condition \eqref{susy:GCG_integ_cond_proj} behaves under the $b$-transformation \eqref{susy:b-transform}, with a non-closed $b$. Let $\mathbbm{L}_+$ again be the $+\i$ eigenbundle of the $H$-twisted generalized complex structure $\genJ$. An element $(X+\xi)\in\mathbbm{L}_+$ is $b$-transformed to $\e^b(X+\xi)$, which is in the $+\i$ eigenbundle of $\e^b\genJ\e^{-b}$, since $\e^b\genJ\e^{-b}\e^{b}(X+\xi)=+\i\e^b(X+\xi)$. Denote the $+\i$ eigenbundle of $\e^b\genJ\e^{-b}$ by $\hat{\mathbbm{L}}_+$. Since $\e^b$ is invertible any element in $\hat{\mathbbm{L}}_+$ can be written as $\e^b(X+\xi)$ for some $(X+\xi)\in\mathbbm{L}_+$. Now consider the $(H-\d b)$-twisted bracket between any two sections of $\hat{\mathbbm{L}}_+$, then \eqref{susy:b-transformed_HCourant} gives $[\e^b(X+\xi),\e^b(Y+\eta)]_{H-\d b} = \e^b[X+\xi,Y+\eta]_H$. Since $\mathbbm{L}_+$ is involutive with respect to the $H$-twisted Courant bracket this implies that $\hat{\mathbbm{L}}_+$ is involutive with respect to the $(H-\d b)$-twisted Courant bracket. The same line of reasoning holds for the $-\i$ eigenbundle. We conclude that, given a H-twisted generalized complex structure $\genJ$, the $b$-transformed structure $\e^b \genJ \e^{-b}$ is integrable with respect to the $(H-\d b)$-twisted Courant bracket. In particular note that if $\d b=0$, the integrability condition remains invariant under the $b$-transformation.

As in the case with the ordinary complex structures the integrability condition \eqref{susy:GCG_integ_cond_proj} can be written in an equivalent form, resembling the definition of the Nijenhuis tensor
\begin{eqnarray}\label{susy:courant_integrability}
&&\hspace*{-1cm}[X+\xi,Y+\eta]_H - [\genJ(X+\xi), \genJ(Y+\eta)]_H \cr 
&&+ \genJ[\genJ(X+\xi), Y+\eta]_H 
+ \genJ[X+\xi, \genJ(Y+\eta)]_H = 0.
\end{eqnarray}

By writing the $H$-twisted generalized complex structure $\genJ$ in local coordinates the above conditions translate into conditions for the components of $\genJ$. In local coordinates $(\partial_\mu,\d x^\mu)$ $\genJ$ reads
\begin{eqnarray}
\genJ = \left(\begin{array}{cc}J & P \\ L & K \end{array}\right),
\end{eqnarray}
where $J = J^\mu_\nu \d x^\nu \otimes \partial_\mu$, $P=P^{\mu\nu}
\partial_\mu\otimes\partial_\nu$, $L= L_{\mu\nu}\d x^\mu\otimes\d x^\nu$ and
$K= K^\mu_\nu \d x^\nu\otimes\partial_\mu$. The condition $\genJ^2=-1$ translates 
into
\begin{align}
J^\mu_\nu J^\nu_\lambda + P^{\mu\nu}L_{\nu\lambda} =& -\delta^\mu_\lambda, \\
J^\mu_\nu P^{\nu\lambda} + P^{\mu\nu}K_\nu^\lambda =& 0, \\
K_\mu^\nu L_{\nu\lambda} + L_{\mu\nu}J^\nu_\lambda =& 0, \\
K_\mu^\nu K_\nu^\lambda + L_{\mu\nu}P^{\nu\lambda} =& -\delta^\lambda_\mu.
\end{align}
Next, the condition of the metric $\genI$ being Hermitean with respect to $\genJ$, $\genJ^t \genI \genJ = \genI$, is in local coordinates equivalent to
\begin{align}
J^\mu_\nu + K^\mu_\nu = 0, \;\;\; P^{\mu\nu} = -P^{\nu\mu}, \;\;\; L_{\mu\nu}= -L_{\nu\mu}.
\end{align}

The condition \eqref{susy:courant_integrability} can be written in local coordinates using \eqref{susy:courant_bracket}. The integrability conditions for the $H$-twisted generalized complex structure, in terms of its components, read \cite{Lindstrom:2004iw}
\begin{align}
0=&\; J^\nu_{[\lambda}J^\mu_{\rho],\nu} + J^\mu_\nu J^\nu_{[\lambda,\rho]} 
+ P^{\mu\nu}\left(L_{[\lambda\rho,\nu]} + J^\sigma_{[\lambda}H_{\rho]\sigma\nu}\right), \label{susy:J_comp_integrability_1}\\
0=&\; P^{[\mu|\nu}P^{|\lambda\rho]}_{\ph{|\lambda\rho]},\nu},\\
0=&\; J^\mu_{\nu,\rho}P^{\rho\lambda} + P^{\rho,\lambda}_{\ph{\rho,\lambda},\nu}J^\mu_\rho
-J^\lambda_{\rho,\nu}P^{\mu\rho} \cr
&+J^\lambda_{\nu,\rho}P^{\mu\rho} - P^{\mu,\lambda}_{\ph{\mu,\lambda},\rho}J^\rho_\nu 
- P^{\lambda\sigma}P^{\mu\rho}H_{\sigma\rho\nu},\\
0=&\; J^\lambda_\nu L_{[\lambda\rho,\gamma]} + L_{\nu\lambda}J^\lambda_{[\gamma,\rho]}
+J^{\lambda}_{\rho}L_{\gamma\nu,\lambda}+ J^\lambda_{\gamma}L_{\nu\rho,\lambda}\cr 
&+L_{\lambda\rho}J^\lambda_{\gamma,\nu} + J^\lambda_\rho L_{\lambda\gamma,\nu}
+ H_{\rho\gamma\nu} - J^\lambda_{[\rho}J^\sigma_\gamma H_{\nu]\lambda\sigma}.\label{susy:J_comp_integrability_4}
\end{align}
Here we have used the convention that the combinatorial factor is not included
in the antisymmetrization, \eg $A_{[\mu\nu\gamma]} = A_{\mu\nu\gamma} +A_{\nu\gamma\mu} +A_{\gamma\mu\nu} 
-A_{\nu\mu\gamma}-A_{\mu\gamma\nu}-A_{\gamma\nu\mu}$.
To get the integrability conditions based on the non-twisted Courant bracket
we simply set $H=0$ in the above equations.

Note that all above is built in the same way as in the standard case of complex geometry, the only difference is that we here have objects with more components and a different bracket.

There is an equivalent definition of a $H$-twisted generalized complex structure entirely given in terms of geometrical objects.  For this we need to introduce some definitions.
Consider a subbundle $\mathbbm{L} \subset (TM\oplus T^*M)\otimes\mathbbm{C}$. The subbundle $\mathbbm{L}$ is called isotropic if
\begin{align}
\langle A,B \rangle = 0, \;\;\;\;\; \forall A,B\in \Gamma(\mathbbm{L}).
\end{align}
Further, $\mathbbm{L}$ is called maximally isotropic if 
\begin{align}
\langle A,B \rangle = 0,\;\;\; \forall A\in \Gamma(\mathbbm{L})
\;\;\;\;\; \mbox{implies}\;\;\; B\in\Gamma(\mathbbm{L}).
\end{align}
The subbundle $\mathbbm{L}$ is called $H$-twisted Courant involutive if for any $A,B\in\Gamma(\mathbbm{L})$ the bracket $[A,B]_H$ is in $\Gamma(\mathbbm{L})$. 

An $H$-twisted complex Dirac structure is an $H$-twisted Courant involutive maximally isotropic subbundle of $(TM\oplus T^*M)\otimes\mathbbm{C}$.

And finally, an $H$-twisted generalized complex structure is a $H$-twisted complex Dirac structure $\mathbbm{L}_+\subset (TM\oplus T^*M)\otimes\mathbbm{C}$ such that $\mathbbm{L}_+\cap\mathbbm{L}_-=\{0\}$, where $\mathbbm{L}_-$ is the complex conjugate of $\mathbbm{L}_+$.

We now turn to two examples of generalized complex structures which shows that standard complex geometry and symplectic geometry has a natural place in generalized complex geometry.
The endomorphism of $(TM\oplus T^*M)\otimes\mathbbm{C}$ 
\begin{align}
\genJ = \gcgMatrix{-J&0\\0&J^t}
\end{align}
is a generalized complex structure if and only if $J$ is a complex structure \cite{Gualtieri:2003dx}. Similarly, the endomorphism of $(TM\oplus T^*M)\otimes\mathbbm{C}$
\begin{align}
\genJ = \gcgMatrix{0&-\omega^{-1}\\ \omega&0}
\end{align}
is a generalized complex structure if and only if $\omega$ is a closed non-degenerate two-form, \ie a symplectic structure \cite{Gualtieri:2003dx}.

Next, we define a $H$-twisted generalized K\"{a}hler structure as $(\genJ_1,\genJ_2)$, where $\genJ_1$ and $\genJ_2$ are two commuting $H$-twisted generalized complex structures such that $\genG= -\genJ_1\genJ_2$ is a positive definite metric on $TM\oplus T^*M$. $\genG$ is sometimes referred to as the generalized metric. 

As in the case of a generalized complex structure, it is possible to give an equivalent  definition of an $H$-twisted generalized K\"{a}hler structure in terms of subbundles of $(TM\oplus T^*M)\otimes\mathbbm{C}$ \cite{Gualtieri:2003dx}.


A geometry equipped with a $H$-twisted generalized K\"{a}hler structure, \ie $H$-twisted generalized K\"{a}hler geometry, is equivalent to bi-Hermitean geometry \cite{Gualtieri:2003dx}. The explicit map between bi-Hermitean geometry, defined by $(g,B,J^{(+)}, J^{(-)})$, and $H$-twisted generalized Kähler geometry is given by
\begin{align}
\genJ_{1,2} = \frac{1}{2} \matrix{cc}{
     J^{(+)}\pm J^{(-)} & -(\omega_+^{-1} \mp \omega^{-1}_-)\\
     \omega_+ \mp \omega_- & -(J^{(+)t} \pm J^{(-)t}) 
},\label{susy:relation_GHR_GKG}
\end{align}
where $\omega_\pm = g J^{(\pm)}$. Further, $\genJ_{1,2}$ satisfy \eqref{susy:cond_aGCS} and the integrability conditions \eqref{susy:GCG_integ_cond_proj} with $H=\d B$. Equation \eqref{susy:relation_GHR_GKG} defines two $H$-twisted generalized complex structures which together form an $H$-twisted K\"{a}hler structure.

\section{Hamiltonian formulation}\label{susy:Hamiltonian_formulation}
We have seen that bi-Hermitean geometry arises out of the conditions on the $\mathcal{N}=(1,1)$ supersymmetric sigma model to have extended $\mathcal{N}=(2,2)$ supersymmetry. We have also seen that bi-Hermitean geometry is equivalent to $H$-twisted generalized K\"{a}hler geometry. A question that arises is whether there is any way of obtaining $H$-twisted generalized K\"{a}hler geometry, formulated in terms of the generalized structures, directly from the sigma model. This question is answered in paper [IV], the relation between bi-Hermitean geometry and generalized K\"{a}hler geometry corresponds to going from the Lagrangian to the Hamiltonian formulation of the sigma model.

Thus, to see how $H$-twisted generalized K\"{a}hler geometry arises out of the $\mathcal{N}=(1,1)$ sigma model we need the phase space, or Hamiltonian formulation of the model.
We begin by defining the phase space for a closed string as in \cite{Zabzine:2005qf}. To find the phase space we need the world-sheet to be given by $\Sigma= S^{1,1}\times\mathbbm{R}$, where $S^{1,1}$ is a supercircle with bosonic coordinate $\sigma$ and Grassmann odd coordinate $\theta$. The superloop space $\mathcal{L}M$ is the space of maps from the supercircle to the target space $M$, \ie $\mathcal{L}M= \{\Phi:S^{1,1}\rightarrow M\}$. The phase space for the closed string is now given by the cotangent bundle of the superloop space, $\Pi T^*\mathcal{L}M$. The $\Pi$ denotes that the fibers of the bundle has reversed parity, \ie are Grassmann odd. Note that since the world-sheet is parametrized by $\sigma$, $\theta$ and one more bosonic coordinate, parametrizing the $\mathbbm{R}$ part, this formulation has room only for manifest $\mathcal{N}=1$ supersymmetry. Local coordinates on the phase space are given by the $\mathcal{N}=1$ fields $\phi^\mu$ and $S_\mu$, where $\phi^\mu$ is the coordinate fields and $S_\mu$ the conjugate momenta.

As an aside, the phase space with coordinates being the fields $X^\mu$ and and the conjugate momenta $p_\mu$, \ie not superfields, is obtained by the same construction as above but with the difference that the world-sheet is considered as a product of the circle and the real line, $S^1\times \mathbbm{R}$ \cite{Alekseev:2004np}. Here however, we will consider only the phase space with $\mathcal{N}=1$ coordinates.

To find the phase space formulation of the manifestly $\mathcal{N}=(1,1)$ action \eqref{susy:S_second_order} we define the new Grassmann odd coordinates  and covariant derivatives as
\begin{align}
\theta^0 = \frac{1}{\sqrt{2}}\left(\theta^+ - \i\theta^-\right),\;\;\;\;\;
\theta^1 = \frac{1}{\sqrt{2}}\left(\theta^+ + \i\theta^-\right).
\end{align}
Next, we consider $\sigma^0$ as the coordinate that parametrize the $\mathbbm{R}$ part of the world-sheet. Further, we let $\sigma^1$ and $\theta^1$ parametrize the supercircle $S^{1,1}$. Since the $\mathcal{N}=(1,1)$ world-sheet is parametrized by two bosonic and two Grassmann odd coordinates, after the above assignments we still have one Grassmann odd coordinate left, $\theta^0$. To get rid of this unwanted coordinate we reduce the action to a manifestly $\mathcal{N}=1$ form by integrating over $\theta^0$.

To this end we introduce the derivatives
\begin{align}
D_0 = \frac{1}{\sqrt{2}}\left(D_+ + \i D_-\right), \;\;\;\;\;
D_1 = \frac{1}{\sqrt{2}}\left(D_+ - \i D_-\right), \label{susy:def_N1_cov_derivatives}
\end{align}
which satisfy $D^2_0=\i\partial_1$, $D^2_1=\i\partial_1$ and $\{D_0,D_1\}=2\i\partial_0$. To conform with the notation of [IV] we introduce an overall factor $1/2$ in the action \eqref{susy:S_second_order} and write it as
\begin{align}
S = \frac{1}{2} \int\d^2\s\d^2\theta \; (g_{\mu\nu} + B_{\mu\nu})D_+\Phi^\mu D_- \Phi^\nu,\label{susy:N11_start_action}
\end{align}
where we once again has defined the metric and the $B$-field as the symmetric and antisymmetric part of the $E^{-1}$ field. Using \eqref{susy:def_N1_cov_derivatives} we may write the action as
\begin{align}
S= -\frac{1}{4}\int \d^2\sigma &\d\theta^1\d\theta^0\;\Big\{
  2D_0\Phi^\mu D_1\Phi^\nu g_{\mu\nu} \cr
  &+ \big(D_1\Phi^\mu D_1\Phi^\nu - D_0\Phi^\mu
D_0\Phi^\nu\big)B_{\mu\nu} \Big\}. \label{susy:action_used_for_reduction}
\end{align}
Further, we introduce the $\mathcal{N}=1$ fields
\begin{align}
  \phi^\mu \equiv \Phi^\mu|_{\theta^0=0}, \;\;\;\;
  S_\mu \equiv g_{\mu\nu}D_0 \Phi^\nu|_{\theta^0=0}, \label{susy:def_N1_fields}
\end{align}
and use the definitions $D\equiv D_1|_{\theta^0=0}$,  $\partial \equiv \partial_1$ and $\theta \equiv \theta^1$. We perform the integration over $\theta^0$ by use of $\int\d\theta^0(\cdot)=D_0(\cdot)|_{\theta^0=0}$ and obtain the reduced phase space action as
\begin{align}
S = \int\d\s^0 \left(
        \left\{\int\d\s\d\theta \; \i(S_\mu-B_{\mu\nu}D\phi^\nu)\partial_0\phi^\mu \right\}
          - \mathcal{H}\right), \label{susy:phase_space_action}
\end{align}
where the Hamiltonian is given by
\begin{align}
\mathcal{H}=
\frac{1}{2}\int\d\s\d\theta \;\Big(
&   \i\partial \phi^\mu D\phi^\nu g_{\mu\nu}
  + S_\mu DS_\nu g^{\mu\nu}
  + S_\mu D\phi^\rho S_\nu g^{\nu\s}\Gamma^{(0)\mu}_{\s\rho}\cr
& + D\phi^\mu D\phi^\nu S_\rho H_{\mu\nu}{}^\rho
  -\frac{1}{3} S_\mu S_\nu S_\rho H^{\mu\nu\rho} \Big),\label{susy:Hamiltonian}
\end{align}
where locally $H=\d B$ and we have used the metric $g$ to raise the indices of $H$. For a mathematical rigorous derivation of this Hamiltonian see \cite{Malikov:2006rm}.

The Liouville form is obtained from the first term in the phase space action \eqref{susy:phase_space_action} and is given by
\begin{align}
\Theta = \i\int\d\s\d\theta \; (S_\mu-b_{\mu\nu}D\phi^\nu)\delta\phi^\mu
\end{align}
where $\delta$ is the de Rham differential on $\Pi T^* \mathcal{L}M$.
This gives the symplectic structure on the phase space as $\omega=\delta\Theta$,
\begin{align}
\omega = \i \int \d\sigma\d\theta\; \left(
\delta S_\mu \wedge \delta\phi^\mu - H_{\mu\nu\rho}D\phi^\mu \delta\phi^\nu\wedge\delta\phi^\rho
\right).\label{susy:phasespace_symplectic_str}
\end{align}
Next, we define the left and right functional derivatives of a smooth functional $F(S,\phi)$ by 
\begin{align}
\delta F =& \int\d\s\d\theta \left(
   \frac{F \overleftarrow{\delta}}{\delta S_\a} \delta S_\a 
+  \frac{F \overleftarrow{\delta}}{\delta\phi^\a} \delta\phi^\a
\right)\cr
=&\int\d\s\d\theta \left(
  \delta S_\a \frac{\overrightarrow{\delta}F}{\delta S_\a} 
   + \delta\phi^\a \frac{\overrightarrow{\delta} F}{\delta\phi^\a}
\right).
\end{align}
Using this definition, from \eqref{susy:phasespace_symplectic_str} we find the Poisson bracket between two smooth functionals on $\Pi T^*\mathcal{L}M$ as 
\begin{align}
\{F,G\}_H = \i \int \d\s\d\theta \Bigg(&
  \frac{F \overleftarrow{\delta}}{\delta S_\a}
  \frac{\overrightarrow{\delta}G}{\delta\phi^\a}
 -\frac{F\overleftarrow{\delta}}{\delta\phi^\a}
  \frac{\overrightarrow{\delta}G}{\delta S_\a}\cr
&+ 2 \frac{F \overleftarrow{\delta}}{\delta S_\b} H_{\a\b\g}D\phi^\a \frac{\overrightarrow{\delta} G}{\delta S_\g}
\Bigg). 
\end{align}
Because of the reversed parity of the fibers $C^\infty(\Pi T^* \mathcal{L}M)$, \ie the space of smooth functionals on the phase space, has a natural $\mathbbm{Z}_2$ grading, $|\cdot|$, defined as 
\begin{align}
|F| = \left\{\begin{array}{lll}
0&\;\;\;\;&\mbox{if $F$ is Grassmann even, \ie bosonic.}\\
1&&\mbox{if $F$ is Grassmann odd, \ie fermionic.} 
\end{array}\right.
\end{align}
The Poisson bracket satisfy the following graded identities
\begin{align}
&\{F,G\}_H = - (-1)^{|F||G|}\{G,F\}_H, \label{susy:superPoisson1}\\
&\{F,GK\}_H = \{F,G\}_H K + (-1)^{|F||G|} G\{F,K\}_H,\\
&0 = (-1)^{|K||F|}\{F,\{G,K\}_H\}_H 
   + (-1)^{|F||G|}\{G,\{K,F\}_H\}_H\cr
&\;\;\;\;\;\;\;\;+ (-1)^{|G||K|}\{K,\{F,G\}_H\}_H.\label{susy:superPoisson3}
\end{align}
where $F,G,K \in C^\infty(\Pi T^* \mathcal{L}M)$.
These are the graded versions of antisymmetry, of the Leibniz rule and of the Jacobi identity respectively. Since the Poisson bracket $\{\cdot,\cdot\}_H$ satisfies \eqref{susy:superPoisson1}-\eqref{susy:superPoisson3}, the space $C^\infty(\Pi T^* \mathcal{L}M)$ with $\{\cdot,\cdot\}_H$ is a superPoisson algebra.

The complete set of canonical transformations on the phase space is given by the diffeomorphisms of $M$ and the b-transforms \cite{Zabzine:2005qf} 
\begin{align}
\phi^\mu \rightarrow \phi^\mu, \;\;\;\;\;\; S_\mu\rightarrow S_\mu - b_{\mu\nu}D\phi^\nu,
\end{align}
where $b$ is a closed two-form on $M$, \ie $b\in\Omega^2_{closed}(M)$.

On the supercircle $S^{1,1}$ there exists two natural operators, the above $D$ derivative, here written in terms of the coordinates $\sigma$ and $\theta$,
\begin{align}
D = \frac{\partial}{\partial\theta} +\i\theta\partial
\end{align}
and the $\mathcal{N}=1$ supersymmetry generator
\begin{align}
Q = \frac{\partial}{\partial\theta} -\i\theta\partial.
\end{align}
These operators satisfy
\begin{align}
D^2 = \i\partial,\;\;\;\;\;
Q^2 = -\i\partial,\;\;\;\;\;
\{D,Q\} = DQ+QD = 0.
\end{align}

The supersymmetry generator on the phase space has the form
\begin{align}
\mathbf{Q}_1(\epsilon) = -\int_{S^{1,1}} \d\s\d\theta \;
\epsilon \left(S_\mu - B_{\mu\nu}D\phi^\nu\right)Q\phi^\mu,
\end{align}
where $\epsilon$ is a Grassmann odd parameter. This generator obeys the algebra
\begin{align}
\{\mathbf{Q}_1(\epsilon_1), \mathbf{Q}_1(\epsilon_2)\}_H = \mathbf{P}(2\epsilon_1\epsilon_2) \label{susy:N1_manifest_susy_generator}
\end{align} 
where 
\begin{align}
\mathbf{P}(a) = \int_{S^{1,1}}\d\s\d\theta\; 
a \left(S_\mu\partial\phi^\mu + \i H_{\mu\nu\rho} D\phi^\mu Q\phi^\nu Q\phi^\rho \right) \label{susy:translation_op}
\end{align}
is the generator of translations along the $\sigma$ direction of $S^{1,1}$. The parameter $a$ is Grassmann even.

A general generator $\mathbf{Q}$ generates transformations of sections of the phase space by use of the Poisson bracket as $\delta(\epsilon)(\cdot) = \{\,\cdot\, ,\mathbf{Q}(\epsilon)\}_H$. We thus find the transformations of the coordinate fields $\phi^\mu$ and $S_\mu$ generated by the above $\mathbf{Q}_1(\epsilon)$ as
\begin{align}
\delta_1(\epsilon)\phi^\mu &=\{\phi^\mu,\mathbf{Q}_1(\epsilon)\}_H = -\i\epsilon Q \phi^\mu,\label{susy:N1_manifest_transf1}\\
\delta_1(\epsilon)\S_\mu &=\{S_\mu,\mathbf{Q}_1(\epsilon)\}_H =-\i\epsilon Q S_\mu.\label{susy:N1_manifest_transf2}
\end{align}

Remember that the action \eqref{susy:S_second_order} is invariant under the manifest $\mathcal{N}=(1,1)$ supersymmetry transformations \eqref{susy:N11_manifest_transfn}. By introducing
\begin{align}
Q_0 = \frac{1}{\sqrt{2}}\left(Q_+ +\i Q_-\right),\;\; & \;\;\;
Q_1 = \frac{1}{\sqrt{2}}\left(Q_+ -\i Q_-\right),\\
\epsilon^0 = \frac{1}{\sqrt{2}}\left(\epsilon_+ -\i \epsilon_-\right),\;\; & \;\;\;
\epsilon^1 = \frac{1}{\sqrt{2}}\left(\epsilon_+ +\i \epsilon_-\right),
\end{align}
we can rewrite the transformations \eqref{susy:N11_manifest_transfn} as
\begin{align}
\delta\Phi^\mu = -\i\epsilon^0 Q_0 \Phi^\mu -\i\epsilon^1 Q_1 \Phi^\mu.
\end{align}
Reducing to the $\mathcal{N}=1$ fields as defined in \eqref{susy:def_N1_fields}
and identifying $Q\equiv Q_1|_{\theta^0=0}$ and $\epsilon\equiv\epsilon^1$
we find exactly the transformations \eqref{susy:N1_manifest_transf1}-\eqref{susy:N1_manifest_transf2} generated by $\mathbf{Q}_1(\epsilon)$. Further, we also find a non-manifest supersymmetry given by
\begin{align}
\tilde{\delta}_1 \phi^\mu = & \epsilon g^{\mu\nu} S_\nu, \label{susy:non-manifest_ham_susy_1}\\
\tilde{\delta}_1 S_\mu = & \i\epsilon g_{\mu\nu} \partial\phi^\nu + \epsilon S_\lambda S_\sigma g^{\lambda\rho}\Gamma^{(0)\sigma}_{\mu\rho}.\label{susy:non-manifest_ham_susy_2}
\end{align}
To find this form of the non-manifest supersymmetry, we need to drop terms containing $\partial_0\phi^\mu$ and $\partial_0 S_\mu$. The motivation for this is that these terms corresponds to time evolution and further, that the transformations \eqref{susy:non-manifest_ham_susy_1}-\eqref{susy:non-manifest_ham_susy_2} 
commute to a translation along the $\sigma$ direction of $S^{1,1}$ and is an invariance of the Hamiltonian \eqref{susy:Hamiltonian}.

Extended supersymmetry on the phase space is given by generators of the type
\cite{Zabzine:2005qf}
\begin{align}
\mathbf{Q}_2(\epsilon) = -\frac{1}{2} \int_{S^{1,1}} \d\s\d\theta \;
\epsilon \langle \Lambda, \genJ \Lambda \rangle \label{susy:Q_2_compact}
\end{align}
where we have used the natural pairing introduced in \eqref{susy:GCG_pairing}, and the definitions
\begin{align}
\genJ = \matrix{cc}{-J&P\\L&J^t},\;\;\;\;\;
\Lambda = \matrix{c}{D\phi\\S}, \label{susy:Q_2_ingredients}
\end{align} 
where $\Lambda \in \Gamma( X^*\Pi((TM\oplus T^*M)\otimes\mathbbm{C}))$, \ie a section of the pull back to the world-sheet by $X$, the bosonic component of $\phi$, of the bundle $(TM\oplus T^*M)\otimes\mathbbm{C}$ with reversed parity on the fibers. This form of $\mathbf{Q}_2$ is the most general ansatz for a supersymmetry generator that does not include any dimensionful parameters. The use of the pairing and the matrix $\genJ$ anticipates that this will have something to do with generalized complex geometry, which is indeed the case. Written out explicitly, $\mathbf{Q}_2$ is given by
\begin{align}
\mathbf{Q}_2(\epsilon) = -\frac{1}{2} \int_{S^{1,1}} \d\s\d\theta \;\epsilon
\left(
2D\phi^\mu S_\nu J^\nu_\mu + D\phi^\mu D\phi^\nu L_{\mu\nu} + S_\mu S_\nu P^{\mu\nu}
\right).
\end{align}

The transformations of the coordinate fields generated by $\mathbf{Q}_2$ are given by
\begin{align}
\delta_2(\epsilon)\phi^\mu =& 
\i\epsilon D\phi^\nu J^\mu_\nu -\i\epsilon S_\nu P^{\mu\nu},\label{susy:Q2_transf_phi}\\
\delta_2(\epsilon)\S_\mu =&
\i\epsilon D(S_\nu J^\nu_\mu) -\frac{\i}{2}\epsilon S_\nu S_\rho P^{\nu\rho}{}_{,\mu}
+\i\epsilon D(D\phi^\nu L_{\mu\nu}) \cr
&+\i\epsilon S_\nu D\phi^\rho J^\nu_{\rho,\mu}
-\frac{\i}{2}D\phi^\nu D\phi^\rho L_{\nu\rho,\mu}.\label{susy:Q2_transf_S}
\end{align}

For $\mathbf{Q}_2$ to be a supersymmetry generator we must demand that the algebra 
\begin{align}
\{\mathbf{Q}_1(\epsilon_1),\mathbf{Q}_2(\epsilon_2)\}_H = 0, \;\;\;\;\;
\{\mathbf{Q}_2(\epsilon_1),\mathbf{Q}_2(\epsilon_2)\}_H = \mathbf{P}(2\epsilon_1\epsilon_2),
\label{susy:Hamiltonian_susy_algebra} 
\end{align}
is fulfilled. Using the ansatz \eqref{susy:Q_2_compact} in this algebra give us the following conditions \cite{Zabzine:2005qf}
\begin{align}
\genJ^2 = -1, \;\;\;\;\; \Pi_\mp [\Pi_\pm(X+\xi), \Pi_\pm(Y+\eta)]_H = 0.
\end{align}
The first condition implies that $\genJ$ must be a generalized almost complex structure. The second condition should be interpreted in terms of the components of $\genJ$ and thus translates into the integrability conditions given in \eqref{susy:J_comp_integrability_1}-\eqref{susy:J_comp_integrability_4}. This implies that $\genJ$ is a $H$-twisted generalized complex structure. Hence, we conclude that when the target manifold $M$ is $H$-twisted generalized complex, the phase space $\Pi T^* \mathcal{L}M$ admits extended supersymmetry. 

Note that demanding the $\mathbf{Q}_2$ to satisfy the supersymmetry algebra \eqref{susy:Hamiltonian_susy_algebra} only impose conditions on the target space. Hence, no conditions on the Hamiltonian of the model arise. In this sense the statement that $\mathbf{Q}_2$ is a supersymmetry generator is model independent. However, when we specify the model we must verify that $\mathbf{Q}_2$ generates a symmetry of the Hamiltonian.

As an example of this we consider briefly the WZ-Poisson sigma model Hamiltonian written in terms of $\mathcal{N}=1$ superfields \cite{Calvo:2005ww}
\begin{align}
\mathcal{H}_{WZP} = \int_{S^{1,1}} \d\s\d\theta\; F_\mu \left(D\phi^\mu - \Pi^{\mu\nu}S_\nu\right),
\end{align}
where the fields $F_\mu$ act as a Lagrange multipliers. $\Pi$ is a $H$-twisted Poisson structure satisfying the modified Jacobi-identity, 
\begin{align}
\Pi^{\mu\rho}\partial_\rho \Pi^{\nu\sigma} 
+ \Pi^{\nu\rho}\partial_\rho \Pi^{\sigma\mu} 
+ \Pi^{\sigma\rho}\partial_\rho \Pi^{\mu\nu} 
= \Pi^{\mu\alpha}\Pi^{\nu\beta}\Pi^{\sigma\gamma} H_{\alpha\beta\gamma} ,
\end{align}
where $H$ is a closed three-form. The above $\mathbf{Q}_2$ in \eqref{susy:Q_2_compact} is an extended supersymmetry of the model if the target space admits an $H$-twisted generalized complex structure and if $\mathbf{Q}_2$ generates a symmetry of $\mathcal{H}_{WZP}$. The condition for the latter is 
\begin{align}
\{\mathbf{Q}_2(\epsilon), \mathcal{H}_{WZP}\}_H=0. \label{susy:WZP_inv_cond}
\end{align}
It is possible to rewrite this condition in terms of an $H$-twisted Dirac structure, defined as
\begin{align}
\mathbbm{L}_\Pi = \{(\Pi^\# \xi, \xi)\in TM\oplus T^*M |\xi\in T^* M\},
\end{align}
where $\Pi$ is an $H$-twisted Poisson structure. In local coordinates $(\Pi^\# \xi)^\mu = \Pi^{\nu\mu} \xi_\nu$. The condition \eqref{susy:WZP_inv_cond} becomes \cite{Calvo:2005ww}
\begin{align}
\genJ(\mathbbm{L}_\Pi) \subset \mathbbm{L}_\Pi. \label{susy:WZP_inv_cond_mod}
\end{align}
Which means that the $H$-twisted Dirac structure is closed under the action of the H-twisted generalized complex structure. If \eqref{susy:WZP_inv_cond_mod} is satisfied the model has $\mathcal{N}=2$ supersymmetry, one supersymmetry being the manifest one and the other being the one generated by $\mathbf{Q}_2$. Once again we find that demanding extended supersymmetry gives conditions on the target space geometry.

\subsection{Generalized K\"{a}hler from a sigma model}\label{GKG_from_sigmamodel}
We now return to our main track, \ie to find the conditions for extended supersymmetry of the sigma model \eqref{susy:S_second_order}. Instead of directly demanding that the supersymmetry generator $\mathbf{Q}_2$ generates an invariance of the Hamiltonian \eqref{susy:Hamiltonian}, we construct one more supersymmetry generator $\mathbf{Q}^{(2)}_2$ of the same form as the first generator $\mathbf{Q}_2$, which we now will call $\mathbf{Q}^{(1)}_2$. The two generators are given by
\begin{align}
\mathbf{Q}^{(1)}_2(\epsilon) =& - \frac{1}{2} \int \d\s \d\theta\; 
\epsilon\langle \Lambda , \genJ^{(1)} \Lambda \rangle, \\
\mathbf{Q}^{(2)}_2(\epsilon) =& -\frac{1}{2}\int \d\s\d\theta\;
\epsilon \langle \Lambda , \genJ^{(2)} \Lambda \rangle.
\end{align}
To be supersymmetry generators these should both satisfy the algebra \eqref{susy:Hamiltonian_susy_algebra}. Implying that $\genJ^{(i)}$, $i=1,2$, now are two $H$-twisted generalized complex structures. Further these should obey
\begin{align}
\{\mathbf{Q}^{(1)}_2(\epsilon_1), \mathbf{Q}^{(2)}_2(\epsilon_2)\}_H = 2\i\epsilon_1\epsilon_2 \mathcal{H}, \label{susy:Hamiltonian_extended_susy_algebra}
\end{align}
where $\mathcal{H}$ is given in \eqref{susy:Hamiltonian}. This means that the two extended supersymmetry generators apart from commuting with the manifest supersymmetry and commuting to a translation along the $\sigma$ direction of $S^{1,1}$ must also commute, as in \eqref{susy:Hamiltonian_extended_susy_algebra}, to a translation along the time direction of the world-sheet, a transformation generated by the Hamiltonian. Imposing the condition \eqref{susy:Hamiltonian_extended_susy_algebra} gives that the $H$-twisted generalized complex structures must obey [VI]
\begin{align}
-\genJ^{(1)} \genJ^{(2)} = \genG \equiv \matrix{cc}{0&g^{-1}\\g&0}. \label{susy:genG_def}
\end{align}
Here we have introduced the positive definite generalized metric $\genG$, that obeys $\genG^2 = \mathbbm{1}_{2d}$. Note that 
\begin{align}
\mathbbm{1}_{2d} = \genG^2 = \genJ^{(1)} \genJ^{(2)} \genJ^{(1)} \genJ^{(2)} = 
(\genJ^{(1)})^2 (\genJ^{(2)})^2 - \genJ^{(1)} [\genJ^{(1)},\genJ^{(2)}] \genJ^{(2)},
\end{align} 
which implies $[\genJ^{(1)},\genJ^{(2)}] =0$, the two $H$-twisted generalized complex structures commute. Together they form an $H$-twisted generalized K\"{a}hler structure $(\genJ^{(1)}, \genJ^{(2)})$. Further, the graded Jacobi identity \eqref{susy:superPoisson3} implies that 
\begin{align}
0=\{\mathbf{Q}^{(i)}_2(\epsilon_1), \{\mathbf{Q}_1(\epsilon_2), \mathbf{Q}_1(\epsilon_3)\}_H\}_H 
 = \{\mathbf{Q}^{(i)}_2(\epsilon_1),P(2\epsilon_2\epsilon_3)\}_H
\end{align}
for $i=1,2$, and
\begin{align}
0=&\{\mathbf{Q}^{(j)}_2(\epsilon_1),\{\mathbf{Q}^{(i)}_2(\epsilon_2),\mathbf{Q}^{(i)}_2(\epsilon_3)\}_H \}_H \cr
&+ \{\mathbf{Q}^{(i)}_2(\epsilon_2),\{\mathbf{Q}^{(i)}_2(\epsilon_3),\mathbf{Q}^{(j)}_2(\epsilon_1)\}_H\}_H\cr
& +\{\mathbf{Q}^{(i)}_2(\epsilon_3),\{\mathbf{Q}^{(j)}_2(\epsilon_1),\mathbf{Q}^{(i)}_2(\epsilon_2)\}_H\}_H
\end{align}
for $i,j =1,2$. For $i\neq j$ the last equation implies $\{\mathbf{Q}^{(i)}_2(\epsilon), \mathcal{H}\}_H=0$ for $i=1,2$. This means that both the generators $\mathbf{Q}^{(i)}_2$ generate symmetries of the Hamiltonian.

We conclude that, the sigma model defined by \eqref{susy:S_second_order} with Hamiltonian \eqref{susy:Hamiltonian} admits two extended supersymmetries, \ie $\mathcal{N}=(2,2)$ supersymmetry,  if the target space is an $H$-twisted generalized K\"{a}hler manifold with $(\genJ^{(1)}, \genJ^{(2)})$ being the $H$-twisted generalized K\"{a}hler structure.

It is straight forward to relate the extended supersymmetry transformations \eqref{susy:Q2_transf_phi}-\eqref{susy:Q2_transf_S} of the phase space fields to the extended supersymmetry transformations \eqref{susy:N11_extended_susy_transfn} of the $\mathcal{N}=(1,1)$ fields. We write the transformations \eqref{susy:N11_extended_susy_transfn} as one combined transformation
\begin{align}
\delta \Phi^\mu = \epsilon^+ D_+ \Phi^\nu J^{(+)\mu}_\nu + \epsilon^- D_- \Phi^\nu  J^{(-)\mu}_\nu \label{susy:GHR_transformation}
\end{align}
and reduce it to $\mathcal{N}=1$ phase space fields by defining $\epsilon^+\equiv -\frac{\i}{\sqrt{2}}(\epsilon^1+\epsilon^2)$, $\epsilon^-\equiv \frac{1}{\sqrt{2}}(\epsilon^2-\epsilon^1)$, using \eqref{susy:def_N1_cov_derivatives} and the definitions \eqref{susy:def_N1_fields}. We find the transformation of $\phi^\mu$ as
\begin{align}
\delta\phi^\mu &= \frac{\i}{2}\epsilon^1\left(
 - D \phi^\nu \left(J^{(+)\mu}_\nu + J^{(-)\mu}_\nu\right) 
 + S_\nu \Big((\omega^{-1}_+)^{\mu\nu} - (\omega^{-1}_-)^{\mu\nu}\Big) 
 \right)\cr
&+\frac{\i}{2}\epsilon^2 \left(
- D \phi^\nu \left(J^{(+)\mu}_\nu - J^{(-)\mu}_\nu\right) 
+ S_\nu \Big((\omega^{-1}_+)^{\mu\nu} + (\omega^{-1}_-)^{\mu\nu}\Big)
\right),
\end{align}
where $(\omega^{-1}_{\pm})^{\mu\nu}= g^{\mu\rho} J^{(\pm)\nu}_\rho$. The transformation $\delta S_\mu$ is found in a similar manner but with the additional complication that we find terms containing the factor $\partial_0\phi^\mu$. To eliminate this factor from the terms we use the equation of motion
\begin{align}
\i\partial_0\phi^\mu  =& DS_\nu g^{\nu\mu} 
 + S_\rho D\F^\lambda  g^{\mu\nu}\Gamma^\rho_{\lambda\nu} \cr
 &+\frac{1}{8}D\phi^\rho D\phi^\lambda H_{\rho\lambda}{}^{\mu}
 -\frac{1}{8}S_\rho S_\lambda H^{\rho\lambda\mu},
\end{align}
found by varying the action \eqref{susy:action_used_for_reduction} and reducing to $\mathcal{N}=1$ fields. Using this equation we ``go on-shell'', which is not strange since the transformations \eqref{susy:GHR_transformation} in general only close on-shell while the transformations \eqref{susy:Q2_transf_phi}-\eqref{susy:Q2_transf_S} close off-shell. The end result is that the transformation of $S_\mu$ can be written in the form \eqref{susy:Q2_transf_S}.

Next, we compare the transformations obtained from reducing \eqref{susy:GHR_transformation} with the transformations generated by $\mathbf{Q}^{(i)}_2$ with $i=1,2$, of the form \eqref{susy:Q2_transf_phi}-\eqref{susy:Q2_transf_S}. In this way we identify the components of the two $H$-twisted generalized complex structures, defining the $\mathbf{Q}^{(i)}_2$'s, as 
\begin{align}
\genJ^{(1,2)} = \frac{1}{2} \matrix{cc}{J^{(+)} \pm J^{(-)} & -(\omega^{-1}_+ \mp \omega_-^{-1})\\
                                         \omega_+ \mp \omega_- & -( J^{(+)t} \pm J^{(-)t})}.
\end{align}
This is the relation \eqref{susy:relation_GHR_GKG} between $H$-twisted generalized K\"{a}hler geometry and bi-Hermitean geometry. We have thus seen that this relation corresponds to going from the Lagrangian to the Hamiltonian, or phase space, formulation of the sigma model.

\subsection{Generalized hyperK\"{a}hler from a sigma model}\label{GHKG_from_sigmamodel}
It is possible to explore under what circumstances the sigma model described by the Hamiltonian \eqref{susy:Hamiltonian} admits even more supersymmetry. We briefly review the results of \cite{Bredthauer:2006sz}. We introduce three generators of extended supersymmetry on the phase space of the form \eqref{susy:Q_2_compact}, 
\begin{align}
\mathbf{Q}^{(1)}_{2,i}(\epsilon) = -\frac{1}{2}\int\d\s\d\theta\; \epsilon\langle\Lambda, \genJ^{(1)}_i \Lambda\rangle, \;\;\;\;\; i=1,2,3.
\end{align}
Demanding that these operators each satisfy the algebra \eqref{susy:Hamiltonian_susy_algebra}, again demands that the three $J^{(1)}_i$'s are $H$-twisted generalized complex structures. Further, demanding that the operators should generate symmetries of the Hamiltonian \eqref{susy:Hamiltonian} 
\begin{align}
\{\mathbf{Q}^{(1)}_{2,i}(\epsilon),\mathcal{H} \}_H=0, \;\;\;\;\; i=1,2,3,
\end{align}
implies that the $\genJ^{(1)}_i$'s commute with the generalized metric $\genG$, introduced in \eqref{susy:genG_def}, \ie $[\genJ^{(1)}_i, \genG ]=0$ for $i=1,2,3$. This makes it possible to introduce three new $H$-twisted generalized complex structures as $\genJ^{(2)}_i = \genG\genJ^{(1)}_i$, $i=1,2,3$. Each pair $(\genJ^{(1)}_i,\genJ^{(2)}_i)$ is a $H$-twisted generalized K\"{a}hler structure. The $\genJ^{(2)}_{i}$'s define supersymmetry generators $\mathbf{Q}^{(2)}_{2,i}$ via \eqref{susy:Q_2_compact}. We further demand that the new generators should commute with their respective $\mathbf{Q}^{(1)}_{2,i}$ to a time translation. The complete set of the six generators must obey the algebra
\begin{align}
\{\mathbf{Q}^{(1)}_{2,i}(\epsilon_1),\mathbf{Q}^{(2)}_{2,j}(\epsilon_2) \}_H
=& 2\i\delta_{ij} \epsilon_1\epsilon_2\mathcal{H},\\
\{\mathbf{Q}^{(n)}_{2,i}(\epsilon_1),\mathbf{Q}^{(n)}_{2,j}(\epsilon_2) \}_H
=& \delta_{ij}\mathbf{P}(2\epsilon_1\epsilon_2),\;\;\;\;\; n = 1,2.
\end{align} 
This algebra implies that all six $H$-twisted generalized complex structures are integrable and satisfy the algebra of bi-quaternions $Cl_{2,1}(\mathbbm{R})$,
\begin{align}
\{\genJ^{(1)}_i,\genJ^{(1)}_j\} &= -\delta_{ij}\mathbbm{1}_{2d} + \varepsilon_{ijk}\genJ^{(1)}_k \label{susy:bi-quaternion_algebra_1}\\
\{\genJ^{(1)}_i,\genJ^{(2)}_j\} &= -\delta_{ij}\genG + \varepsilon_{ijk}\genJ^{(2)}_k\\
\{\genJ^{(2)}_i,\genJ^{(2)}_j\} &= -\delta_{ij}\mathbbm{1}_{2d} + \varepsilon_{ijk}\genJ^{(1)}_k\\
\{\genJ^{(2)}_i,\genJ^{(1)}_j\} &= -\delta_{ij}\genG + \varepsilon_{ijk}\genJ^{(2)}_k. \label{susy:bi-quaternion_algebra_2}
\end{align}
A manifold equipped with six $H$-twisted generalized complex structures that satisfy the algebra \eqref{susy:bi-quaternion_algebra_1}-\eqref{susy:bi-quaternion_algebra_2} is called an $H$-twisted generalized hyperK\"{a}hler manifold.

Hence, we conclude that the sigma model with Hamiltonian \eqref{susy:Hamiltonian} admits $\mathcal{N}=(4,4)$ supersymmetry if the target space has an $H$-twisted generalized hyperK\"{a}hler manifold. To summarize, the supersymmetries present are: one manifest supersymmetry generated by $\mathbf{Q}_1$ \eqref{susy:N1_manifest_susy_generator}, one non-manifest supersymmetry \eqref{susy:non-manifest_ham_susy_1}-\eqref{susy:non-manifest_ham_susy_2}, three supersymmetries generated by the $\mathbf{Q}^{(1)}_{2,i}$'s and three supersymmetries generated by the $\mathbf{Q}^{(2)}_{2,i}$'s. In total eight supersymmetries.

\section{Manifest $\mathcal{N}=(2,2)$ formulation} \label{susy:manifest_form}
As mentioned at the end of section \ref{susy:geom_interpret} it is possible to write the action of the sigma model with extended $\mathcal{N}=(2,2)$ supersymmetry in terms of $\mathcal{N}=(2,2)$ superfields, making the extended supersymmetry manifest. 

To this end we introduce the two-dimensional $\mathcal{N}=(2,2)$ superspace by defining the coordinates as $(\sigma^\ppmm, \theta^\pm, \bar{\theta}^\pm)$. Further, we introduce the covariant derivatives $\mathbbm{D}_\pm$ and $\bar{\mathbbm{D}}_\pm$ that satisfy
\begin{align}
\{\mathbbm{D}_\pm,\bar{\mathbbm{D}}_\pm\}= \i\partial_{\ppmm}. \label{susy:N22_cov_algebra}
\end{align}
All other anti-commutators between the $\mathbbm{D}_\pm$'s and $\bar{\mathbbm{D}}_\pm$'s vanish.

In \cite{Gates:1984nk} it was shown that when $[J^{(+)},J^{(-)}]=0$ the target space is parametrized by chiral $\varphi$ and anti-chiral $\bar{\varphi}$ superfields, defined by
\begin{align}
\bar{\mathbbm{D}}_\pm\varphi &= 0 \label{susy:chiral_cond1}\\
\mathbbm{D}_\pm \bar{\varphi} &=0,\label{susy:chiral_cond2}
\end{align}
and twisted chiral $\chi$ and twisted anti-chiral $\bar{\chi}$ superfields, defined by
\begin{align}
\mathbbm{D}_+ \chi &= \bar{\mathbbm{D}}_-\chi =0 \label{susy:twisted_chiral_cond1}\\
\bar{\mathbbm{D}}_+\bar{\chi} &= \mathbbm{D}_- \bar{\chi} = 0.\label{susy:twisted_chiral_cond2}
\end{align}

For the more general case where $[J^{(+)},J^{(-)}]\neq0$ we also need left and right semi-chiral $\mathbbm{X}_{L,R}$ superfields and left and right semi-anti-chiral $\bar{\mathbbm{X}}_{L,R}$ superfields \cite{Buscher:1987uw, Sevrin:1996jr} defined by
\begin{align}
\bar{\mathbbm{D}}_+\mathbbm{X}_L &= \mathbbm{D}_+ \bar{\mathbbm{X}}_L = 0 \label{susy:semi_chiral_cond1}\\
\bar{\mathbbm{D}}_-\mathbbm{X}_R &= \mathbbm{D}_- \bar{\mathbbm{X}}_R = 0. \label{susy:semi_chiral_cond2}
\end{align}
to completely parametrize the target space \cite{Lindstrom:2004hi,Lindstrom:2005zr}. 

In the case when there are multiplets of each type of fields we introduce the following indices 
\begin{align}
\begin{array}{lll}
\varphi^\alpha, \bar{\varphi}^{\bar{\alpha}},&\hspace{0.5cm}& \alpha,\bar{\alpha} = 1,...,d_c\\
\chi^{\alpha'}, \bar{\chi}^{\bar{\alpha}'}, && \alpha',\bar{\alpha}' = 1,...,d_t\\
\mathbbm{X}_L^{a}, \bar{\mathbbm{X}}_L^{\bar{a}}, && a,\bar{a}=1,...,d_{L}\\
\mathbbm{X}_R^{a'}, \bar{\mathbbm{X}}_R^{\bar{a}'}, && a',\bar{a}'=1,...,d_{R}.
\end{array}
\end{align}
where indices without a bar are holomorphic indices and the once with a bar are anti-holomorphic indices. Further, to simplify notation we introduce the multi-indices $\mathcal{A}=(\alpha,\bar{\alpha})$, $\mathcal{A}'=(\alpha',\bar{\alpha}')$  and $A=(a,\bar{a})$ and $A'=(a',\bar{a}')$. For convenience we also introduce four complex structures of dimensions $2d_c$, $2d_t$, $2d_{L}$ and $2d_{R}$ of the form
\begin{align}
J = \matrix{cc}{i&0\\0&-i}.
\end{align}


In \cite{Lindstrom:2005zr} it is shown that chiral, twisted chiral and semi-chiral superfields are enough to describe the full generalized K\"{a}hler geometry and in \cite{Maes:2006bm} it is shown that no other types of $\mathcal{N}=(2,2)$ superfields can appear in a manifest $\mathcal{N}=(2,2)$ sigma model. The general action is given in terms of the generalized K\"{a}hler potential $K$ as
\begin{align}
S = \int \d^2\sigma\d^2\theta\d^2\bar{\theta}\; K(\varphi^{\mathcal{A}},  \chi^{\mathcal{A}'},\mathbbm{X}_L^{A}, \mathbbm{X}_R^{A'}). \label{susy:N22_manifest_action}
\end{align}
The metric and $B$-field of the target space are given in terms of the generalized K\"{a}hler potential \cite{Lindstrom:2005zr}.

To reduce the above action to $\mathcal{N}=(1,1)$ form we introduce the $\mathcal{N}=(1,1)$ operators
\begin{align}
D_\pm = \mathbbm{D}_\pm + \bar{\mathbbm{D}}_\pm, \;\;\;\;\;
\hat{Q}_\pm =\i\left( \mathbbm{D}_\pm - \bar{\mathbbm{D}}_\pm \right).
\end{align}
When reducing, two of the manifest supersymmetries become non-manifest and are generated by the operators $\hat{Q}_\pm$. The $D_\pm$ are the covariant derivatives of the remaining manifest $\mathcal{N}=(1,1)$ supersymmetry.

The $\mathcal{N}=(1,1)$ field content of the above $\mathcal{N}=(2,2)$ fields is given by
\begin{align}
\begin{array}{lll}
\varphi^{\mathcal{A}}\equiv \varphi^{\mathcal{A}}|, &\hspace{1.5cm}& 
\chi^{\mathcal{A}'}\equiv\chi^{\mathcal{A}'}|,\\
X_L^A \equiv \mathbbm{X}_L^A|, &&
X_R^{A'} \equiv \mathbbm{X}_R^{A'}|,\\
\Psi_{L-}^A \equiv \hat{Q}_- \mathbbm{X}_L^A|, &&
\Psi_{R+}^{A'} \equiv \hat{Q}_+ \mathbbm{X}_R^{A'}|,
\end{array}
\end{align}
where $|$ denotes formally setting the odd (complex) coordinate $\theta_2 \propto \theta - \bar{\theta}$ to zero. The conditions for the chiral superfields (\ref{susy:chiral_cond1}-\ref{susy:chiral_cond2}) give the non-manifest supersymmetry transformation
\begin{align}
\hat{Q}_\pm\varphi^{\mathcal{A}} = J^\mathcal{A}_\mathcal{B} D_\pm \varphi^\mathcal{B}.
\end{align}
From the twisted chiral conditions \eqref{susy:twisted_chiral_cond1}-\eqref{susy:twisted_chiral_cond2} the transformation
\begin{align}
\hat{Q}_\pm \chi^{\mathcal{A}'} = \mp J^{\mathcal{A}'}_{\mathcal{B}'} D_\pm \chi^{\mathcal{B}'}
\end{align}
follows. Using \eqref{susy:N22_cov_algebra}, the conditions on the semi-chiral fields \eqref{susy:semi_chiral_cond1}-\eqref{susy:semi_chiral_cond2} give rise to the transformations
\begin{align}
\begin{array}{lll}
\hat{Q}_+ X_L^A = J^A_B D_+ X_L^B, & \hspace{1.5cm}& \hat{Q}_+ \Psi^A_{L-} = J^A_B D_+ \Psi^B_{L-},\\
\hat{Q}_- X_L^A = \Psi^A_{L-}, && \hat{Q}_- \Psi^A_{L-} = \i\partial_\mm X^A_{L},\\
\hat{Q}_+ X_R^{A'} = \Psi^{A'}_{R+}, && \hat{Q}_+ \Psi^{A'}_{R+} = \i\partial_\pp X^A_{R},\\
\hat{Q}_- X_R^{A'} = J^{A'}_{B'} D_- X_R^{B'}, && \hat{Q}_- \Psi^{A'}_{R+} = J^{A'}_{B'} D_- \Psi^{B'}_{R+}.
\end{array}
\end{align}

Reducing the action to its $\mathcal{N}=(1,1)$ form is done by integrating over the coordinates $\bar{\theta}^\pm$ as
\begin{align}
\int\d^2\sigma\d^2\theta\d^2\bar{\theta}\; K =& 
\int\d^2\sigma\mathbbm{D}_+\mathbbm{D}_-\bar{\mathbbm{D}}_+ \bar{\mathbbm{D}}_-\; K| \cr
=& -\frac{\i}{4}\int\d^2\sigma D_+ D_- \hat{Q}_+ \hat{Q}_-\; K|\cr
=& -\frac{\i}{4} \int\d^2\sigma\d^2\theta \hat{Q}_+ \hat{Q}_- \;K|.
\end{align}
Hence, up to the overall factor we obtain the $\mathcal{N}=(1,1)$ Lagrangian as $L(\varphi^{\mathcal{A}},\chi^{\mathcal{A}'}, X_L^A, X_R^{A'}, \Psi_{L-}^A, \Psi_{R+}^{A'}) = \hat{Q}_+ \hat{Q}_- \;K(\varphi^{\mathcal{A}},  \chi^{\mathcal{A}'}, \mathbbm{X}_L^A, \mathbbm{X}_R^{A'})|$. By proper redefinitions of the fields the $\mathcal{N}=(1,1)$ action may be brought into a first order form, expressed in terms of the $\mathcal{N}=(1,1)$ fields $\Phi^\mu$ and auxiliary fields $\Psi_\mu$. This is the action \eqref{1st:S_first_order}, as we will discuss in chapter \ref{1storder_susy}. When we have the same number of left- and right-semi-chiral multiplets, \ie $d_R=d_L$, and the $E$-field in \eqref{1st:S_first_order}, obtained from the generalized K\"{a}hler potential $K$ \cite{Lindstrom:2005zr}, is invertible we may integrate out the auxiliary spinorial fields and obtain a second order action of the form \eqref{susy:S_second_order}.

        \chapter{T-duality and extended supersymmetry}\label{T-dual:chapter}

T-duality, or target space duality, is an equivalence between two string theories living on two, possibly, different space times. The two T-dual theories describe the same physics but in different ways.

The easiest example of T-duality is given by bosonic closed string theory compactified on a circle with radius $R$ by the identification $X^{25} \sim X^{25}+ 2\pi R$. The string can wind around the compact dimension and the number of times it does this is given by the winding number $m$. Since the $X^{25}$ direction is compact the momentum of the string in this direction will be discrete and take the values $n/R$, where $n\in\mathbbm{Z}$. The spectrum of the theory, as observed by an observer in the 25 non-compact dimensions, is given by, see \eg\cite{Johnson:2003gi},
\begin{align}
M^2 = \frac{n^2}{R^2} + \frac{w^2 R^2}{\alpha^{\prime 2}} 
+ \frac{2}{\alpha'} \left( \sum_{n=1}^{\infty}\alpha_{-n}^\mu\alpha_{n \mu}  
+\sum_{n=1}^{\infty}\bar{\alpha}_{-n}^\mu \bar{\alpha}_{n\mu}
- 2 \right),
\end{align}
where we denote the left- and right-moving mode operators by $\alpha$ and $\bar{\alpha}$ respectively. This spectrum should be compared to the spectrum obtained in the non-compactified theory \eqref{intro:closed_string_spectrum}, the two first terms are new and arise from making the $X^{25}$ direction compact. Further, the level matching condition, \cf \eqref{intro:level_matching_cond}, reads
\begin{align}
\sum_{n=1}^{\infty}\alpha_{-n}^\mu\alpha_{n \mu}  
- \sum_{n=1}^{\infty}\bar{\alpha}_{-n}^\mu \bar{\alpha}_{n\mu} = nw.
\end{align}
The interesting thing about this is that both the spectrum and level matching condition are invariant under the transformation
\begin{align}
\left\{
\begin{array}{rl}
R\longrightarrow& \tilde{R}= \alpha'/R,\\
n\longrightarrow& \tilde{n}=w,\\
w\longrightarrow& \tilde{w}=n,
\end{array}
\right.
\end{align}
which hints that the original theory, compactified on a circle of radius R, is equivalent to a theory compactified on a circle with radius $\alpha'/R$, with the momentum quantum number and the winding number exchanged. It has been shown \cite{Nair:1986zn} that T-duality holds to all orders in string perturbation theory, meaning that the above invariance of the spectrum is not a coincidence. We say that the bosonic closed string theory compactified on a circle of radius $R$ is T-dual to the same theory compactified on a circle of radius $\alpha'/R$. Note that at the radius $R=\sqrt{\alpha'}$ the theory is dual to itself.

For the superstring, T-duality transforms a theory into a different one. It has been shown \cite{Dine:1989vu,Dai:1989ua} that Type IIA string theory compactified on a circle of radius $R$ is T-dual to Type IIB compactified on a circle of radius $\alpha'/R$. Moreover, Heterotic $SO(32)$ string theory compactified on a circle of radius $R$ is T-dual to Heterotic $E_8\times E_8$ theory compactified on a circle of radius $\alpha'/R$ \cite{Narain:1985jj,Narain:1986am, Ginsparg:1986bx}.

In this chapter we will only consider classical T-duality and study the implications for extended supersymmetry on the world-sheet.

For the case where the string lives in a curved space-time with a background $B$-field, the Buscher's rules \cite{Buscher:1987sk, Buscher:1987qj} specify the map between the background of the string and its T-dual background. One way of deriving Buscher's rules is to consider the isometry of the background and notice that the action is invariant under this isometry. Gauging the isometry produces a, so called, ``parent'' action and fixing the gauge in an appropriate manner produces the T-dual model with the correct background \cite{Hull:1985pq, Rocek:1991ps}.

Perhaps the easiest way to perform a T-duality and to find Buscher's rules in the case of the bosonic string is to go to the phase space formulation, perform a specific canonical transformation and then return to the Lagrangian formulation. This procedure gives the T-dual action \cite{Alvarez:1994wj}. Since in the last chapter we developed the phase space formulation of the string sigma model in terms of $\mathcal{N}=1$ phase space fields, we have the necessary tools to generalize this result to find the corresponding T-duality transformation in terms of the $\mathcal{N}=1$ fields. This was done by the author in article [V].

\section{T-duality as a canonical transformation}
Before turning to the supersymmetric case we briefly review the derivation of T-duality in the bosonic string case, as presented in \cite{Alvarez:1994wj}. For this we start with the action for the bosonic string given by
\begin{align}
S_{bos}=\frac{1}{2}\int\d^2\s \; (g_{\mu\nu}+B_{\mu\nu}) \partial_\pp X^\mu \partial_\mm X^\mu, \label{T-dual:bos_start_action}
\end{align}
with $\sigma_\ppmm=\frac{1}{2}(\sigma^0\pm\sigma^1)$. We now assume that there is an isometry generated by the Killing vector field $k$ such that $\mathcal{L}_k g_{\mu\nu}=0$ and $i_k H = -\d v$ for some one-form $v$ and $H=\d B$. This implies that $\mathcal{L}_k B=\d (i_k B - v)$ and since we assume that there is only one Killing vector $k$ we may choose a gauge where $\mathcal{L}_k B=0$. This further implies that there exists a coordinate system on the target space, $x^\mu = (x^0, x^a)$ where $a=1,...,D-1$, adapted to the isometry such that $k = \frac{\partial}{\partial x^0}$. In this system the metric and $B$-field are independent of the $x^0$ coordinate.

Writing the action in the adapted coordinates and using $\dot{X}^0=\partial_0 X^0$ and $X^{\prime 0}=\partial_1 X^0$ we find 
\begin{align}
S_{bos} = \frac{1}{2} \int \d^2\s\;\Big\{ &
   g_{00}\left((\dot{X}^0)^2 - (X^{\prime 0})^2 \right)\cr
& + (g_{0a} + B_{0a})\partial_\mm X^a(\dot{X}^0 + X^{\prime 0}) \cr
& + (g_{0a} - B_{0a})\partial_\pp X^a(\dot{X}^0 - X^{\prime 0}) \cr
& + (g_{ab} + B_{ab})\partial_\pp X^a \partial_\mm X^b
\Big\}.\label{T-dual:adapted_coord_bosonic_action}
\end{align}
It follows that the momenta conjugate to $X^0$ is
\begin{align}
P_0 = g_{00} \dot{X}^0 
+ \frac{1}{2}(g_{0a}-B_{0a})\partial_\pp X^a
+ \frac{1}{2}(g_{0a}+B_{0a})\partial_\mm X^a.
\end{align}
By performing a Legendre transformation for the coordinate $x^0$ we obtain the Hamiltonian as $\mathcal{H}_{bos} = P_0 \dot{X}^0 - L_{bos}$, where $L_{bos}$ is the Lagrangian defined by \eqref{T-dual:adapted_coord_bosonic_action}. We consider the components of the fields transverse to the isometry direction, \ie $X^a$ as spectator fields and in this sense $\mathcal{H}_{bos}$ is a Hamiltonian for the components of the fields in the $x^0$ direction. By performing the canonical transformation 
\begin{align}
\left\{
\begin{array}{l}
P_0 = - \tilde{X}^{\prime 0}\\
X^{\prime 0} = -\tilde{P}_0 
\end{array}
\right.\label{T-dual:bos_canonical_transf}
\end{align}
the Hamiltonian $\mathcal{H}_{bos}$ transforms into its T-dual $\tilde{\mathcal{H}}_{bos}$. Next we Legendre transform the T-dual Hamiltonian to the T-dual action with Lagrangian $\tilde{L}_{bos}= \tilde{P}_0 \dot{\tilde{X}}^{0} - \tilde{\mathcal{H}}_{bos}$. Explicitly the T-dual action is given by
\begin{align}
\tilde{S}_{bos} = \frac{1}{2} \int \d^2\s\;\Bigg\{&
\frac{1}{g_{00}}((\dot{\tilde{X}}^0)^2 - (\tilde{X}^{\prime 0})^2) \cr
&+ \frac{1}{g_{00}} (g_{0a} - B_{0a})\partial_\pp X^a (\dot{\tilde{X}}^0 - \tilde{X}^{\prime 0})\cr
&- \frac{1}{g_{00}} (g_{0a} + B_{0a})\partial_\mm X^a (\dot{\tilde{X}}^0 + \tilde{X}^{\prime 0})\cr
&- \frac{1}{g_{00}} (g_{0a} - B_{0a})(g_{0a} + B_{0a})\partial_\pp X^a \partial_\mm X^b\cr
&+(g_{ab} + B_{ab})\partial_\pp X^a \partial_\mm X^b  
\Bigg\}.
\end{align}
By writing this action in the same form as \eqref{T-dual:adapted_coord_bosonic_action} we identify the T-dual metric and $B$-field as
\begin{align}
\tilde{g}_{00} =& \frac{1}{g_{00}},\;\;\;\;\;\;\;\;
\tilde{g}_{0a} = -\frac{B_{0a}}{g_{00}},\;\;\;\;\;\;\;\;
\tilde{g}_{ab} = g_{ab} - \frac{g_{0a}g_{0b}}{g_{00}} + \frac{B_{0a}B_{0b}}{g_{00}},\cr
&\tilde{B}_{0a} = -\frac{g_{0a}}{g_{00}}, \;\;\;\;\;\;\;\;
\tilde{B}_{ab} = B_{ab} - \frac{g_{0a}B_{0b}}{g_{00}} + \frac{B_{0a}g_{0b}}{g_{00}}. \label{T-dual:Buschers_rules}
\end{align}
These are the so called Buscher's rules \cite{Buscher:1987sk, Buscher:1987qj}.
The inverse of the T-dual metric is found to be
\begin{align}
\tilde{g}^{00} = g_{00} + g^{ab}B_{0a}B_{0b},\;\;\;\;\;\;\;\;
\tilde{g}^{0a} = g^{ab}B_{0b},\;\;\;\;\;\;\;\;
\tilde{g}^{ab} = g^{ab}.
\end{align}

Since \eqref{T-dual:adapted_coord_bosonic_action} is nothing but the action \eqref{T-dual:bos_start_action} we can write the T-dual action in the same form as 
\eqref{T-dual:bos_start_action} but with the coordinate fields given by $\tilde{X}^\mu = (\tilde{X}^0, X^a)$ and the background given by the Buscher's rules \eqref{T-dual:Buschers_rules}.

In conclusion, since a canonical transformation does not change the physics of the system, the transformation \eqref{T-dual:Buschers_rules} give us two equivalent descriptions of the same physics. We say that the theory living on the background $(g,B)$ and the theory living on the background $(\tilde{g}, \tilde{B})$ are target space dual, or T-dual for short. 

In the following we will construct the analogue of the canonical transformation \eqref{T-dual:bos_canonical_transf} in the $\mathcal{N}=1$ phase space formulation, but first we will consider an alternative way of performing T-duality.

\section{T-duality as a gauging of the isometry}
An alternative derivation of the Buscher's rules is to note that when the isometry is also a symmetry of the action we may gauge the symmetry, making it local. The gauging introduces gauge fields and we require these to be pure gauge by introducing a term with a Lagrange multiplier. Then the original action or the T-dual action may be obtained by integrating out fields and fixing a specific gauge.

Here we follow the lines presented in \cite{Albertsson:2004gr} to review this procedure for the case of the manifest $\mathcal{N}=(1,1)$ action \eqref{susy:N11_start_action}
\begin{align}
S = \frac{1}{2} \int\d^2\s\d^2\theta \; (g_{\mu\nu} + B_{\mu\nu})D_+\Phi^\mu D_- \Phi^\nu. \label{T-dual:N11_action}
\end{align}
The isometry, given by the Killing vector field $k$, induces a transformation on the fields as
\begin{align}
\delta_i(a) \Phi^\mu = a k^\mu(\Phi), \label{T-dual:N11_isometry}
\end{align}
where $a$ is a constant bosonic parameter. The action \eqref{T-dual:N11_action} is invariant under this transformation if $\mathcal{L}_k g = 0$ and $\mathcal{L}_k B = 0$. Which is what we assumed in the previous section to have an adapted coordinate system. We now go to the adapted coordinates, such that $k$ is parallel with $\Phi^0$. We introduce the Grassmann odd gauge fields $A_\pm$ and the field $\tilde{\Phi}^0$ acting as a Lagrange multiplier and write the action as
\begin{align}
S_{parent}= \frac{1}{2}\int\d^2\s\d^2\theta \Big\{& (D_+\Phi^0 + A_+) (D_-\Phi^0 + A_-)g_{00}\cr
&+ (D_+\Phi^0 + A_+)D_- \Phi^a (g_{0a}+B_{0a})\cr
&+ D_+\Phi^a (D_-\Phi^0 + A_-) (g_{0a}-B_{0a})\cr
&+ D_+\Phi^a D_-\Phi^b (g_{ab}+B_{ab})\cr
&- \tilde{\Phi}^0 (D_+ A_- + D_- A_+) \Big\}.\label{T-dual:parent_action}
\end{align}
This action is called the parent action, since as we shall see it gives rise to both the original action \eqref{T-dual:N11_action} and its T-dual partner. 
The gauging makes the isometry transformation local and in the adapted coordinates the gauge transformation reads
\begin{align}
\delta_i(a) \Phi^0 = a, \;\;\;\;\;\;\;
\delta_i(a) A_{\pm} = - D_{\pm} a,\label{T-dual:gauge_transfn}
\end{align}
where now $a=a(\sigma^\alpha)$. Using $\delta_i(D_\pm \Phi^0 + A_\pm)=0$ and the property \eqref{susy:N11_odd_derivative} of the spinor derivatives, we find immediately that the parent action \eqref{T-dual:parent_action} is invariant under this local gauge transformation.

The equation of motion that follow from \eqref{T-dual:parent_action} by varying $\tilde{\Phi}^0$ is
\begin{align}
0 =& D_+ A_- + D_- A_+. \label{T-dual:eomA}
\end{align}
This equation implies that, for some scalar superfield $\lambda$, the gauge fields is given by $A_\pm = D_\pm \lambda$. Comparing to \eqref{T-dual:gauge_transfn} we see that the gauge fields can be set to zero by a gauge transformation, thus the gauge fields are pure gauge. 

By using the equation of motion \eqref{T-dual:eomA} in the parent action \eqref{T-dual:parent_action} and choosing the gauge $\lambda=0$ we recover the original action \eqref{T-dual:N11_action}. Thus, we find that the parent action is indeed a generalization of the action \eqref{T-dual:N11_action}.

Variation of the $A_{\pm}$'s in the parent action \eqref{T-dual:parent_action} give the following equations of motion
\begin{align}
0 =& D_+ \tilde{\Phi}^0 + \left(D_+ \Phi^0 + A_+ \right)g_{00} + D_+\Phi^a (g_{0a}-B_{0a}), \label{T-dual:eom+}\\
0 =& - D_- \tilde{\Phi}^0 +\left(D_- \Phi^0 + A_- \right)g_{00} + D_-\Phi^a (g_{0a}+B_{0a}).\label{T-dual:eom-}
\end{align}
Note that these equations relate the Lagrange multiplier field $\tilde{\Phi}^0$ and the coordinate field $\Phi^0$. This will be important in the next section when we formulate the T-duality transformation in phase space. 

We now integrate out $A_\pm$ from the parent action \eqref{T-dual:parent_action} using \eqref{T-dual:eom+} and \eqref{T-dual:eom-}. Next we fix the gauge as $\lambda=-\Phi^0$. This effectively removes the original coordinate field $\Phi^0$ and $\tilde{\Phi}^0$ becomes the new coordinate field in the T-dual model. The end result of this procedure is 
\begin{align}
\tilde{S} = \frac{1}{2}\int\d^2\s\d^2\theta (\tilde{g}_{\mu\nu} + \tilde{B}_{\mu\nu})D_+\tilde{\Phi}^\mu D_-\tilde{\Phi}^\nu, \label{T-dual:T-dual_action}
\end{align}
which is the T-dual action where we have grouped together the new coordinate $\tilde{\Phi}^0$ with the remaining original coordinates as $\tilde{\Phi}^\mu=(\tilde{\Phi}^0, \Phi^a)$. The T-dual metric and $B$-field are given by Buscher's rules \eqref{T-dual:Buschers_rules}.

Once again we have found a T-dual action as an equivalent description of the physics in the model, but on a different background. Here however we have achieved this in a manifestly $\mathcal{N}=(1,1)$ description. The above procedure can be summarized by the picture
\begin{align}
S(g,B) \longleftrightarrow S_{parent} \longrightarrow \tilde{S}(\tilde{g},\tilde{B}).
\end{align}

\section{T-duality as a symplectomorphism}

We now turn to the the question of the T-duality in the $\mathcal{N}=1$ phase space formulation. We want to find the transformation in this setting that corresponds to the canonical transformation \eqref{T-dual:bos_canonical_transf} that gave the T-duality in the bosonic model. Here we review the results of [V].
 
To this end we consider the equations of motion \eqref{T-dual:eom+} and \eqref{T-dual:eom-} that arose from the parent action. Since we now know that the $\tilde{\Phi}^0$ is the T-dual coordinate, these equations relate the coordinates in the two dual models. Accompanying these relations is 
\begin{align}
\Phi^a=\tilde{\Phi}^a,\label{T-dual:coordinate-realtion0}
\end{align}
found by noting how the T-dual coordinate fields are defined in the T-dual action \eqref{T-dual:T-dual_action}.  Remember that when we obtained the original action from the parent action we used the gauge $\lambda = 0$, further, in this gauge the equations \eqref{T-dual:eom+} and \eqref{T-dual:eom-} simplify. We will use this gauge in the following. 

We will now reduce the \eqref{T-dual:eom+} and \eqref{T-dual:eom-} to the $\mathcal{N}=1$ form used in the phase space formulation. For this we use the odd derivatives defined in \eqref{susy:def_N1_cov_derivatives} to rewrite the equations as
\begin{align}
D_0\tilde{\Phi}^0 &= B_{0a} D_0\Phi^a - g_{0\mu} D_1\Phi^\mu, \label{T-dual:coordinate-relation1}\\
D_1\tilde{\Phi}^0 &= B_{0a} D_1\Phi^a - g_{0\mu} D_0\Phi^\mu. \label{T-dual:coordinate-relation2}
\end{align}

When reducing we look for a relation between the T-dual phase space coordinates $(\tilde{\phi}^\mu,\tilde{S}_\mu)$ and the original phase space coordinates $(\phi^\mu, S_\mu)$. We use the definitions \eqref{susy:def_N1_fields} of the original $\mathcal{N}=1$ phase space fields and the definitions
\begin{align}
\tilde{\phi}^\mu \equiv \tilde{\Phi}^\mu|_{\theta^0=0}, \;\;\;\;\;\;\;
\tilde{S}_\mu \equiv (\tilde{g}_{\mu\nu} D_0\tilde{\Phi}^\mu)|_{\theta^0=0}
\end{align}
of the T-dual $\mathcal{N}=1$ fields.

The relation \eqref{T-dual:coordinate-realtion0} and the equation \eqref{T-dual:coordinate-relation2} reduce straightforwardly to
\begin{align}
\tilde{\phi}^a &= \phi^a,\\
D\tilde{\phi}^0 &= - S_0 + B_{0a}D\phi^a.
\end{align}
To find the relation between the $\tilde{S}_\mu$ fields and the fields $\phi^\mu$ and $S_\mu$ is slightly more involved. We need to consider the cases $\mu=0$ and $\mu=a$ separately. We also use the definitions of the T-dual metric in terms of the original one \eqref{T-dual:Buschers_rules} and the equation \eqref{T-dual:coordinate-relation1}. This produces the relations
\begin{align}
\tilde{S}_0 &= -D\phi^0 - \frac{g_{0a}}{g_{00}}D\phi^a,\\
\tilde{S}_a &= S_a -\frac{g_{0a}}{g_{00}}S_0 + B_{0a}D\phi^0 + \frac{B_{0a}g_{0b}}{g_{00}}D\phi^b.
\end{align}
Altogether, the T-duality transformation may be written as
\begin{align}
\left\{
\begin{array}{rl}
D\tilde{\phi}^0 =& - S_0 + B_{0a} D\phi^a \\
D\tilde{\phi}^a =&   D\phi^a\\
\tilde{S}_0   =& - D\phi^0 - \frac{g_{0a}}{g_{00}}D\phi^a \\
\tilde{S}_a   =&   S_a - \frac{g_{0a}}{g_{00}} S_0 
                 + B_{0a} D\phi^0 + B_{0a}\frac{g_{0b}}{g_{00}} D\phi^b.
\end{array}
\right.\label{T-dual:Tduality-transform}
\end{align}
Further, it is straightforward to invert \eqref{T-dual:Tduality-transform} to find the inverse T-duality transformation.

Define $X^*$ as the pull back by the bosonic component of $\phi$ to the world-sheet and $\tilde{X}^*$ is the pull back by the bosonic component of $\tilde{\phi}$ to the T-dual world-sheet. Using this notation, the transformation \eqref{T-dual:Tduality-transform} defines how an object in the bundle $X^*\Pi((TM\oplus T^*M)\otimes{C})$ maps into an object in the T-dual bundle $\tilde{X}^*\Pi((T\tilde{M}\oplus T^*\tilde{M})\otimes{C})$. Such a map is called a bundle morphism. 

It is now straightforward to investigate how the Hamiltonian \eqref{susy:Hamiltonian} transforms under the T-duality. Using \eqref{T-dual:Tduality-transform} in \eqref{susy:Hamiltonian} gives the T-dual Hamiltonian as
\begin{align}
\tilde{\mathcal{H}}=
\frac{1}{2}\int\d\s\d\theta \;\Big(
&   \i\partial\tilde{\phi}^\mu D\tilde{\phi}^\nu \tilde{g}_{\mu\nu}
  + \tilde{S}_\mu D\tilde{S}_\nu \tilde{g}^{\mu\nu}
  + \tilde{S}_\mu D\tilde{\phi}^\rho \tilde{S}_\nu \tilde{g}^{\nu\s}\tilde{\Gamma}^{(0)\mu}_{\s\rho}\cr
& + D\tilde{\phi}^\mu D\tilde{\phi}^\nu \tilde{S}_\rho \tilde{H}_{\mu\nu}{}^\rho
  -\frac{1}{3} \tilde{S}_\mu \tilde{S}_\nu \tilde{S}_\rho \tilde{H}^{\mu\nu\rho} \Big), \label{T-dual:dual_hamiltonian}
\end{align}
where we have used the T-dual metric to raise the indices of the $\tilde{H}$-field. Explicitly the components of $\tilde{H}$ are given by
\begin{align}
\tilde{H}_{0ab} &= \frac{1}{2}\left(\partial_a\left(\frac{g_{0b}}{g_{00}}\right) 
           -\partial_b\left(\frac{g_{0a}}{g_{00}}\right)\right), \label{T-dual:Hdual0ab}\\
\tilde{H}_{abc} &= H_{abc} - B_{0[a|}\tilde{H}_{0|bc]} - \frac{1}{g_{00}}g_{0[a|} H_{0|bc]},
\end{align}
in the second relation we have used $\tilde{H}_{0ab}$ as a convenient abbreviation and the anti-symmetrization is taken without a combinatorial factor.

The Hamiltonian \eqref{T-dual:dual_hamiltonian} is what we expect to find, since it follows from the T-dual action \eqref{T-dual:T-dual_action} in the same way as the Hamiltonian \eqref{susy:Hamiltonian} was derived from the action \eqref{susy:N11_start_action} in section \ref{susy:Hamiltonian_formulation}.

An alternative way of deriving the T-duality transformations \eqref{T-dual:Tduality-transform} is to start with the ansatz 
\begin{align}
\left\{
\begin{array}{rl}
D\phi^0 =& A \tilde{S}_0 + C^{0a} \tilde{S}_a + F D\tilde{\phi}^0  + K^0_a D\tilde{\phi}^a\\
D\phi^a =& D\tilde{\phi}^a\\
S_0 =& L \tilde{S}_0 + M_0^a \tilde{S}_a + N D\tilde{\phi}^0 + P^0_a D\tilde{\phi}^a \\
S_a =& Q^0_{a} \tilde{S}_0 + R^b_a \tilde{S}_b + T_{a0} D\tilde{\phi}^0 + U_{ab} D\tilde{\phi}^b 
\end{array}
\right.\label{T-dual:ansatz}
\end{align}
and require that the Hamiltonian \eqref{susy:Hamiltonian} transforms into \eqref{T-dual:dual_hamiltonian}. The reason for not allowing the $\phi^a$ field to transform is that we do not want the complication that the derivatives of the metric and $B$-field to transform, hence the ansatz \eqref{T-dual:ansatz} is not the most general. By considering the special case where the background fields are constant we find that all the unknown parameters in the ansatz are fixed up to two independent signs. Considering the complete case of non-constant background fields fixes the signs and we find the transformation \eqref{T-dual:Tduality-transform}.

To explore the properties of the T-duality transformation \eqref{T-dual:Tduality-transform} we examine how the first term in the phase space action \eqref{susy:phase_space_action} transforms. It follows that 
\begin{align}
\int\d\s\d\theta \; \i(S_\mu-B_{\mu\nu}D\phi^\nu)\partial_0\phi^\mu  \longrightarrow
\int\d\s\d\theta \; \i(\tilde{S}_\mu-\tilde{B}_{\mu\nu}D\tilde{\phi}^\nu)\partial_0\tilde{\phi}^\mu, \label{T-dual:transf_of_Liouville_gen_term}
\end{align}
where we have dropped a term that is a total derivative with respect to $\sigma^0$ and hence, does not affect the Hamiltonian equations of motion. Next, recall that the transformed term defines the Liouville form, which in turn defines the symplectic structure and the Poisson bracket on the phase space. From \eqref{T-dual:transf_of_Liouville_gen_term} we find that the T-duality transformation changes the $H$-twisted Poisson bracket into the $\tilde{H}$-twisted Poisson bracket, which is indeed the one that arises in the derivation of the T-dual Hamiltonian \eqref{T-dual:dual_hamiltonian} from the T-dual action \eqref{T-dual:T-dual_action}. 

\begin{figure}
\begin{centering}
\input{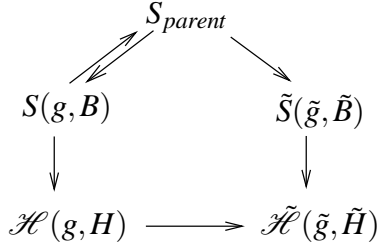}
\caption{Relations between the different formulations of the sigma model. Using the T-duality transformation \eqref{T-dual:transf_of_Liouville_gen_term} the diagram commutes in the sense that if we start from $S(g,B)$ we can choose any of the two paths that takes us to $\tilde{\mathcal{H}}(\tilde{g},\tilde{H})$ and obtain the same result. }
\label{T-dual:T-duality_relations}
\end{centering}
\end{figure}

We conclude that the T-duality transformation \eqref{T-dual:Tduality-transform} is a symplectomorphism that takes the Hamiltonian into the correct T-dual Hamiltonian and the Poisson bracket into the correct T-dual bracket. The complete situation can be summarized as in figure \ref{T-dual:T-duality_relations}.

\section{Extended supersymmetry in the T-dual model}
In the previous chapter we presented ways of introducing extended supersymmetry in the sigma model. Here we address the question if the amount of extended supersymmetry survives the T-duality transformation. In the process we will find how a generalized complex structure transforms under T-duality. 

Recall that in phase space the original model has one manifest supersymmetry \eqref{susy:N1_manifest_transf1}-\eqref{susy:N1_manifest_transf2} and one non-manifest supersymmetry \eqref{susy:non-manifest_ham_susy_1}-\eqref{susy:non-manifest_ham_susy_2}. Since the T-dual Hamiltonian arise from the T-dual action, which has exactly the same form as the original one, in phase space the T-dual model will also have one manifest supersymmetry and one non-manifest supersymmetry. These are of the same form as for the original model but with the fields replaced by their T-dual partners.

We will now assume that the original model has an extended supersymmetry, generated by a generator $\mathbf{Q}_2$ as in \eqref{susy:Q_2_compact}. For readability we repeat the definition, 
\begin{align}
\mathbf{Q}_2(\epsilon) = -\frac{1}{2} \int_{S^{1,1}} \d\s\d\theta \;
\epsilon \langle \Lambda, \genJ \Lambda \rangle. \label{T-dual:ext_generator}
\end{align}
In particular, remember that the $\Lambda$ is a section of $X^*\Pi((TM\oplus T^*M)\otimes\mathbbm{C})$ and that the transformation \eqref{T-dual:Tduality-transform} tells us how such an object transforms under T-duality. However, to apply \eqref{T-dual:Tduality-transform} we use the adapted coordinates and write the objects appearing in \eqref{T-dual:ext_generator} as, \cf \eqref{susy:GCG_pairing}, \eqref{susy:Q_2_ingredients},
\begin{align}
\genI = \Tpairing{a}{b},\;\;\;
\genJ = \TgenJ{a}{b},\;\;\;
\Lambda = \tLambda{a}.
\end{align}
Next, we define the section $\tilde{\Lambda}$ of the T-dual bundle $\tilde{X}^*\Pi((T\tilde{M}\oplus T^*\tilde{M})\otimes\mathbbm{C})$ by
$\tilde{\Lambda}^t =(D\tilde{\f}^0, D\tilde{\f}^a, \tilde{S}_0,\tilde{S}_a)$. 
We also define the bundle morphism $T$ and its inverse $T^{-1}$,
\begin{align}
T:(TM\oplus T^*M)\otimes\mathbbm{C}\longrightarrow(T\tilde{M}\oplus T^*\tilde{M})\otimes\mathbbm{C},\\
T^{-1}:(T\tilde{M}\oplus T^*\tilde{M})\otimes\mathbbm{C} \longrightarrow (TM\oplus T^*M)\otimes\mathbbm{C},
\end{align}
by
\begin{align}
T = \Tdual{a}{b} , \;\;\;\;
T^{-1} = \invTdual{a}{b}.
\end{align}
The T-duality transformation \eqref{T-dual:Tduality-transform} is now written as
$\tilde{\Lambda} = T\Lambda$ and its inverse as $\Lambda = T^{-1}\tilde{\Lambda}$.
It is straightforward to show that 
\begin{align}
\genI = T^t \genI T, \label{T-dual:T_in_Odd}
\end{align} which imply that the T-duality transformation preserves the natural pairing \eqref{susy:GCG_pairing} and hence is an element of $O(2d,2d)$.

We now have the necessary tools to find the T-dual partner to the generator of the extended supersymmetry $\mathbf{Q}_2$ in \eqref{T-dual:ext_generator}. For this we express the fields in $\mathbf{Q}_2$ in their T-dual partners as
\begin{align}
\langle \Lambda, \genJ \Lambda \rangle 
&= \langle T^{-1}\tilde{\Lambda}, \genJ T^{-1}\tilde{\Lambda} \rangle 
= \tilde{\Lambda}^t (T^{-1})^t \genI \genJ T^{-1}\tilde{\Lambda}\cr
&= \tilde{\Lambda}^t \genI T \genJ T^{-1}\tilde{\Lambda}
= \langle \tilde{\Lambda}, T \genJ T^{-1}\tilde{\Lambda} \rangle
\end{align}
where we have used that $(T^{-1})^t \genI = \genI T$ which follows from \eqref{T-dual:T_in_Odd}. This defines a generator $\tilde{Q}_2$ on the T-dual phase space as
\begin{align}
\tilde{\mathbf{Q}}_2(\epsilon) = -\frac{1}{2}\int_{S^{1,1}}\d\s\d\theta\; \epsilon
\langle \tilde{\Lambda}, \tilde{\genJ} \tilde{\Lambda} \rangle. \label{T-dual:T-dual_generator}
\end{align}
with $\tilde{\genJ}= T\genJ T^{-1}$. To be a generator of extended supersymmetry in the T-dual phase space, the map $\tilde{\genJ}$ in $\tilde{\mathbf{Q}}_2$ is required to be a $\tilde{H}$-twisted generalized complex structure. Using that $\genJ$ is a $H$-twisted generalized complex structure it follows that $\tilde{\genJ}^2 = -\mathbbm{1}$ and $\tilde{\genJ}^t\genI\tilde{\genJ} = \genI$,
meaning that $\tilde{\genJ}$ satisfies the conditions \eqref{susy:cond_aGCS} for being a generalized almost complex structure on the space $(T\tilde{M}\oplus T^* \tilde{M})\otimes\mathbbm{C}$. 

One way to verify that the generalized almost complex structure $\tilde{\genJ}$ is integrable, is to write its components in terms of the components of $\genJ$ and verifying that the T-dual components obey the equations \eqref{susy:J_comp_integrability_1} to \eqref{susy:J_comp_integrability_4}, with $H$ replaced by $\tilde{H}$. To verify that $\tilde{\genJ}$ satisfies these equation we need to use that $\genJ$ is a $H$-twisted generalized complex structure. The calculation is very lengthy but straightforward.

Another way to verify that $\tilde{\genJ}$ is integrable is to note that when transforming the supersymmetry algebra \eqref{susy:Hamiltonian_susy_algebra} of the generator $\mathbf{Q}_2$ to its T-dual, as a result of the T-duality transformation being a symplectomorphism, the algebra does not change. Explicitly,
\begin{align}
\{\mathbf{Q}_1(\epsilon_1), \mathbf{Q}_2(\epsilon_2)\}_H = 0 
&\longrightarrow 
\{\tilde{\mathbf{Q}}_1(\epsilon_1), \tilde{\mathbf{Q}}_2(\epsilon_2)\}_{\tilde{H}} = 0, \label{T-dual:T-dual_susy_algebra1}\\
\{\mathbf{Q}_2(\epsilon_1), \mathbf{Q}_2(\epsilon_2)\}_H = \mathbf{P}(2\epsilon_1\epsilon_2)
&\longrightarrow 
\{\tilde{\mathbf{Q}}_2(\epsilon_1), \tilde{\mathbf{Q}}_2(\epsilon_2)\}_{\tilde{H}} = \tilde{\mathbf{P}}(2\epsilon_1\epsilon_2),\label{T-dual:T-dual_susy_algebra2}
\end{align}
where $\tilde{\mathbf{Q}}_1$ generates the manifest supersymmetry in the T-dual phase space and $\tilde{\mathbf{P}}$ is the translation generator on the T-dual phase space, given by \eqref{susy:translation_op} with the fields replaced by their T-duals. That the T-dual of the translation operator $\mathbf{P}$, defined in \eqref{susy:translation_op}, takes this form is verified by a straightforward calculation if we note that $\int_{S^{1,1}}\d\s\d\theta H_{\mu\nu\rho}D\f^\mu Q\f^\nu Q\f^\rho = -\frac{1}{3}\int_{S^{1,1}}\d\s\d\theta H_{\mu\nu\rho}D\f^\mu D\f^\nu D\f^\rho$.

Since the generator $\tilde{\mathbf{Q}}_2$ satisfies the algebra \eqref{T-dual:T-dual_susy_algebra1}-\eqref{T-dual:T-dual_susy_algebra2} it is a supersymmetry generator. This implies that the map $\tilde{\genJ}$, defining $\tilde{\mathbf{Q}}_2$ through \eqref{T-dual:T-dual_generator}, is a $\tilde{H}$-twisted generalized complex structure \cite{Zabzine:2005qf}. Thus we have proved integrability of the T-dual $\tilde{\genJ}$.

In article [V] the integrability of $\tilde{\genJ}$ is shown in a third way based on modifying a proof given by Cavalcanti in \cite{Cavalcanti:2005hq}. In the proof in article [V] the target space and the T-dual target space are considered as trivial fiber bundles with the fibers given by $S^1$'s, the isometry directions. By specifying the connections on the fiber bundles it is possible to write the $H$ and $\tilde{H}$ fields in a convenient way. Further, in the article the T-duality transformation of differential forms is considered and the transformation \eqref{T-dual:Tduality-transform} of sections of the bundle is reformulated in the new language using the connections. A useful relation between the T-duality transformations of the differential forms and of the sections is found using the new expressions of $H$ and $\tilde{H}$. Together with this relation, an expression for how the H-twisted Courant bracket acts on differential forms, given in \cite{Gualtieri:2003dx, Cavalcanti:2005hq}, is used to show that the T-dual of the $+i$ eigenspace of a $H$-twisted generalized complex structure is involutive under the $\tilde{H}$-twisted Courant bracket. This shows that the T-dual $\tilde{\genJ}$ is integrable. However, this proof requires that $\mathcal{L}_k\genJ = 0$. That $\genJ$ satisfies this condition is a consequence of requiring that the isometry transformation in phase space, \eqref{T-dual:N11_isometry} reduced to $\mathcal{N}=1$ fields, should commute with the extended supersymmetry generated by $\mathbf{Q}_2$.

We summarize the above discussion of integrability: Given a $H$-twisted generalized complex structure $\genJ$ its T-dual is a $\tilde{H}$-twisted generalized complex structure given by $\tilde{\genJ}=T\genJ T^{-1}$.

Since the T-dual $\tilde{\genJ}$ is integrable it defines an extended supersymmetry generator $\mathbf{Q}_2$ as in \eqref{T-dual:T-dual_generator}. Hence, given an extended supersymmetry in the original model there is an extended supersymmetry in the T-dual model, the amount of extended supersymmetry is preserved under the T-duality transformation. 

We now relate this result to the discussions on extended supersymmetry and geometry in sections \ref{GKG_from_sigmamodel} and \ref{GHKG_from_sigmamodel}. The sigma model with $\mathcal{N}=(2,2)$ extended supersymmetry has as target space a $H$-twisted generalized K\"{a}hler geometry. The corresponding T-dual sigma model also has $\mathcal{N}=(2,2)$ extended supersymmetry and the target space is required to be $\tilde{H}$-twisted generalized K\"{a}hler. The reason for the $\tilde{H}$-twist is that the T-dual $\tilde{\genJ}^{(i)}$'s, defining the two generators of extended supersymmetry, are $\tilde{H}$-twisted generalized complex structures. Further, the sigma model with $\mathcal{N}=(4,4)$ extended supersymmetry has as target space a $H$-twisted generalized hyperK\"{a}hler geometry. Its T-dual sigma model thus has $\mathcal{N}=(4,4)$ extended supersymmetry and its target space is required to be $\tilde{H}$-twisted generalized hyperK\"{a}hler.  
        \chapter{First order sigma model and extended supersymmetry}\label{1storder_susy}

It is natural to look for alternative ways in which generalized complex geometry naturally arise out of demanding the sigma model to have extended supersymmetry. 

In \cite{Lindstrom:2004iw} several types of supersymmetric sigma
models are studied. It is found that a topological model admits $\mathcal{N}=(2,0)$
supersymmetry if the target space is a generalized complex manifold, where the
closure of the supersymmetry algebra is assured by the integrability with
respect to the (untwisted) Courant bracket of a generalized complex structure.
It is also found that adding a term containing a $B$-field to the topological
action restricts the geometry of the target space to that of a generalized
complex manifold where the integrability for the generalized complex structure
now is given in terms of the $H$-twisted Courant bracket.

In this chapter we study a supersymmetric sigma model written in a first order form, introduced in \cite{Lindstrom:2004eh}, and investigate the implications of off-shell extended $\mathcal{N}=(2,2)$ supersymmetry on the target space. We find a structure that seems to contain more than generalized complex geometry. It is described in terms of two copies of the cotangent bundle instead of one. Further, we present an example that shows how generalized complex geometry and ordinary complex geometry is contained in the new framework.

\section{The sigma model}
The action that we will consider is a first order action. It introduces Grassmann-odd valued auxiliary $\mathcal{N}=(1,1)$ superfields that are present to give the possibility that the extended $\mathcal{N}=(2,2)$ supersymmetry closes off shell. The action we will study is given by \cite{Lindstrom:2004eh} 
\begin{align}\label{1st:S_first_order}
S_{1st} = \int \d^2 \sigma \d^2 \theta\left\{
         \Psi_{(+|\mu} D_{-)}\Phi^\mu + \Psi_{+\mu}\Psi_{-\nu}E^{\mu\nu}
                           \right\},
\end{align}
where $\Phi$, $\Psi_{\pm}$ are $\mathcal{N}=(1,1)$ superfields and $E=E(\Phi^\mu)$. $E$ can be thought of as the inverse of $g_{\mu\nu}+B_{\mu\nu}$. The equations of motion that follow from variation of this action are
\begin{align}
\delta \Psi_{+\mu} \Longrightarrow & \;\;\; D_-\Phi^\mu + \Psi_{-\nu} E^{\mu\nu} =0, \label{1st:eom_psi+}\\
\delta \Psi_{-\nu} \Longrightarrow & \;\;\; D_+\Phi^\mu - \Psi_{+\nu} E^{\nu\mu} =0, \label{1st:eom_psi-}\\
\delta \Phi^\mu \Longrightarrow & \;\;\; D_{(+} \Psi_{-)\mu} + 
\Psi_{+\nu}\Psi_{-\gamma}E^{\nu\gamma}_{\ph{\nu\gamma},\mu} = 0,
\end{align}
where we, as before, use the notation $E^{\nu\gamma}_{\ph{\nu\gamma},\mu}=\partial_\mu E^{\nu\gamma}$. Note that the first order action \eqref{1st:S_first_order} does not require that the $E$-tensor is invertible. This feature makes this formulation suitable for models where the target space is assumed to be a Poisson manifold. In such manifolds the metric is absent and the symplectic structure is in general not invertible.

However, if we assume that the $E$-tensor is invertible and integrate out the $S_{\pm\mu}$-fields from the action \eqref{1st:S_first_order}, using the equations of motion \eqref{1st:eom_psi+} and \eqref{1st:eom_psi-}, we obtain the second order action \eqref{susy:S_second_order} with $(E^{-1})_{\mu\nu} = g_{\mu\nu}+B_{\mu\nu}$. This shows the equivalence of the two actions under the assumption that $E$ is invertible.

In \cite{Lindstrom:2004eh} extended $\mathcal{N}=(2,2)$ supersymmetry of the action \eqref{1st:S_first_order} was studied on $\Psi$-shell, \ie by using the equations of motion \eqref{1st:eom_psi+} and \eqref{1st:eom_psi-}, and requiring a metric to be present. The supersymmetry transformations found was later shown to hold in the case where $E$ is a non-invertible Poisson tensor \cite{Bergamin:2004sk}. In article [III] we study the off-shell extended supersymmetry of the sigma model, which is what we turn to next.

\section{The supersymmetry transformation}
The sigma model \eqref{1st:S_first_order} is written in terms of $\mathcal{N}=(1,1)$ superfields. This means that it is invariant under the manifest supersymmetry. Further this supersymmetry does not pose any conditions on the target manifold. To study the conditions for extended supersymmetry in this setting we make an ansatz and demand that the left and right moving extended supersymmetry obey the standard conditions similar to \eqref{susy:12_comm:1} - \eqref{susy:2pp_comm},
\begin{align}
\left[\delta_1^\pm(\epsilon_1^\pm), \delta_2^\pm(\epsilon_2^\pm)\right] A &= 0, \label{1st:12_comm:1}\\
\left[\delta_1^\pm(\epsilon_1^\pm), \delta_2^\mp(\epsilon_2^\mp)\right] A &= 0, \label{1st:12_comm:2}\\
\left[\delta_2^\pm(\epsilon^\pm_1), \delta_2^\mp(\epsilon^\mp_2)\right] A &= 0, \label{1st:2pm_comm}\\
\left[\delta_2^\pm(\epsilon^\pm_1), \delta_2^\pm(\epsilon^\pm_2)\right] A 
      &= -2\i\epsilon^\pm_1 \epsilon^\pm_2 \partial_\ppmm A,\label{1st:2pp_comm}
\end{align}
with $A$ being a generic field representing $\Phi^\mu$, $\Psi_{+\mu}$ or $\Psi_{-\mu}$. As before, these conditions mean that the extended supersymmetry must commute with the manifest supersymmetry, the right moving commutes with the left moving supersymmetry and they commute to a world-sheet translation of the field. We will write the ansatz in terms of $\mathcal{N}=(1,1)$ fields so that the conditions \eqref{1st:12_comm:1} and \eqref{1st:12_comm:2} are automatically satisfied. Hence we will have to investigate the twelve conditions arising from \eqref{1st:2pm_comm} and \eqref{1st:2pp_comm} with $A=\Phi^\mu, \Psi_{+\mu}, \Psi_{-\mu}$. 


The most general ansatz for the extended supersymmetry that does not include any dimensionful parameters is given by \cite{Lindstrom:2004eh}
\begin{align}
  \delta_2^{(\pm)}\Phi^\mu =&
     \eps^{\pm}\left(D_{\pm}\Phi^\nu
     J^{(\pm)\mu}_{~~\nu}-\Psi_{\pm\nu}P^{(\pm)\mu\nu}\right),
     \label{1st:del+-Phi:general}\\
  \delta_2^{(\pm)} \Psi_{\pm\mu} =&
    \eps^{\pm}\left(D_\pm^2\Phi^\nu
    L^{(\pm)}_{\mu\nu}-D_{\pm}\Psi_{\pm\nu}K^{(\pm) \nu}_{\mu}
    +\Psi_{\pm\nu}\Psi_{\pm\sigma}N^{(\pm)\nu\sigma}_{\mu}\right.\nonumber \\
  &\left. \qquad \qquad +D_{\pm}\Phi^\nu D_{\pm}\Phi^\rho
    M^{(\pm)}_{\mu\nu\rho}+D_{\pm}\Phi^\nu \Psi_{\pm\sigma}
    Q^{(\pm)\sigma}_{\mu\nu}\right), \label{1st:del+-S+-:general} \\
  \delta_2^{(\pm)} \Psi_{\mp\mu} =&
    \eps^{\pm}\left(D_{\pm}\Psi_{\mp\nu}R^{(\pm)\nu}_{\mu}+
    D_{\mp}\Psi_{\pm\nu}Z^{(\pm)\nu}_{\mu}+D_{\pm}D_{\mp}\Phi^\nu
      T^{(\pm)}_{\mu\nu}\right.
    \nonumber \\
  &\left. \qquad \qquad +\Psi_{\pm\rho}D_{\mp}\Phi^\nu U^{(\pm)\rho}_{\mu\nu}
    +D_{\pm}\Phi^\nu \Psi_{\mp\rho} V^{(\pm)\rho}_{\mu\nu}\right.\nonumber \\
  &\left. \qquad \qquad +D_{\pm}\Phi^\nu D_{\mp}\Phi^\rho X^{(\pm)}_{\mu\nu\rho}
    +\Psi_{\pm\nu}\Psi_{\mp\rho}Y^{(\pm)\nu\rho}_{\mu}\right).
  \label{1st:del+-S-+:general}
\end{align}
Here, all the tensors in general depend on $\Phi$ and are at this stage arbitrary. We will determine them by the requiring that the transformation is an extended supersymmetry. Using this ansatz we calculate the commutators in \eqref{1st:2pm_comm} and \eqref{1st:2pp_comm}. The results of the calculation are given in appendices \ref{Appendix:n21} and \ref{Appendix:n22}. Requiring that the commutators should obey the algebra \eqref{1st:2pm_comm}-\eqref{1st:2pp_comm} give us about one hundred coupled non-linear tensor equations.

To examine the conditions that arise it is convenient to introduce the space $\mathbbm{E}=(TM\oplus T^* M_+ \oplus T^* M_-)\otimes\mathbbm{C}$, which is the complexified direct sum of the tangent bundle of the $d$-dimensional target space $M$ and two copies of the cotangent space of $M$. To motivate the introduction of this space recall that when we considered generalized complex geometry in chapter \ref{extended_susy} we had only one field, the $S_\mu$ field, living on $T^* M$, which together with $\phi^\mu$ gave us a natural basis of the phase space. Here we have the field $\Phi^\mu$ living on $TM$ and the fields $\Psi_{\pm\mu}$ both living on $T^* M$. For this reason we introduce two copies $T^* M_\pm$, corresponding to the two Grassmann directions on the world-sheet and use $(\Phi^\mu, \Psi_{+\mu}, \Psi_{-\mu})$ as local coordinates on $\mathbbm{E}$. Objects living on $\mathbbm{E}$ was first considered in connection to supersymmetry by \cite{Lindstrom:2004iw}.

We define the $6d\times6d$ matrices $\GgenJ{(\pm)}$ acting on $\mathbbm{E}$ by
\begin{align}
\GgenJ{(+)} = \Litenmatrix{ 
                        \ph{-}J^{(+)} & \ph{-}P^{(+)}&0\\
                        -L^{(+)}& \ph{-}K^{(+)}&0 \\
                        \ph{-}T^{(+)}& -Z^{(+)}& \ph{-}R^{(+)}
                        },\;\;\;
\GgenJ{(-)} = \Litenmatrix{
                      \ph{-}J^{(-)} & \ph{-}0 & - P^{(-)}\\
                      \ph{-}T^{(-)} & \ph{-}R^{(-)} & -Z^{(-)}\\
                      -L^{(-)} & \ph{-}0 & \ph{-} K^{(-)}
                       }.
\end{align}
It turns out that the algebraic conditions, \ie those that do not involve any derivatives of the tensors, arising from the commutators \eqref{1st:2pm_comm} and \eqref{1st:2pp_comm} can be written naturally in terms of $\GgenJ{(\pm)}$ as
\begin{align}
\GgenJ{(\pm)2} = -\mathbbm{1}, \;\;\;\;\;\;\;\;
[\GgenJ{(+)},\GgenJ{(-)}]=0.\label{1st:SUSY_algebra_algebraic_cond}
\end{align}
This implies that $\GgenJ{(\pm)}$ are two commuting almost complex structures on $\mathbbm{E}$.

Next we must ensure that the transformation \eqref{1st:del+-Phi:general} - \eqref{1st:del+-S-+:general} is an invariance of the action \eqref{1st:S_first_order}. The result of the variation of the action \eqref{1st:S_first_order} is presented in appendix \ref{Appendix:var_action} and demanding invariance impose eleven further conditions on the tensors in the ansatz. The algebraic conditions that arise can nicely be written with the help of the $6d\times6d$ matrix
\begin{align}
\GgenG = \matrix{ccc}{0&1&-1\\1&0&E\\-1&E^t&0}.
\end{align}
They become
\begin{align}
\GgenJ{(\pm)t}\GgenG\GgenJ{(\pm)} = \GgenG, \label{1st:action_inv_algebraic_cond}
\end{align}
which implies that $\GgenG$ is Hermitean with respect to both $\GgenJ{\pm}$. However, since $\GgenG$ in general is not invertible, it can not be considered as a metric on $\mathbbm{E}$.

The simplicity of the conditions \eqref{1st:SUSY_algebra_algebraic_cond} and \eqref{1st:action_inv_algebraic_cond} suggest that the differential conditions on the tensors might be written equally simple in terms of objects acting on $\mathbbm{E}$. One way to proceed is to try to extend the $H$-twisted Courant bracket to the space $\mathbbm{E}$ so that an integrability condition similar to \eqref{susy:GCG_integ_cond_proj} reproduces all the differential conditions on the tensors. Unfortunately there is no natural and unique extension of the $H$-twisted Courant bracket to include two copies of $T^* M$. In spite of the beauty of this approach it is rather lengthy to obtain the integrability conditions arising from any given extended bracket and a appropriate bracket that reproduce the equations for the most general case has not been found.

Instead, to study the differential conditions that arise from the algebra and the invariance of the action, we make two further simplifying assumptions, we assume that the $P^{\pm}$ tensors are invertible and that the background field $E$ is a Poisson tensor $\Pi^{\mu\nu}$ of full rank. This means that $\Pi^{-1}$ is a symplectic structure, \ie a closed non-degenerate two-form, and the sigma model under consideration is the first order form of a symplectic sigma model. 

Before considering the differential conditions we briefly investigate what this symplectic sigma model describes. For this we integrate out the auxiliary $\Psi_{\pm}$ fields and end up with an action of the form \eqref{susy:S_2nd_order_explicit} with $g=0$ and $B=\Pi^{-1}$. Since $\Pi^{-1}$ is closed the three-form field $H$ vanishes and hence, the sigma model is trivial. Nevertheless, in article [III] this symplectic sigma model is used as a toy model for exploring the implications of extended supersymmetry in the setting of the sigma model with auxiliary fields.

We use the conditions arising from the invariance of the action to relate the tensors. Then we use the obtained relations to simplify the conditions from the algebra. In particular this gives us that $J^{(\pm)}$, $K^{(\pm)}$ and $R^{(\pm)}$ are six complex structures that are covariantly constant, with the covariant derivative defined by the torsionfree connections $\Gamma^{(J)}$, $\Gamma^{(K)}$ and $\Gamma^{(R)}$ respectively. These connections turn out to be related by the differential conditions arising from the supersymmetry algebra.

We use the above connections to introduce the connection matrix
\begin{align}
\GGamma = \mbox{diag}\left(\Gamma^{(J)},-\Gamma^{(K)},-\Gamma^{(R)}\right)
\end{align}
to formulate the differential conditions in terms of objects acting on $\mathbbm{E}$. The conditions now imply that the almost complex structures $\GgenJ{(\pm)}$ are covariantly constant
\begin{align}
\nabla \GgenJ{(\pm)} \equiv \partial \GgenJ{(\pm)} - \GgenJ{(\pm)}\cdot\GGamma + \GGamma\cdot\GgenJ{(\pm)} =0, \label{1st:cov_const_ggenJ}
\end{align}
and that the generalized Riemann tensor constructed with $\GGamma$ vanishes,
\begin{align}
\GR \equiv \d\GGamma - \GGamma \circ \GGamma = 0.
\end{align}
For K\"{a}hler geometries the Nijenhuis tensor may be expressed in terms of the covariant derivative of the complex structure, further if the complex structure is covariantly constant the Nijenhuis tensor vanishes. This suggests that the condition \eqref{1st:cov_const_ggenJ} could be considered as an integrability condition for $\GgenJ{(\pm)}$. However, as mentioned above it is not clear under which bracket the eigenspaces of $\GgenJ{(\pm)}$ should be involutive. 

Remember that in section \ref{susy:GCG} we found that in generalized complex geometry the $b$-transformation, with $\d b =0$, leaves the integrability condition invariant. Hence, to further study the condition \eqref{1st:cov_const_ggenJ} we generalize the $b$-transformation \eqref{susy:b-transform} to act on sections of $\mathbbm{E}$ by defining it as 
\begin{align}
U = \matrix{ccc}{1&0&0\\b&1&0\\b&0&1}, \label{1st:gen_b-transform}
\end{align}
with $b$ being a closed two-form. If we now consider $U$ as a gauge transformation on $\mathbbm{E}$, the $\GGamma$ transforms as a connection and the almost complex structures as $\GgenJ{(\pm)}\rightarrow U\GgenJ{(\pm)}U^{-1}$. This implies that the condition $\nabla\GgenJ{(\pm)}=0$ remains invariant. Hence, the condition \eqref{1st:cov_const_ggenJ} behaves as we would like a condition for integrability to behave under a $b$-transformation. This further motivates the interpretation of \eqref{1st:cov_const_ggenJ} as a condition for integrability of $\GgenJ{(\pm)}$.

The relations found above from the differential conditions are nicely expressed in terms of geometrical objects. Even though it should be emphasized that they are valid only for the symplectic first order sigma model, the relations indicate that objects living on $\mathbbm{E}$ might be useful to describe the target space of a first order sigma model with a more general background field $E$. They also hint that a richer structure might arise out of requiring extended supersymmetry of the first order sigma model \eqref{1st:S_first_order} compared to what arises from the same requirement on the corresponding second order sigma model \eqref{susy:S_second_order}, which has a target space that is bi-Hermitean, or equivalently generalized K\"{a}hler.

In article [III] we elaborate on the relation to the formulation of the sigma model in terms of $\mathcal{N}=(2,2)$ superfields and find a relation to generalized complex geometry. For this we consider a special case of \eqref{susy:N22_manifest_action} given by 
\begin{align}
S = \int \d^2\s\d^2\theta\d^2\bar{\theta} \;\left\{\mathbbm{X}^A B_{AB'}\mathbbm{Y}^{B'}\right\}, \label{1st:N22_toy_action}
\end{align}
where $B_{AB'}$ is a constant antisymmetric matrix of full rank, $\mathbbm{X}$ is a left semi-chiral and $\mathbbm{Y}$ a right semi-anti-chiral superfield, \cf section \ref{susy:manifest_form}. Being a special case of \eqref{susy:N22_manifest_action} this action has a generalized K\"{a}hler geometry as target space \cite{Lindstrom:2005zr} and hence we know that it incorporates generalized complex geometry. To explore this in detail we reduce the action to its $\mathcal{N}=(1,1)$ form. Doing this we find that it lacks some fields to be written in a form that can be related to \eqref{1st:S_first_order}. For this, we we need to introduce the extra auxiliary fields $\Psi_{+A'}$ and $\Psi_{-A}$ and require that
\begin{align}
\Psi_{+A'} = 0, \;\;\;\;\;\; \Psi_{-A} = 0. \label{1st:half-Psi-constraints}
\end{align}
Defining the indices as $\mu=(A,A')$ makes it possible to write the reduced version of \eqref{1st:N22_toy_action} as 
\begin{align}
S = \int \d^2\s\d^2\theta \;\left\{
\Psi_{+\mu}(B^{-1})^{\mu\nu}\Psi_{-\nu} + D_+\Phi^\mu B_{\mu\nu} D_-\Phi^\nu
\right\}. \label{1st:N11_reduced_action}
\end{align}
The background is given by the constant antisymmetric $B$, which implies that the second term vanishes. Here we have kept the second term since when we perform $b$-transformation, \eqref{1st:gen_b-transform} with $b=-B$, we obtain the action \eqref{1st:S_first_order} with the constant $E=B^{-1}$. Further, we note that the equations of motion that follow from \eqref{1st:N11_reduced_action} for half of the $\Psi_{\pm\mu}$-fields are equivalent to the constraints \eqref{1st:half-Psi-constraints}. Thus, the action \eqref{1st:N11_reduced_action} has more fields than can be incorporated in the manifest $\mathcal{N}=(2,2)$ formulation. 

From the opposite point of view, to write \eqref{1st:N11_reduced_action} in terms of $\mathcal{N}=(2,2)$ fields we need to integrate out half of the auxiliary fields. It is then consistent to simply remove the corresponding components in the almost complex structures $\GgenJ{(\pm)}$. Doing this we find that the reduction of our almost complex structures on $\mathbbm{E}$ match the two commuting generalized complex structures found in \cite{Lindstrom:2004hi}. Further, integrating out the remaining half of the auxiliary fields give us the corresponding second order action, \eqref{susy:S_second_order} with $E^{-1}=B$, and hence the geometry is described by two ordinary commuting complex structures. In summary we have the following diagram
\begin{align}
\GgenJ{(\pm)}\xrightarrow{\Psi_{+A'},\Psi_{-A}=0} 
\genJ^{(\pm)}
\xrightarrow{\Psi_{+A},\Psi_{-A'}=0} 
J^{(\pm)}.
\end{align}
The above example indicates that to completely describe the geometry arising from 
demanding extended supersymmetry in the sigma model given by \eqref{1st:S_first_order} we may need a geometry beyond generalized complex geometry.

To conclude this chapter we mention that attempts to find a supersymmetry transformation for the sigma model with a general $E$-field have been made by the authors of [III]. However, we have not been able to find the most general answer and thus, the most general $\mathcal{N}=(2,2)$ supersymmetry transformations of the sigma model \eqref{1st:S_first_order} with a general background is yet to be found.

        \chapter{Epilogue}

In this thesis we have considered some aspects of string theory. After a general introduction to the subject we, by considering the tensionless limit of the string, constructed a IIB supergravity background generated by a tensionless string. The background had the characteristics of a gravitational shock-wave. We then went on to consider quantization of the tensile and tensionless string in a pp-wave background. We found, due to the special type of background, that if we quantize the theory before the tensionless limit is taken we find the same result as if we quantize after taking the limit. The last part of the thesis considers different aspects of extended world-sheet supersymmetry of sigma models. We described how complex geometry and generalized complex geometry naturally arise out of the requirement of extended world-sheet supersymmetry. Further, the Hamiltonian formulation of the sigma model was constructed and a physical explanation to the equivalence of the bi-Hermitean geometry and generalized K\"{a}hler geometry was found. This equivalence follows from the equivalence of the Lagrangian- and Hamiltonian formulations of the sigma model. Then, T-duality in the Hamiltonian formulation of the sigma model was studied. The outcome was that, when we consider all fields being independent of the direction in which the T-duality is performed, the amount of extended supersymmetry is preserved under the T-duality transformation.

The work presented in this thesis has given contributions to our present understanding of some aspects of string theory. In particular, it contributes to the understanding of the tensionless string and it provides insight into the classification of the different target spaces required by different models of the string.

Despite the amount of research put into string theory, to date we do not have a complete understanding of it and there is still a lot to be investigated. It will be very exciting to follow the future developments of the field and to see what the theory will teach us about the world we live in. In particular, it will be interesting to see when, and if, experimental tests of string theory will be constructed and what they will tell us. To conclude, it will be interesting to find out what the future holds for string theory and whether the smallest constituents of matter really are vibrating strings or perhaps something even more fundamental.
        \chapter*{Acknowledgments\markboth{Acknowledgments}{Acknowledgments}} 
\addcontentsline{toc}{chapter}{Acknowledgments}

First of all, I would like to thank my supervisor Ulf Lindstr\"{o}m for giving me the opportunity to study string theory, for rewarding collaborations and for enlightening discussions and guidance. I would also like to thank Maxim Zabzine for fruitful discussions and cooperation. Thanks also to my collaborators Andreas Bredthauer and Linus Wulff for stimulating discussions and great fun.

Many thanks to Rolf Paulsson for organizing wonderful ski-conferences and for giving me the opportunity to teach. Thanks also to Ulf Danielsson and Staffan Yngve for enjoyable teaching collaborations and to Rabab Elkarib for all the help with administrative issues.

I would like to thank all of the past and present members of the Department of Theoretical Physics for providing a stimulating environment and for all the fun during my time as a Ph.D. student. 

During the preparation of this thesis I have benefited much from discussions with several people. In particular, I am grateful to Thomas Klose and Hans Johansson for comments and feedback on the manuscript. Thanks also to Roberth Asplund for entertaining physics discussions and for reading and commenting on the first draft, to Joel Ekstrand for useful comments and help with the typesetting and to Magdalena Larfors, Niklas Johansson and Valentina Giangreco Puletti for helpful comments.

I would also like to thank Andreas H\"{o}glund, Bjarte Mohn and Ola Wessely for pleasant physics discussions and a lot of fun. Thanks also to Martin L\"{u}bcke for support, good food and bad movies.

I also wish to express my gratitude to my parents and to my and \rund{A}sa's families for all support during my time as a Ph.D. student. 

Last, but certainly not least, I would like to express my deepest gratitude to \rund{A}sa for all love, patience and support.

        \appendix
        \chapter*{Appendix A: Conditions from the supersymmetry algebra}
\addcontentsline{toc}{chapter}{Appendix A: Conditions from the supersymmetry algebra}
\setcounter{chapter}{1}

In this appendix we state the results needed when using the most general ansatz for extended supersymmetry given by \eqref{1st:del+-Phi:general} - \eqref{1st:del+-S-+:general}. We first give the explicit outcome of using the ansatz to perform the variation of the action \eqref{1st:S_first_order}. We then give the results of the calculations of the commutators needed for the ansatz to be a supersymmetry.

\section{Variation of the action}\label{Appendix:var_action}
Using the most general transformations \eqref{1st:del+-Phi:general} - \eqref{1st:del+-S-+:general} and performing the variation of the action \eqref{1st:S_first_order} we obtain
{\footnotesize
\begin{align*}
\delta_2^{(+)}S =& \int \d^2\s\d^2\theta \epsilon^+  \Big\{
     \smallhalf D_+^2\Phi^\a D_-\Phi^\beta \big( 
            T^{(+)}_{(\a\b)} 
          + L^{(+)}_{(\b\a)} 
          \big) \label{eqn:action:full-variation}\\
& + \Psi_{-\a}D_+^2\Phi^\b \big(
            J^{(+)\a}_\b 
          + L^{(+)}_{\mu\b}E^{\mu\a}
          + R^{(+)\a}_\b 
          \big) \\
& + \Psi_{+\a}D_+D_-\Phi^\b \big(
          - K^{(+)\a}_\b
          - Z^{(+)\a}_\b
          - E^{\a\mu}T^{(+)}_{\mu\b}
          - J^{(+)\a}_\b
          \big) \\
& + \Psi_{+\a}D_+\Psi_{-\b}\big(
          - K^{(+)\a}_\mu E^{\mu\b}
          - E^{\a\mu} R^{(+)\b}_\mu
          - P^{(+)\b\a}
          \big) \\
& - \smallhalf \Psi_{+\a}D_-\Psi_{+\b}\big(
           E^{(\a|\mu}Z^{(+)\b)}_\mu
          + P^{(+)(\a\b)}
          \big) \\
& + \smallhalf \Psi_{+\a}\Psi_{+\b}D_-\Phi^\r \big(
           N^{(+)[\a\b]}_{\ph{(+)[\a\b]}\r}
          + U^{(+)[\a}_{\mu\r}E^{\b]\mu} \nonumber \\
& \ph{+ \smallhalf \Psi_{+\a}\Psi_{+b}D_-\Phi^\r \big(}
          - \smallhalf[ E^{[\a|\mu}Z^{\p\b]}_\mu
             - P^{\p[\a\b]} ]_\r
          \big) \\
& + \smallhalf \Psi_{-\a}D_+\Phi^\b D_+\Phi^\g\big(
          - J^{(+)\a}_{[\b,\g]}
          + E^{\mu\a}M^{\p}_{\mu[\b\g]} 
          + V^{(+)\a}_{[\b\g]}
          - R^{(+)\a}_{[\b,\g]}
          \big)\\ 
& + \Psi_{+\a}D_+\Phi^\b D_-\Phi^\g\big(
          - J^{(+)\a}_{\b\g}
          - Q^{(+)\a}_{\g\b}
          - U^{(+)\a}_{\b\g} \cr
&\ph{+ \Psi_{+\a}D_+\Phi^\b D_-\Phi^\g\big(}
          - E^{\a\mu}X^{(+)}_{\mu\b\g}
          - K^{(+)\a}_{\g,\b}
          - Z^{(+)\a}_{\b,\g}
          \big) \\
& + \smallhalf \Psi_{+\a}\Psi_{+\b}\Psi_{-\g}\big(
            N^{(+)[\a\b]}_\mu E^{\mu\g}
          + E^{[\b|\mu} Y^{(+)\a]\g}_{\mu} \cr
&\ph{+ \smallhalf \Psi_{+\a}\Psi_{+\b}\Psi_{-\g}\big(}
          - P^{(+)\mu[\a}E^{\b]\g}_{\ph{\b]\g},\mu}
          \big) \\
& + \smallhalf D_+\Phi^\a D_+\Phi^\b D_-\Phi^\g \big(
            M^{(+)}_{\g[\a\b]}
          + X^{(+)}_{[\a\b]\g}
          - T^{(+)}_{[\a|\g|\b]} \nonumber \\
& \ph{ + \smallhalf D_+\Phi^\a D_+\Phi^\b D_-\Phi^\g \big(}
          + \smallhalf [T^\p_{[\b\a\g]}+L^\p_{[\a\b\g]}]
          \big) \\
& + \Psi_{+\a}\Psi_{-\b}D_+\Phi^\g \big(
            Q^{(+)\a}_{\mu\g} E^{\mu\b}
          + E^{\a\mu}V^{(+)\b}_{\mu\g}
          + Y^{(+)\a\b}_{\g}  \cr
&\ph{+ \Psi_{+\a}\Psi_{-\b}D_+\Phi^\g \big(}
          + E^{\a\b}_{\ph{\a\b},\mu} J^{(+)\mu}_{\g}
          + (K^{(+)\a}_{\mu}E^{\mu\b})_{,\g}
          \big)
\Big\}.
\end{align*} 
}
Demanding that the action \eqref{1st:S_first_order} is invariant under the transformation gives us eleven conditions on the tensors in the transformation.

\section{$\mathcal{N}=(2,1)$ commutators}\label{Appendix:n21}
To have $\mathcal{N}=(2,1)$ supersymmetry, \ie one extended supersymmetry in the left or right moving sector, we need the following commutators to obey the supersymmetry algebra \eqref{1st:2pp_comm}.

{\footnotesize
\begin{eqnarray*}
[&&\hspace{-18pt}\delta_2^{+}(\eps^+_1), \delta_2^{+}(\eps^+_2)]\Phi^\mu = \cr
 && \ph{+}2\eps_1^{+}\eps_2^{+} D^2_{+}\Phi^\alpha \Big(
          J^{(+)\nu}_{~~\alpha} J^{(+)\mu}_{~~\nu}
          +L^{(+)}_{\nu\alpha} P^{(+)\mu\nu}
          \Big)\cr
 && + 2\eps_1^{+}\eps_2^{+} D_{+}\Psi_{+\alpha} \Big(
         -P^{(+)\nu\alpha} J^{(+)\mu}_{~~\nu}
         -K^{(+) \alpha}_{\nu} P^{(+)\mu\nu}
         \Big)\cr
\cr 
 && + 2\eps_1^{+}\eps_2^{+} D_{+}\Phi^\alpha D_{+}\Phi^\beta \Big(
         -J^{(+)\nu}_{~~\alpha,\beta} J^{(+)\mu}_{~~\nu}
         +J^{(+)\mu}_{~~\alpha\rho} J^{(+)\rho}_{~~\beta}
         +M^{(+)}_{\nu\alpha\beta}P^{(+)\mu\nu}
         \Big) \cr
 && + 2\eps_1^{+}\eps_2^{+} \Psi_{+\alpha}D_+\Phi^\beta \Big(
         P^{(+)\nu\alpha}_{\ph{(+)\nu\alpha},\beta}J^{(+)\mu}_{~~\nu}
         +J^{(+)\mu}_{~~\beta\rho} P^{(+)\rho\alpha}\cr
 &&\ph{+ 2\eps_1^{+}\eps_2^{+} \Psi_{+\alpha}D_+\Phi^\beta \Big(}
         -Q^{(+)\alpha}_{\nu\beta} P^{(+)\mu\nu}
         -P^{(+)\mu\alpha}_{\ph{(+)\mu\alpha},\rho} J^{(+)\rho}_{~~\beta}
         \Big)\cr
 && + 2\eps_1^{+}\eps_2^{+} \Psi_{+\alpha}\Psi_{+\beta} \Big(
         N^{(+)\alpha\beta}_{\nu} P^{(+)\mu\nu}
         +P^{(+)\mu\alpha}_{\ph{(+)\mu\alpha},\rho}P^{(+)\rho\beta}
         \Big)\cr
\end{eqnarray*}
}
{\footnotesize
\begin{eqnarray*}
[&&\hspace{-18pt}\delta_2^+(\eps^+_1),\delta_2^+(\eps^+_2)] \Psi_{+\mu} = \cr 
 &&   \ph{+}  2\eps_1^{+}\eps_2^{+} D^3_{+}\Phi^\alpha \Big(
            - J^{(+)\nu}_{~~\alpha} L^{(+)}_{\mu\nu} 
            - L^{(+)}_{\nu\alpha} K^{(+) \nu}_{\mu}
            \Big)\cr
&&  +2\eps_1^{+}\eps_2^{+} D^2_{+}\Psi_{+\alpha}\Big(
              P^{(+)\nu\alpha} L^{(+)}_{\mu\nu}
            + K^{(+) \alpha}_{\nu} K^{(+) \nu}_{\mu}
            \Big)\cr
\cr 
&&  +2\eps_1^{+}\eps_2^{+} D_{+}\Phi^\alpha D^2_{+}\Phi^\beta \Big(
            - J^{(+)\nu}_{~~\alpha,\beta} L^{(+)}_{\mu\nu}
            - J^{(+)\rho}_{~~\alpha} L^{(+)}_{\mu\beta,\rho} 
            - L^{(+)}_{\nu\beta,\alpha}K^{(+) \nu}_{\mu}   \cr            
&& \ph{+2\eps_1^{+}\eps_2^{+} D_{+}\Phi^\alpha D^2_{+}\Phi^\beta \Big(}
            + M^{(+)}_{\nu[\alpha\beta]}K^{(+) \nu}_{\mu}
            + J^{(+)\nu}_{~~\beta} M^{(+)}_{\mu[\nu\alpha]}
            + L^{(+)}_{\s\beta}Q^{(+)\s}_{\mu\alpha}
            \Big)\cr
&&  +2\eps_1^{+}\eps_2^{+} \Psi_{+\alpha}D^2_{+}\Phi^\beta \Big(
              P^{(+)\nu\alpha}_{\ph{(+)\nu\alpha},\beta} L^{(+)}_{\mu\nu} 
            + P^{(+)\rho\alpha} L^{(+)}_{\mu\beta,\rho} \cr
&& \ph{+2\eps_1^{+}\eps_2^{+} \Psi_{+\alpha}D^2_{+}\Phi^\beta \Big(}
            - Q^{(+)\alpha}_{\nu\beta}K^{(+) \nu}_{\mu}   
            - L^{(+)}_{\nu\beta}N^{(+)[\nu\alpha]}_{\mu}
            + J^{(+)\nu}_{~~\beta} Q^{(+)\alpha}_{\mu\nu}
            \Big)\cr
&& +2\eps_1^{+}\eps_2^{+} D_{+}\Psi_{+\alpha} D_{+}\Phi^\beta \Big(
              K^{(+) \alpha}_{\nu,\beta}K^{(+) \nu}_{\mu}
            + Q^{(+)\alpha}_{\nu\beta}K^{(+) \nu}_{\mu} \cr        
&& \ph{+2\eps_1^{+}\eps_2^{+} D_{+}\Psi_{+\alpha} D_{+}\Phi^\beta \Big(}
            + J^{(+)\rho}_{~~\beta}K^{(+) \alpha}_{\mu,\rho}
            - P^{(+)\nu\alpha} M^{(+)}_{\mu[\nu\beta]} 
            - K^{(+) \alpha}_{\s}Q^{(+)\s}_{\mu\beta}
            \Big)\cr
&& +2\eps_1^{+}\eps_2^{+} \Psi_{+\alpha}D_{+}\Psi_{+\beta} \Big(
            - N^{(+)[\beta\alpha]}_{\nu}K^{(+) \nu}_{\mu}
            - P^{(+)\rho\alpha}K^{(+) \beta}_{\mu,\rho} \cr
&&\ph{+2\eps_1^{+}\eps_2^{+} \Psi_{+\alpha}D_{+}\Psi_{+\beta} \Big(}
            + K^{(+) \beta}_{\nu} N^{(+)[\nu\alpha]}_{\mu}
            - P^{(+)\nu\beta} Q^{(+)\alpha}_{\mu\nu}
            \Big)\cr
\cr 
&& +2\eps_1^{+}\eps_2^{+} \Psi_{+\alpha}\Psi_{+\beta}D_{+}\Phi^\gamma \Big(
            - N^{(+)\alpha\beta}_{\nu,\gamma}K^{(+) \nu}_{\mu}
            - Q^{(+)\alpha}_{\nu\gamma}N^{(+)[\nu\beta]}_{\mu}\cr
&& \ph{+2\eps_1^{+}\eps_2^{+} \Psi_{+\alpha}\Psi_{+\beta}D_{+}\Phi^\gamma \Big(}
            - J^{(+)\rho}_{~~\gamma} N^{(+)\alpha\beta}_{\mu,\rho} 
            - P^{(+)\nu\alpha}_{\ph{(+)\nu\alpha},\gamma}Q^{(+)\beta}_{\mu\nu}\cr
&& \ph{+2\eps_1^{+}\eps_2^{+} \Psi_{+\alpha}\Psi_{+\beta}D_{+}\Phi^\gamma \Big(}
            + N^{(+)\alpha\beta}_{\s}Q^{(+)\s}_{\mu\gamma}
            + P^{(+)\rho\beta}Q^{(+)\alpha}_{\mu\gamma,\rho}
            \Big)\cr
&& +2\eps_1^{+}\eps_2^{+} D_{+}\Phi^\alpha D_{+}\Phi^\beta D_{+}\Phi^\gamma \Big(
            - M^{(+)}_{\nu\alpha\beta,\gamma}K^{(+) \nu}_{\mu}
            - J^{(+)\nu}_{~~\alpha,\beta} M^{(+)}_{\mu[\nu\gamma]} \cr
&& \ph{+2\eps_1^{+}\eps_2^{+} D_{+}\Phi^\alpha D_{+}\Phi^\beta D_{+}\Phi^\gamma \Big(}
            - J^{(+)\rho}_{~~\gamma}M^{(+)}_{\mu\alpha\beta,\rho}
            + M^{(+)}_{\s\beta\gamma}Q^{(+)\s}_{\mu\alpha}
            \Big)\cr
&& +2\eps_1^{+}\eps_2^{+} \Psi_{+\alpha}D_{+}\Phi^\beta D_{+}\Phi^\gamma \Big(
               Q^{(+)\alpha}_{\nu\beta,\gamma}K^{(+) \nu}_{\mu}         
             - M^{(+)}_{\nu\beta\gamma}N^{(+)[\nu\alpha]}_{\mu} \cr
&& \ph{+2\eps_1^{+}\eps_2^{+} \Psi_{+\alpha}D_{+}\Phi^\beta D_{+}\Phi^\gamma}
             + P^{(+)\nu\alpha}_{\ph{(+)\nu\alpha},\beta} M^{(+)}_{\mu[\nu\gamma]}
             + P^{(+)\tau\alpha}M^{(+)}_{\mu\beta\gamma,\tau}   \cr
&& \ph{+2\eps_1^{+}\eps_2^{+} \Psi_{+\alpha}D_{+}\Phi^\beta D_{+}\Phi^\gamma}
             - J^{(+)\nu}_{~~\beta,\gamma}Q^{(+)\alpha}_{\mu\nu}
             + Q^{(+)\alpha}_{\s\gamma}Q^{(+)\s}_{\mu\beta}
             + J^{(+)\rho}_{~~\gamma}Q^{(+)\alpha}_{\mu\beta,\rho}
             \Big)\cr
&& +2\eps_1^{+}\eps_2^{+} \Psi_{+\alpha}\Psi_{+\beta}\Psi_{+\gamma} \Big(
             - N^{(+)\alpha\beta}_{\nu}N^{(+)[\nu\gamma]}_{\mu}     
             + P^{(+)\rho\gamma}N^{(+)\alpha\beta}_{\mu,\rho}
             \Big) 
\end{eqnarray*}
}
{\footnotesize
\begin{eqnarray*}
[&&\hspace{-18pt}\delta_2^+(\eps^+_1),\delta_2^+(\eps^+_2)]\Psi_{-\mu} = \cr 
 &&  \ph{+}2\eps_1^{+}\eps_2^{+} D^2_{+}\Psi_{-\a} \Big(R^{(+)\a}_{\nu} R^{(+)\nu}_{\mu} \Big)   \cr
 && + 2\eps_1^{+}\eps_2^{+} D_{+}D_{-}\Psi_{+\a} \Big(
           Z^{(+)\a}_{\nu} R^{(+)\nu}_{\mu} 
         + K^{(+) \a}_{\nu}Z^{(+)\nu}_{\mu}
         + P^{(+)\nu\a} T^{(+)}_{\mu\nu}
         \Big)\cr
 && +2\eps_1^{+}\eps_2^{+} D^2_{+}D_{-}\Phi^\a \Big(
           T^{(+)}_{\nu\a}R^{(+)\nu}_{\mu}         
         + L^{(+)}_{\nu\a}Z^{(+)\nu}_{\mu}         
         + J^{(+)\nu}_{~~\a} T^{(+)}_{\mu\nu}      
         \Big)\cr
\cr 
 &&  + 2\eps_1^{+}\eps_2^{+} D_{+}\Psi_{-\a} D_{+}\Phi^\b \Big(
           R^{(+)\a}_{\nu,\b} R^{(+)\nu}_{\mu}
         - V^{(+)\a}_{\nu\b}R^{(+)\nu}_{\mu}   \cr   
 &&  \ph{+ 2\eps_1^{+}\eps_2^{+} D_{+}\Psi_{-\a} D_{+}\Phi^\b \Big(}
         - J^{(+)\r}_{\b}R^{(+)\a}_{\mu,\r}
         + R^{(+)\a}_{\r} V^{(+)\r}_{\mu\b} 
         \Big)\cr
 && + 2\eps_1^{+}\eps_2^{+} D_{-}\Psi_{+\a} D_{+}\Phi^\b \Big(
           Z^{(+)\a}_{\nu,\b} R^{(+)\nu}_{\mu}     
         - Q^{(+)\a}_{\nu\b}Z^{(+)\nu}_{\mu}       
         - J^{(+)\r}_{\b} Z^{(+)\a}_{\mu\r} \cr    
 && \ph{+ 2\eps_1^{+}\eps_2^{+} D_{-}\Psi_{+\a} D_{+}\Phi^\b \Big(}
         + P^{(+)\nu\a}_{\ph{(+)\nu\a},\b} T^{(+)}_{\mu\nu}        
         + Z^{(+)\a}_{\r} V^{(+)\r}_{\mu\b}        
         + P^{(+)\r\a} X^{(+)}_{\mu\b\r}           
         \Big)\cr 
 && +2\eps_1^{+}\eps_2^{+} D_{+}\Phi^\a D_{+}D_{-}\Phi^\b \Big(
           T^{(+)}_{\nu\b,\a}R^{(+)\nu}_{\mu}      
         - X^{(+)}_{\nu\a\b} R^{(+)\nu}_{\mu}      
         - M^{(+)}_{\nu[\b\a]}Z^{(+)\nu}_{\mu} \cr 
 &&\ph{+2\eps_1^{+}\eps_2^{+} D_{+}\Phi^\a}
         + J^{(+)\nu}_{[\b,\a]} T^{(+)}_{\mu\nu}   
         - J^{(+)\t}_{~\a} T^{(+)}_{\mu\b,\t}       
         + T^{(+)}_{\r\b} V^{(+)\r}_{\mu\a}        
         + J^{(+)\r}_{~~\b} X^{(+)}_{\mu\a\r}        
         \Big)\cr
 && +2\eps_1^{+}\eps_2^{+} D_{+}\Psi_{+\a}D_{-}\Phi^\b \Big(
           U^{(+)\a}_{\nu\b} R^{(+)\nu}_{\mu}      
         - K^{(+) \a}_{\nu,\b}Z^{(+)\nu}_{\mu} \cr 
 && \ph{+2\eps_1^{+}\eps_2^{+} D_{+}\Psi_{+\a}D_{-}\Phi^\b \Big(}
         - P^{(+)\nu\a}_{\ph{(+)\nu\a},\b} T^{(+)}_{\mu\nu}        
         + K^{(+)\a}_{\r} U^{(+)\r}_{\mu\b}        
         - P^{(+)\nu\a} X^{(+)}_{\mu\nu\b}         
         \Big)\cr
 && +2\eps_1^{+}\eps_2^{+} \Psi_{+\a}D_{+}D_{-}\Phi^\b \Big(
         - U^{(+)\a}_{\nu\b} R^{(+)\nu}_{\mu}      
         - Q^{(+)\a}_{\nu\b}Z^{(+)\nu}_{\mu}       
         + P^{(+)\nu\a}_{\ph{(+)\nu\a},\b} T^{(+)}_{\mu\nu} \cr    
 && \ph{+2\eps_1^{+}\eps_2^{+} \Psi_{+\a}D_{+}D_{-}\Phi^\b \Big(}
         + P^{(+)\t\a}T^{(+)}_{\mu\b,\t}           
         + J^{(+)\nu}_{~~\b} U^{(+)\a}_{\mu\nu}    
         + T^{(+)}_{\r\b} Y^{(+)\a\r}_{\mu}        
         \Big)\cr
 && +2\eps_1^{+}\eps_2^{+} \Psi_{-\a} D^2_{+}\Phi^\b \Big(
           V^{(+)\a}_{\nu\b}R^{(+)\nu}_{\mu}       
         + J^{(+)\nu}_{~~\b} V^{(+)\a}_{\mu\nu}    
         - L^{(+)}_{\nu\b} Y^{(+)\nu\a}_{\mu}      
         \Big)\cr
 && +2\eps_1^{+}\eps_2^{+} D^2_{+}\Phi^\a D_{-}\Phi^\b \Big( 
           X^{(+)}_{\nu\a\b}R^{(+)\nu}_{\mu}       
         + L^{(+)}_{\nu\a,\b}Z^{(+)\nu}_{\mu} \cr  
 && \ph{+2\eps_1^{+}\eps_2^{+} D^2_{+}\Phi^\a D_{-}\Phi^\b \Big(}
         + J^{(+)\nu}_{~~\a,\b} T^{(+)}_{\mu\nu}   
         - L^{(+)}_{\r\a} U^{(+)\r}_{\mu\b}        
         + J^{(+)\nu}_{~~\a} X^{(+)}_{\mu\nu\b}    
         \Big)\cr
 && +2\eps_1^{+}\eps_2^{+} \Psi_{-\a} D_{+}\Psi_{+\b} \Big(
           Y^{(+)\b\a}_{\nu} R^{(+)\nu}_{\mu}      
         - P^{(+)\nu\b} V^{(+)\a}_{\mu\nu}         
         + K^{(+)\b}_{\nu} Y^{(+)\nu\a}_{\mu}      
         \Big)\cr
 && +2\eps_1^{+}\eps_2^{+} \Psi_{+\a} D_{+}\Psi_{-\b} \Big(
         - Y^{(+)\a\b}_{\nu} R^{(+)\nu}_{\mu}     
         + P^{(+)\r\a}R^{(+)\b}_{\mu,\r}          
         + R^{(+)\b}_{\r}Y^{(+)\a\r}_{\mu}        
         \Big)\cr
 && +2\eps_1^{+}\eps_2^{+} \Psi_{+\a}D_{-}\Psi_{+\b} \Big(
           N^{(+)[\b\a]}_{\nu}Z^{(+)\nu}_{\mu}     
         + P^{(+)\r\a} Z^{(+)\b}_{\mu\r}           
         + P^{(+)\nu\b} U^{(+)\a}_{\mu\nu}      \cr
 &&\ph{+2\eps_1^{+}\eps_2^{+} \Psi_{+\a}D_{-}\Psi_{+\b} \Big(}
         + Z^{(+)\b}_{\r}Y^{(+)\a\r}_{\mu}         
         \Big)\cr
\cr 
 && +2\eps_1^{+}\eps_2^{+} \Psi_{+\a} D_{+}\Phi^\b D_{-}\Phi^\g \Big(
         - U^{(+)\a}_{\nu\g,\b}R^{(+)\nu}_{\mu}    
         - Q^{(+)\a}_{\nu\b,\g}Z^{(+)\nu}_{\mu}    
         + P^{(+)\nu\a}_{\ph{(+)\nu\a},\g\b} T^{(+)}_{\mu\nu}  \cr      
 && \ph{+2\eps_1^{+}\eps_2^{+} \Psi_{+\a} D_{+}\Phi^\b D_{-}\Phi^\g \Big(}
         + Q^{(+)\a}_{\r\b}U^{(+)\r}_{\mu\g}       
         + J^{(+)\nu}_{~~\b,\g} U^{(+)\a}_{\mu\nu} 
         + J^{(+)\t}_{~~\b}U^{(+)\a}_{\mu\g,\t} \cr
 && \ph{+2\eps_1^{+}\eps_2^{+} \Psi_{+\a} D_{+}\Phi^\b D_{-}\Phi^\g \Big(}
         - U^{(+)\a}_{\r\g} V^{(+)\r}_{\mu\b}      
         + P^{(+)\nu\a}_{\ph{(+)\nu\a},\b} X^{(+)}_{\mu\nu\g}    
         + P^{(+)\r\a}_{\ph{(+)\r\a},\g} X^{(+)}_{\mu\b\r}   \cr 
 && \ph{+2\eps_1^{+}\eps_2^{+} \Psi_{+\a} D_{+}\Phi^\b D_{-}\Phi^\g \Big(}
         + P^{(+)\t\a}X^{(+)}_{\mu\b\g,\t}         
         + X^{(+)}_{\r\b\g}Y^{(+)\a\r}_{\mu}       
         \Big)\cr
 && +2\eps_1^{+}\eps_2^{+} \Psi_{-\a} D_{+}\Phi^\b D_{+}\Phi^\g \Big(
           V^{(+)\a}_{\nu\g,\b}R^{(+)\nu}_{\mu}    
         - J^{(+)\nu}_{~~\b,\g} V^{(+)\a}_{\mu\nu} \cr 
 && \ph{+2\eps_1^{+}\eps_2^{+} \Psi_{-\a} D_{+}\Phi^\b D_{+}\Phi^\g \Big(}
         + V^{(+)\a}_{\r\g} V^{(+)\r}_{\mu\b}      
         + J^{(+)\t}_{~~\g} V^{(+)\a}_{\mu\b,\t}   
         - M^{(+)}_{\nu\b\g}Y^{(+)\nu\a}_{\mu}     
         \Big)\cr
 && +2\eps_1^{+}\eps_2^{+} D_{+}\Phi^\a D_{+}\Phi^\b D_{-}\Phi^\g \Big(
         - X^{(+)}_{\nu\a\g,\b}R^{(+)\nu}_{\mu}    
         + M^{(+)}_{\nu\a\b,\g}Z^{(+)\nu}_{\mu}    
         - J^{(+)\nu}_{~~\a,\g\b} T^{(+)}_{\mu\nu} \cr
 && \ph{+2\eps_1^{+}\eps_2^{+} D_{+}\Phi^\a D_{+}\Phi^\b D_{-}\Phi^\g \Big(}
         - M^{(+)}_{\r\a\b}U^{(+)\r}_{\mu\g}       
         + X^{(+)}_{\r\b\g} V^{(+)\r}_{\mu\a}      
         - J^{(+)\nu}_{~~\a,\b} X^{(+)}_{\mu\nu\g} \cr 
 && \ph{+2\eps_1^{+}\eps_2^{+} D_{+}\Phi^\a D_{+}\Phi^\b D_{-}\Phi^\g \Big(}
         + J^{(+)\r}_{~~\b,\g} X^{(+)}_{\mu\a\r}   
         + J^{(+)\t}_{~~\b}X^{(+)}_{\mu\a\g,\t}    
         \Big)\cr
 && +2 \eps_1^{+}\eps_2^{+} \Psi_{+\a} \Psi_{-\b} D_{+}\Phi^\g \Big(
           Y^{(+)\a\b}_{\nu,\g}R^{(+)\nu}_{\mu}               
         - P^{(+)\nu\a}_{\ph{(+)\nu\s},\g} V^{(+)\b}_{\mu\nu} 
         + Y^{(+)\a\b}_{\r} V^{(+)\r}_{\mu\g}             \cr 
 &&\ph{+2 \eps_1^{+}\eps_2^{+} \Psi_{+\a} \Psi_{-\b} D_{+}\Phi^\g \Big(}
         - P^{(+)\t\a} V^{(+)\b}_{\mu\g,\t}                   
         - Q^{(+)\a}_{\nu\g} Y^{(+)\nu\b}_{\mu}               
         - V^{(+)\b}_{\r\g}Y^{(+)\a\r}_{\mu}  \cr             
 && \ph{+2 \eps_1^{+}\eps_2^{+} \Psi_{+\a} \Psi_{-\b} D_{+}\Phi^\g \Big(}
         - J^{(+)\t}_{~~\g} Y^{(+)\a\b}_{\mu,\t}              
         \Big)\cr
 && +2\eps_1^{+}\eps_2^{+} \Psi_{+\a}\Psi_{+\b}D_{-}\Phi^\g \Big(
           N^{(+)\a\b}_{\nu,\g}Z^{(+)\nu}_{\mu}                   
         - N^{(+)\a\b}_{\r} U^{(+)\r}_{\mu\g}                     
         - P^{(+)\nu\b}_{\ph{(+)\nu\b},\g} U^{(+)\a}_{\mu\nu} \cr 
 && \ph{+2\eps_1^{+}\eps_2^{+} \Psi_{+\a}\Psi_{+\b}D_{-}\Phi^\g \Big(}
         - P^{(+)\t\b} U^{(+)\a}_{\mu\g,\t}                       
         + U^{(+)\b}_{\r\g} Y^{(+)\a\r}_{\mu}                     
         \Big)\cr
 && +2\eps_1^{+}\eps_2^{+} \Psi_{+\a} \Psi_{+\b} \Psi_{-\g} \Big(
         - N^{(+)\a\b}_{\nu}Y^{(+)\nu\g}_{\mu}                    
         + Y^{(+)\b\g}_{\r} Y^{(+)\a\r}_{\mu}                     
         - P^{(+)\t\b} Y^{(+)\a\g}_{\mu,\t}                       
         \Big)
\end{eqnarray*}
}

\section{$\mathcal{N}=(2,2)$ commutators}\label{Appendix:n22}
To have $\mathcal{N}=(2,2)$ supersymmetry we need the three commutators given in the previous section and the following two commutators to obey the supersymmetry algebra \eqref{1st:12_comm:2}-\eqref{1st:2pp_comm}. The other commutators needed are obtained by replacing $+ \leftrightarrow -$ in all five commutators.
{\footnotesize
\begin{eqnarray*}
[&&\hspace{-18pt}\delta_2^+(\eps^+_1),\delta_2^-(\eps^-_2)]\Phi^\mu =\cr 
&& \ph{+}\eps_1^{+}\eps_2^{-} D_{+}D_{-}\Phi^\alpha \Big( 
              - J^{(+)\nu}_{~~\alpha} J^{(-)\mu}_{~~\nu}
              + T^{(+)}_{\nu\alpha} P^{(-)\mu\nu} \cr
&& \ph{\eps_1^{+}\eps_2^{-} D_{+}D_{-}\Phi^\alpha \Big(}
              + J^{(-)\nu}_{~~\alpha} J^{(+)\mu}_{~~\nu} 
              - T^{(-)}_{\nu\alpha} P^{(+)\mu\nu} 
              \Big) \cr
&& +\eps_1^{+}\eps_2^{-} D_{-}\Psi_{+\alpha}\Big(
               -P^{(+)\nu\alpha} J^{(-)\mu}_{~~\nu}
               +Z^{(+)\alpha}_{\nu} P^{(-)\mu\nu}
               +R^{(-)\alpha}_{\nu} P^{(+)\mu\nu}
              \Big)\cr
&& + \eps_1^{+}\eps_2^{-} D_{+}\Psi_{-\alpha} \Big(
                 R^{(+)\alpha}_{\nu} P^{(-)\mu\nu}
               - P^{(-)\nu\alpha} J^{(+)\mu}_{~~\nu}
               + Z^{(-)\alpha}_{\nu} P^{(+)\mu\nu}
               \Big)\cr
\cr 
&& + \eps_1^{+}\eps_2^{-} D_{+}\Phi^\alpha D_{-}\Phi^\beta \Big(
              -J^{(+)\nu}_{~~\alpha,\beta} J^{(-)\mu}_{~~\nu}
              -J^{(+)\rho}_{~~\alpha} J^{(-)\mu}_{~~\beta,\rho}
              + X^{(+)}_{\nu\alpha\beta} P^{(-)\mu\nu} \cr
&& \ph{+ \eps_1^{+}\eps_2^{-} D_{+}\Phi^\alpha D_{-}\Phi^\beta \Big(}
              +J^{(-)\nu}_{~~\beta,\alpha} J^{(+)\mu}_{~~\nu}
              +J^{(-)\rho}_{~~\beta}J^{(+)\mu}_{~~\alpha,\rho}
              - X^{(-)}_{\nu\beta\alpha} P^{(+)\mu\nu}
              \Big)\cr 
&& +\eps_1^{+}\eps_2^{-} \Psi_{+\alpha} D_{-}\Phi^\beta \Big(
                 P^{(+)\nu\alpha}_{\ph{(+)\nu\alpha},\beta} J^{(-)\mu}_{~~\nu}
               + P^{(+)\rho\alpha} J^{(-)\mu}_{~~\beta,\rho} \cr
&&\ph{+\eps_1^{+}\eps_2^{-} \Psi_{+\alpha} D_{-}\Phi^\beta \Big(}
               + U^{(+)\alpha}_{\nu\beta} P^{(-)\mu\nu}
               - V^{(-)\alpha}_{\nu\beta} P^{(+)\mu\nu}
               - J^{(-)\rho}_{~~\beta}P^{(+)\mu\alpha}_{\ph{(+)\mu\alpha},\rho}
               \Big)\cr
&& + \eps_1^{+}\eps_2^{-} \Psi_{-\alpha}D_{+}\Phi^\beta \Big(
               - V^{(+)\alpha}_{\nu\beta} P^{(-)\mu\nu}
               - J^{(+)\rho}_{~~\beta}P^{(-)\mu\alpha}_{\ph{(-)\mu\alpha},\rho} \cr
&& \ph{+ \eps_1^{+}\eps_2^{-} \Psi_{-\alpha}D_{+}\Phi^\beta \Big(}
               + P^{(-)\nu\alpha}_{\ph{(-)\nu\alpha},\beta} J^{(+)\mu}_{~~\nu}
               + P^{(-)\rho\alpha}J^{(+)\mu}_{~~\beta,\rho} 
               + U^{(-)\alpha}_{\nu\beta} P^{(+)\mu\nu}
               \Big)\cr 
&& + \eps_1^{+}\eps_2^{-} \Psi_{+\alpha}\Psi_{-\beta} \Big(
                 Y^{(+)\alpha\beta}_{\nu} P^{(-)\mu\nu}
               - P^{(+)\rho\alpha}P^{(-)\mu\beta}_{\ph{(-)\mu\beta},\rho} 
               - Y^{(-)\beta\alpha}_{\nu} P^{(+)\mu\nu}\cr
&& \ph{+ \eps_1^{+}\eps_2^{-} \Psi_{+\alpha}\Psi_{-\beta} \Big(}
               + P^{(-)\rho\beta}P^{(+)\mu\alpha}_{\ph{(+)\mu\alpha},\rho}
               \Big)
\end{eqnarray*}
}
{\footnotesize
\begin{eqnarray*}
[&&\hspace{-18pt}\delta_2^+(\eps^+_1),\delta_2^-(\eps^-_2)] \Psi_{+\mu} = \cr 
&& \ph{+}\eps_1^{+}\eps_2^{-} D^2_{+}D_{-}\Phi^\a \Big(
            L^{(+)}_{\nu\a}R^{(-)\nu}_{\mu}        
          + T^{(+)}_{\nu\a} Z^{(-)\nu}_{\mu}       
          - J^{(+)\nu}_{~~\a} T^{(-)}_{\mu\nu} \cr 
&&\ph{\eps_1^{+}\eps_2^{-} D^2_{+}D_{-}\Phi^\a \Big(}
          - J^{(-)\nu}_{~~\a} L^{(+)}_{\mu\nu}          
          + T^{(-)}_{\nu\a} K^{(+) \nu}_{\mu}           
          \Big)\cr
&& +\eps_1^{+}\eps_2^{-} D_{+}D_{-}\Psi_{+\a} \Big(
            K^{(+)\a}_{\nu} R^{(-)\nu}_{\mu}     
          + Z^{(+)\a}_{\nu} Z^{(-)\nu}_{\mu}     
          - P^{(+)\nu\a} T^{(-)}_{\mu\nu}   \cr  
&& \ph{+\eps_1^{+}\eps_2^{-} D_{+}D_{-}\Psi_{+\a} \Big(}
              - R^{(-)\a}_{\nu} K^{(+)\nu}_{\mu}            
          \Big)\cr
&& + \eps_1^{+}\eps_2^{-} D^2_{+}\Psi_{-\a} \Big(
            R^{(+)\a}_{\nu} Z^{(-)\nu}_{\mu}                     
              + P^{(-)\nu\a} L^{(+)}_{\mu\nu}               
              - Z^{(-)\a}_{\nu} K^{(+) \nu}_{\mu}           
          \Big)\cr
\cr
&& +\eps_1^{+}\eps_2^{-} D_{-}\Phi^\a D^2_{+}\Phi^\b \Big(
            L^{(+)}_{\nu\b,\a} R^{(-)\nu}_{\mu}   
          + X^{(+)}_{\nu\b\a} Z^{(-)\nu}_{\mu}    
          - J^{(+)\nu}_{~~\b,\a} T^{(-)}_{\mu\nu} \cr 
&&\ph{+\eps_1^{+}\eps_2^{-} D_{-}\Phi^\a D^2_{+}\Phi^\b \Big(}
          + L^{(+)}_{\r\b} V^{(-)\r}_{\mu\a}      
          - J^{(+)\r}_{~~\b} X^{(-)}_{\mu\a\r}    
              - J^{(-)\nu}_{~~\a,\b} L^{(+)}_{\mu\nu}   \cr 
&&\ph{+\eps_1^{+}\eps_2^{-} D_{-}\Phi^\a D^2_{+}\Phi^\b \Big(}
              - J^{(-)\r}_{~~\a}L^{(+)}_{\mu\b,\r}          
              + X^{(-)}_{\nu\a\b} K^{(+)\nu}_{\mu}          
          \Big)\cr
&& +\eps_1^{+}\eps_2^{-} D_{+}\Psi_{+\a} D_{-}\Phi^\b \Big(
          - K^{(+)\a}_{\nu,\b} R^{(-)\nu}_{\mu}  
          + U^{(+)\a}_{\nu\b} Z^{(-)\nu}_{\mu}   
          + P^{(+)\nu\a}_{\ph{(+)\nu\a},\b} T^{(-)}_{\mu\nu} \cr 
&& \ph{+\eps_1^{+}\eps_2^{-} D_{+}\Psi_{+\a} D_{-}\Phi^\b \Big(}
          - K^{(+)\a}_{\r} V^{(-)\r}_{\mu\b}     
          + P^{(+)\r\a} X^{(-)}_{\mu\b\r}        
             + V^{(-)\a}_{\nu\b} K^{(+)\nu}_{\mu}   \cr 
&& \ph{+\eps_1^{+}\eps_2^{-} D_{+}\Psi_{+\a} D_{-}\Phi^\b \Big(}
             + J^{(-)\r}_{~~\b} K^{(+) \a}_{\mu,\r}    
          \Big)\cr
&& +\eps_1^{+}\eps_2^{-} \Psi_{+\a} D_{-}\Psi_{+\b} \Big(
            N^{(+)[\b\a]}_{\nu} R^{(-)\nu}_{\mu}  
          + P^{(+)\r\a}R^{(-)\b}_{\mu,\r}         
          - P^{(+)\nu\b} V^{(-)\a}_{\mu\nu}    \cr
&&\ph{+\eps_1^{+}\eps_2^{-} \Psi_{+\a} D_{-}\Psi_{+\b} \Big(}
          - Z^{(+)\b}_{\nu} Y^{(-)\nu\a}_{\mu}    
             + R^{(-)\b}_{\s} N^{(+)[\a\s]}_{\mu}       
          \Big)\cr
&& +\eps_1^{+}\eps_2^{-} D_{+}\Phi^\a D_{+}D_{-}\Phi^\b \Big(
          - M^{(+)}_{\nu[\b\a]} R^{(-)\nu}_{\mu}    
          + T^{(+)}_{\nu\b,\a} Z^{(-)\nu}_{\mu}     
          - X^{(+)}_{\nu\a\b} Z^{(-)\nu}_{\mu} \cr  
&&\ph{+\eps_1^{+}\eps_2^{-} D_{+}\Phi^\a D_{+}D_{-}\Phi^\b \Big(}
          + J^{(+)\nu}_{~~[\a,\b]} T^{(-)}_{\mu\nu} 
          + J^{(+)\r}_{~~\a} T^{(-)}_{\mu\b,\r}     
          - T^{(+)}_{\r\b} U^{(-)\r}_{\mu\a}     \cr
&&\ph{+\eps_1^{+}\eps_2^{-} D_{+}\Phi^\a D_{+}D_{-}\Phi^\b \Big(}
          - J^{(+)\nu}_{~~\b} X^{(-)}_{\mu\nu\a}    
          + T^{(-)}_{\nu\b,\a} K^{(+) \nu}_{\mu}        
          - X^{(-)}_{\nu\b\a} K^{(+)\nu}_{\mu}    \cr   
&&\ph{+\eps_1^{+}\eps_2^{-} D_{+}\Phi^\a D_{+}D_{-}\Phi^\b \Big(}
          - J^{(-)\r}_{~~\b} M^{(+)}_{\mu[\a\r]}        
          - T^{(-)}_{\s\b} Q^{(+)\s}_{\mu\a}            
          \Big)\cr
&& +\eps_1^{+}\eps_2^{-} \Psi_{+\a} D_{+}D_{-}\Phi^\b \Big(
          - Q^{(+)\a}_{\nu\b} R^{(-)\nu}_{\mu}                   
          - U^{(+)\a}_{\nu\b} Z^{(-)\nu}_{\mu}                   
          - P^{(+)\nu\a}_{\ph{(+)\nu\a},\b} T^{(-)}_{\mu\nu} \cr 
&&\ph{+\eps_1^{+}\eps_2^{-} \Psi_{+\a} D_{+}D_{-}\Phi^\b \Big(}
          - P^{(+)\r\a}T^{(-)}_{\mu\b,\r}                        
          - J^{(+)\nu}_{~~\b} V^{(-)\a}_{\mu\nu}                 
          - T^{(+)}_{\nu\b} Y^{(-)\nu\a}_{\mu}              \cr  
&&\ph{+\eps_1^{+}\eps_2^{-} \Psi_{+\a} D_{+}D_{-}\Phi^\b \Big(}
              - V^{(-)\a}_{\nu\b} K^{(+)\nu}_{\mu}    
              - T^{(-)}_{\s\b} N^{(+)[\a\s]}_{\mu}    
              + J^{(-)\nu}_{~~\b} Q^{(+)\a}_{\mu\nu}  
          \Big)\cr
&& +\eps_1^{+}\eps_2^{-} D_{-}\Psi_{+\a} D_{+}\Phi^\b \Big(
          - Q^{(+)\a}_{\nu\b}R^{(-)\nu}_{\mu}                    
          - J^{(+)\r}_{~~\b} R^{(-)\a}_{\mu,\r}                  
          + Z^{(+)\a}_{\nu,\b} Z^{(-)\nu}_{\mu}              \cr 
&&\ph{+\eps_1^{+}\eps_2^{-} D_{-}\Psi_{+\a} D_{+}\Phi^\b \Big(}
          - P^{(+)\nu\a}_{\ph{(+)\nu\a},\b} T^{(-)}_{\mu\nu}     
          - Z^{(+)\a}_{\r} U^{(-)\r}_{\mu\b}                     
          - P^{(+)\nu\a} X^{(-)}_{\mu\nu\b}                 \cr  
&&\ph{+\eps_1^{+}\eps_2^{-} D_{-}\Psi_{+\a} D_{+}\Phi^\b \Big(}
              - R^{(-)\a}_{\nu,\b} K^{(+) \nu}_{\mu}        
              + R^{(-)\a}_{\s} Q^{(+)\s}_{\mu\b}            
          \Big)\cr
&& + \eps_1^{+}\eps_2^{-} D_{+}\Psi_{-\a}D_{+}\Phi^\b \Big(
            R^{(+)\a}_{\nu,\b} Z^{(-)\nu}_{\mu}                  
          - V^{(+)\a}_{\nu\b} Z^{(-)\nu}_{\mu}                   
          - J^{(+)\r}_{~~\b}Z^{(-)\a}_{\mu,\r}              \cr  
&&\ph{+ \eps_1^{+}\eps_2^{-} D_{+}\Psi_{-\a}D_{+}\Phi^\b \Big(}
          - R^{(+)\a}_{\r} U^{(-)\r}_{\mu\b}                     
              - Z^{(-)\a}_{\nu,\b} K^{(+)\nu}_{\mu}         
              - U^{(-)\a}_{\nu\b} K^{(+) \nu}_{\mu}     \cr 
&&\ph{+ \eps_1^{+}\eps_2^{-} D_{+}\Psi_{-\a}D_{+}\Phi^\b \Big(}
              + P^{(-)\r\a} M^{(+)}_{\mu[\b\r]}             
              + Z^{(-)\a}_{\s} Q^{(+)\s}_{\mu\b}            
          \Big)\cr
&& + \eps_1^{+}\eps_2^{-} \Psi_{-\a} D^2_{+}\Phi^\b \Big(
            V^{(+)\a}_{\nu\b} Z^{(-)\nu}_{\mu}                   
          - J^{(+)\nu}_{~~\b} U^{(-)\a}_{\mu\nu}                 
          + L^{(+)}_{\r\b} Y^{(-)\a\r}_{\mu}            \cr      
&&\ph{+ \eps_1^{+}\eps_2^{-} \Psi_{-\a} D^2_{+}\Phi^\b \Big(}
              + P^{(-)\nu\a}_{\ph{(-)\nu\a},\b} L^{(+)}_{\mu\nu} 
              + P^{(-)\r\a}L^{(+)}_{\mu\b,\r}                    
              + U^{(-)\a}_{\nu\b} K^{(+) \nu}_{\mu}              
          \Big)\cr
&& + \eps_1^{+}\eps_2^{-} \Psi_{-\a} D_{+}\Psi_{+\b} \Big(
            Y^{(+)\b\a}_{\nu} Z^{(-)\nu}_{\mu}                   
          + P^{(+)\nu\b} U^{(-)\a}_{\mu\nu}                      
          - K^{(+)\b}_{\r} Y^{(-)\a\r}_{\mu}          \cr        
&&\ph{+\eps_1^{+}\eps_2^{-} \Psi_{-\a} D_{+}\Psi_{+\b} \Big(}
              + Y^{(-)\a\b}_{\nu} K^{(+) \nu}_{\mu}      
              - P^{(-)\r\a} K^{(+)\b}_{\mu,\r}           
          \Big)\cr
&& + \eps_1^{+}\eps_2^{-} \Psi_{+\a} D_{+}\Psi_{-\b} \Big(
          - Y^{(+)\a\b}_{\nu} Z^{(-)\nu}_{\mu}                   
          + P^{(+)\r\a} Z^{(-)\b}_{\mu,\r}                       
          - R^{(+)\b}_{\nu} Y^{(-)\nu\a}_{\mu}       \cr         
&&\ph{+ \eps_1^{+}\eps_2^{-} \Psi_{+\a} D_{+}\Psi_{-\b} \Big(}
              - Y^{(-)\b\a}_{\nu} K^{(+) \nu}_{\mu}      
              + Z^{(-)\b}_{\s} N^{(+)[\a\s]}_{\mu}       
              - P^{(-)\nu\b} Q^{(+)\a}_{\mu\nu}          
          \Big)\cr
\cr 
&& +\eps_1^{+}\eps_2^{-} \Psi_{+\a} \Psi_{+\b} D_{-}\Phi^\g \Big(
            N^{(+)\a\b}_{\nu,\g} R^{(-)\nu}_{\mu}   
          - P^{(+)\nu\a}_{\ph{(+)\nu\a},\g} V^{(-)\b}_{\mu\nu}         
          + N^{(+)\a\b}_{\r} V^{(-)\r}_{\mu\g}  \cr 
&&\ph{+\eps_1^{+}\eps_2^{-} \Psi_{+\a} \Psi_{+\b} D_{-}\Phi^\g \Big(}
          + P^{(+)\t\b}V^{(-)\a}_{\mu\g,\t}         
          + U^{(+)\a}_{\nu\g} Y^{(-)\nu\b}_{\mu}    
              - V^{(-)\b}_{\s\g} N^{(+)[\a\s]}_{\mu} \cr 
&&\ph{+\eps_1^{+}\eps_2^{-} \Psi_{+\a} \Psi_{+\b} D_{-}\Phi^\g \Big(}
              - J^{(-)\r}_{~~\g} N^{(+)\a\b}_{\mu,\r}    
          \Big)\cr
&& +\eps_1^{+}\eps_2^{-} D_{+}\Phi^\a D_{+}\Phi^\b D_{-}\Phi^\g \Big(
            M^{(+)}_{\nu\a\b,\g} R^{(-)\nu}_{\mu}       
          - X^{(+)}_{\nu\a\g,\b} Z^{(-)\nu}_{\mu}       
          + J^{(+)\nu}_{~~\a,\b\g} T^{(-)}_{\mu\nu} \cr 
&&\ph{+\eps_1^{+}\eps_2^{-} D_{+}\Phi^\a D_{+}\Phi^\b D_{-}\Phi^\g \Big(}
          + X^{(+)}_{\r\a\g} U^{(-)\r}_{\mu\b}          
          + M^{(+)}_{\r\a\b} V^{(-)\r}_{\mu\g}          
          + J^{(+)\nu}_{~~\a,\g} X^{(-)}_{\mu\nu\b} \cr 
&&\ph{+\eps_1^{+}\eps_2^{-} D_{+}\Phi^\a D_{+}\Phi^\b D_{-}\Phi^\g \Big(}
          + J^{(+)\r}_{~~\a,\b} X^{(-)}_{\mu\g\r}       
          - J^{(+)\t}_{~~\b} X^{(-)}_{\mu\g\a,\t}       
              - X^{(-)}_{\nu\g\a,\b} K^{(+)\nu}_{\mu}     \cr 
&&\ph{+\eps_1^{+}\eps_2^{-} D_{+}\Phi^\a D_{+}\Phi^\b D_{-}\Phi^\g \Big(}
              - J^{(-)\r}_{~~\g,\b} M^{(+)}_{\mu[\a\r]}       
              - J^{(-)\t}_{~~\g} M^{(+)}_{\mu\a\b,\t}         
              - X^{(-)}_{\s\g\b} Q^{(+)\s}_{\mu\a}            
          \Big)\cr
&& +\eps_1^{+}\eps_2^{-} \Psi_{+\a}D_{+}\Phi^\b D_{-}\Phi^\g \Big(
          - Q^{(+)\a}_{\nu\b,\g} R^{(-)\nu}_{\mu}                  
          - U^{(+)\a}_{\nu\g,\b} Z^{(-)\nu}_{\mu}                  
          - P^{(+)\nu\a}_{\ph{(+)\nu\a},\b\g} T^{(-)}_{\mu\nu} \cr 
&&\ph{+\eps_1^{+}\eps_2^{-} \Psi_{+\a}D_{+}\Phi^\b D_{-}\Phi^\g \Big(}
          + U^{(+)\a}_{\r\g} U^{(-)\r}_{\mu\b}                     
          - J^{(+)\nu}_{~~\b,\g} V^{(-)\a}_{\mu\nu}                
          - Q^{(+)\a}_{\r\b} V^{(-)\r}_{\mu\g}               \cr   
&&\ph{+\eps_1^{+}\eps_2^{-} \Psi_{+\a}D_{+}\Phi^\b D_{-}\Phi^\g \Big(}
          - J^{(+)\t}_{~~\b} V^{(-)\a}_{\mu\g,\t}                  
          - P^{(+)\nu\a}_{\ph{(+)\nu\t},\g} X^{(-)}_{\mu\nu\b}     
          - P^{(+)\r\a}_{\ph{(+)\r\t},\b} X^{(-)}_{\mu\g\r}   \cr  
&&\ph{+\eps_1^{+}\eps_2^{-} \Psi_{+\a}D_{+}\Phi^\b D_{-}\Phi^\g \Big(}
          - P^{(+)\t\a} X^{(-)}_{\mu\g\b,\t}                     
          - X^{(+)}_{\nu\b\g} Y^{(-)\nu\a}_{\mu}        \cr      
&&\ph{+\eps_1^{+}\eps_2^{-} \Psi_{+\a}D_{+}\Phi^\b D_{-}\Phi^\g \Big(}
              - V^{(-)\a}_{\nu\g,\b} K^{(+) \nu}_{\mu}        
              - X^{(-)}_{\s\g\b} N^{(+)[\a\s]}_{\mu}          
              + J^{(-)\nu}_{~~\g,\b} Q^{(+)\a}_{\mu\nu}   \cr 
&&\ph{+\eps_1^{+}\eps_2^{-} \Psi_{+\a} D_{+}\Phi^\b D_{-}\Phi^\g \Big(}
              + V^{(-)\a}_{\s\g} Q^{(+)\s}_{\mu\b}            
              + J^{(-)\t}_{~~\g} Q^{(+)\a}_{\mu\b,\t}         
          \Big)\cr
&& + \eps_1^{+}\eps_2^{-} \Psi_{-\a} D_{+}\Phi^\b D_{+}\Phi^\g \Big(
          - V^{(+)\a}_{\nu\b,\g} Z^{(-)\nu}_{\mu}                
          + V^{(+)\a}_{\r\b} U^{(-)\r}_{\mu\g}                   
          + J^{(+)\nu}_{~~\b,\g} U^{(-)\a}_{\mu\nu}          \cr 
&&\ph{+ \eps_1^{+}\eps_2^{-} \Psi_{-\a} D_{+}\Phi^\b D_{+}\Phi^\g\Big(}
          - J^{(+)\t}_{~~\g} U^{(-)\a}_{\mu\b,\t}                
          + M^{(+)}_{\r\b\g} Y^{(-)\a\r}_{\mu}                   
              - U^{(-)\a}_{\nu\b,\g} K^{(+) \nu}_{\mu}       \cr 
&&\ph{+ \eps_1^{+}\eps_2^{-} \Psi_{-\a} D_{+}\Phi^\b D_{+}}
              + P^{(-)\r\a}_{\ph{(-)\r\t},\g} M^{(+)}_{\mu[\b\r]} 
              + P^{(-)\t\a} M^{(+)}_{\mu\b\g,\t}                  
              - U^{(-)\a}_{\s\g} Q^{(+)\s}_{\mu\b}                
          \Big)\cr
&& + \eps_1^{+}\eps_2^{-} \Psi_{+\a} \Psi_{-\b} D_{+}\Phi^\g \Big(
            Y^{(+)\a\b}_{\nu,\g} Z^{(-)\nu}_{\mu}                
          - Y^{(+)\a\b}_{\r} U^{(-)\r}_{\mu\g}                   
          + P^{(+)\nu\a}_{\ph{(+)\nu\a},\g} U^{(-)\b}_{\mu\nu} \cr
&&\ph{+ \eps_1^{+}\eps_2^{-} \Psi_{+\a} \Psi_{-\b} D_{+}\Phi^\g \Big(}
          + P^{(+)\t\a} U^{(-)\b}_{\mu\g,\t}                     
          + V^{(+)\b}_{\nu\g} Y^{(-)\nu\a}_{\mu}                 
          + Q^{(+)\a}_{\r\g} Y^{(-)\b\r}_{\mu}               \cr 
&&\ph{+ \eps_1^{+}\eps_2^{-} \Psi_{+\a} \Psi_{-\b} D_{+}\Phi^\g \Big(}
          + J^{(+)\t}_{~~\g} Y^{(-)\b\a}_{\mu,\t}                
              + Y^{(-)\b\a}_{\nu,\g} K^{(+)\nu}_{\mu}                  
              + U^{(-)\b}_{\s\g} N^{(+)[\a\s]}_{\mu}            \cr    
&&\ph{+ \eps_1^{+}\eps_2^{-} \Psi_{+\a} \Psi_{-\b} D_{+}\Phi^\g \Big(}
              + P^{(-)\nu\b}_{\ph{(-)\nu\t},\g} Q^{(+)\a}_{\mu\nu}     
              - Y^{(-)\b\a}_{\s} Q^{(+)\s}_{\mu\g}                     
              + P^{(-)\t\b} Q^{(+)\a}_{\mu\g,\t}                       
          \Big)\cr
&& + \eps_1^{+}\eps_2^{-} \Psi_{+\a} \Psi_{+\b} \Psi_{-\g} \Big(
            Y^{(+)\a\g}_{\nu} Y^{(-)\nu\b}_{\mu}                 
          + N^{(+)\a\b}_{\r} Y^{(-)\g\r}_{\mu}                   
          + P^{(+)\t\b} Y^{(-)\g\a}_{\mu,\t}                 \cr 
&&\ph{+ \eps_1^{+}\eps_2^{-} \Psi_{+\a} \Psi_{+\b} \Psi_{-\g} \Big(}
              - Y^{(-)\g\b}_{\s} N^{(+)[\a\s]}_{\mu}     
              + P^{(-)\r\g} N^{(+)\a\b}_{\mu,\r}         
          \Big)
\end{eqnarray*}
}

    \backmatter
    \nocite{*} 
    \bibliographystyle{alpha}
    \bibliography{References}

\newcommand{\etalchar}[1]{$^{#1}$}
\begin{thebibliography}{BFOHP02b}

\bibitem[AAGL94]{Alvarez:1994wj}
Enrique Alvarez, Luis Alvarez-Gaume, and Yolanda Lozano.
\newblock A canonical approach to duality transformations.
\newblock {\em Phys. Lett.}, B336:183--189, 1994.

\bibitem[AK88]{Amati:1988ww}
D.~Amati and C.~Klimcik.
\newblock Strings in a shock wave background and generation of curved geometry
  from flat space string theory.
\newblock {\em Phys. Lett.}, B210:92, 1988.

\bibitem[ALZ04]{Albertsson:2004gr}
Cecilia Albertsson, Ulf Lindström, and Maxim Zabzine.
\newblock {T}-duality for the sigma model with boundaries.
\newblock {\em JHEP}, 12:056, 2004.

\bibitem[AS71]{Aichelburg:1970dh}
P.~C. Aichelburg and R.~U. Sexl.
\newblock On the gravitational field of a massless particle.
\newblock {\em Gen. Rel. Grav.}, 2:303--312, 1971.

\bibitem[AS05]{Alekseev:2004np}
Anton Alekseev and Thomas Strobl.
\newblock Current algebra and differential geometry.
\newblock {\em JHEP}, 03:035, 2005.

\bibitem[BDVH76]{Brink:1976sc}
L.~Brink, P.~Di~Vecchia, and Paul~S. Howe.
\newblock A locally supersymmetric and reparametrization invariant action for
  the spinning string.
\newblock {\em Phys. Lett.}, B65:471--474, 1976.

\bibitem[Ber05]{Bergamin:2004sk}
L.~Bergamin.
\newblock Generalized complex geometry and the {P}oisson sigma model.
\newblock {\em Mod. Phys. Lett.}, A20:985--996, 2005.

\bibitem[BFOHP02a]{Blau:2001ne}
Matthias Blau, Jose Figueroa-O'Farrill, Christopher Hull, and George
  Papadopoulos.
\newblock A new maximally supersymmetric background of {IIB} superstring
  theory.
\newblock {\em JHEP}, 01:047, 2002.

\bibitem[BFOHP02b]{Blau:2002dy}
Matthias Blau, Jose Figueroa-O'Farrill, Christopher Hull, and George
  Papadopoulos.
\newblock Penrose limits and maximal supersymmetry.
\newblock {\em Class. Quant. Grav.}, 19:L87--L95, 2002.

\bibitem[BFOP02]{Blau:2002mw}
Matthias Blau, Jose Figueroa-O'Farrill, and George Papadopoulos.
\newblock Penrose limits, supergravity and brane dynamics.
\newblock {\em Class. Quant. Grav.}, 19:4753, 2002.

\bibitem[BLR88]{Buscher:1987uw}
T.~Buscher, U.~Lindström, and M.~Rocek.
\newblock New supersymmetric sigma models with {W}ess-{Z}umino terms.
\newblock {\em Phys. Lett.}, B202:94, 1988.

\bibitem[BMN02]{Berenstein:2002jq}
David Berenstein, Juan~M. Maldacena, and Horatiu Nastase.
\newblock Strings in flat space and pp waves from {N} = 4 super {Y}ang {M}ills.
\newblock {\em JHEP}, 04:013, 2002.

\bibitem[Bre06]{Bredthauer:2006sz}
Andreas Bredthauer.
\newblock Generalized hyperkaehler geometry and supersymmetry.
\newblock 2006.

\bibitem[BRS75]{Becchi:1974md}
C.~Becchi, A.~Rouet, and R.~Stora.
\newblock Renormalization of the abelian higgs-kibble model.
\newblock {\em Commun. Math. Phys.}, 42:127--162, 1975.

\bibitem[Bus87]{Buscher:1987sk}
T.~H. Buscher.
\newblock A symmetry of the string background field equations.
\newblock {\em Phys. Lett.}, B194:59, 1987.

\bibitem[Bus88]{Buscher:1987qj}
T.~H. Buscher.
\newblock Path integral derivation of quantum duality in nonlinear sigma
  models.
\newblock {\em Phys. Lett.}, B201:466, 1988.

\bibitem[Cal06]{Calvo:2005ww}
Ivan Calvo.
\newblock Supersymmetric {WZ-Poisson} sigma model and twisted generalized
  complex geometry.
\newblock {\em Lett. Math. Phys.}, 77:53--62, 2006.

\bibitem[Cav05]{Cavalcanti:2005hq}
Gil~R. Cavalcanti.
\newblock New aspects of the ddc-lemma.
\newblock 2005.

\bibitem[CM67]{Coleman:1967ad}
Sidney~R. Coleman and J.~Mandula.
\newblock All possible symmetries of the {S} matrix.
\newblock {\em Phys. Rev.}, 159:1251--1256, 1967.

\bibitem[CMPF85]{Callan:1985ia}
C.~G. Callan, E.~J. Martinec, M.~J. Perry, and D.~Friedan.
\newblock Strings in background fields.
\newblock {\em Nucl. Phys.}, B262:593, 1985.

\bibitem[Cou90]{courant}
T.~Courant.
\newblock Dirac manifolds.
\newblock {\em Trans. Amer. Math. Soc.}, 319 no. 2:631--661, 1990.

\bibitem[dAS96]{deAlwis:1996ze}
S.~P. de~Alwis and K.~Sato.
\newblock D-strings and f-strings from string loops.
\newblock {\em Phys. Rev.}, D53:7187--7196, 1996.

\bibitem[DHS89]{Dine:1989vu}
Michael Dine, Patrick~Y. Huet, and N.~Seiberg.
\newblock Large and small radius in string theory.
\newblock {\em Nucl. Phys.}, B322:301, 1989.

\bibitem[Dir50]{Dirac:1950pj}
Paul A.~M. Dirac.
\newblock Generalized hamiltonian dynamics.
\newblock {\em Can. J. Math.}, 2:129--148, 1950.

\bibitem[DLP89]{Dai:1989ua}
Jin Dai, R.~G. Leigh, and Joseph Polchinski.
\newblock New connections between string theories.
\newblock {\em Mod. Phys. Lett.}, A4:2073--2083, 1989.

\bibitem[dVGN95]{deVega:1994hu}
H.~J. de~Vega, I.~Giannakis, and A.~Nicolaidis.
\newblock String quantization in curved space-times: Null string approach.
\newblock {\em Mod. Phys. Lett.}, A10:2479--2484, 1995.

\bibitem[DZ76]{Deser:1976rb}
S.~Deser and B.~Zumino.
\newblock A complete action for the spinning string.
\newblock {\em Phys. Lett.}, B65:369--373, 1976.

\bibitem[FP67]{Faddeev:1967fc}
L.~D. Faddeev and V.~N. Popov.
\newblock Feynman diagrams for the yang-mills field.
\newblock {\em Phys. Lett.}, B25:29--30, 1967.

\bibitem[GG75]{Gurses:1974cm}
Metin Gurses and Feza Gursey.
\newblock Derivation of the string equation of motion in general relativity.
\newblock {\em Phys. Rev.}, D11:967, 1975.

\bibitem[GGRS83]{Gates:1983nr}
S.~J. Gates, Marcus~T. Grisaru, M.~Rocek, and W.~Siegel.
\newblock Superspace, or one thousand and one lessons in supersymmetry.
\newblock {\em Front. Phys.}, 58:1--548, 1983.

\bibitem[GHMR85]{Gross:1985fr}
David~J. Gross, Jeffrey~A. Harvey, Emil~J. Martinec, and Ryan Rohm.
\newblock Heterotic string theory. 1. the free heterotic string.
\newblock {\em Nucl. Phys.}, B256:253, 1985.

\bibitem[GHR84]{Gates:1984nk}
Jr. Gates, S.~J., C.~M. Hull, and M.~Rocek.
\newblock Twisted multiplets and new supersymmetric nonlinear sigma models.
\newblock {\em Nucl. Phys.}, B248:157, 1984.

\bibitem[Gin87]{Ginsparg:1986bx}
Paul~H. Ginsparg.
\newblock Comment on toroidal compactification of heterotic superstrings.
\newblock {\em Phys. Rev.}, D35:648, 1987.

\bibitem[GKP98]{Gubser:1998bc}
S.~S. Gubser, I.~R. Klebanov, and A.~M. Polyakov.
\newblock Gauge theory correlators from non-critical string theory.
\newblock {\em Phys. Lett.}, B428:105--114, 1998.

\bibitem[GLS{\etalchar{+}}95]{Gustafsson:1994kr}
H.~Gustafsson, U.~Lindström, P.~Saltsidis, B.~Sundborg, and R.~van Unge.
\newblock Hamiltonian brst quantization of the conformal string.
\newblock {\em Nucl. Phys.}, B440:495--520, 1995.

\bibitem[Got71]{Goto:1971ce}
Tetsuo Goto.
\newblock Relativistic quantum mechanics of one-dimensional mechanical
  continuum and subsidiary condition of dual resonance model.
\newblock {\em Prog. Theor. Phys.}, 46:1560--1569, 1971.

\bibitem[Gro88]{Gross:1988ue}
David~J. Gross.
\newblock High-energy symmetries of string theory.
\newblock {\em Phys. Rev. Lett.}, 60:1229, 1988.

\bibitem[GS84a]{Green:1984sg}
Michael~B. Green and John~H. Schwarz.
\newblock Anomaly cancellation in supersymmetric d=10 gauge theory and
  superstring theory.
\newblock {\em Phys. Lett.}, B149:117--122, 1984.

\bibitem[GS84b]{Green:1983wt}
Michael~B. Green and John~H. Schwarz.
\newblock Covariant description of superstrings.
\newblock {\em Phys. Lett.}, B136:367--370, 1984.

\bibitem[GSO77]{Gliozzi:1976qd}
F.~Gliozzi, Joel Scherk, and David~I. Olive.
\newblock Supersymmetry, supergravity theories and the dual spinor model.
\newblock {\em Nucl. Phys.}, B122:253--290, 1977.

\bibitem[GSW87]{Green:1987sp}
Michael~B. Green, J.~H. Schwarz, and Edward Witten.
\newblock Superstring theory.
\newblock 1987.
\newblock Cambridge, Uk: Univ. Pr. (1987) 469 p + 596 p. (Cambridge Monographs
  On Mathematical Physics).

\bibitem[Gua03]{Gualtieri:2003dx}
Marco Gualtieri.
\newblock Generalized complex geometry.
\newblock 2003.

\bibitem[Gue00]{Gueven:2000ru}
R.~Gueven.
\newblock Plane wave limits and {T}-duality.
\newblock {\em Phys. Lett.}, B482:255--263, 2000.

\bibitem[Hit03]{Hitchin:2004ut}
Nigel Hitchin.
\newblock Generalized {Calabi-Yau} manifolds.
\newblock {\em Quart. J. Math. Oxford Ser.}, 54:281--308, 2003.

\bibitem[HKLR86]{Hull:1985pq}
C.~M. Hull, A.~Karlhede, U.~Lindström, and M.~Rocek.
\newblock Nonlinear sigma models and their gauging in and out of superspace.
\newblock {\em Nucl. Phys.}, B266:1, 1986.

\bibitem[HLS75]{Haag:1974qh}
Rudolf Haag, Jan~T. Lopuszanski, and Martin Sohnius.
\newblock All possible generators of supersymmetries of the {S} matrix.
\newblock {\em Nucl. Phys.}, B88:257, 1975.

\bibitem[HS91]{Horowitz:1991cd}
Gary~T. Horowitz and Andrew Strominger.
\newblock Black strings and p-branes.
\newblock {\em Nucl. Phys.}, B360:197--209, 1991.

\bibitem[Hul95]{Hull:1995nu}
C.~M. Hull.
\newblock String-string duality in ten-dimensions.
\newblock {\em Phys. Lett.}, B357:545--551, 1995.

\bibitem[ILS92]{Isberg:1992ia}
J.~Isberg, U.~Lindström, and B.~Sundborg.
\newblock Space-time symmetries of quantized tensionless strings.
\newblock {\em Phys. Lett.}, B293:321--326, 1992.

\bibitem[ILST94]{Isberg:1993av}
J.~Isberg, U.~Lindström, B.~Sundborg, and G.~Theodoridis.
\newblock Classical and quantized tensionless strings.
\newblock {\em Nucl. Phys.}, B411:122--156, 1994.

\bibitem[Joh03]{Johnson:2003gi}
C.~V. Johnson.
\newblock D-branes.
\newblock 2003.
\newblock Cambridge, USA: Univ. Pr. (2003) 548 p.

\bibitem[KL86]{Karlhede:1986wb}
A.~Karlhede and U.~Lindström.
\newblock The classical bosonic string in the zero tension limit.
\newblock {\em Class. Quant. Grav.}, 3:L73--L75, 1986.

\bibitem[Lin02]{Lindstrom:2002ph}
U.~Lindström.
\newblock Supersymmetry, a biased review.
\newblock 2002.

\bibitem[Lin04]{Lindstrom:2004eh}
Ulf Lindström.
\newblock Generalized {N} = (2,2) supersymmetric non-linear sigma models.
\newblock {\em Phys. Lett.}, B587:216--224, 2004.

\bibitem[LMTZ05]{Lindstrom:2004iw}
Ulf Lindström, Ruben Minasian, Alessandro Tomasiello, and Maxim Zabzine.
\newblock Generalized complex manifolds and supersymmetry.
\newblock {\em Commun. Math. Phys.}, 257:235--256, 2005.

\bibitem[LRvUZ05]{Lindstrom:2004hi}
Ulf Lindström, Martin Rocek, Rikard von Unge, and Maxim Zabzine.
\newblock Generalized kaehler geometry and manifest {N} = (2,2) supersymmetric
  nonlinear sigma-models.
\newblock {\em JHEP}, 07:067, 2005.

\bibitem[LRvUZ07]{Lindstrom:2005zr}
Ulf Lindström, Martin Rocek, Rikard von Unge, and Maxim Zabzine.
\newblock Generalized {K}aehler manifolds and off-shell supersymmetry.
\newblock {\em Commun. Math. Phys.}, 269:833--849, 2007.

\bibitem[LST91]{Lindstrom:1990qb}
U.~Lindström, B.~Sundborg, and G.~Theodoridis.
\newblock The zero tension limit of the superstring.
\newblock {\em Phys. Lett.}, B253:319--323, 1991.

\bibitem[LZ02]{Lyakhovich:2002kc}
Simon Lyakhovich and Maxim Zabzine.
\newblock Poisson geometry of sigma models with extended supersymmetry.
\newblock {\em Phys. Lett.}, B548:243--251, 2002.

\bibitem[LZ04]{Lindstrom:2003mg}
Ulf Lindström and Maxim Zabzine.
\newblock Tensionless strings, wzw models at critical level and massless higher
  spin fields.
\newblock {\em Phys. Lett.}, B584:178--185, 2004.

\bibitem[Mal98]{Maldacena:1997re}
Juan~M. Maldacena.
\newblock The large {N} limit of superconformal field theories and
  supergravity.
\newblock {\em Adv. Theor. Math. Phys.}, 2:231--252, 1998.

\bibitem[Mal06]{Malikov:2006rm}
Fyodor Malikov.
\newblock Lagrangian approach to sheaves of vertex algebras.
\newblock 2006.

\bibitem[Met02]{Metsaev:2001bj}
R.~R. Metsaev.
\newblock Type {IIB Green-Schwarz} superstring in plane wave {R}amond-{R}amond
  background.
\newblock {\em Nucl. Phys.}, B625:70--96, 2002.

\bibitem[MN05]{Marshakov:2005fn}
Andrei Marshakov and Antti~J. Niemi.
\newblock Yang-mills, complex structures and chern's last theorem.
\newblock {\em Mod. Phys. Lett.}, A20:2583--2600, 2005.

\bibitem[MS06]{Maes:2006bm}
Joris Maes and Alexander Sevrin.
\newblock A note on {N} = (2,2) superfields in two dimensions.
\newblock {\em Phys. Lett.}, B642:535--539, 2006.

\bibitem[MT02]{Metsaev:2002re}
R.~R. Metsaev and A.~A. Tseytlin.
\newblock Exactly solvable model of superstring in plane wave {R}amond-{R}amond
  background.
\newblock {\em Phys. Rev.}, D65:126004, 2002.

\bibitem[Nak90]{Nakahara:1990th}
M.~Nakahara.
\newblock Geometry, topology and physics.
\newblock 1990.
\newblock Bristol, UK: Hilger (1990) 505 p. (Graduate student series in
  physics).

\bibitem[Nam70]{Nambu:1970}
Y.~Nambu.
\newblock Duality and hydrodynamics.
\newblock 1970.
\newblock Lectures at the {Copenhagen} symposium.

\bibitem[Nar86]{Narain:1985jj}
K.~S. Narain.
\newblock New heterotic string theories in uncompactified dimensions < 10.
\newblock {\em Phys. Lett.}, B169:41, 1986.

\bibitem[NSSW87]{Nair:1986zn}
V.~P. Nair, A.~D. Shapere, A.~Strominger, and F.~Wilczek.
\newblock Compactification of the twisted heterotic string.
\newblock {\em Nucl. Phys.}, B287:402, 1987.

\bibitem[NSW87]{Narain:1986am}
K.~S. Narain, M.~H. Sarmadi, and Edward Witten.
\newblock A note on toroidal compactification of heterotic string theory.
\newblock {\em Nucl. Phys.}, B279:369, 1987.

\bibitem[Pen76]{Penrose:1976}
R.~Penrose.
\newblock Any space-time has a plane wave as a limit.
\newblock {\em Differential Geometry and relativity, Reidel, Dordrecht}, pages
  271--275, 1976.

\bibitem[Ple04]{Plefka:2003nb}
Jan~Christoph Plefka.
\newblock Lectures on the plane-wave string / gauge theory duality.
\newblock {\em Fortsch. Phys.}, 52:264--301, 2004.

\bibitem[Pol81a]{Polyakov:1981rd}
Alexander~M. Polyakov.
\newblock Quantum geometry of bosonic strings.
\newblock {\em Phys. Lett.}, B103:207--210, 1981.

\bibitem[Pol81b]{Polyakov:1981re}
Alexander~M. Polyakov.
\newblock Quantum geometry of fermionic strings.
\newblock {\em Phys. Lett.}, B103:211--213, 1981.

\bibitem[Pol95]{Polchinski:1995mt}
Joseph Polchinski.
\newblock Dirichlet-branes and ramond-ramond charges.
\newblock {\em Phys. Rev. Lett.}, 75:4724--4727, 1995.

\bibitem[Pol98]{Polchinski:1998rq}
J.~Polchinski.
\newblock String theory.
\newblock 1998.
\newblock Cambridge, UK: Univ. Pr. (1998) 402 p + 531 p.

\bibitem[PS95]{Peskin:1995ev}
Michael~E. Peskin and D.~V. Schroeder.
\newblock An introduction to quantum field theory.
\newblock 1995.
\newblock Reading, USA: Addison-Wesley (1995) 842 p.

\bibitem[RV92]{Rocek:1991ps}
Martin Rocek and Erik~P. Verlinde.
\newblock Duality, quotients, and currents.
\newblock {\em Nucl. Phys.}, B373:630--646, 1992.

\bibitem[Sch77]{Schild:1976vq}
Alfred Schild.
\newblock Classical null strings.
\newblock {\em Phys. Rev.}, D16:1722, 1977.

\bibitem[Sch95]{Schwarz:1995dk}
John~H. Schwarz.
\newblock An {SL(2,Z)} multiplet of type {IIB} superstrings.
\newblock {\em Phys. Lett.}, B360:13--18, 1995.

\bibitem[Sen02a]{Sen:2002nu}
Ashoke Sen.
\newblock Rolling tachyon.
\newblock {\em JHEP}, 04:048, 2002.

\bibitem[Sen02b]{Sen:2002in}
Ashoke Sen.
\newblock Tachyon matter.
\newblock {\em JHEP}, 07:065, 2002.

\bibitem[SS74]{Scherk:1974ca}
Joel Scherk and John~H. Schwarz.
\newblock Dual models for nonhadrons.
\newblock {\em Nucl. Phys.}, B81:118--144, 1974.

\bibitem[SSJ04]{Sadri:2003pr}
Darius Sadri and Mohammad~M. Sheikh-Jabbari.
\newblock The plane-wave / super yang-mills duality.
\newblock {\em Rev. Mod. Phys.}, 76:853, 2004.

\bibitem[ST97]{Sevrin:1996jr}
Alexander Sevrin and Jan Troost.
\newblock Off-shell formulation of {N} = 2 non-linear sigma-models.
\newblock {\em Nucl. Phys.}, B492:623--646, 1997.

\bibitem[Sun01]{Sundborg:2000wp}
Bo~Sundborg.
\newblock Stringy gravity, interacting tensionless strings and massless higher
  spins.
\newblock {\em Nucl. Phys. Proc. Suppl.}, 102:113--119, 2001.

\bibitem[Tyu75]{Tyutin:1975qk}
I.~V. Tyutin.
\newblock Gauge invariance in field theory and statistical physics in operator
  formalism.
\newblock 1975.
\newblock LEBEDEV-75-39.

\bibitem[Ven68]{Veneziano:1968yb}
G.~Veneziano.
\newblock Construction of a crossing - symmetric, regge behaved amplitude for
  linearly rising trajectories.
\newblock {\em Nuovo. Cim.}, A57:190--197, 1968.

\bibitem[Wit95a]{Witten:1995zh}
Edward Witten.
\newblock Some comments on string dynamics.
\newblock 1995.

\bibitem[Wit95b]{Witten:1995ex}
Edward Witten.
\newblock String theory dynamics in various dimensions.
\newblock {\em Nucl. Phys.}, B443:85--126, 1995.

\bibitem[Wit98]{Witten:1998qj}
Edward Witten.
\newblock Anti-de {S}itter space and holography.
\newblock {\em Adv. Theor. Math. Phys.}, 2:253--291, 1998.

\bibitem[Yon74]{Yoneya:1974jg}
T.~Yoneya.
\newblock Connection of dual models to electrodynamics and gravidynamics.
\newblock {\em Prog. Theor. Phys.}, 51:1907--1920, 1974.

\bibitem[Zab06a]{Zabzine:2005qf}
Maxim Zabzine.
\newblock Hamiltonian perspective on generalized complex structure.
\newblock {\em Commun. Math. Phys.}, 263:711--722, 2006.

\bibitem[Zab06b]{Zabzine:2006uz}
Maxim Zabzine.
\newblock Lectures on generalized complex geometry and supersymmetry.
\newblock 2006.

\bibitem[Zwi04]{Zwiebach:2004tj}
B.~Zwiebach.
\newblock A first course in string theory.
\newblock 2004.
\newblock Cambridge, UK: Univ. Pr. (2004) 558 p.

\end{thebibliography}
\end{document}